\documentclass[aps,reprint,twocolumn,superscriptaddress]{revtex4-1}
\usepackage[colorlinks,bookmarks=true,citecolor=blue,linkcolor=red,urlcolor=blue]{hyperref}
\usepackage{amsmath,amsfonts,amssymb,mathtools}
\usepackage{mathrsfs}
\usepackage{graphicx,float}
\usepackage{multirow}
\usepackage{makecell}
\usepackage[ruled,vlined]{algorithm2e}
\usepackage{algorithmic}
\usepackage{color}
\usepackage{dcolumn}
\usepackage{bm}
\usepackage{subfiles}
\usepackage{verbatim} 
\usepackage{tikz}
\usepackage{longtable}
\usepackage{rotating} 
\usepackage{float}
\usepackage[title]{appendix}
\usepackage{url}
\begin{document}

\title{Topological classification and diagnosis in magnetically ordered electronic materials}
\author{Bingrui Peng}
\thanks{These authors contributed equally to this study.}
\affiliation{Beijing National Laboratory for Condensed Matter Physics and Institute of Physics, Chinese Academy of Sciences, Beijing 100190, China}
\affiliation{University of Chinese Academy of Sciences, Beijing 100049, China}
\author{Yi Jiang}
\thanks{These authors contributed equally to this study.}
\affiliation{Beijing National Laboratory for Condensed Matter Physics and Institute of Physics, Chinese Academy of Sciences, Beijing 100190, China}
\affiliation{University of Chinese Academy of Sciences, Beijing 100049, China}
\author{Zhong Fang}
\affiliation{Beijing National Laboratory for Condensed Matter Physics and Institute of Physics, Chinese Academy of Sciences, Beijing 100190, China}
\author{Hongming Weng}
\affiliation{Beijing National Laboratory for Condensed Matter Physics and Institute of Physics, Chinese Academy of Sciences, Beijing 100190, China}
\affiliation{Songshan Lake Materials Laboratory, Dongguan, Guangdong 523808, China}
\author{Chen Fang}
\email{cfang@iphy.ac.cn}
\affiliation{Beijing National Laboratory for Condensed Matter Physics and Institute of Physics, Chinese Academy of Sciences, Beijing 100190, China}
\affiliation{Songshan Lake Materials Laboratory, Dongguan, Guangdong 523808, China}
\affiliation{Kavli Institute for Theoretical Sciences, Chinese Academy of Sciences, Beijing 100190, China}

\begin{abstract}
We show that compositions of time-reversal and spatial symmetries, also known as the magnetic-space-group symmetries, protect topological invariants as well as surface states that are distinct from those of all preceding topological states.
We obtain, by explicit and exhaustive construction, the topological classification of electronic band insulators that are magnetically ordered for each one of the 1421 magnetic space groups in three dimensions.
We have also computed the symmetry-based indicators for each nontrivial class, and, by doing so, establish the complete mapping from symmetry representations to topological invariants.
\end{abstract}
\maketitle

\section{Introduction}
Magnetic space groups (MSGs)\cite{bradley1968magnetic} describe the symmetry of lattices where spins are magnetically ordered.
Magnetic ordering necessarily breaks time-reversal symmetry, and, as the order parameter is a vector, usually also breaks some point-group symmetries (rotation and reflection).
In many magnetically ordered materials, specially those having antiferromagnetism, there are a special type of composite symmetries: a group element $m=g\cdot{T}$ is the composition of a space-group (SG) symmetry $g$ and time-reversal symmetry $T$.
Consider for example an anti-ferromagnetic Neel order polarized along $z$ with propagation vector $\mathbf{Q}=(\pi/a,0,0)$, where $a$ is the lattice constant.
The lattice translation by one unit cell along $x$, $\{E|100\}$, is broken, and time reversal, $T$, is also broken, while their composition $\{E|100\}\cdot{T}$ remains a symmetry.
(Here we use $E$ to represent the identity $3\times3$ matrix, representing the trivial element of $O(3)$.)
According to the types of $g$ in $m=g\cdot{T}$, the MSGs are classified into four types:
if $g$ does not exist (so that $m$ does not exist), the magnetic group is type-I; if $g$ is identity, $\{E|000\}$, type-II; if $g$ involves a nontrivial point group operation, type-III; and if $g$ is a pure lattice translation, $\{E|klm\}$, the group is type-IV.
By this definition, type-II MSGs contain time reversal, and hence describe nonmagnetic materials.
In this work, we focus on type-I, III, IV MSGs\footnote{the ``MSGs'' we use in this work means type-1, 3, and 4 MSGs, and the non-magnetic type-2 MSGs are excluded.}.

MSGs are also the symmetries of the effective Hamiltonians describing elementary excitations, such as magnons and electrons, that move within a magnetically ordered lattice.
In this work, we focus on magnetic materials in which coherent quasiparticle fermion excitations form band(s) within a finite range of the Fermi energy, and study the band topology of these fermions.
Our theory can in principle be applied to any magnetic materials where the notion of electron-like quasiparticles is valid, at least near the Fermi energy.
Itinerant magnets\cite{shimizu1981itinerant}, heavy-fermion metals\cite{stewart1984heavy}, and doped Mott insulators\cite{lee2006doping} that maintain a magnetically ordered state, are considered to belong to this large class of materials.

The interplay between symmetry and topology has been a focus of modern condensed-matter research\cite{hasan2010colloquium,qi2011topological,chiu2016classification,armitage2018weyl}.
For a given symmetry group and a nonzero gap, all band Hamiltonians are grouped into \emph{equivalence classes}, where two Hamiltonians in the same (different) class(es) can(not) be smoothly deformed into each other, while maintaining both the gap and the symmetry.
Each equivalence class is denoted by a unique set of integers called the \emph{topological invariants}\cite{thouless1982quantized, haldane1983nonlinear, affleck1987rigorous, wen1990topological, kitaev2001unpaired, kane2005z, moore2007topological, fu2007topological}, the forms and types of which only depend on the symmetry group and dimensionality.
How many distinct equivalence classes exist for a given symmetry group in a given dimension, and what are the topological invariants for each class?
This is the question we call the problem of \emph{topological classification}.
The theory of topological classification for time-reversal and particle-hole symmetries has been done in all dimensions using the K-theory\cite{schnyder2008classification, kitaev2009periodic, ryu2010topological}; the classification problem for a single spatial symmetry plus time reversal in three dimensions has been solved, usually heuristically, for several symmetries\cite{fu2007topological3d,hsieh2012topological,teo2008surface,wang2016hourglass,shiozaki2014topology,song2017d,fang2019new,turner2010entanglement,hughes2011inversion,song2018quantitative,khalaf2018symmetry}; and the classification problem for arbitrary SGs plus time reversal in three dimensions have been attempted using either the ``real-space recipe'' argument\cite{song2018quantitative,song2019topological,song2020real} or the ``double-strong-topological-insulator construction''\cite{fang2019new,khalaf2018symmetry} argument more recently.

The classification problem for MSGs begins with the theory of axion insulators protected by space-inversion symmetry without time reversal\cite{qi2008topological,li2010dynamical,hughes2011inversion,qi2011topological,turner2012quantized,yue2019symmetry,zhang2019topological,otrokov2019prediction, rienks2019large}, followed by the theory of antiferromagnetic topological insulators\cite{mong2010antiferromagnetic,fang2013topological,liu2013antiferromagnetic,zhang2015topological}, again followed by the discovery of several topological invariants protected by wallpaper groups\cite{liu2014topological,shiozaki2017topological}.
A more systematic attempt is made in Ref.\cite{elcoro2020magnetic}, where the layer-construction (LC) method reveals a number of new topological states.
In this work, we use the \emph{real-space recipe} method developed in \cite{song2019topological,song2020real}.
Effectively, this method converts the problem of topological classification into a ``LEGO puzzle'', where one tries to find distinct ways one can build an ``edgeless'' construction using some given pieces.
This ``LEGO puzzle'' is then further transformed into finding all independent integer solutions of a set of linear equations on a $\mathbb{Z}_n$ ring.
This method not only yields a complete topological classification of gapped bands in each of the 1421 groups, but also gives us, for each nontrivial equivalence class, one explicit and microscopic construction that we call the ``topological crystal'' (TC)\cite{song2019topological}.
We emphasize that our method finds more topological classes than the layer-construction method, because LCs are a type of TCs, while not all TCs are LCs.
(In fact, we show that at least 553 of the 1421 MSGs have at least one topological state that cannot be layer-constructed.)

If the classification gives us the labels (topological invariants) for the equivalence classes into which gapped states are put in, the \emph{topological diagnosis} then tells us to which equivalence class a specific, given material (Hamiltonian) belongs.
Ideally, a diagnosis scheme computes the topological invariants of that Hamiltonian, and by comparing these values with the labels on the equivalence classes, one puts the Hamiltonian into the right one.
However, topological invariants are notoriously difficult to compute\cite{fang2015new,wang2016hourglass}, and for some, we do not even have the explicit expressions in terms of the wave functions of the bands\cite{song2017d,fang2019new}.
Fortunately, if we relax the requirement of ideal diagnosis to approximate diagnosis, the story is completely changed.
An approximate diagnosis uses partial information on the wave function, and in return gives us partial information on the topological invariants, not invariants themselves.
For example, an approximate diagnosis for systems with $n$-fold rotation symmetry yields the invariant (Chern number) modulo $n$ by using only the rotation eigenvalues at several high-symmetry momenta\cite{hughes2011inversion,fang2012bulk}.
Recently, the theory of symmetry-based indicators (SIs)\cite{po2017symmetry} and that of topological quantum chemistry\cite{bradlyn2017topological}, enhanced by the full mapping from indicators to topological invariants\cite{song2018quantitative}, give birth to a fast approximate diagnosis scheme.
This fast-diagnosis scheme has been applied to a large number of non-magnetic materials\cite{zhang2019catalogue,vergniory2019complete,tang2019comprehensive}.

In this work, we extend the above diagnosis scheme to magnetic materials that can be well-characterized in terms of band structures. We find the explicit formulas for all SIs in terms of band representations (which in part differ from previous works\cite{ono2018unified, elcoro2020magnetic}); and we calculate the values of indicators for each TC in every MSG.
Since each gapped state can be adiabatically continued to a TC, this result in fact yields the mapping from invariants to indicators.
In this calculation, we find that certain indicator values are never taken in any TC, and hence can only indicate nodal band structures
\cite{fang2003anomalous, wan2011topological, xu2011chern, wang2016time, kubler2016weyl, chang2016room, yang2017topological, liu2018giant, wang2018large, belopolski2019discovery, morali2019fermi, liu2019magnetic, nie2017topological, tang2016dirac, hua2018dirac, kim2018large, nie2019topological, zou2019study}.
These nodes are ``evasive'' as they are away from any high-symmetry points or lines.
In fact, we show that all these nodal indicators indicate Weyl nodes at generic momenta.
In a numerical diagnosis, the indicators are by far easier to obtain than the invariants, because the former only depend on the band representations at fewer than or equal to eight momenta and the latter depend on the valence-band wave functions in the entire Brillouin zone (BZ).
Therefore, an inverse mapping from indicators to invariants/nodes is generated using a script, provided along with the paper at \href{https://github.com/yjiang-iop/MSG_mapping}{https://github.com/yjiang-iop/MSG\_mapping}.

\section{Classification}

\subsection{General scheme}
The basic idea of the real-space recipe is that all topological crystalline insulators (TCIs) can be adiabatically deformed into a special form of real-space constructions called the TCs, such that classifying TCIs is equivalent to classifying TCs.
TCs are real-space patterns built from topological pieces in lower dimensions, which were first applied to non-magnetic SGs in order to obtain the full classifications of non-magnetic TCIs\cite{song2019topological}.
In this work, we apply the real-space recipe for the construction and the classification of all gapped topological states protected by MSGs.

To start with, we build a structure of cell complex by using MSG symmetries, including both unitary and anti-unitary ones, and partition the 3D space into finite 3D regions called asymmetric units (AUs) that fill the whole 3D space without overlaps.
In fact, all AUs are symmetry-related and can be generated by choosing one AU and then copying it using MSG operations.
AUs are also called 3-cells, and the 2D faces where they meet are 2-cells.
Similarly, 1D lines where 2-cells intersect are 1-cells, and the endpoints of 1-cells are 0-cells.

To construct TCs in 3D, we should take account of all $d$-dimensional topological building blocks with $d \leq 3$.
For each cell, its local symmetry group is defined as the collection of symmetries that keep every point of the cell unchanged.
The local symmetry group of a cell determines the onsite symmetry class (Altland-Zirnbauer class\cite{altland1997nonstandard}) of the Hamiltonian on that cell.
For MSGs, the effective symmetry class of a cell is always class A or class AI.
Note that although some cells have mirror plane as onsite symmetry, the states on them can be divided into two sectors by mirror eigenvalues, each of which belongs to class A.
A 2-cell may have local symmetry group generated by $M\cdot T$, and a 1-cell local symmetry group generated by $C_2\cdot T$.
Since $(M\cdot T)^2=(C_2\cdot T)^2=+1$, those cells belong to symmetry class AI.
According to the tenfold way results\cite{ryu2010topological}, systems of class AI have trivial classification in all dimension $\le 3$, and systems of class A have nontrivial classification in 2D and trivial classification in 0D, 1D, and 3D, which means only 2D topological building blocks need to be considered in the construction of TCs for MSG.
Furthermore, there are two types of 2-cells depending on whether they coincide with mirror planes. If coincide, they can be decorated with mirror Chern insulators characterized by two $\mathbb{Z}$ numbers, i.e., two mirror Chern numbers for $\pm i$ mirror sectors\footnote{Due to the absence of TRS in MSGs, the ``mirror Chern insulator'' we use here do not require two mirror Chern numbers to be opposite with each other, i.e., $C^{+}_m \neq C^{-}_m$ generically, which are different with the mirror Chern insulators in non-magnetic SGs with $C^{+}_m=-C^{-}_m$\cite{teo2008surface,hsieh2012topological}.}, and if not, with Chern insulators characterized by one $\mathbb{Z}$ number, i.e., the Chern number.
Therefore, there are only two types of building blocks for our real-space recipe, i.e., Chern insulators and mirror Chern insulators.

Having building blocks in hand, we next enumerate all topological inequivalent decorations on the cell complexes in MSGs.
Before proceeding, note that the building blocks themselves form a finitely generated Abelian group, e.g., $\mathbb{Z}$ for Chern insulators and $\mathbb{Z}^2$ for mirror Chern insulators.
Therefore, the TCs built from them form a linear space with integer coefficients, such that two TCs can be ``added'' to obtain another TC, and there exists a maximal set of linearly independent TCs (the generators) for each MSG.
As a result, one just needs to obtain the generators to describe the full set of TCs.
As TCs are supposed to be fully gapped topological states, all the boundary states contributed by 2D building blocks (Chern insulators and mirror Chern insulators) should cancel with each other on each 1-cell, leading to fully gapped states inside the bulk, a condition known as the ``gluing condition''\cite{song2019topological} or ``no-open-edge condition''\cite{song2020real}.
After this procedure, we obtain a set of generators that form an Abelian group $\mathbb{Z}^n$.
However, this is generally not the final classification, because some generators will reduce from $\mathbb{Z}$-type to $\mathbb{Z}_2$-type after a process of subtracting topological trivial elements called ``bubble-equivalence''\cite{song2019topological,song2020real}.
The final classification can be expressed as a quotient group Ker/Img, a structure resembling group (co)homology, where Ker stands for the linear space of TCs satisfying no-open-edge condition and Img for the space of bubble-equivalence.
These final classifications of MSGs have the form $\mathbb{Z}^n\times\mathbb{Z}_2^l$ and can be found in Appendix.\ref{AppendixN}.

Lastly, we compare the real-space constructions for non-magnetic SGs and MSGs.
First, observe that their building blocks are different.
In non-magnetic SGs, due to the time-reversal symmetry (TRS), the building blocks are two-dimensional topological insulators and mirror Chern insulators with $C^{+}_m = - C^{-}_m$, while in MSGs, the building blocks are Chern insulators and mirror Chern insulators with independent $C^{+}_m$ and $C^{-}_m$\footnote{in the following context we use the notation $C^{\pm}_{m}$ for both real-space mirror Chern numbers and momentum-space mirror Chern numbers, and the specific meaning of $C^{\pm}_m$ can be inferred from the context}.
More significantly, for non-magnetic SGs, one can obtain most of the TCs by LCs, which involve only layered 2D TIs and mirror Chern insulators as building blocks\cite{song2018quantitative}, with only 12 non-magnetic SGs having states beyond LCs\cite{song2019topological}.
However, non-layer constructions (non-LCs) exist widely in MSGs, and constructing with layers only may lose a number of TCs.
By contrast, the real-space recipe employing the structure of the cell complex automatically includes both LCs and non-LCs, giving the complete collection of TCs, hence the complete classification of gapped topological states.

\begin{figure*}
	\centering
	\includegraphics[width=1\textwidth]{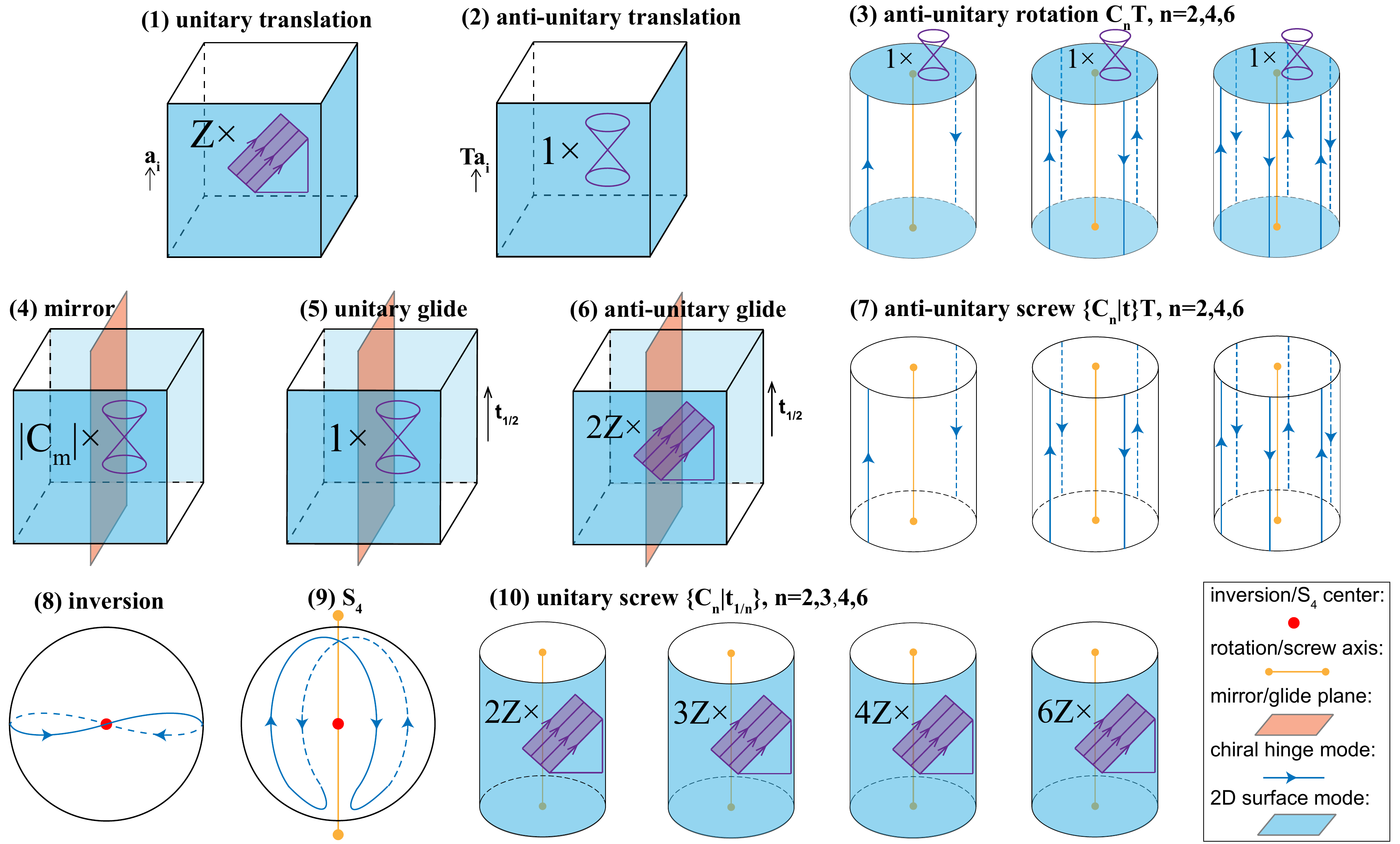}
	\caption{\label{surface_states}Surface states of symmetries with non-trivial topological invariants in MSGs, with the surface terminations preserving corresponding symmetries. These surface states can be 1D chiral hinge modes and 2D surface modes, with 2D modes being either slope-like chiral surface modes in (1),(6),(10), or Dirac cones in (2),(3),(4),(5). More details can be found in Appendix.\ref{AppendixC}.}
\end{figure*}

\subsection{Topological invariants}
TCIs are characterized by crystalline-symmetry-protected topological invariants.
Previous works have looked into some of the topological invariants protected by magnetic crystalline symmetries, with the earliest one being the axion insulators with inversion invariant\cite{turner2012quantized,hughes2011inversion}, followed by the anti-ferromagnetic topological insulators protected by anti-unitary translations\cite{mong2010antiferromagnetic,fang2013topological}.
In this work, we exhaustively enumerate all topological invariants in MSGs.

Formally, having nontrivial symmetry-protected topological invariants means a topological state cannot be smoothly deformed into a trivial state when the symmetry is preserved.
Specially, if a TCI can not be adiabatically connected to an atomic insulator with the symmetry operation $g$ preserved, we say it has nontrivial $g$-invariant.
As all TCIs are adiabatically connected to TCs, we can utilize TCs to derive all invariants protected by MSGs symmetries.
In our real-space recipe, given a symmetry operation $g$ alone, if nontrivial TCs compatible with $g$ can be constructed, then we say $g$ can protect nontrivial topological invariants.
More specifically, each independent TC corresponds to an independent invariant, and they have the same group structure, e.g., a $\mathbb{Z}_2$ TC owns a $\mathbb{Z}_2$ invariant.

To find which symmetry operations in MSGs can protect nontrivial invariants and what kinds of the invariants they protect, we take account of all of MSG symmetries one by one, including translation, rotation, inversion, mirror, rotoinversion ($S_n$), screw, glide, and those combined with TRS such as $C_n \cdot T$, etc.
We consider when each of them is present alone, what TCs can be constructed.
If no TC exists, this symmetry has only trivial invariant; if there exist TCs, this symmetry must protect nontrivial topological invariants, and we further decide how many independent TCs and whether they are $\mathbb{Z}$-type or $\mathbb{Z}_n$ type by checking if the TCs can be smoothly deformed into trivial states after multiplying $n$ times.
In this way, we know all the MSG symmetries that can singly protect nontrivial topological invariants, with each symmetry hosting at most one invariant, being either $\mathbb{Z}_2$-type or $\mathbb{Z}$-type.
We summarize the results in Table.\ref{table_invariants}, and classify these invariants into three types:
\begin{itemize}
    \item Trivial invariants: $C_n$ rotations and anti-unitary improper point group symmetries in MSGs cannot host nontrivial decorations.
    Take $C_2$ for example.
    When there is only a $C_2$ symmetry alone, consider a 2D plane that passes the $C_2$-axis, which is partitioned into two 2-cells by the axis.
    Chern insulators cannot be decorated on the two 2-cells, because their chiral edge modes on the $C_2$-axis are $C_2$-related, and as such are in the same direction and cannot gap out each other.
    This is different from the non-magnetic case where $C_2$ has a $\mathbb{Z}_2$ invariant\cite{song2019topological, song2018quantitative}, as the building blocks in non-magnetic SGs are 2D TIs and two $C_2$-related helical edge states can cancel with each other.

    \item $\mathbb{Z}$ invariants: Unlike non-magnetic MSGs where only mirror symmetries have $\mathbb{Z}$ invariant, here in MSGs, as the building blocks are both $\mathbb{Z}$-type, we have many other $\mathbb{Z}$ invariants, including the invariant of unitary translation, unitary screw, and anti-unitary glide.

    \item $\mathbb{Z}_2$ invariants:
    A large proportion of MSG symmetries protect $\mathbb{Z}_2$ invariants, with many of them unique to MSGs, such as anti-unitary translations/rotations/screws. 
    In Appendix.\ref{AppendixD}, we define a special type of decoration called ``$Z_2$ decoration'' which has zero weak invariants and all $\mathbb{Z}_2$ invariants bond together and equal to 1 (for mirror Chern numbers, we can define $C_m^++C_m^-\bmod 2$ as a $\mathbb{Z}_2$ invariant), and can be seen as an axion insulator with the axion angle $\theta=\pi$. For $Z_2$ decorations, all these $\mathbb{Z}_2$ invariants merge into one $\mathbb{Z}_2$ invariant, i.e., the ``axion invariant'', whose definition can be taken as the 3D magnetoelectric polarization $P_3$ according to Ref.\cite{qi2008topological, qi2011topological,fang2012bulk}.
\end{itemize}

\begin{table}
	\centering
	\begin{tabular}{cc}
\hline\hline	
\textbf{MSG Symmetries} & \textbf{Invariant type}
\\
\hline
unitary rotation $ C_n $& \multirow{3}{*}{Trivial} \\
\makecell[c]{anti-unitary improper point group \\
symmetries $P\cdot T$, $M\cdot T$, $S_n\cdot T$} &
\\
\hline
unitary improper $ S_n$, $P$, $\{M|\frac{1}{2}\}$ & \multirow{3}{*}{$\mathbb{Z}_2$} \\
anti-unitary translation $ \{ E|t\}\cdot T$ &  \\
anti-unitary proper $C_n\cdot T$, $\{C_n|t\}\cdot T$ & \\
\hline
unitary translation $\{ E | R\}$ & \multirow{4}{*}{$\mathbb{Z}$} \\
unitary mirror $M$ & \\
unitary screw $ \{ C_n | t\} $ &  \\
anti-unitary glide $ \{ M | \frac{1}{2} \}\cdot T$ & \\
\hline\hline			
	\end{tabular}
\caption{\label{table_invariants} MSG symmetries and their corresponding invariant types. We use the Seitz Symbol $\{O|t\}$ to represent symmetries with nonzero translations, where $R$ denotes a lattice translation and $t$ a fractional translation. The invariants of the unitary screw and anti-unitary glide are bound to the translation invariant, with their values being $\frac{1}{n}$ and $\frac{1}{2}$ of the translation invariant, respectively.}
\end{table}

Each nontrivial invariant has its distinct anomalous surface state due to the bulk-boundary correspondence.
Because the topological invariants together with their surface states can be superimposed, we only need to derive the surface state for every single invariant, and the surface states of all TCIs can be readily known from their invariants.
Among these surface states in MSGs, some have already been discussed in previous works, including those protected by $C_2 \cdot T$
\cite{shiozaki2014topology,fang2015new, ahn2019symmetry}, $C_4\cdot T$\cite{schindler2018higher}, glide\cite{fang2015new,shiozaki2015z,shiozaki2016topology,kim2019glide,kim2020glide}, and anti-unitary translation\cite{mong2010antiferromagnetic,fang2013topological}, etc. (for spinless invariants including $C_n\cdot T$ and anti-unitary translation, see Ref.\cite{liu2013antiferromagnetic,zhang2015topological}), while some are first proposed in this work, such as the one protected by anti-unitary glide.
We plot these surface states in Fig.\ref{surface_states}, and more details can be found in Appendix.\ref{AppendixC}.
As an example shown in Fig.\ref{surface_states}(3), on a cylinder geometry, the surface states protected by $C_n \cdot T, n = 2,4,6$ have a single Dirac cone on the top surface, and $n$ chiral hinge modes related to each other under $ C_n\cdot T$ on the side surface.

Although we find all the topological invariants protected by MSG symmetries, we have not derived their explicit formulas in k-space, which could be complicated.
Instead, we compute these topological invariants in real-space for the TCs we built using a unified method as follows\cite{song2019topological}.
Given an MSG $M$ and an element $g\in M$, first choose a generic point $r$ inside an arbitrary AU, then draw a path connecting $r$ and its image point $g \cdot r$, under the only constraint that the path does not cross any 1-cells and 0-cells. The corresponding invariant $\delta(g)$ is determined by the decorated 2-cells that the path crosses, i.e., the total Chern number or mirror Chern number accumulated through the path. The invariants thus calculated are well-defined and do not depend on the choice of the generic point $r$ or the path.
The full listing of topological invariants of the TCs will prove to be useful in the second part of this work, where we use SIs to diagnose topological states, and TCs function as an intermediate to connect SIs and invariants.

We take one example to show the correspondence between invariants and surface states for a given MSG.
The type-III MSG 3.3 $P2'$ has three independent generators of TCs.
One of them is protected by $C_{2y}\cdot T$ with nonzero invariant $\delta(C_{2y}\cdot T) = 1$, while the other two are protected by the translation symmetries in $x$ and $z$ directions, with nonzero weak invariant $\delta( \{ E | 100 \} ) = 1$ and $\delta{\{ E | 001\} } = 1 $, respectively.
For the first decoration, the surface state protected by $C_{2y}\cdot T$ has been described before, as shown in Fig.\ref{surface_states}(3), while the other two translation decorations have one chiral surface mode on the 2D surface preserving the translation symmetry in $x/z$ direction, as shown in Fig.\ref{surface_states}(1).

\subsection{Non-layer constructions}

The abundance of non-LCs distinguishes the TCs in MSGs from those in non-magnetic SGs, where most of the decorations are LCs.
In fact, LCs are just a special type of TCs where 2D planes are uniformly decorated, while non-LCs contain non-uniform decorations or incomplete 2D planes.

\begin{figure}
	\centering
	\includegraphics[width=0.2\textwidth]{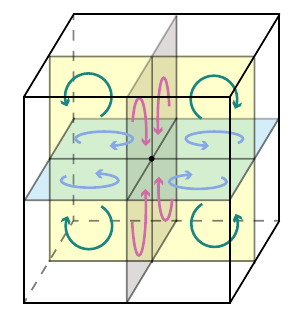}
	\caption{\label{Pmmm_nonLC}The non-LC of $Pmmm$, where the arrows denote the directions of the chiral edge modes on the 2-cells. Note we only plot the 2-cells around the origin point inside the unit cell, and omit the six side-surfaces for simplicity, which are in fact all decorated. Assume the 8 inversion centers have coordinate $\bm{r}=\frac{1}{2}\mathbf{a}_1\delta_1+\frac{1}{2}\mathbf{a}_2\delta_2+\frac{1}{2}\mathbf{a}_3\delta_3$, where $\delta_i=0,1$ and $\mathbf{a}_i$ is the lattice vector. The 2-cells around an inversion center have the same decoration as shown in the figure if $\sum_i\delta_i$ is even, while the 2-cells have opposite directional edge modes if $\sum_i\delta_i$ is odd.}
\end{figure}

We use type-1 MSG $Pmmm$ as a representative to show the characters of non-LCs.
$Pmmm$ has three orthogonal mirrors $M_x, M_y$, and $M_z$, which do not commute with each other due to the spin rotation.
Placing mirror Chern insulator layers with  $ ( C^{+}_m, C^{-}_m ) = ( 1,-1 ) $ on any of the six mirror planes, i.e., $x,y,z = 0,\frac{1}{2}$, forms 6 independent LCs.
However, one can still construct a distinct non-layer decoration falling outside the linear space of these 6 LCs.
As shown in Fig.\ref{Pmmm_nonLC}, this non-layer
decoration is constructed by sewing small patches of mirror Chern insulators with $ ( C^{+}_m, C^{-}_m ) = (1, 0)$ or $ ( 0, -1 ) $ on each mirror 2-cell, with adjacent patches having opposite mirror Chern numbers, making each mirror plane non-uniform.
Each 1-cell is shared by 4 patches, with patches in different directions contributing opposite chiral edge modes, canceling with each other and satisfying the no-open-edge condition.
This non-LC is distinct from LCs, where each mirror plane is decorated with a uniform, infinite-sized mirror Chern insulator.
The decorated 2-cells of this non-LC are all pinned on the mirror planes, preventing it to be deformed into an LC.

Although this non-LC seems complicated, we observe that it can be connected to the time-reversal strong TI (STI) from the perspective of surface states.
On the one hand, this non-LC has mirror Chern numbers $(C_{m,k_i=0}^+,C_{m,k_i=0}^-,C_{m,k_i=\pi}^+,C_{m,k_i=\pi}^-)=(1,-1,0,0), i=x,y,z$, which lead to a nontrivial mirror-protected surface state with a single Dirac cone on $k_i= 0$ in the 2D surface BZ.
On the other hand, an STI put on $Pmmm$ lattice also has a single Dirac cone on each surface.
By the principle of bulk-edge correspondence, we conclude that the STI with $Pmmm$ symmetry can be adiabatically deformed into this non-LC by breaking the TRS while preserving crystalline symmetries.
Note that for $Pmmm$, despite the absence of TRS, the three mirrors anti-commuting with each other also enforce the energy bands to form Kramer-like pairs, i.e., twofold degeneracy with the same parity and opposite mirror eigenvalues.
Therefore, the three mirror symmetries still pin the surface Dirac cone on the mirror invariant lines when the TRS is broken.
As a result, we can use the tight-binding models of STIs, for example, the 3D Bernevig-Hughes-Zhang (BHZ) model\cite{bernevig2006quantum}, to describe this non-LC.

As mentioned before, unlike non-magnetic SGs where only 12 SGs have non-LCs\cite{song2019topological}, non-LCs exist widely in MSGs.
For instance, all the supergroups of MSG 47.249 $Pmmm$ can host a non-LC as one of its TCI classification generators, and similar for MSG 25.57 $Pmm2$, 84.51 $P4_2/m$, 10.44 $P2'/m$, 75.3 $P4'$, 6.21 $P_am$, and 75.5 $P_C4$, thus the total number of MSGs that have non-LCs is at least 553, i.e., the number of the supergroups of these MSGs, including MSGs with different Bravais lattices.
As argued in Ref.\cite{song2019topological}, STI is compatible with all crystalline symmetries, thus the abundance of non-LCs in MSGs can be understood in a way that they can be obtained by inducing magnetism in STIs in different lattices. 

We remark that the non-LCs in these MSGs, when being a ``$Z_2$ decoration'', are all axion insulators. As STIs are also axion insulators, when the TRS is broken, there still exist other unitary improper or anti-unitary proper symmetries that have non-trivial $\mathbb{Z}_2$ invariants, preserving the quantized $\pi$ axion angle.


\section{Diagnosis}
Symmetry-based indicator is a powerful tool for diagnosing topological states\cite{bradlyn2017topological, po2017symmetry, kruthoff2017topological} and has been applied in both non-magnetic SGs\cite{song2018quantitative, song2018diagnosis, khalaf2018symmetry} and MSGs\cite{ono2018unified, watanabe2018structure, elcoro2020magnetic, bouhon2020topological}, which leads to the discovery of a significant number of new topological materials\cite{zhang2019catalogue, vergniory2019complete, tang2019comprehensive, xu2020high}.
SI theory in MSGs has been partly tackled in previous works, with the SI group structures and part of the SI expressions given.

However, the mappings between SIs to topological invariants in MSGs have not been fully investigated.
In this work, we derive the explicit formulas of all indicators, and their quantitative mappings to topological invariants if correspond to gapped states, and possible Weyl point configurations if correspond to gapless states.
These two different correspondences to gapped and gapless states are found thanks to the TCI classifications, which include all the gapped TCI states, and states having nonzero SI while not belonging to any gapped state must be Weyl semimetals.

\subsection{Explicit expression of indicators}

Among 1421 MSGs, 688 of them have nontrivial SIs.
Despite the seemingly large number, SIs in all MSGs can be induced from 16 generating MSGs, which we list in Table.\ref{generatingSI}, with only 10 corner cases to be discussed later.

\begin{table}
	\centering
	\begin{tabular}{c|c|c}
		\hline\hline
		\textbf{MSG} & $\bm{X}_{\text{BS}}$ & \textbf{SI} \\\hline
		$P\overline{1}$  & $\mathbb{Z}_{2,2,2,4}$ & $z_{2P,1},z_{2P,2},z_{2P,3},z_{4P}$ \\\hline
		$Pmmm$ & $\mathbb{Z}_{2,2,2,4}$ & $z_{2P,1}',z_{2P,2}',z_{2P,3}',z_{4P}'$ \\\hline
		$Pn,n=2,3,4,6$ & $\mathbb{Z}_n$ & $z_{nC}$ \\\hline
		$Pn/m,n=2,3,4,6$ & $\mathbb{Z}_{n,n,n}$ & $z_{nm,0}^+,z_{nm,0}^-,z_{nm,\pi}^+$ \\\hline
		$P\overline{4}$ & $\mathbb{Z}_{2,2,4}$ & $z_{2,S_4}, z_{2,\text{Weyl}}, z_{4C}$ \\\hline
		$P4/mmm$ & $\mathbb{Z}_{2,4,8}$ & $z_{2P,1}^\prime, z_{4m,\pi}^+, z_8$ \\\hline
		$P6/mmm$ & $\mathbb{Z}_{6,12}$ & $z_{6m,\pi}^+, z_{12}$ \\\hline
		$Pnc'c',n=2,4,6$ & $\mathbb{Z}_{n}$ & $z_{nC}^\prime$ \\
		\hline\hline
	\end{tabular}
	\caption{\label{generatingSI}Generating SIs and generating MSGs in MSGs. We use a simplified notation to represent the SI group, i.e., $\mathbb{Z}_{n_1,n_2,\cdots}=\mathbb{Z}_{n_1}\times\mathbb{Z}_{n_2}\times\cdots$. Among these generating MSGs, only $Pnc'c'$ are type-3 MSGs, while all the others are type-1. The definition and interpretation of these SIs can be found in Appendix.\ref{AppendixG}.}
\end{table}

Most of the SIs can be expressed in terms of some topological invariants.
In fact, the idea of diagnosing nontrivial topology using only the symmetry data on high-symmetry points (HSPs) in the BZ stems from the Fu-Kane formula\cite{fu2007topological} for non-magnetic SGs with inversion symmetry, and in the following works\cite{hughes2011inversion,turner2012quantized, fang2012bulk} $C_n$ rotation eigenvalues are used to calculate the Chern number modulo $n$, i.e., the $z_{nC}$ indicator in Table.\ref{generatingSI}.
When there exist mirror planes perpendicular to the rotation axis, the $z_{nC}$ indicator can be applied to two mirror sectors which give the $z_{nm,0/\pi}^\pm$ indicator, where $\pm$ represents two mirror sectors.
There are also three type-3 generating MSGs $Pnc'c',n=2,4,6$ which have anti-unitary glide symmetries that make all irreducible corepresentations (coirreps) twofold degenerate (with the same $C_n$ eigenvalues) on the $k_z=\pi$ plane, allowing the definition of a new set of SIs $z_{nC}^\prime$ using the number of degenerate pairs, the value of which corresponds to the Chern number divided by two modulo n of the 2D BZ.
The interpretations of other generating SIs are left in Appendix.\ref{AppendixG}.

The word ``generating'' means the SIs in other MSGs can be expressed in terms of these generating SIs.
Some times, the ``generation'' involves the reduction of the indicator group, e.g. from $\mathbb{Z}_4$ to $\mathbb{Z}_2$ by taking only the even numbers in $\mathbb{Z}_4$.
Moreover, the interpretation of the SI, generated from the same indicator, can be different in different MSGs.
For example, the $z_{nC}$ indicator originally represents the Berry phase of a closed loop in the 2D BZ as shown in Appendix.\ref{AppendixK}, and it becomes the Chern number when the state is gapped.
However, as shown later, in some other groups, this indicator if nonzero indicates gapless topological states, and its value represents the number of Weyl points.
For another example, the $z_{4P}^\prime$ indicator is adopted in some MSGs where the coirreps are not doubly degenerate with the same parity.
In these cases, the interpretation of these indicators may change which takes case-by-case examination, and the SI expressions are only effective and may become invalid when, for example, the origin point is changed.
Thus we fix the coordinate system to avoid these problems by adopting the Bilbao convention\cite{aroyo2006bilbao1,aroyo2006bilbao2,aroyo2011crystallography} when calculating the indicators.

To sum up, the indicator formulas in generating SIs can be used to express all SIs in other MSGs, except 10 corner-case MSGs which have SI formulas given independently using coirreps in Appendix.\ref{AppendixH}.
Here we show an example of MSG 42.222 $Fm'm'2$, which has a $\mathbb{Z}_2$ indicator with expression:
\begin{equation}
z_{2,42.222} = N(\overline{\Gamma}_3) - N(\overline{A}_3) \bmod 2
\end{equation}
where $N(\bar{K}_i)$ represents the number of coirrep $\bar{K}_i$.
This MSG has TCI classification $\mathbb{Z}$, the generator of which has weak invariants $\delta_w=(1,1,0)$ and all other invariants equal zero.
The SI $z_{2,42.222}=1$ represents this generator as well as all the odd number copies of it.
Note that this indicator is different from the choice in Ref.\cite{elcoro2020magnetic}.

\subsection{Computation of indicators for topological crystals}

In this section, we show how one can calculate the SIs for a given TC.
As most of the SIs correspond to some topological invariants, including $z_{2P,i},z_{4P},z_{2P}^\prime,z_{nC},z_{nm,0/\pi}^\pm,z_{2,S_4}$, and $z_{nC}^\prime$, their values can be determined directly.
For example, the $\mathbb{Z}_{2}\times\mathbb{Z}_{2}\times\mathbb{Z}_{2}\times\mathbb{Z}_{4}$ indicators defined in $P\bar{1}$ have straightforward meanings:
\begin{equation}
	z_{2P,i}=\delta_{w,i}\bmod 2,\ \ z_{4P}=2\delta(P)
\end{equation}
where $\delta_{w,i}(i=1,2,3)$ is the weak invariant and $\delta(P)$ is the inversion invariant.
Note that $z_{4P}=1,3$ represent Weyl semimetals\cite{hughes2011inversion, turner2012quantized}.

However, there also exist many SIs that cannot be read directly from the invariants, including $z_{4P}^\prime$, $z_8$, and $z_{12}$.
To determine their value for a TC, we need to find a compatible set of coirreps for the TC and then calculate the SIs of the coirrep set.
The result does not, however, depend on which particular set we choose as long as it is compatible with the TC.
This is because a given set of invariants, that is, a given TC, can only correspond to one possible set of SIs.
Therefore, we only need to find one compatible set of coirreps for the TC and calculate its SI.

Finding compatible coirreps for an LC is simple, but to a non-LC TC can be challenging.
Fortunately, as shown before that many of the non-LCs in MSGs are adiabatically connected to STIs, the BHZ model describing STIs can be used, which has symmetries of SG 221 plus the TRS, a supergroup to many MSGs.
As a result, we can adjust the model parameters such that the BHZ model and the non-LCs share the same set of invariants, and then compute the indicators for the BHZ model, with details in Appendix.\ref{AppendixL}.

The mappings between invariants and SIs only need to be derived for the generating MSGs shown in Table.\ref{generatingSI}.
Other MSGs, except the corner cases and Weyl states, either are the supergroups of these generating MSGs or have a different Bravais lattice, and their mappings can be induced from the mappings of the generating MSGs.
For corner cases, we derive their mappings case-by-case in Appendix.\ref{AppendixH}.

\subsection{The mapping from indicators to invariants}

So far we have obtained the mapping from invariants to indicators for topological gapped states protected by all MSGs, listed in Appendix.\ref{AppendixO}.
A practically more useful mapping is its inverse, one from indicators to invariants.
This is because, on one hand, the indicators only depend on the coirreps at HSPs, and as such are far easier to obtain in first-principles calculations.
On the other hand, the invariants, as shown in the previous section, are directly related to the topological surface states that may be detected experimentally.
A mapping from indicator to invariants hence links first-principles calculation to experimental observables.
These mappings from SIs to invariants are ``one-to-many'' in general, in which a given SI set may be mapped to multiple invariant sets.

Assume an MSG has $m$ linearly independent TCs, which have invariant sets $V_1, V_2, ..., V_m$ and SI sets $S_1, S_2, ..., S_m$.
Given an arbitrary SI set $S$ from the SI group, to find all the possible invariant sets corresponding to it, one needs to solve the linear equation $\sum_{i=1}^m x_i S_i=S$, where $x_i$ are coefficients to be determined.
The invariant sets corresponding to $S$ can be obtained using the solved $x_i$.
This inhomogeneous linear equation can be solved by first finding a special solution and then adding all the general solutions to its homogeneous counterpart, i.e., the solutions of the corresponding homogeneous linear equation, $\sum_{i=1}^m x_i S_i=0$, which gives the invariant sets that have zero SIs.

Following this line, we have developed an algorithm that automatically computes all possible sets of invariants for a given set of indicators.
The code and the full results which have been entered into a large table may be downloaded from 
\href{https://github.com/yjiang-iop/MSG_mapping}{https://github.com/yjiang-iop/MSG\_mapping}.

\subsection{Weyl semimetal indicators}

Unlike non-magnetic spinful SGs\cite{song2018quantitative} where all SIs indicate topological gapped states, we find many SIs in MSGs correspond to gapless Weyl semimetals.
As the full mapping from TCs to SIs has been found for each MSG, indicator values that do not belong to the image of this mapping necessarily correspond to gapless states.
Although these Weyl semimetals have Weyl points at generic momenta in the BZ, the symmetries of MSGs require their creation or annihilation to happen at HSPs, which makes it possible to detect them using SIs.
These Weyl semimetals can be classified into two types, with one being inter-plane Weyl points that lie between $k_i=0$ and $k_i=\pi$ plane, and the other being in-plane Weyl points that lie on $k_i=0$ or $\pi$ plane, where $i$ denotes the main rotation axis of the MSG.
We leave the full discussion of Weyl semimetals to Appendix.\ref{AppendixI}, and use one example of type-4 MSG 81.37 $P_C\overline{4}$ to illustrate the basic idea.

\begin{figure}[H]
	\centering
	\includegraphics[width=0.2\textwidth]{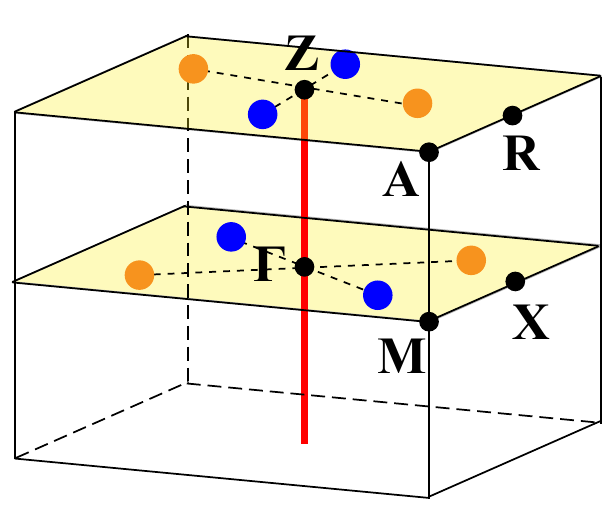}
	\caption{\label{81_37weyl}The Weyl point configuration of MSG 81.37, where the blue and orange dots represent Weyl points of opposite chiralities, the red line represents the $S_4$ axis, and two yellow planes represent $k_z=0,\pi$ planes.}
\end{figure}

MSG 81.37 can host a multiple of 4 Weyl points on both $k_z=0/\pi$ plane, with Weyl points connected by $S_4$ symmetry having opposite chirality, as shown in Fig.\ref{81_37weyl}. This MSG has TCI classification $\mathbb{Z}_2$ and SI group $\mathbb{Z}_2\times\mathbb{Z}_2$.
The SI group is larger than the TCI group.
One of the $\mathbb{Z}_2$ indicator can be chosen as $z_{2,S_4}$, the odd value of which corresponds to the decoration with $S_4$ invariant $\delta(S_4)=1$, while the other $\mathbb{Z}_2$ indicator can be chosen as $z_{4C}/2$, the odd value of which represents the Weyl states.
Notice that the Weyl points are $S_4$-symmetric, and they can only be annihilated at HSPs on $k_z=0/\pi$ plane.
Moreover, $z_{4C,k_z=0}=z_{4C,k_z=\pi}=0$ or $2$ always holds, which means the Weyl points will appear simultaneously on $k_z=0,\pi$ planes.
The in-plane nature of this type of Weyl states is enforced by the $C_2\cdot T$ symmetry.

\bibliography{ms}

\begin{thebibliography}{94}
\expandafter\ifx\csname natexlab\endcsname\relax\def\natexlab#1{#1}\fi
\expandafter\ifx\csname bibnamefont\endcsname\relax
  \def\bibnamefont#1{#1}\fi
\expandafter\ifx\csname bibfnamefont\endcsname\relax
  \def\bibfnamefont#1{#1}\fi
\expandafter\ifx\csname citenamefont\endcsname\relax
  \def\citenamefont#1{#1}\fi
\expandafter\ifx\csname url\endcsname\relax
  \def\url#1{\texttt{#1}}\fi
\expandafter\ifx\csname urlprefix\endcsname\relax\def\urlprefix{URL }\fi
\providecommand{\bibinfo}[2]{#2}
\providecommand{\eprint}[2][]{\url{#2}}

\bibitem[{\citenamefont{Bradley and Davies}(1968)}]{bradley1968magnetic}
\bibinfo{author}{\bibfnamefont{C.~J.} \bibnamefont{Bradley}} \bibnamefont{and}
  \bibinfo{author}{\bibfnamefont{B.~L.} \bibnamefont{Davies}},
  \bibinfo{journal}{Rev. Mod. Phys.} \textbf{\bibinfo{volume}{40}},
  \bibinfo{pages}{359} (\bibinfo{year}{1968}),
  \urlprefix\url{https://link.aps.org/doi/10.1103/RevModPhys.40.359}.

\bibitem[{Note1()}]{Note1}
Note1, \bibinfo{note}{the ``MSGs'' we use in this work means type-1, 3, and 4
  MSGs, and the non-magnetic type-2 MSGs are excluded.}

\bibitem[{\citenamefont{Moriya and Takahashi}(1984)}]{shimizu1981itinerant}
\bibinfo{author}{\bibfnamefont{T.}~\bibnamefont{Moriya}} \bibnamefont{and}
  \bibinfo{author}{\bibfnamefont{Y.}~\bibnamefont{Takahashi}},
  \bibinfo{journal}{Annual Review of Materials Science}
  \textbf{\bibinfo{volume}{14}}, \bibinfo{pages}{1} (\bibinfo{year}{1984}),
  \eprint{https://doi.org/10.1146/annurev.ms.14.080184.000245},
  \urlprefix\url{https://doi.org/10.1146/annurev.ms.14.080184.000245}.

\bibitem[{\citenamefont{Stewart}(1984)}]{stewart1984heavy}
\bibinfo{author}{\bibfnamefont{G.~R.} \bibnamefont{Stewart}},
  \bibinfo{journal}{Rev. Mod. Phys.} \textbf{\bibinfo{volume}{56}},
  \bibinfo{pages}{755} (\bibinfo{year}{1984}),
  \urlprefix\url{https://link.aps.org/doi/10.1103/RevModPhys.56.755}.

\bibitem[{\citenamefont{Lee et~al.}(2006)\citenamefont{Lee, Nagaosa, and
  Wen}}]{lee2006doping}
\bibinfo{author}{\bibfnamefont{P.~A.} \bibnamefont{Lee}},
  \bibinfo{author}{\bibfnamefont{N.}~\bibnamefont{Nagaosa}}, \bibnamefont{and}
  \bibinfo{author}{\bibfnamefont{X.-G.} \bibnamefont{Wen}},
  \bibinfo{journal}{Rev. Mod. Phys.} \textbf{\bibinfo{volume}{78}},
  \bibinfo{pages}{17} (\bibinfo{year}{2006}),
  \urlprefix\url{https://link.aps.org/doi/10.1103/RevModPhys.78.17}.

\bibitem[{\citenamefont{Hasan and Kane}(2010)}]{hasan2010colloquium}
\bibinfo{author}{\bibfnamefont{M.~Z.} \bibnamefont{Hasan}} \bibnamefont{and}
  \bibinfo{author}{\bibfnamefont{C.~L.} \bibnamefont{Kane}},
  \bibinfo{journal}{Rev. Mod. Phys.} \textbf{\bibinfo{volume}{82}},
  \bibinfo{pages}{3045} (\bibinfo{year}{2010}),
  \urlprefix\url{https://link.aps.org/doi/10.1103/RevModPhys.82.3045}.

\bibitem[{\citenamefont{Qi and Zhang}(2011)}]{qi2011topological}
\bibinfo{author}{\bibfnamefont{X.-L.} \bibnamefont{Qi}} \bibnamefont{and}
  \bibinfo{author}{\bibfnamefont{S.-C.} \bibnamefont{Zhang}},
  \bibinfo{journal}{Rev. Mod. Phys.} \textbf{\bibinfo{volume}{83}},
  \bibinfo{pages}{1057} (\bibinfo{year}{2011}),
  \urlprefix\url{https://link.aps.org/doi/10.1103/RevModPhys.83.1057}.

\bibitem[{\citenamefont{Chiu et~al.}(2016)\citenamefont{Chiu, Teo, Schnyder,
  and Ryu}}]{chiu2016classification}
\bibinfo{author}{\bibfnamefont{C.-K.} \bibnamefont{Chiu}},
  \bibinfo{author}{\bibfnamefont{J.~C.~Y.} \bibnamefont{Teo}},
  \bibinfo{author}{\bibfnamefont{A.~P.} \bibnamefont{Schnyder}},
  \bibnamefont{and} \bibinfo{author}{\bibfnamefont{S.}~\bibnamefont{Ryu}},
  \bibinfo{journal}{Rev. Mod. Phys.} \textbf{\bibinfo{volume}{88}},
  \bibinfo{pages}{035005} (\bibinfo{year}{2016}),
  \urlprefix\url{https://link.aps.org/doi/10.1103/RevModPhys.88.035005}.

\bibitem[{\citenamefont{Armitage et~al.}(2018)\citenamefont{Armitage, Mele, and
  Vishwanath}}]{armitage2018weyl}
\bibinfo{author}{\bibfnamefont{N.~P.} \bibnamefont{Armitage}},
  \bibinfo{author}{\bibfnamefont{E.~J.} \bibnamefont{Mele}}, \bibnamefont{and}
  \bibinfo{author}{\bibfnamefont{A.}~\bibnamefont{Vishwanath}},
  \bibinfo{journal}{Rev. Mod. Phys.} \textbf{\bibinfo{volume}{90}},
  \bibinfo{pages}{015001} (\bibinfo{year}{2018}),
  \urlprefix\url{https://link.aps.org/doi/10.1103/RevModPhys.90.015001}.

\bibitem[{\citenamefont{Thouless et~al.}(1982)\citenamefont{Thouless, Kohmoto,
  Nightingale, and den Nijs}}]{thouless1982quantized}
\bibinfo{author}{\bibfnamefont{D.~J.} \bibnamefont{Thouless}},
  \bibinfo{author}{\bibfnamefont{M.}~\bibnamefont{Kohmoto}},
  \bibinfo{author}{\bibfnamefont{M.~P.} \bibnamefont{Nightingale}},
  \bibnamefont{and} \bibinfo{author}{\bibfnamefont{M.}~\bibnamefont{den Nijs}},
  \bibinfo{journal}{Phys. Rev. Lett.} \textbf{\bibinfo{volume}{49}},
  \bibinfo{pages}{405} (\bibinfo{year}{1982}),
  \urlprefix\url{https://link.aps.org/doi/10.1103/PhysRevLett.49.405}.

\bibitem[{\citenamefont{Haldane}(1983)}]{haldane1983nonlinear}
\bibinfo{author}{\bibfnamefont{F.~D.~M.} \bibnamefont{Haldane}},
  \bibinfo{journal}{Phys. Rev. Lett.} \textbf{\bibinfo{volume}{50}},
  \bibinfo{pages}{1153} (\bibinfo{year}{1983}),
  \urlprefix\url{https://link.aps.org/doi/10.1103/PhysRevLett.50.1153}.

\bibitem[{\citenamefont{Affleck et~al.}(1987)\citenamefont{Affleck, Kennedy,
  Lieb, and Tasaki}}]{affleck1987rigorous}
\bibinfo{author}{\bibfnamefont{I.}~\bibnamefont{Affleck}},
  \bibinfo{author}{\bibfnamefont{T.}~\bibnamefont{Kennedy}},
  \bibinfo{author}{\bibfnamefont{E.~H.} \bibnamefont{Lieb}}, \bibnamefont{and}
  \bibinfo{author}{\bibfnamefont{H.}~\bibnamefont{Tasaki}},
  \bibinfo{journal}{Phys. Rev. Lett.} \textbf{\bibinfo{volume}{59}},
  \bibinfo{pages}{799} (\bibinfo{year}{1987}),
  \urlprefix\url{https://link.aps.org/doi/10.1103/PhysRevLett.59.799}.

\bibitem[{\citenamefont{Wen}(1990)}]{wen1990topological}
\bibinfo{author}{\bibfnamefont{X.-G.} \bibnamefont{Wen}},
  \bibinfo{journal}{International Journal of Modern Physics B}
  \textbf{\bibinfo{volume}{4}}, \bibinfo{pages}{239} (\bibinfo{year}{1990}),
  \urlprefix\url{https://www.worldscientific.com/doi/abs/10.1142/S0217979290000139}.

\bibitem[{\citenamefont{Kitaev}(2001)}]{kitaev2001unpaired}
\bibinfo{author}{\bibfnamefont{A.~Y.} \bibnamefont{Kitaev}},
  \bibinfo{journal}{Physics-Uspekhi} \textbf{\bibinfo{volume}{44}},
  \bibinfo{pages}{131} (\bibinfo{year}{2001}),
  \urlprefix\url{https://iopscience.iop.org/article/10.1070/1063-7869/44/10S/S29}.

\bibitem[{\citenamefont{Kane and Mele}(2005)}]{kane2005z}
\bibinfo{author}{\bibfnamefont{C.~L.} \bibnamefont{Kane}} \bibnamefont{and}
  \bibinfo{author}{\bibfnamefont{E.~J.} \bibnamefont{Mele}},
  \bibinfo{journal}{Phys. Rev. Lett.} \textbf{\bibinfo{volume}{95}},
  \bibinfo{pages}{146802} (\bibinfo{year}{2005}),
  \urlprefix\url{https://link.aps.org/doi/10.1103/PhysRevLett.95.146802}.

\bibitem[{\citenamefont{Moore and Balents}(2007)}]{moore2007topological}
\bibinfo{author}{\bibfnamefont{J.~E.} \bibnamefont{Moore}} \bibnamefont{and}
  \bibinfo{author}{\bibfnamefont{L.}~\bibnamefont{Balents}},
  \bibinfo{journal}{Phys. Rev. B} \textbf{\bibinfo{volume}{75}},
  \bibinfo{pages}{121306} (\bibinfo{year}{2007}),
  \urlprefix\url{https://link.aps.org/doi/10.1103/PhysRevB.75.121306}.

\bibitem[{\citenamefont{Fu and Kane}(2007)}]{fu2007topological}
\bibinfo{author}{\bibfnamefont{L.}~\bibnamefont{Fu}} \bibnamefont{and}
  \bibinfo{author}{\bibfnamefont{C.~L.} \bibnamefont{Kane}},
  \bibinfo{journal}{Phys. Rev. B} \textbf{\bibinfo{volume}{76}},
  \bibinfo{pages}{045302} (\bibinfo{year}{2007}),
  \urlprefix\url{https://link.aps.org/doi/10.1103/PhysRevB.76.045302}.

\bibitem[{\citenamefont{Schnyder et~al.}(2008)\citenamefont{Schnyder, Ryu,
  Furusaki, and Ludwig}}]{schnyder2008classification}
\bibinfo{author}{\bibfnamefont{A.~P.} \bibnamefont{Schnyder}},
  \bibinfo{author}{\bibfnamefont{S.}~\bibnamefont{Ryu}},
  \bibinfo{author}{\bibfnamefont{A.}~\bibnamefont{Furusaki}}, \bibnamefont{and}
  \bibinfo{author}{\bibfnamefont{A.~W.~W.} \bibnamefont{Ludwig}},
  \bibinfo{journal}{Phys. Rev. B} \textbf{\bibinfo{volume}{78}},
  \bibinfo{pages}{195125} (\bibinfo{year}{2008}),
  \urlprefix\url{https://link.aps.org/doi/10.1103/PhysRevB.78.195125}.

\bibitem[{\citenamefont{Kitaev}(2009)}]{kitaev2009periodic}
\bibinfo{author}{\bibfnamefont{A.}~\bibnamefont{Kitaev}}, in
  \emph{\bibinfo{booktitle}{AIP conference proceedings}}
  (\bibinfo{organization}{American Institute of Physics},
  \bibinfo{year}{2009}), vol. \bibinfo{volume}{1134}, pp.
  \bibinfo{pages}{22--30},
  \urlprefix\url{https://aip.scitation.org/doi/abs/10.1063/1.3149495}.

\bibitem[{\citenamefont{Ryu et~al.}(2010)\citenamefont{Ryu, Schnyder, Furusaki,
  and Ludwig}}]{ryu2010topological}
\bibinfo{author}{\bibfnamefont{S.}~\bibnamefont{Ryu}},
  \bibinfo{author}{\bibfnamefont{A.~P.} \bibnamefont{Schnyder}},
  \bibinfo{author}{\bibfnamefont{A.}~\bibnamefont{Furusaki}}, \bibnamefont{and}
  \bibinfo{author}{\bibfnamefont{A.~W.} \bibnamefont{Ludwig}},
  \bibinfo{journal}{New Journal of Physics} \textbf{\bibinfo{volume}{12}},
  \bibinfo{pages}{065010} (\bibinfo{year}{2010}),
  \urlprefix\url{https://iopscience.iop.org/article/10.1088/1367-2630/12/6/065010/meta}.

\bibitem[{\citenamefont{Fu et~al.}(2007)\citenamefont{Fu, Kane, and
  Mele}}]{fu2007topological3d}
\bibinfo{author}{\bibfnamefont{L.}~\bibnamefont{Fu}},
  \bibinfo{author}{\bibfnamefont{C.~L.} \bibnamefont{Kane}}, \bibnamefont{and}
  \bibinfo{author}{\bibfnamefont{E.~J.} \bibnamefont{Mele}},
  \bibinfo{journal}{Phys. Rev. Lett.} \textbf{\bibinfo{volume}{98}},
  \bibinfo{pages}{106803} (\bibinfo{year}{2007}),
  \urlprefix\url{https://link.aps.org/doi/10.1103/PhysRevLett.98.106803}.

\bibitem[{\citenamefont{Hsieh et~al.}(2012)\citenamefont{Hsieh, Lin, Liu, Duan,
  Bansil, and Fu}}]{hsieh2012topological}
\bibinfo{author}{\bibfnamefont{T.~H.} \bibnamefont{Hsieh}},
  \bibinfo{author}{\bibfnamefont{H.}~\bibnamefont{Lin}},
  \bibinfo{author}{\bibfnamefont{J.}~\bibnamefont{Liu}},
  \bibinfo{author}{\bibfnamefont{W.}~\bibnamefont{Duan}},
  \bibinfo{author}{\bibfnamefont{A.}~\bibnamefont{Bansil}}, \bibnamefont{and}
  \bibinfo{author}{\bibfnamefont{L.}~\bibnamefont{Fu}},
  \bibinfo{journal}{Nature communications} \textbf{\bibinfo{volume}{3}},
  \bibinfo{pages}{1} (\bibinfo{year}{2012}),
  \urlprefix\url{https://www.nature.com/articles/ncomms1969}.

\bibitem[{\citenamefont{Teo et~al.}(2008)\citenamefont{Teo, Fu, and
  Kane}}]{teo2008surface}
\bibinfo{author}{\bibfnamefont{J.~C.~Y.} \bibnamefont{Teo}},
  \bibinfo{author}{\bibfnamefont{L.}~\bibnamefont{Fu}}, \bibnamefont{and}
  \bibinfo{author}{\bibfnamefont{C.~L.} \bibnamefont{Kane}},
  \bibinfo{journal}{Phys. Rev. B} \textbf{\bibinfo{volume}{78}},
  \bibinfo{pages}{045426} (\bibinfo{year}{2008}),
  \urlprefix\url{https://link.aps.org/doi/10.1103/PhysRevB.78.045426}.

\bibitem[{\citenamefont{Wang et~al.}(2016{\natexlab{a}})\citenamefont{Wang,
  Alexandradinata, Cava, and Bernevig}}]{wang2016hourglass}
\bibinfo{author}{\bibfnamefont{Z.}~\bibnamefont{Wang}},
  \bibinfo{author}{\bibfnamefont{A.}~\bibnamefont{Alexandradinata}},
  \bibinfo{author}{\bibfnamefont{R.~J.} \bibnamefont{Cava}}, \bibnamefont{and}
  \bibinfo{author}{\bibfnamefont{B.~A.} \bibnamefont{Bernevig}},
  \bibinfo{journal}{Nature} \textbf{\bibinfo{volume}{532}},
  \bibinfo{pages}{189} (\bibinfo{year}{2016}{\natexlab{a}}),
  \urlprefix\url{https://doi.org/10.1038/nature17410}.

\bibitem[{\citenamefont{Shiozaki and Sato}(2014)}]{shiozaki2014topology}
\bibinfo{author}{\bibfnamefont{K.}~\bibnamefont{Shiozaki}} \bibnamefont{and}
  \bibinfo{author}{\bibfnamefont{M.}~\bibnamefont{Sato}},
  \bibinfo{journal}{Phys. Rev. B} \textbf{\bibinfo{volume}{90}},
  \bibinfo{pages}{165114} (\bibinfo{year}{2014}),
  \urlprefix\url{https://link.aps.org/doi/10.1103/PhysRevB.90.165114}.

\bibitem[{\citenamefont{Song et~al.}(2017)\citenamefont{Song, Fang, and
  Fang}}]{song2017d}
\bibinfo{author}{\bibfnamefont{Z.}~\bibnamefont{Song}},
  \bibinfo{author}{\bibfnamefont{Z.}~\bibnamefont{Fang}}, \bibnamefont{and}
  \bibinfo{author}{\bibfnamefont{C.}~\bibnamefont{Fang}},
  \bibinfo{journal}{Phys. Rev. Lett.} \textbf{\bibinfo{volume}{119}},
  \bibinfo{pages}{246402} (\bibinfo{year}{2017}),
  \urlprefix\url{https://link.aps.org/doi/10.1103/PhysRevLett.119.246402}.

\bibitem[{\citenamefont{Fang and Fu}(2019)}]{fang2019new}
\bibinfo{author}{\bibfnamefont{C.}~\bibnamefont{Fang}} \bibnamefont{and}
  \bibinfo{author}{\bibfnamefont{L.}~\bibnamefont{Fu}},
  \bibinfo{journal}{Science advances} \textbf{\bibinfo{volume}{5}},
  \bibinfo{pages}{eaat2374} (\bibinfo{year}{2019}),
  \urlprefix\url{https://advances.sciencemag.org/content/5/12/eaat2374}.

\bibitem[{\citenamefont{Turner et~al.}(2010)\citenamefont{Turner, Zhang, and
  Vishwanath}}]{turner2010entanglement}
\bibinfo{author}{\bibfnamefont{A.~M.} \bibnamefont{Turner}},
  \bibinfo{author}{\bibfnamefont{Y.}~\bibnamefont{Zhang}}, \bibnamefont{and}
  \bibinfo{author}{\bibfnamefont{A.}~\bibnamefont{Vishwanath}},
  \bibinfo{journal}{Phys. Rev. B} \textbf{\bibinfo{volume}{82}},
  \bibinfo{pages}{241102} (\bibinfo{year}{2010}),
  \urlprefix\url{https://link.aps.org/doi/10.1103/PhysRevB.82.241102}.

\bibitem[{\citenamefont{Hughes et~al.}(2011)\citenamefont{Hughes, Prodan, and
  Bernevig}}]{hughes2011inversion}
\bibinfo{author}{\bibfnamefont{T.~L.} \bibnamefont{Hughes}},
  \bibinfo{author}{\bibfnamefont{E.}~\bibnamefont{Prodan}}, \bibnamefont{and}
  \bibinfo{author}{\bibfnamefont{B.~A.} \bibnamefont{Bernevig}},
  \bibinfo{journal}{Phys. Rev. B} \textbf{\bibinfo{volume}{83}},
  \bibinfo{pages}{245132} (\bibinfo{year}{2011}),
  \urlprefix\url{https://link.aps.org/doi/10.1103/PhysRevB.83.245132}.

\bibitem[{\citenamefont{Song et~al.}(2018{\natexlab{a}})\citenamefont{Song,
  Zhang, Fang, and Fang}}]{song2018quantitative}
\bibinfo{author}{\bibfnamefont{Z.}~\bibnamefont{Song}},
  \bibinfo{author}{\bibfnamefont{T.}~\bibnamefont{Zhang}},
  \bibinfo{author}{\bibfnamefont{Z.}~\bibnamefont{Fang}}, \bibnamefont{and}
  \bibinfo{author}{\bibfnamefont{C.}~\bibnamefont{Fang}},
  \bibinfo{journal}{Nature communications} \textbf{\bibinfo{volume}{9}},
  \bibinfo{pages}{1} (\bibinfo{year}{2018}{\natexlab{a}}),
  \urlprefix\url{https://www.nature.com/articles/s41467-018-06010-w}.

\bibitem[{\citenamefont{Khalaf et~al.}(2018)\citenamefont{Khalaf, Po,
  Vishwanath, and Watanabe}}]{khalaf2018symmetry}
\bibinfo{author}{\bibfnamefont{E.}~\bibnamefont{Khalaf}},
  \bibinfo{author}{\bibfnamefont{H.~C.} \bibnamefont{Po}},
  \bibinfo{author}{\bibfnamefont{A.}~\bibnamefont{Vishwanath}},
  \bibnamefont{and} \bibinfo{author}{\bibfnamefont{H.}~\bibnamefont{Watanabe}},
  \bibinfo{journal}{Phys. Rev. X} \textbf{\bibinfo{volume}{8}},
  \bibinfo{pages}{031070} (\bibinfo{year}{2018}),
  \urlprefix\url{https://link.aps.org/doi/10.1103/PhysRevX.8.031070}.

\bibitem[{\citenamefont{Song et~al.}(2019)\citenamefont{Song, Huang, Qi, Fang,
  and Hermele}}]{song2019topological}
\bibinfo{author}{\bibfnamefont{Z.}~\bibnamefont{Song}},
  \bibinfo{author}{\bibfnamefont{S.-J.} \bibnamefont{Huang}},
  \bibinfo{author}{\bibfnamefont{Y.}~\bibnamefont{Qi}},
  \bibinfo{author}{\bibfnamefont{C.}~\bibnamefont{Fang}}, \bibnamefont{and}
  \bibinfo{author}{\bibfnamefont{M.}~\bibnamefont{Hermele}},
  \bibinfo{journal}{Science advances} \textbf{\bibinfo{volume}{5}},
  \bibinfo{pages}{eaax2007} (\bibinfo{year}{2019}),
  \urlprefix\url{https://advances.sciencemag.org/content/5/12/eaax2007}.

\bibitem[{\citenamefont{Song et~al.}(2020)\citenamefont{Song, Fang, and
  Qi}}]{song2020real}
\bibinfo{author}{\bibfnamefont{Z.}~\bibnamefont{Song}},
  \bibinfo{author}{\bibfnamefont{C.}~\bibnamefont{Fang}}, \bibnamefont{and}
  \bibinfo{author}{\bibfnamefont{Y.}~\bibnamefont{Qi}},
  \bibinfo{journal}{Nature communications} \textbf{\bibinfo{volume}{11}},
  \bibinfo{pages}{1} (\bibinfo{year}{2020}),
  \urlprefix\url{https://www.nature.com/articles/s41467-020-17685-5}.

\bibitem[{\citenamefont{Qi et~al.}(2008)\citenamefont{Qi, Hughes, and
  Zhang}}]{qi2008topological}
\bibinfo{author}{\bibfnamefont{X.-L.} \bibnamefont{Qi}},
  \bibinfo{author}{\bibfnamefont{T.~L.} \bibnamefont{Hughes}},
  \bibnamefont{and} \bibinfo{author}{\bibfnamefont{S.-C.} \bibnamefont{Zhang}},
  \bibinfo{journal}{Phys. Rev. B} \textbf{\bibinfo{volume}{78}},
  \bibinfo{pages}{195424} (\bibinfo{year}{2008}),
  \urlprefix\url{https://link.aps.org/doi/10.1103/PhysRevB.78.195424}.

\bibitem[{\citenamefont{Li et~al.}(2010)\citenamefont{Li, Wang, Qi, and
  Zhang}}]{li2010dynamical}
\bibinfo{author}{\bibfnamefont{R.}~\bibnamefont{Li}},
  \bibinfo{author}{\bibfnamefont{J.}~\bibnamefont{Wang}},
  \bibinfo{author}{\bibfnamefont{X.-L.} \bibnamefont{Qi}}, \bibnamefont{and}
  \bibinfo{author}{\bibfnamefont{S.-C.} \bibnamefont{Zhang}},
  \bibinfo{journal}{Nature Physics} \textbf{\bibinfo{volume}{6}},
  \bibinfo{pages}{284} (\bibinfo{year}{2010}),
  \urlprefix\url{https://www.nature.com/articles/nphys1534}.

\bibitem[{\citenamefont{Turner et~al.}(2012)\citenamefont{Turner, Zhang, Mong,
  and Vishwanath}}]{turner2012quantized}
\bibinfo{author}{\bibfnamefont{A.~M.} \bibnamefont{Turner}},
  \bibinfo{author}{\bibfnamefont{Y.}~\bibnamefont{Zhang}},
  \bibinfo{author}{\bibfnamefont{R.~S.~K.} \bibnamefont{Mong}},
  \bibnamefont{and}
  \bibinfo{author}{\bibfnamefont{A.}~\bibnamefont{Vishwanath}},
  \bibinfo{journal}{Phys. Rev. B} \textbf{\bibinfo{volume}{85}},
  \bibinfo{pages}{165120} (\bibinfo{year}{2012}),
  \urlprefix\url{https://link.aps.org/doi/10.1103/PhysRevB.85.165120}.

\bibitem[{\citenamefont{Yue et~al.}(2019)\citenamefont{Yue, Xu, Song, Weng, Lu,
  Fang, and Dai}}]{yue2019symmetry}
\bibinfo{author}{\bibfnamefont{C.}~\bibnamefont{Yue}},
  \bibinfo{author}{\bibfnamefont{Y.}~\bibnamefont{Xu}},
  \bibinfo{author}{\bibfnamefont{Z.}~\bibnamefont{Song}},
  \bibinfo{author}{\bibfnamefont{H.}~\bibnamefont{Weng}},
  \bibinfo{author}{\bibfnamefont{Y.-M.} \bibnamefont{Lu}},
  \bibinfo{author}{\bibfnamefont{C.}~\bibnamefont{Fang}}, \bibnamefont{and}
  \bibinfo{author}{\bibfnamefont{X.}~\bibnamefont{Dai}},
  \bibinfo{journal}{Nature Physics} \textbf{\bibinfo{volume}{15}},
  \bibinfo{pages}{577} (\bibinfo{year}{2019}),
  \urlprefix\url{https://www.nature.com/articles/s41567-019-0457-0}.

\bibitem[{\citenamefont{Zhang et~al.}(2019{\natexlab{a}})\citenamefont{Zhang,
  Shi, Zhu, Xing, Zhang, and Wang}}]{zhang2019topological}
\bibinfo{author}{\bibfnamefont{D.}~\bibnamefont{Zhang}},
  \bibinfo{author}{\bibfnamefont{M.}~\bibnamefont{Shi}},
  \bibinfo{author}{\bibfnamefont{T.}~\bibnamefont{Zhu}},
  \bibinfo{author}{\bibfnamefont{D.}~\bibnamefont{Xing}},
  \bibinfo{author}{\bibfnamefont{H.}~\bibnamefont{Zhang}}, \bibnamefont{and}
  \bibinfo{author}{\bibfnamefont{J.}~\bibnamefont{Wang}},
  \bibinfo{journal}{Phys. Rev. Lett.} \textbf{\bibinfo{volume}{122}},
  \bibinfo{pages}{206401} (\bibinfo{year}{2019}{\natexlab{a}}),
  \urlprefix\url{https://link.aps.org/doi/10.1103/PhysRevLett.122.206401}.

\bibitem[{\citenamefont{Otrokov et~al.}(2019)\citenamefont{Otrokov,
  Klimovskikh, Bentmann, Estyunin, Zeugner, Aliev, Ga{\ss}, Wolter, Koroleva,
  Shikin et~al.}}]{otrokov2019prediction}
\bibinfo{author}{\bibfnamefont{M.~M.} \bibnamefont{Otrokov}},
  \bibinfo{author}{\bibfnamefont{I.~I.} \bibnamefont{Klimovskikh}},
  \bibinfo{author}{\bibfnamefont{H.}~\bibnamefont{Bentmann}},
  \bibinfo{author}{\bibfnamefont{D.}~\bibnamefont{Estyunin}},
  \bibinfo{author}{\bibfnamefont{A.}~\bibnamefont{Zeugner}},
  \bibinfo{author}{\bibfnamefont{Z.~S.} \bibnamefont{Aliev}},
  \bibinfo{author}{\bibfnamefont{S.}~\bibnamefont{Ga{\ss}}},
  \bibinfo{author}{\bibfnamefont{A.}~\bibnamefont{Wolter}},
  \bibinfo{author}{\bibfnamefont{A.}~\bibnamefont{Koroleva}},
  \bibinfo{author}{\bibfnamefont{A.~M.} \bibnamefont{Shikin}},
  \bibnamefont{et~al.}, \bibinfo{journal}{Nature}
  \textbf{\bibinfo{volume}{576}}, \bibinfo{pages}{416} (\bibinfo{year}{2019}),
  \urlprefix\url{https://doi.org/10.1038/s41586-019-1840-9}.

\bibitem[{\citenamefont{Rienks et~al.}(2019)\citenamefont{Rienks, Wimmer
  et~al.}}]{rienks2019large}
\bibinfo{author}{\bibfnamefont{E.~D.} \bibnamefont{Rienks}},
  \bibinfo{author}{\bibfnamefont{S.}~\bibnamefont{Wimmer}},
  \bibnamefont{et~al.}, \bibinfo{journal}{Nature}
  \textbf{\bibinfo{volume}{576}}, \bibinfo{pages}{423} (\bibinfo{year}{2019}),
  \urlprefix\url{https://doi.org/10.1038/s41586-019-1826-7}.

\bibitem[{\citenamefont{Mong et~al.}(2010)\citenamefont{Mong, Essin, and
  Moore}}]{mong2010antiferromagnetic}
\bibinfo{author}{\bibfnamefont{R.~S.~K.} \bibnamefont{Mong}},
  \bibinfo{author}{\bibfnamefont{A.~M.} \bibnamefont{Essin}}, \bibnamefont{and}
  \bibinfo{author}{\bibfnamefont{J.~E.} \bibnamefont{Moore}},
  \bibinfo{journal}{Phys. Rev. B} \textbf{\bibinfo{volume}{81}},
  \bibinfo{pages}{245209} (\bibinfo{year}{2010}),
  \urlprefix\url{https://link.aps.org/doi/10.1103/PhysRevB.81.245209}.

\bibitem[{\citenamefont{Fang et~al.}(2013)\citenamefont{Fang, Gilbert, and
  Bernevig}}]{fang2013topological}
\bibinfo{author}{\bibfnamefont{C.}~\bibnamefont{Fang}},
  \bibinfo{author}{\bibfnamefont{M.~J.} \bibnamefont{Gilbert}},
  \bibnamefont{and} \bibinfo{author}{\bibfnamefont{B.~A.}
  \bibnamefont{Bernevig}}, \bibinfo{journal}{Phys. Rev. B}
  \textbf{\bibinfo{volume}{88}}, \bibinfo{pages}{085406}
  (\bibinfo{year}{2013}),
  \urlprefix\url{https://link.aps.org/doi/10.1103/PhysRevB.88.085406}.

\bibitem[{\citenamefont{Liu}(2013)}]{liu2013antiferromagnetic}
\bibinfo{author}{\bibfnamefont{C.-X.} \bibnamefont{Liu}},
  \bibinfo{journal}{arXiv preprint arXiv:1304.6455}  (\bibinfo{year}{2013}),
  \urlprefix\url{https://arxiv.org/abs/1304.6455}.

\bibitem[{\citenamefont{Zhang and Liu}(2015)}]{zhang2015topological}
\bibinfo{author}{\bibfnamefont{R.-X.} \bibnamefont{Zhang}} \bibnamefont{and}
  \bibinfo{author}{\bibfnamefont{C.-X.} \bibnamefont{Liu}},
  \bibinfo{journal}{Phys. Rev. B} \textbf{\bibinfo{volume}{91}},
  \bibinfo{pages}{115317} (\bibinfo{year}{2015}),
  \urlprefix\url{https://link.aps.org/doi/10.1103/PhysRevB.91.115317}.

\bibitem[{\citenamefont{Liu et~al.}(2014)\citenamefont{Liu, Zhang, and
  VanLeeuwen}}]{liu2014topological}
\bibinfo{author}{\bibfnamefont{C.-X.} \bibnamefont{Liu}},
  \bibinfo{author}{\bibfnamefont{R.-X.} \bibnamefont{Zhang}}, \bibnamefont{and}
  \bibinfo{author}{\bibfnamefont{B.~K.} \bibnamefont{VanLeeuwen}},
  \bibinfo{journal}{Phys. Rev. B} \textbf{\bibinfo{volume}{90}},
  \bibinfo{pages}{085304} (\bibinfo{year}{2014}),
  \urlprefix\url{https://link.aps.org/doi/10.1103/PhysRevB.90.085304}.

\bibitem[{\citenamefont{Shiozaki et~al.}(2017)\citenamefont{Shiozaki, Sato, and
  Gomi}}]{shiozaki2017topological}
\bibinfo{author}{\bibfnamefont{K.}~\bibnamefont{Shiozaki}},
  \bibinfo{author}{\bibfnamefont{M.}~\bibnamefont{Sato}}, \bibnamefont{and}
  \bibinfo{author}{\bibfnamefont{K.}~\bibnamefont{Gomi}},
  \bibinfo{journal}{Phys. Rev. B} \textbf{\bibinfo{volume}{95}},
  \bibinfo{pages}{235425} (\bibinfo{year}{2017}),
  \urlprefix\url{https://link.aps.org/doi/10.1103/PhysRevB.95.235425}.

\bibitem[{\citenamefont{Elcoro et~al.}(2020)\citenamefont{Elcoro, Wieder, Song,
  Xu, Bradlyn, and Bernevig}}]{elcoro2020magnetic}
\bibinfo{author}{\bibfnamefont{L.}~\bibnamefont{Elcoro}},
  \bibinfo{author}{\bibfnamefont{B.~J.} \bibnamefont{Wieder}},
  \bibinfo{author}{\bibfnamefont{Z.}~\bibnamefont{Song}},
  \bibinfo{author}{\bibfnamefont{Y.}~\bibnamefont{Xu}},
  \bibinfo{author}{\bibfnamefont{B.}~\bibnamefont{Bradlyn}}, \bibnamefont{and}
  \bibinfo{author}{\bibfnamefont{B.~A.} \bibnamefont{Bernevig}},
  \bibinfo{journal}{arXiv preprint arXiv:2010.00598}  (\bibinfo{year}{2020}),
  \urlprefix\url{https://arxiv.org/abs/2010.00598}.

\bibitem[{\citenamefont{Fang and Fu}(2015)}]{fang2015new}
\bibinfo{author}{\bibfnamefont{C.}~\bibnamefont{Fang}} \bibnamefont{and}
  \bibinfo{author}{\bibfnamefont{L.}~\bibnamefont{Fu}}, \bibinfo{journal}{Phys.
  Rev. B} \textbf{\bibinfo{volume}{91}}, \bibinfo{pages}{161105}
  (\bibinfo{year}{2015}),
  \urlprefix\url{https://link.aps.org/doi/10.1103/PhysRevB.91.161105}.

\bibitem[{\citenamefont{Fang et~al.}(2012)\citenamefont{Fang, Gilbert, and
  Bernevig}}]{fang2012bulk}
\bibinfo{author}{\bibfnamefont{C.}~\bibnamefont{Fang}},
  \bibinfo{author}{\bibfnamefont{M.~J.} \bibnamefont{Gilbert}},
  \bibnamefont{and} \bibinfo{author}{\bibfnamefont{B.~A.}
  \bibnamefont{Bernevig}}, \bibinfo{journal}{Phys. Rev. B}
  \textbf{\bibinfo{volume}{86}}, \bibinfo{pages}{115112}
  (\bibinfo{year}{2012}),
  \urlprefix\url{https://link.aps.org/doi/10.1103/PhysRevB.86.115112}.

\bibitem[{\citenamefont{Po et~al.}(2017)\citenamefont{Po, Vishwanath, and
  Watanabe}}]{po2017symmetry}
\bibinfo{author}{\bibfnamefont{H.~C.} \bibnamefont{Po}},
  \bibinfo{author}{\bibfnamefont{A.}~\bibnamefont{Vishwanath}},
  \bibnamefont{and} \bibinfo{author}{\bibfnamefont{H.}~\bibnamefont{Watanabe}},
  \bibinfo{journal}{Nature communications} \textbf{\bibinfo{volume}{8}},
  \bibinfo{pages}{1} (\bibinfo{year}{2017}),
  \urlprefix\url{https://www.nature.com/articles/s41467-017-00133-2}.

\bibitem[{\citenamefont{Bradlyn et~al.}(2017)\citenamefont{Bradlyn, Elcoro,
  Cano, Vergniory, Wang, Felser, Aroyo, and Bernevig}}]{bradlyn2017topological}
\bibinfo{author}{\bibfnamefont{B.}~\bibnamefont{Bradlyn}},
  \bibinfo{author}{\bibfnamefont{L.}~\bibnamefont{Elcoro}},
  \bibinfo{author}{\bibfnamefont{J.}~\bibnamefont{Cano}},
  \bibinfo{author}{\bibfnamefont{M.}~\bibnamefont{Vergniory}},
  \bibinfo{author}{\bibfnamefont{Z.}~\bibnamefont{Wang}},
  \bibinfo{author}{\bibfnamefont{C.}~\bibnamefont{Felser}},
  \bibinfo{author}{\bibfnamefont{M.}~\bibnamefont{Aroyo}}, \bibnamefont{and}
  \bibinfo{author}{\bibfnamefont{B.~A.} \bibnamefont{Bernevig}},
  \bibinfo{journal}{Nature} \textbf{\bibinfo{volume}{547}},
  \bibinfo{pages}{298} (\bibinfo{year}{2017}),
  \urlprefix\url{https://www.nature.com/articles/nature23268}.

\bibitem[{\citenamefont{Zhang et~al.}(2019{\natexlab{b}})\citenamefont{Zhang,
  Jiang, Song, Huang, He, Fang, Weng, and Fang}}]{zhang2019catalogue}
\bibinfo{author}{\bibfnamefont{T.}~\bibnamefont{Zhang}},
  \bibinfo{author}{\bibfnamefont{Y.}~\bibnamefont{Jiang}},
  \bibinfo{author}{\bibfnamefont{Z.}~\bibnamefont{Song}},
  \bibinfo{author}{\bibfnamefont{H.}~\bibnamefont{Huang}},
  \bibinfo{author}{\bibfnamefont{Y.}~\bibnamefont{He}},
  \bibinfo{author}{\bibfnamefont{Z.}~\bibnamefont{Fang}},
  \bibinfo{author}{\bibfnamefont{H.}~\bibnamefont{Weng}}, \bibnamefont{and}
  \bibinfo{author}{\bibfnamefont{C.}~\bibnamefont{Fang}},
  \bibinfo{journal}{Nature} \textbf{\bibinfo{volume}{566}},
  \bibinfo{pages}{475} (\bibinfo{year}{2019}{\natexlab{b}}),
  \urlprefix\url{https://www.nature.com/articles/s41586-019-0944-6}.

\bibitem[{\citenamefont{Vergniory et~al.}(2019)\citenamefont{Vergniory, Elcoro,
  Felser, Regnault, Bernevig, and Wang}}]{vergniory2019complete}
\bibinfo{author}{\bibfnamefont{M.}~\bibnamefont{Vergniory}},
  \bibinfo{author}{\bibfnamefont{L.}~\bibnamefont{Elcoro}},
  \bibinfo{author}{\bibfnamefont{C.}~\bibnamefont{Felser}},
  \bibinfo{author}{\bibfnamefont{N.}~\bibnamefont{Regnault}},
  \bibinfo{author}{\bibfnamefont{B.~A.} \bibnamefont{Bernevig}},
  \bibnamefont{and} \bibinfo{author}{\bibfnamefont{Z.}~\bibnamefont{Wang}},
  \bibinfo{journal}{Nature} \textbf{\bibinfo{volume}{566}},
  \bibinfo{pages}{480} (\bibinfo{year}{2019}),
  \urlprefix\url{https://www.nature.com/articles/s41586-019-0954-4}.

\bibitem[{\citenamefont{Tang et~al.}(2019)\citenamefont{Tang, Po, Vishwanath,
  and Wan}}]{tang2019comprehensive}
\bibinfo{author}{\bibfnamefont{F.}~\bibnamefont{Tang}},
  \bibinfo{author}{\bibfnamefont{H.~C.} \bibnamefont{Po}},
  \bibinfo{author}{\bibfnamefont{A.}~\bibnamefont{Vishwanath}},
  \bibnamefont{and} \bibinfo{author}{\bibfnamefont{X.}~\bibnamefont{Wan}},
  \bibinfo{journal}{Nature} \textbf{\bibinfo{volume}{566}},
  \bibinfo{pages}{486} (\bibinfo{year}{2019}),
  \urlprefix\url{https://www.nature.com/articles/s41586-019-0937-5}.

\bibitem[{\citenamefont{Ono and Watanabe}(2018)}]{ono2018unified}
\bibinfo{author}{\bibfnamefont{S.}~\bibnamefont{Ono}} \bibnamefont{and}
  \bibinfo{author}{\bibfnamefont{H.}~\bibnamefont{Watanabe}},
  \bibinfo{journal}{Phys. Rev. B} \textbf{\bibinfo{volume}{98}},
  \bibinfo{pages}{115150} (\bibinfo{year}{2018}),
  \urlprefix\url{https://link.aps.org/doi/10.1103/PhysRevB.98.115150}.

\bibitem[{\citenamefont{Fang et~al.}(2003)\citenamefont{Fang, Nagaosa,
  Takahashi, Asamitsu, Mathieu, Ogasawara, Yamada, Kawasaki, Tokura, and
  Terakura}}]{fang2003anomalous}
\bibinfo{author}{\bibfnamefont{Z.}~\bibnamefont{Fang}},
  \bibinfo{author}{\bibfnamefont{N.}~\bibnamefont{Nagaosa}},
  \bibinfo{author}{\bibfnamefont{K.~S.} \bibnamefont{Takahashi}},
  \bibinfo{author}{\bibfnamefont{A.}~\bibnamefont{Asamitsu}},
  \bibinfo{author}{\bibfnamefont{R.}~\bibnamefont{Mathieu}},
  \bibinfo{author}{\bibfnamefont{T.}~\bibnamefont{Ogasawara}},
  \bibinfo{author}{\bibfnamefont{H.}~\bibnamefont{Yamada}},
  \bibinfo{author}{\bibfnamefont{M.}~\bibnamefont{Kawasaki}},
  \bibinfo{author}{\bibfnamefont{Y.}~\bibnamefont{Tokura}}, \bibnamefont{and}
  \bibinfo{author}{\bibfnamefont{K.}~\bibnamefont{Terakura}},
  \bibinfo{journal}{Science} \textbf{\bibinfo{volume}{302}},
  \bibinfo{pages}{92} (\bibinfo{year}{2003}),
  \urlprefix\url{https://science.sciencemag.org/content/302/5642/92.abstract}.

\bibitem[{\citenamefont{Wan et~al.}(2011)\citenamefont{Wan, Turner, Vishwanath,
  and Savrasov}}]{wan2011topological}
\bibinfo{author}{\bibfnamefont{X.}~\bibnamefont{Wan}},
  \bibinfo{author}{\bibfnamefont{A.~M.} \bibnamefont{Turner}},
  \bibinfo{author}{\bibfnamefont{A.}~\bibnamefont{Vishwanath}},
  \bibnamefont{and} \bibinfo{author}{\bibfnamefont{S.~Y.}
  \bibnamefont{Savrasov}}, \bibinfo{journal}{Phys. Rev. B}
  \textbf{\bibinfo{volume}{83}}, \bibinfo{pages}{205101}
  (\bibinfo{year}{2011}),
  \urlprefix\url{https://link.aps.org/doi/10.1103/PhysRevB.83.205101}.

\bibitem[{\citenamefont{Xu et~al.}(2011)\citenamefont{Xu, Weng, Wang, Dai, and
  Fang}}]{xu2011chern}
\bibinfo{author}{\bibfnamefont{G.}~\bibnamefont{Xu}},
  \bibinfo{author}{\bibfnamefont{H.}~\bibnamefont{Weng}},
  \bibinfo{author}{\bibfnamefont{Z.}~\bibnamefont{Wang}},
  \bibinfo{author}{\bibfnamefont{X.}~\bibnamefont{Dai}}, \bibnamefont{and}
  \bibinfo{author}{\bibfnamefont{Z.}~\bibnamefont{Fang}},
  \bibinfo{journal}{Phys. Rev. Lett.} \textbf{\bibinfo{volume}{107}},
  \bibinfo{pages}{186806} (\bibinfo{year}{2011}),
  \urlprefix\url{https://link.aps.org/doi/10.1103/PhysRevLett.107.186806}.

\bibitem[{\citenamefont{Wang et~al.}(2016{\natexlab{b}})\citenamefont{Wang,
  Vergniory, Kushwaha, Hirschberger, Chulkov, Ernst, Ong, Cava, and
  Bernevig}}]{wang2016time}
\bibinfo{author}{\bibfnamefont{Z.}~\bibnamefont{Wang}},
  \bibinfo{author}{\bibfnamefont{M.~G.} \bibnamefont{Vergniory}},
  \bibinfo{author}{\bibfnamefont{S.}~\bibnamefont{Kushwaha}},
  \bibinfo{author}{\bibfnamefont{M.}~\bibnamefont{Hirschberger}},
  \bibinfo{author}{\bibfnamefont{E.~V.} \bibnamefont{Chulkov}},
  \bibinfo{author}{\bibfnamefont{A.}~\bibnamefont{Ernst}},
  \bibinfo{author}{\bibfnamefont{N.~P.} \bibnamefont{Ong}},
  \bibinfo{author}{\bibfnamefont{R.~J.} \bibnamefont{Cava}}, \bibnamefont{and}
  \bibinfo{author}{\bibfnamefont{B.~A.} \bibnamefont{Bernevig}},
  \bibinfo{journal}{Phys. Rev. Lett.} \textbf{\bibinfo{volume}{117}},
  \bibinfo{pages}{236401} (\bibinfo{year}{2016}{\natexlab{b}}),
  \urlprefix\url{https://link.aps.org/doi/10.1103/PhysRevLett.117.236401}.

\bibitem[{\citenamefont{K{\"u}bler and Felser}(2016)}]{kubler2016weyl}
\bibinfo{author}{\bibfnamefont{J.}~\bibnamefont{K{\"u}bler}} \bibnamefont{and}
  \bibinfo{author}{\bibfnamefont{C.}~\bibnamefont{Felser}},
  \bibinfo{journal}{EPL (Europhysics Letters)} \textbf{\bibinfo{volume}{114}},
  \bibinfo{pages}{47005} (\bibinfo{year}{2016}),
  \urlprefix\url{https://iopscience.iop.org/article/10.1209/0295-5075/114/47005/meta}.

\bibitem[{\citenamefont{Chang et~al.}(2016)\citenamefont{Chang, Xu, Zheng,
  Singh, Hsu, Bian, Alidoust, Belopolski, Sanchez, Zhang
  et~al.}}]{chang2016room}
\bibinfo{author}{\bibfnamefont{G.}~\bibnamefont{Chang}},
  \bibinfo{author}{\bibfnamefont{S.-Y.} \bibnamefont{Xu}},
  \bibinfo{author}{\bibfnamefont{H.}~\bibnamefont{Zheng}},
  \bibinfo{author}{\bibfnamefont{B.}~\bibnamefont{Singh}},
  \bibinfo{author}{\bibfnamefont{C.-H.} \bibnamefont{Hsu}},
  \bibinfo{author}{\bibfnamefont{G.}~\bibnamefont{Bian}},
  \bibinfo{author}{\bibfnamefont{N.}~\bibnamefont{Alidoust}},
  \bibinfo{author}{\bibfnamefont{I.}~\bibnamefont{Belopolski}},
  \bibinfo{author}{\bibfnamefont{D.~S.} \bibnamefont{Sanchez}},
  \bibinfo{author}{\bibfnamefont{S.}~\bibnamefont{Zhang}},
  \bibnamefont{et~al.}, \bibinfo{journal}{Scientific reports}
  \textbf{\bibinfo{volume}{6}}, \bibinfo{pages}{1} (\bibinfo{year}{2016}),
  \urlprefix\url{https://doi.org/10.1038/srep38839}.

\bibitem[{\citenamefont{Yang et~al.}(2017)\citenamefont{Yang, Sun, Zhang, Shi,
  Parkin, and Yan}}]{yang2017topological}
\bibinfo{author}{\bibfnamefont{H.}~\bibnamefont{Yang}},
  \bibinfo{author}{\bibfnamefont{Y.}~\bibnamefont{Sun}},
  \bibinfo{author}{\bibfnamefont{Y.}~\bibnamefont{Zhang}},
  \bibinfo{author}{\bibfnamefont{W.-J.} \bibnamefont{Shi}},
  \bibinfo{author}{\bibfnamefont{S.~S.} \bibnamefont{Parkin}},
  \bibnamefont{and} \bibinfo{author}{\bibfnamefont{B.}~\bibnamefont{Yan}},
  \bibinfo{journal}{New Journal of Physics} \textbf{\bibinfo{volume}{19}},
  \bibinfo{pages}{015008} (\bibinfo{year}{2017}),
  \urlprefix\url{https://iopscience.iop.org/article/10.1088/1367-2630/aa5487/meta}.

\bibitem[{\citenamefont{Liu et~al.}(2018)\citenamefont{Liu, Sun, Kumar,
  Muechler, Sun, Jiao, Yang, Liu, Liang, Xu et~al.}}]{liu2018giant}
\bibinfo{author}{\bibfnamefont{E.}~\bibnamefont{Liu}},
  \bibinfo{author}{\bibfnamefont{Y.}~\bibnamefont{Sun}},
  \bibinfo{author}{\bibfnamefont{N.}~\bibnamefont{Kumar}},
  \bibinfo{author}{\bibfnamefont{L.}~\bibnamefont{Muechler}},
  \bibinfo{author}{\bibfnamefont{A.}~\bibnamefont{Sun}},
  \bibinfo{author}{\bibfnamefont{L.}~\bibnamefont{Jiao}},
  \bibinfo{author}{\bibfnamefont{S.-Y.} \bibnamefont{Yang}},
  \bibinfo{author}{\bibfnamefont{D.}~\bibnamefont{Liu}},
  \bibinfo{author}{\bibfnamefont{A.}~\bibnamefont{Liang}},
  \bibinfo{author}{\bibfnamefont{Q.}~\bibnamefont{Xu}}, \bibnamefont{et~al.},
  \bibinfo{journal}{Nature physics} \textbf{\bibinfo{volume}{14}},
  \bibinfo{pages}{1125} (\bibinfo{year}{2018}),
  \urlprefix\url{https://doi.org/10.1038/s41567-018-0234-5}.

\bibitem[{\citenamefont{Wang et~al.}(2018)\citenamefont{Wang, Xu, Lou, Liu, Li,
  Huang, Shen, Weng, Wang, and Lei}}]{wang2018large}
\bibinfo{author}{\bibfnamefont{Q.}~\bibnamefont{Wang}},
  \bibinfo{author}{\bibfnamefont{Y.}~\bibnamefont{Xu}},
  \bibinfo{author}{\bibfnamefont{R.}~\bibnamefont{Lou}},
  \bibinfo{author}{\bibfnamefont{Z.}~\bibnamefont{Liu}},
  \bibinfo{author}{\bibfnamefont{M.}~\bibnamefont{Li}},
  \bibinfo{author}{\bibfnamefont{Y.}~\bibnamefont{Huang}},
  \bibinfo{author}{\bibfnamefont{D.}~\bibnamefont{Shen}},
  \bibinfo{author}{\bibfnamefont{H.}~\bibnamefont{Weng}},
  \bibinfo{author}{\bibfnamefont{S.}~\bibnamefont{Wang}}, \bibnamefont{and}
  \bibinfo{author}{\bibfnamefont{H.}~\bibnamefont{Lei}},
  \bibinfo{journal}{Nature communications} \textbf{\bibinfo{volume}{9}},
  \bibinfo{pages}{1} (\bibinfo{year}{2018}),
  \urlprefix\url{https://doi.org/10.1038/s41467-018-06088-2}.

\bibitem[{\citenamefont{Belopolski et~al.}(2019)\citenamefont{Belopolski,
  Manna, Sanchez, Chang, Ernst, Yin, Zhang, Cochran, Shumiya, Zheng
  et~al.}}]{belopolski2019discovery}
\bibinfo{author}{\bibfnamefont{I.}~\bibnamefont{Belopolski}},
  \bibinfo{author}{\bibfnamefont{K.}~\bibnamefont{Manna}},
  \bibinfo{author}{\bibfnamefont{D.~S.} \bibnamefont{Sanchez}},
  \bibinfo{author}{\bibfnamefont{G.}~\bibnamefont{Chang}},
  \bibinfo{author}{\bibfnamefont{B.}~\bibnamefont{Ernst}},
  \bibinfo{author}{\bibfnamefont{J.}~\bibnamefont{Yin}},
  \bibinfo{author}{\bibfnamefont{S.~S.} \bibnamefont{Zhang}},
  \bibinfo{author}{\bibfnamefont{T.}~\bibnamefont{Cochran}},
  \bibinfo{author}{\bibfnamefont{N.}~\bibnamefont{Shumiya}},
  \bibinfo{author}{\bibfnamefont{H.}~\bibnamefont{Zheng}},
  \bibnamefont{et~al.}, \bibinfo{journal}{Science}
  \textbf{\bibinfo{volume}{365}}, \bibinfo{pages}{1278} (\bibinfo{year}{2019}),
  \urlprefix\url{https://science.sciencemag.org/content/365/6459/1278.abstract}.

\bibitem[{\citenamefont{Morali et~al.}(2019)\citenamefont{Morali, Batabyal,
  Nag, Liu, Xu, Sun, Yan, Felser, Avraham, and Beidenkopf}}]{morali2019fermi}
\bibinfo{author}{\bibfnamefont{N.}~\bibnamefont{Morali}},
  \bibinfo{author}{\bibfnamefont{R.}~\bibnamefont{Batabyal}},
  \bibinfo{author}{\bibfnamefont{P.~K.} \bibnamefont{Nag}},
  \bibinfo{author}{\bibfnamefont{E.}~\bibnamefont{Liu}},
  \bibinfo{author}{\bibfnamefont{Q.}~\bibnamefont{Xu}},
  \bibinfo{author}{\bibfnamefont{Y.}~\bibnamefont{Sun}},
  \bibinfo{author}{\bibfnamefont{B.}~\bibnamefont{Yan}},
  \bibinfo{author}{\bibfnamefont{C.}~\bibnamefont{Felser}},
  \bibinfo{author}{\bibfnamefont{N.}~\bibnamefont{Avraham}}, \bibnamefont{and}
  \bibinfo{author}{\bibfnamefont{H.}~\bibnamefont{Beidenkopf}},
  \bibinfo{journal}{Science} \textbf{\bibinfo{volume}{365}},
  \bibinfo{pages}{1286} (\bibinfo{year}{2019}),
  \urlprefix\url{https://science.sciencemag.org/content/365/6459/1286.abstract}.

\bibitem[{\citenamefont{Liu et~al.}(2019)\citenamefont{Liu, Liang, Liu, Xu, Li,
  Chen, Pei, Shi, Mo, Dudin et~al.}}]{liu2019magnetic}
\bibinfo{author}{\bibfnamefont{D.}~\bibnamefont{Liu}},
  \bibinfo{author}{\bibfnamefont{A.}~\bibnamefont{Liang}},
  \bibinfo{author}{\bibfnamefont{E.}~\bibnamefont{Liu}},
  \bibinfo{author}{\bibfnamefont{Q.}~\bibnamefont{Xu}},
  \bibinfo{author}{\bibfnamefont{Y.}~\bibnamefont{Li}},
  \bibinfo{author}{\bibfnamefont{C.}~\bibnamefont{Chen}},
  \bibinfo{author}{\bibfnamefont{D.}~\bibnamefont{Pei}},
  \bibinfo{author}{\bibfnamefont{W.}~\bibnamefont{Shi}},
  \bibinfo{author}{\bibfnamefont{S.}~\bibnamefont{Mo}},
  \bibinfo{author}{\bibfnamefont{P.}~\bibnamefont{Dudin}},
  \bibnamefont{et~al.}, \bibinfo{journal}{Science}
  \textbf{\bibinfo{volume}{365}}, \bibinfo{pages}{1282} (\bibinfo{year}{2019}),
  \urlprefix\url{https://science.sciencemag.org/content/365/6459/1282.abstract}.

\bibitem[{\citenamefont{Nie et~al.}(2017)\citenamefont{Nie, Xu, Prinz, and
  Zhang}}]{nie2017topological}
\bibinfo{author}{\bibfnamefont{S.}~\bibnamefont{Nie}},
  \bibinfo{author}{\bibfnamefont{G.}~\bibnamefont{Xu}},
  \bibinfo{author}{\bibfnamefont{F.~B.} \bibnamefont{Prinz}}, \bibnamefont{and}
  \bibinfo{author}{\bibfnamefont{S.-c.} \bibnamefont{Zhang}},
  \bibinfo{journal}{Proceedings of the National Academy of Sciences}
  \textbf{\bibinfo{volume}{114}}, \bibinfo{pages}{10596}
  (\bibinfo{year}{2017}).

\bibitem[{\citenamefont{Tang et~al.}(2016)\citenamefont{Tang, Zhou, Xu, and
  Zhang}}]{tang2016dirac}
\bibinfo{author}{\bibfnamefont{P.}~\bibnamefont{Tang}},
  \bibinfo{author}{\bibfnamefont{Q.}~\bibnamefont{Zhou}},
  \bibinfo{author}{\bibfnamefont{G.}~\bibnamefont{Xu}}, \bibnamefont{and}
  \bibinfo{author}{\bibfnamefont{S.-C.} \bibnamefont{Zhang}},
  \bibinfo{journal}{Nature Physics} \textbf{\bibinfo{volume}{12}},
  \bibinfo{pages}{1100} (\bibinfo{year}{2016}),
  \urlprefix\url{https://doi.org/10.1038/nphys3839}.

\bibitem[{\citenamefont{Hua et~al.}(2018)\citenamefont{Hua, Nie, Song, Yu, Xu,
  and Yao}}]{hua2018dirac}
\bibinfo{author}{\bibfnamefont{G.}~\bibnamefont{Hua}},
  \bibinfo{author}{\bibfnamefont{S.}~\bibnamefont{Nie}},
  \bibinfo{author}{\bibfnamefont{Z.}~\bibnamefont{Song}},
  \bibinfo{author}{\bibfnamefont{R.}~\bibnamefont{Yu}},
  \bibinfo{author}{\bibfnamefont{G.}~\bibnamefont{Xu}}, \bibnamefont{and}
  \bibinfo{author}{\bibfnamefont{K.}~\bibnamefont{Yao}},
  \bibinfo{journal}{Phys. Rev. B} \textbf{\bibinfo{volume}{98}},
  \bibinfo{pages}{201116} (\bibinfo{year}{2018}),
  \urlprefix\url{https://link.aps.org/doi/10.1103/PhysRevB.98.201116}.

\bibitem[{\citenamefont{Kim et~al.}(2018)\citenamefont{Kim, Seo, Lee, Ko, Kim,
  Jang, Ok, Lee, Jo, Kang et~al.}}]{kim2018large}
\bibinfo{author}{\bibfnamefont{K.}~\bibnamefont{Kim}},
  \bibinfo{author}{\bibfnamefont{J.}~\bibnamefont{Seo}},
  \bibinfo{author}{\bibfnamefont{E.}~\bibnamefont{Lee}},
  \bibinfo{author}{\bibfnamefont{K.-T.} \bibnamefont{Ko}},
  \bibinfo{author}{\bibfnamefont{B.}~\bibnamefont{Kim}},
  \bibinfo{author}{\bibfnamefont{B.~G.} \bibnamefont{Jang}},
  \bibinfo{author}{\bibfnamefont{J.~M.} \bibnamefont{Ok}},
  \bibinfo{author}{\bibfnamefont{J.}~\bibnamefont{Lee}},
  \bibinfo{author}{\bibfnamefont{Y.~J.} \bibnamefont{Jo}},
  \bibinfo{author}{\bibfnamefont{W.}~\bibnamefont{Kang}}, \bibnamefont{et~al.},
  \bibinfo{journal}{Nature materials} \textbf{\bibinfo{volume}{17}},
  \bibinfo{pages}{794} (\bibinfo{year}{2018}),
  \urlprefix\url{https://doi.org/10.1038/s41563-018-0132-3}.

\bibitem[{\citenamefont{Nie et~al.}(2019)\citenamefont{Nie, Weng, and
  Prinz}}]{nie2019topological}
\bibinfo{author}{\bibfnamefont{S.}~\bibnamefont{Nie}},
  \bibinfo{author}{\bibfnamefont{H.}~\bibnamefont{Weng}}, \bibnamefont{and}
  \bibinfo{author}{\bibfnamefont{F.~B.} \bibnamefont{Prinz}},
  \bibinfo{journal}{Phys. Rev. B} \textbf{\bibinfo{volume}{99}},
  \bibinfo{pages}{035125} (\bibinfo{year}{2019}),
  \urlprefix\url{https://link.aps.org/doi/10.1103/PhysRevB.99.035125}.

\bibitem[{\citenamefont{Zou et~al.}(2019)\citenamefont{Zou, He, and
  Xu}}]{zou2019study}
\bibinfo{author}{\bibfnamefont{J.}~\bibnamefont{Zou}},
  \bibinfo{author}{\bibfnamefont{Z.}~\bibnamefont{He}}, \bibnamefont{and}
  \bibinfo{author}{\bibfnamefont{G.}~\bibnamefont{Xu}}, \bibinfo{journal}{npj
  Computational Materials} \textbf{\bibinfo{volume}{5}}, \bibinfo{pages}{1}
  (\bibinfo{year}{2019}),
  \urlprefix\url{https://doi.org/10.1038/s41524-019-0237-5}.

\bibitem[{\citenamefont{Altland and Zirnbauer}(1997)}]{altland1997nonstandard}
\bibinfo{author}{\bibfnamefont{A.}~\bibnamefont{Altland}} \bibnamefont{and}
  \bibinfo{author}{\bibfnamefont{M.~R.} \bibnamefont{Zirnbauer}},
  \bibinfo{journal}{Phys. Rev. B} \textbf{\bibinfo{volume}{55}},
  \bibinfo{pages}{1142} (\bibinfo{year}{1997}),
  \urlprefix\url{https://link.aps.org/doi/10.1103/PhysRevB.55.1142}.

\bibitem[{Note2()}]{Note2}
Note2, \bibinfo{note}{due to the absence of TRS in MSGs, the ``mirror Chern
  insulator'' we use here do not require two mirror Chern numbers to be
  opposite with each other, i.e., $C^{+}_m \not =C^{-}_m$ generically, which
  are different with the mirror Chern insulators in non-magnetic SGs with
  $C^{+}_m=-C^{-}_m$\cite {teo2008surface,hsieh2012topological}.}

\bibitem[{Note3()}]{Note3}
Note3, \bibinfo{note}{in the following context we use the notation $C^{\pm
  }_{m}$ for both real-space mirror Chern numbers and momentum-space mirror
  Chern numbers, and the specific meaning of $C^{\pm }_m$ can be inferred from
  the context}.

\bibitem[{\citenamefont{Ahn and Yang}(2019)}]{ahn2019symmetry}
\bibinfo{author}{\bibfnamefont{J.}~\bibnamefont{Ahn}} \bibnamefont{and}
  \bibinfo{author}{\bibfnamefont{B.-J.} \bibnamefont{Yang}},
  \bibinfo{journal}{Phys. Rev. B} \textbf{\bibinfo{volume}{99}},
  \bibinfo{pages}{235125} (\bibinfo{year}{2019}),
  \urlprefix\url{https://link.aps.org/doi/10.1103/PhysRevB.99.235125}.

\bibitem[{\citenamefont{Schindler et~al.}(2018)\citenamefont{Schindler, Cook,
  Vergniory, Wang, Parkin, Bernevig, and Neupert}}]{schindler2018higher}
\bibinfo{author}{\bibfnamefont{F.}~\bibnamefont{Schindler}},
  \bibinfo{author}{\bibfnamefont{A.~M.} \bibnamefont{Cook}},
  \bibinfo{author}{\bibfnamefont{M.~G.} \bibnamefont{Vergniory}},
  \bibinfo{author}{\bibfnamefont{Z.}~\bibnamefont{Wang}},
  \bibinfo{author}{\bibfnamefont{S.~S.~P.} \bibnamefont{Parkin}},
  \bibinfo{author}{\bibfnamefont{B.~A.} \bibnamefont{Bernevig}},
  \bibnamefont{and} \bibinfo{author}{\bibfnamefont{T.}~\bibnamefont{Neupert}},
  \bibinfo{journal}{Science Advances} \textbf{\bibinfo{volume}{4}}
  (\bibinfo{year}{2018}),
  \urlprefix\url{https://advances.sciencemag.org/content/4/6/eaat0346}.

\bibitem[{\citenamefont{Shiozaki et~al.}(2015)\citenamefont{Shiozaki, Sato, and
  Gomi}}]{shiozaki2015z}
\bibinfo{author}{\bibfnamefont{K.}~\bibnamefont{Shiozaki}},
  \bibinfo{author}{\bibfnamefont{M.}~\bibnamefont{Sato}}, \bibnamefont{and}
  \bibinfo{author}{\bibfnamefont{K.}~\bibnamefont{Gomi}},
  \bibinfo{journal}{Phys. Rev. B} \textbf{\bibinfo{volume}{91}},
  \bibinfo{pages}{155120} (\bibinfo{year}{2015}),
  \urlprefix\url{https://link.aps.org/doi/10.1103/PhysRevB.91.155120}.

\bibitem[{\citenamefont{Shiozaki et~al.}(2016)\citenamefont{Shiozaki, Sato, and
  Gomi}}]{shiozaki2016topology}
\bibinfo{author}{\bibfnamefont{K.}~\bibnamefont{Shiozaki}},
  \bibinfo{author}{\bibfnamefont{M.}~\bibnamefont{Sato}}, \bibnamefont{and}
  \bibinfo{author}{\bibfnamefont{K.}~\bibnamefont{Gomi}},
  \bibinfo{journal}{Phys. Rev. B} \textbf{\bibinfo{volume}{93}},
  \bibinfo{pages}{195413} (\bibinfo{year}{2016}),
  \urlprefix\url{https://link.aps.org/doi/10.1103/PhysRevB.93.195413}.

\bibitem[{\citenamefont{Kim et~al.}(2019)\citenamefont{Kim, Shiozaki, and
  Murakami}}]{kim2019glide}
\bibinfo{author}{\bibfnamefont{H.}~\bibnamefont{Kim}},
  \bibinfo{author}{\bibfnamefont{K.}~\bibnamefont{Shiozaki}}, \bibnamefont{and}
  \bibinfo{author}{\bibfnamefont{S.}~\bibnamefont{Murakami}},
  \bibinfo{journal}{Phys. Rev. B} \textbf{\bibinfo{volume}{100}},
  \bibinfo{pages}{165202} (\bibinfo{year}{2019}),
  \urlprefix\url{https://link.aps.org/doi/10.1103/PhysRevB.100.165202}.

\bibitem[{\citenamefont{Kim and Murakami}(2020)}]{kim2020glide}
\bibinfo{author}{\bibfnamefont{H.}~\bibnamefont{Kim}} \bibnamefont{and}
  \bibinfo{author}{\bibfnamefont{S.}~\bibnamefont{Murakami}},
  \bibinfo{journal}{Phys. Rev. B} \textbf{\bibinfo{volume}{102}},
  \bibinfo{pages}{195202} (\bibinfo{year}{2020}),
  \urlprefix\url{https://link.aps.org/doi/10.1103/PhysRevB.102.195202}.

\bibitem[{\citenamefont{Bernevig et~al.}(2006)\citenamefont{Bernevig, Hughes,
  and Zhang}}]{bernevig2006quantum}
\bibinfo{author}{\bibfnamefont{B.~A.} \bibnamefont{Bernevig}},
  \bibinfo{author}{\bibfnamefont{T.~L.} \bibnamefont{Hughes}},
  \bibnamefont{and} \bibinfo{author}{\bibfnamefont{S.-C.} \bibnamefont{Zhang}},
  \bibinfo{journal}{science} \textbf{\bibinfo{volume}{314}},
  \bibinfo{pages}{1757} (\bibinfo{year}{2006}),
  \urlprefix\url{https://science.sciencemag.org/content/314/5806/1757}.

\bibitem[{\citenamefont{Kruthoff et~al.}(2017)\citenamefont{Kruthoff, de~Boer,
  van Wezel, Kane, and Slager}}]{kruthoff2017topological}
\bibinfo{author}{\bibfnamefont{J.}~\bibnamefont{Kruthoff}},
  \bibinfo{author}{\bibfnamefont{J.}~\bibnamefont{de~Boer}},
  \bibinfo{author}{\bibfnamefont{J.}~\bibnamefont{van Wezel}},
  \bibinfo{author}{\bibfnamefont{C.~L.} \bibnamefont{Kane}}, \bibnamefont{and}
  \bibinfo{author}{\bibfnamefont{R.-J.} \bibnamefont{Slager}},
  \bibinfo{journal}{Phys. Rev. X} \textbf{\bibinfo{volume}{7}},
  \bibinfo{pages}{041069} (\bibinfo{year}{2017}),
  \urlprefix\url{https://link.aps.org/doi/10.1103/PhysRevX.7.041069}.

\bibitem[{\citenamefont{Song et~al.}(2018{\natexlab{b}})\citenamefont{Song,
  Zhang, and Fang}}]{song2018diagnosis}
\bibinfo{author}{\bibfnamefont{Z.}~\bibnamefont{Song}},
  \bibinfo{author}{\bibfnamefont{T.}~\bibnamefont{Zhang}}, \bibnamefont{and}
  \bibinfo{author}{\bibfnamefont{C.}~\bibnamefont{Fang}},
  \bibinfo{journal}{Phys. Rev. X} \textbf{\bibinfo{volume}{8}},
  \bibinfo{pages}{031069} (\bibinfo{year}{2018}{\natexlab{b}}),
  \urlprefix\url{https://link.aps.org/doi/10.1103/PhysRevX.8.031069}.

\bibitem[{\citenamefont{Watanabe et~al.}(2018)\citenamefont{Watanabe, Po, and
  Vishwanath}}]{watanabe2018structure}
\bibinfo{author}{\bibfnamefont{H.}~\bibnamefont{Watanabe}},
  \bibinfo{author}{\bibfnamefont{H.~C.} \bibnamefont{Po}}, \bibnamefont{and}
  \bibinfo{author}{\bibfnamefont{A.}~\bibnamefont{Vishwanath}},
  \bibinfo{journal}{Science advances} \textbf{\bibinfo{volume}{4}},
  \bibinfo{pages}{eaat8685} (\bibinfo{year}{2018}),
  \urlprefix\url{https://advances.sciencemag.org/content/4/8/eaat8685}.

\bibitem[{\citenamefont{Bouhon et~al.}(2020)\citenamefont{Bouhon, Lange, and
  Slager}}]{bouhon2020topological}
\bibinfo{author}{\bibfnamefont{A.}~\bibnamefont{Bouhon}},
  \bibinfo{author}{\bibfnamefont{G.~F.} \bibnamefont{Lange}}, \bibnamefont{and}
  \bibinfo{author}{\bibfnamefont{R.-J.} \bibnamefont{Slager}},
  \bibinfo{journal}{arXiv preprint arXiv:2010.10536}  (\bibinfo{year}{2020}),
  \urlprefix\url{https://arxiv.org/abs/2010.10536}.

\bibitem[{\citenamefont{Xu et~al.}(2020)\citenamefont{Xu, Elcoro, Song, Wieder,
  Vergniory, Regnault, Chen, Felser, and Bernevig}}]{xu2020high}
\bibinfo{author}{\bibfnamefont{Y.}~\bibnamefont{Xu}},
  \bibinfo{author}{\bibfnamefont{L.}~\bibnamefont{Elcoro}},
  \bibinfo{author}{\bibfnamefont{Z.-D.} \bibnamefont{Song}},
  \bibinfo{author}{\bibfnamefont{B.~J.} \bibnamefont{Wieder}},
  \bibinfo{author}{\bibfnamefont{M.}~\bibnamefont{Vergniory}},
  \bibinfo{author}{\bibfnamefont{N.}~\bibnamefont{Regnault}},
  \bibinfo{author}{\bibfnamefont{Y.}~\bibnamefont{Chen}},
  \bibinfo{author}{\bibfnamefont{C.}~\bibnamefont{Felser}}, \bibnamefont{and}
  \bibinfo{author}{\bibfnamefont{B.~A.} \bibnamefont{Bernevig}},
  \bibinfo{journal}{Nature} \textbf{\bibinfo{volume}{586}},
  \bibinfo{pages}{702} (\bibinfo{year}{2020}),
  \urlprefix\url{https://www.nature.com/articles/s41586-020-2837-0}.

\bibitem[{\citenamefont{Aroyo et~al.}(2006{\natexlab{a}})\citenamefont{Aroyo,
  Perez-Mato, Capillas, Kroumova, Ivantchev, Madariaga, Kirov, and
  Wondratschek}}]{aroyo2006bilbao1}
\bibinfo{author}{\bibfnamefont{M.~I.} \bibnamefont{Aroyo}},
  \bibinfo{author}{\bibfnamefont{J.~M.} \bibnamefont{Perez-Mato}},
  \bibinfo{author}{\bibfnamefont{C.}~\bibnamefont{Capillas}},
  \bibinfo{author}{\bibfnamefont{E.}~\bibnamefont{Kroumova}},
  \bibinfo{author}{\bibfnamefont{S.}~\bibnamefont{Ivantchev}},
  \bibinfo{author}{\bibfnamefont{G.}~\bibnamefont{Madariaga}},
  \bibinfo{author}{\bibfnamefont{A.}~\bibnamefont{Kirov}}, \bibnamefont{and}
  \bibinfo{author}{\bibfnamefont{H.}~\bibnamefont{Wondratschek}},
  \bibinfo{journal}{Zeitschrift f{\"u}r Kristallographie-Crystalline Materials}
  \textbf{\bibinfo{volume}{221}}, \bibinfo{pages}{15}
  (\bibinfo{year}{2006}{\natexlab{a}}),
  \urlprefix\url{https://www.degruyter.com/document/doi/10.1524/zkri.2006.221.1.15/html}.

\bibitem[{\citenamefont{Aroyo et~al.}(2006{\natexlab{b}})\citenamefont{Aroyo,
  Kirov, Capillas, Perez-Mato, and Wondratschek}}]{aroyo2006bilbao2}
\bibinfo{author}{\bibfnamefont{M.~I.} \bibnamefont{Aroyo}},
  \bibinfo{author}{\bibfnamefont{A.}~\bibnamefont{Kirov}},
  \bibinfo{author}{\bibfnamefont{C.}~\bibnamefont{Capillas}},
  \bibinfo{author}{\bibfnamefont{J.}~\bibnamefont{Perez-Mato}},
  \bibnamefont{and}
  \bibinfo{author}{\bibfnamefont{H.}~\bibnamefont{Wondratschek}},
  \bibinfo{journal}{Acta Crystallographica Section A: Foundations of
  Crystallography} \textbf{\bibinfo{volume}{62}}, \bibinfo{pages}{115}
  (\bibinfo{year}{2006}{\natexlab{b}}),
  \urlprefix\url{http://scripts.iucr.org/cgi-bin/paper?S0108767305040286}.

\bibitem[{\citenamefont{Aroyo et~al.}(2011)\citenamefont{Aroyo, Perez-Mato,
  Orobengoa, Tasci, de~la Flor, and Kirov}}]{aroyo2011crystallography}
\bibinfo{author}{\bibfnamefont{M.~I.} \bibnamefont{Aroyo}},
  \bibinfo{author}{\bibfnamefont{J.}~\bibnamefont{Perez-Mato}},
  \bibinfo{author}{\bibfnamefont{D.}~\bibnamefont{Orobengoa}},
  \bibinfo{author}{\bibfnamefont{E.}~\bibnamefont{Tasci}},
  \bibinfo{author}{\bibfnamefont{G.}~\bibnamefont{de~la Flor}},
  \bibnamefont{and} \bibinfo{author}{\bibfnamefont{A.}~\bibnamefont{Kirov}},
  \bibinfo{journal}{Bulg. Chem. Commun} \textbf{\bibinfo{volume}{43}},
  \bibinfo{pages}{183} (\bibinfo{year}{2011}),
  \urlprefix\url{http://bcc.bas.bg/BCC_Volumes/Volume_43_Number_2_2011/Volume_43_Number_2_2011_PDF/2011_43_2_1.pdf}.

\bibitem[{\citenamefont{Fang et~al.}(2016)\citenamefont{Fang, Lu, Liu, and
  Fu}}]{fang2016topological}
\bibinfo{author}{\bibfnamefont{C.}~\bibnamefont{Fang}},
  \bibinfo{author}{\bibfnamefont{L.}~\bibnamefont{Lu}},
  \bibinfo{author}{\bibfnamefont{J.}~\bibnamefont{Liu}}, \bibnamefont{and}
  \bibinfo{author}{\bibfnamefont{L.}~\bibnamefont{Fu}},
  \bibinfo{journal}{Nature Physics} \textbf{\bibinfo{volume}{12}},
  \bibinfo{pages}{936} (\bibinfo{year}{2016}),
  \urlprefix\url{https://www.nature.com/articles/nphys3782}.

\bibitem[{\citenamefont{Qi et~al.}(2006)\citenamefont{Qi, Wu, and
  Zhang}}]{qi2006topological}
\bibinfo{author}{\bibfnamefont{X.-L.} \bibnamefont{Qi}},
  \bibinfo{author}{\bibfnamefont{Y.-S.} \bibnamefont{Wu}}, \bibnamefont{and}
  \bibinfo{author}{\bibfnamefont{S.-C.} \bibnamefont{Zhang}},
  \bibinfo{journal}{Physical Review B} \textbf{\bibinfo{volume}{74}},
  \bibinfo{pages}{085308} (\bibinfo{year}{2006}),
  \urlprefix\url{https://journals.aps.org/prb/abstract/10.1103/PhysRevB.74.085308}.

\bibitem[{\citenamefont{Cano et~al.}(2018)\citenamefont{Cano, Bradlyn, Wang,
  Elcoro, Vergniory, Felser, Aroyo, and Bernevig}}]{cano2018building}
\bibinfo{author}{\bibfnamefont{J.}~\bibnamefont{Cano}},
  \bibinfo{author}{\bibfnamefont{B.}~\bibnamefont{Bradlyn}},
  \bibinfo{author}{\bibfnamefont{Z.}~\bibnamefont{Wang}},
  \bibinfo{author}{\bibfnamefont{L.}~\bibnamefont{Elcoro}},
  \bibinfo{author}{\bibfnamefont{M.~G.} \bibnamefont{Vergniory}},
  \bibinfo{author}{\bibfnamefont{C.}~\bibnamefont{Felser}},
  \bibinfo{author}{\bibfnamefont{M.~I.} \bibnamefont{Aroyo}}, \bibnamefont{and}
  \bibinfo{author}{\bibfnamefont{B.~A.} \bibnamefont{Bernevig}},
  \bibinfo{journal}{Phys. Rev. B} \textbf{\bibinfo{volume}{97}},
  \bibinfo{pages}{035139} (\bibinfo{year}{2018}),
  \urlprefix\url{https://link.aps.org/doi/10.1103/PhysRevB.97.035139}.

\end{thebibliography}

\clearpage
\tableofcontents
\appendix
\appendixpage
\addappheadtotoc

\section{Topological crystals, construction and classification}\label{AppendixA}

In this section, we discuss in detail the complete procedure of constructing topological crystals (TCs) protected by magnetic space group (MSG) symmetries using our real-space recipe, including cell-complex, building blocks, no-open-edge condition, and bubble equivalence. 

\subsection{Cell Complex}
To construct a TC, one first needs to find the geometric structure called cell-complex, which is obtained by using the elements of crystalline symmetry group to cut the 3D space into small regions and lower-dimensional pieces, i.e., 3, 2, 1, and 0-dimensional cells:
\begin{itemize}
	\item 3D regions are called 3-cells. A single 3-cell is also called an asymmetric unit (AU), which fills the whole space without overlap after copied by all symmetry group operations;
	\item 2D planes where two 3-cells meet are called 2-cells;
	\item 1D lines where 2-cells intersect are called 1-cells;
	\item 0D end-points of 1-cells are called 0-cells.
\end{itemize}

One thing worth noting is that, for any d-cell with $d = 0,1,2,3 $, two different points inside it cannot be related by any crystalline symmetries, which means a single d-cell owns no crystalline symmetries itself. However, some symmetries can keep every point of the d-cell invariant, forming its on-site symmetry group.
The property that a single d-cell having no crystalline symmetries is significant, as one can only place topological states protected by onsite symmetries on a d-cell, whose classifications are already known from the ten-fold way results\cite{ryu2010topological}.  

The cell complex for MSGs is similar to that for space groups (SGs) discussed comprehensively in Ref.\cite{song2019topological}, with the only difference being that MSGs can have anti-unitary symmetries, including magnetic half translations, which makes the cell complex more complicated.

\begin{figure}[ht]
	\centering
	\includegraphics[width=0.2\textwidth]{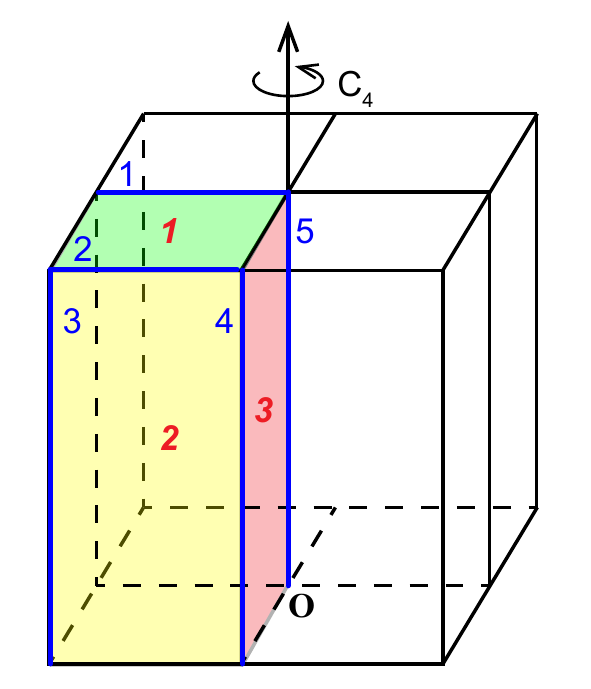}
	\caption{\label{cell_complex_P4}The cell complex of $P4$, where the red number 1-3 mark three 2-cells and blue number 1-5 mark five 1-cells.}
\end{figure}

We show in Fig.\ref{cell_complex_P4} the cell-complex of type-1 MSG 75.1 $P4$. The unit cell is divided into four cuboids by $C_4$ symmetry. The AU, or the symmetry-independent 3-cell, could be chosen as any one of the four cuboids. We choose the 3 colored faces as symmetry-independent 2-cells and the 5 blue lines as symmetry-independent 1-cells.

\subsection{2D building Blocks}
Next, we consider what topological states can be decorated on the cell complex. 
To begin with, we have to decide the effective symmetry classes, i.e., Altland-Zirnbauer classes\cite{altland1997nonstandard}, of d-cells in the cell-complex, which are determined by their on-site symmetry groups.
We notice that for all d-cells with $d=0,1,2,3$ protected by MSG symmetries, their effective symmetry classes are either class A or class A\uppercase\expandafter{\romannumeral1}.
Although some d-cells can have onsite unitary symmetry, denoted by $g$, the states on them can be diagonalized according to the eigenvalues of $g$ into different sectors, with each sector belonging to class A.
Some d-cells belong to class A\uppercase\expandafter{\romannumeral1} since they have onsite anti-unitary symmetries with square $+1$, e.g., the combination of mirror and time-reversal symmetry (TRS), $ M \cdot T $.
Thanks to the well-know result of ten-fold way classifications, the classification of class A is trivial when $d=0,1,3$ and $\mathbb{Z}$ when $d=2$, while class A\uppercase\expandafter{\romannumeral1} have trivial classifications for $d\le 3$.
Therefore, we only need to consider nontrivial decorations on 2-cells, which can be further divided into two cases:
\begin{enumerate}
	\item When the 2-cell does not coincide with any mirror planes, the effective symmetry class is just class A with $\mathbb{Z}$ classification in 2D and the topological state to be decorated on it is nothing but the Chern insulator.
	\item When the 2-cell coincides with a mirror plane, we can divide the states into two mirror $\pm i$ sectors, with each sector's effective symmetry class being A and having $\mathbb{Z}$ classification, which corresponds to the mirror Chern number\cite{hsieh2012topological,teo2008surface}. 
	In this case, the topological state decorated on it is the mirror Chern insulator. 
	Note that here TRS is absent, so the two mirror Chern numbers $C_m^+$ and $C_m^-$ are not necessarily opposite, and the total Chern number $C_m^{+} + C_m^{-}$ are generically nonzero. 
\end{enumerate}

In summary, there are two kinds of 2D building blocks for constructing TCs protected by MSG symmetries: (i) Chern insulators and (ii) mirror Chern insulators.

\subsection{No-open-edge condition}
Given a certain MSG, finding all independent magnetic TCs is equivalent to enumerating all the topologically distinct ways of decorating the cell complex. 
As topological states we construct are fully gapped in the bulk, all the boundary states contributed by the 2-cells must cancel with each other on the 1-cells they meet, a condition known as the 
``non-open-edge condition''.

We discuss the no-open-edge condition for the two types of building blocks separately:
\begin{enumerate}
	\item For non-mirror 2-cells, the no-open-edge condition requires an equal number of opposite directional 1D edge modes such that they cancel with each other and result in a fully gapped state on the 1-cell. 
	
	Below we show an example of a 2D non-mirror plane sliced by $C_2$ rotation into two 2-cells. As shown in Fig.\ref{no_open_edge}(a), when the 2-cells are parallel to the $C_2$ axis, the $C_2$-related edge modes run in the same direction and cannot be canceled. However, when the 2-cells are vertical to the $C_2$ axis shown in Fig.\ref{no_open_edge}(b), the edge modes on the intersecting 1-cell have opposite directions and can be canceled. The no-open-edge condition fails for the first case while holds for the second case.
    \begin{figure}[ht]
	    \centering
	    \includegraphics[width=0.4\textwidth]{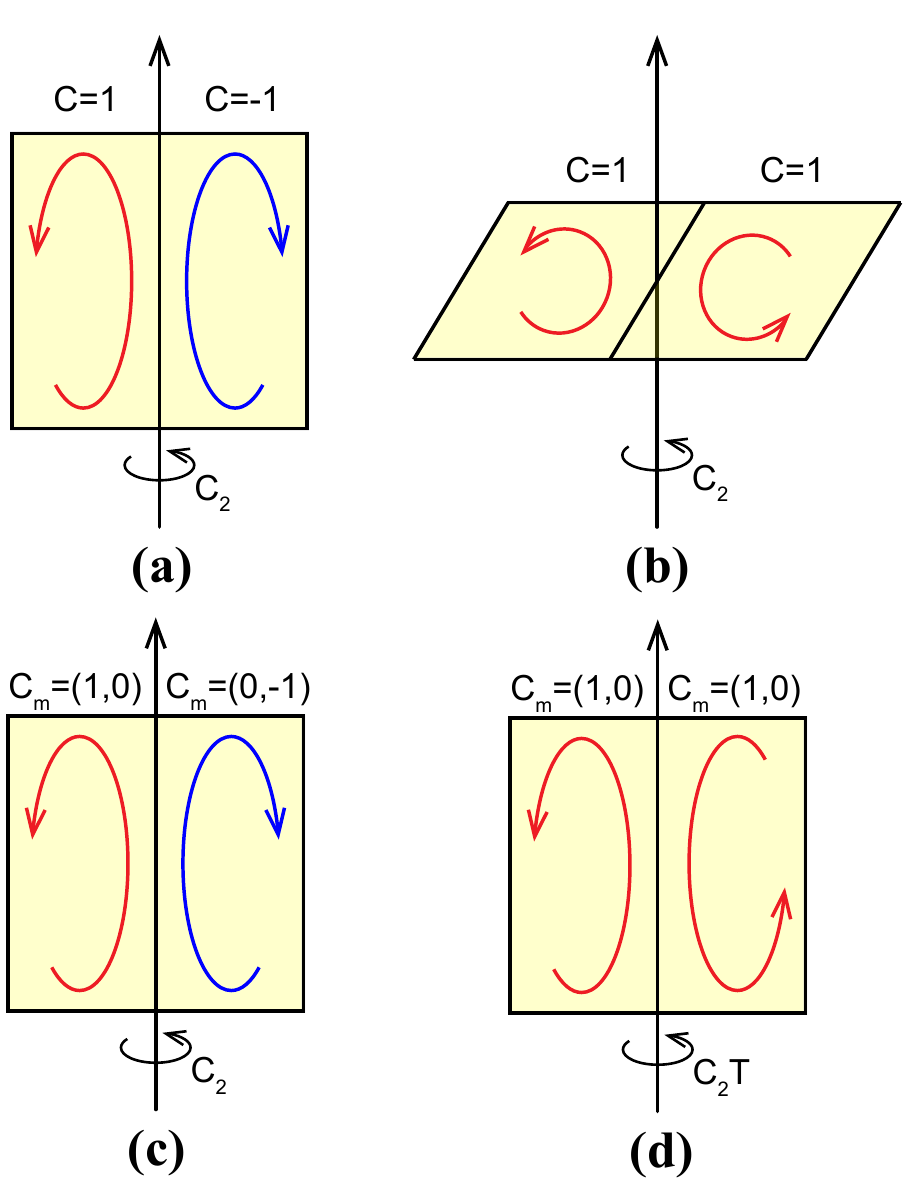}
	    \caption{\label{no_open_edge}No-open-edge conditions. (a). Two $C_2$-related 2-cells that pass the $C_2$ axis fail the no-open-edge condition. (b). Two $C_2$-related 2-cells vertical to the $C_2$ axis satisfy the no-open-edge condition. (c). A mirror plane decorated with mirror Chern insulators that passes the $C_2$ axis fails the no-open-edge condition. (d) A mirror plane decorated with mirror Chern insulators that passes the $C_2\cdot T$ axis satisfies the no-open-edge condition.}
    \end{figure}

	\item For mirror 2-cells, the no-open-edge condition is more complicated because not only the chiralities but also the mirror eigenvalues of the edge modes need to be considered, i.e., the total chiralities of edge modes should be zero for both mirror $\pm i$ sectors. In another world, two edge modes can be gapped out if and only if they have the same mirror eigenvalue and opposite chiralities. 
	
	For example, see a mirror plane passes the $C_2$ or $C_2 \cdot T$ axis, as shown in Fig.\ref{no_open_edge}(c)(d).
	For the $ C_2 $ case, note $C_2$ preserves the direction of edge mode and inverts the mirror sector, as the $C_2$ and mirror axes are perpendicular which leads to $C_2M=-MC_2$. As a result, the two edge states on the 1-cell have opposite mirror eigenvalues and the same chirality, which can not be gapped out.
	
	For the $ C_2 \cdot T $ case, note that $T$ also inverts mirror sector as $ [M,T] = 0 $ and $ T( \pm i ) = \mp i $. The combination of $C_2$ and $T$ keeps the mirror sector unchanged. $T$ also inverts the chirality. Thus the two edge modes on the 1-cell have the same mirror eigenvalue but opposite chiralities, which can be gapped out.

\end{enumerate}             

Practically, we define a ``boundary map'' which maps symmetry-independent 2-cells to symmetry-independent 1-cells.
This boundary map decides the way how boundary states are contributed on each 1-cell by all 2-cells attached to it, 
written in a matrix form of dimension $N_{1\text{-cells}} \times N_{2\text{-cells}}$. 
By calculating the kernel space of this matrix, one obtains the generators of decorations that satisfy the no-open-edge condition.

We use the example of MSG 75.1 $P4$ to illustrate this procedure. 
As shown in Fig.\ref{cell_complex_P4}, the cell complex of $P4$ contains 3 independent 2-cells labeled as $ c_1, c_2, c_3$ and five independent 1-cells labeled as $ b_1, b_2, b_3, b_4, b_5 $. The matrix of boundary map is calculated as
\begin{align}
\begin{pmatrix} 
0 & 0 & 0 \\
0 & 0 & 0 \\
0 & 4 & 0 \\
0 & 2 & -2 \\
0 & 0 & 4 
\end{pmatrix}
\end{align}
where the $i$-th rows represent 1-cell $b_i$ and the $j$-th column represent 2-cell $c_j$.
The $(i,j)$-th entry tells us how many edge modes will be contributed on the $i$-th 1-cell if one puts a unit decoration, i.e., a Chern insulator with Chern number $C=1$, on the $j$-th 2-cell. 
The kernel space of this matrix is spanned by only one vector, $ (1,0,0) $, which means there is only one generator of decorations satisfying the no-open-edge condition. The vector $(1,0,0)$ means only the 2-cell $c_1$ can be decorated with Chern insulators, and after copying $c_1$ using $C_4$ and translation symmetries, this decoration turns out to be a 3D quantum anomalous Hall insulator (QAHI), i.e., Chern insulators stacked along $z$ direction.

\subsection{Bubble equivalence}
\label{section bubble equivalence}
By solving the no-open-edge condition for MSGs, one gets the generators of TCs forming a $\mathbb{Z}^n$ group. 
However, the states generated by superimposing these generators are not all necessarily nontrivial, and one needs to consider the so-called ``bubble equivalence'' process to remove the trivial states and arrive at the finial TCI classification. Sometimes an even number of copies of a generator could be trivialized, resulting in a $\mathbb{Z}_2$ classification.

Specifically, a ``bubble'' represents a topological trivial state in 3D space which has a 2D surface with non-vanishing Chern number, as shown in Fig.\ref{bubble_eq}(a). 
It can be created at a generic point inside an AU and grow larger until it coincides with the AU. Despite its surface with nonzero Chern number, a bubble itself is intrinsically trivial since it can smoothly shrink back into a point without breaking any symmetries. 

When a bubble is created in an AU, there will be symmetry-related bubbles created simultaneously inside all the symmetry-related 3-cells, and the relative signs of surface Chern numbers of these bubbles are also determined by symmetries. 
On the 2-cell where two bubbles meet, the Chern number can be changed by either $\pm2$ or $0$, which can be decided by considering how Berry curvature transforms under the symmetry operation relating these two bubbles. 

Similar to the discussion of the no-open-edge condition, we discuss bubble equivalence for non-mirror 2-cells and mirror 2-cells separately.
\begin{enumerate}
\item For non-mirror 2-cells, we only have to consider the change of Chern number on it by bubble equivalence, which could be either $\pm 2$ or $0$.
When the change of Chern number is $\pm 2$, the TCI classification is affected by bubble equivalence. 

For example, we consider the type-4 MSG 1.3, $P_S 1$, which has an anti-unitary half translation in the $z$ direction aside from three unitary integer translations. Note that the layer generated by anti-unitary half translation has the opposite Chern number from the original layer. 
As shown in Fig.\ref{bubble_eq}(b), the decoration with even Chern numbers on each layer can be trivialized by bubble equivalence, which leads to the $\mathbb{Z}_2$ TCI classification of MSG 1.3, whose generator can be chosen as a layer construction with $C=1$ on integer planes and $C=-1$ on half-integer planes. 

\begin{figure*}
	\centering
	\includegraphics[width=1\textwidth]{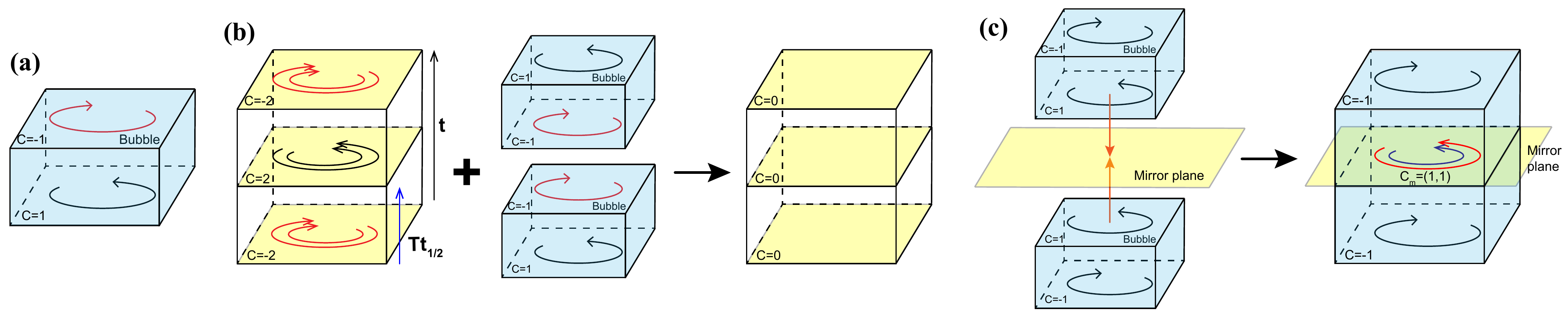}
	\caption{\label{bubble_eq}Bubble equivalence. (a). A simple bubble with surfaces Chern number $C=\pm1$, where we only plot the upper and lower surfaces, and the side surfaces are in fact all have $\pm1$ Chern numbers. (b). A bubble equivalence process that transforms a $C=\pm2$ decoration into a trivial decoration. (c). When two mirror-related bubbles coincide with a mirror plane, they change the mirror Chern number by $(C_m^+,C_m^-)=\pm(1,1)$.}
\end{figure*}

\item For mirror 2-cells, one has to take into account the mirror eigenvalues aside from Chern numbers. 
We show that a minimal bubble equivalence contribute mirror Chern numbers $( C_m^{+}, C_m^{-}) = \pm (1,1) $ on the 2-cell where they meet.
This property is significant and will be used frequently in the following context. 

Without loss of generality, we set the mirror in $z$ direction.  
Note the Berry curvature defined on x-y plane, $(\nabla \times \vec{A})_z=\partial_x A_y-\partial_y A_x$, is a pseudo vector which is unchanged under $M_z$. 
Thus two mirror-symmetric bubbles always modify the total Chern number on the mirror plane by an even number, i.e., the $M_z$-related surfaces of two bubbles have the same Chern number.

Next, we decide the mirror eigenvalues of two $M_z$-related bubbles. Denoting the states on the two bubbles as $\psi_1$ and $\psi_2$. 
They transform to each other under $M_z$ as: $ \psi_1 \rightarrow \zeta \psi_2 $, $ \psi_2 \rightarrow \zeta' \psi_1 $, where $\zeta \cdot \zeta' = -1$ since $M_z^2 =-1$. Thus the matrix representation of $M_z$
is $\begin{pmatrix} 0 & \zeta \\ \zeta' & 0 \end{pmatrix} $, 
which has two eigenvalues $\pm i$.  
One can construct two linear combinations of $ \psi_{1}$ and $\psi_2$,
i.e., $ \frac{1}{\sqrt{1+|\zeta|^2}}(  i\psi_1 + \zeta \psi_2) $ and 
 $ \frac{1}{\sqrt{1+|\zeta'|^2}}(  \zeta' \psi_1 - i \psi_2) $,
as the two eigenstates with mirror eigenvalue $+i$ and $-i$, respectively.
As a result, a minimal bubble equivalence changes the mirror Chern numbers by $( C_m^{+}, C_m^{-}) = \pm (1,1) $, as shown in Fig.\ref{bubble_eq}(c).

\end{enumerate}

\subsection{Decorations, first encounter}
According to previous discussions, there are two types of building blocks, i.e., Chern insulators and mirror Chern insulators, according to their coincidence with mirror planes or not. 
Nevertheless, we can also classify the building blocks in another way for the convenience of the further procedure, that is, whether the net Chern numbers they host are zero or not. 
The decorations with zero net Chern number are called M-building blocks, and if nonzero, C-building blocks.

M-building blocks are only decorated on mirror 2-cells and are actually mirror Chern insulators with zero net Chern numbers, i.e., $  C^{+}_m + C^{-}_m = 0 $. By contrast, C-building blocks can be decorated on both mirror and non-mirror 2-cells, in both cases with nonzero net Chern numbers. In another word, C-building blocks are either Chern insulators or mirror Chern insulators with $ C^{+}_m \neq - C^{-}_m $.

\paragraph{Decorations by M-building blocks}
In practice, for MSGs with mirror symmetries, one does not have to deal with no-open-edge condition for two mirror sectors directly, which could be technically complicated as some symmetry operations can invert the direction of a mirror plane thus convert mirror $+i$ sector to mirror $-i$ sector or vise versa.

As discussed in Ref.\cite{song2019topological}, when decorated with M-building blocks, all 2-cells on the mirror plane can always be ``glued'' together and extend to the whole mirror plane, which means the non-open-edge condition does not need to be considered explicitly, as well as the bubble equivalence which only effects decorations by C-building blocks.
Instead, the number of independent decorations is obtained by enumerating all symmetry-independent mirror planes.
This step, i.e., decorating M-building blocks, results in a $\mathbb{Z}^m$ group as part of the final classification, where $m$ is the number of independent mirror planes.

In a word, decorations by M-building blocks can be seen as the whole collection of decorations with zero net Chern numbers on all 2-cells. 

\paragraph{Decorations by C-building blocks}
To arrive at the final classification, the next step is to find the complement of decorations by M-building blocks, i.e., all the decorations with (at least some) 2-cells hosting nonzero net Chern numbers, which are exactly the decorations by C-building blocks. 

In this step, even for mirror 2-cells, we only have to assign one Chern number, i.e., the total Chern number.
The reason is as following.
Note we already have the decorations by M-building blocks, which can be freely added or subtracted now as they can be regarded as trivial elements in the space of decorations by C-building blocks. 
More specifically, if we assign two mirror Chern numbers, $(C_m^{+}, C_m^{-})$, on a mirror 2-cell, we can add an M-building block with $ ( C_m^{-}, -C_m^{-} ) $ to it, resulting in a state of $ ( C_m^{+} + C_m^{-}, 0 )$, which can be represented by its total Chern number.
With this simplification, we effectively reduce the degree of freedom on each mirror 2-cell from 2 to 1 and only have to deal with the boundary maps together with the bubble equivalence involving only one Chern number (the total Chern number) for each 2-cell. 

In practice, the result of this step can be calculated by the quotient space Ker/Img, where Ker stands for the kernel space of no-open-edge condition and Img stands for the space defined by bubble equivalence. After this step, we get a set of generators like $\mathbb{Z}^l\times\mathbb{Z}_2^n$ by decorating C-building blocks, where $n = 0, 1$, which means there is at most one $\mathbb{Z}_2$ generator and will be explained later.

\subsection{Final TCI classification}

Finally, we combine the results obtained by decorating M-building blocks and C-building blocks into $\mathbb{Z}^m \times \mathbb{Z}^l \times \mathbb{Z}_2^n$. 
However, this could still not be the final classification, because when there are mirror planes, the $\mathbb{Z}_2$ generator obtained in the second step can be absorbed into $\mathbb{Z}^m$ and results in $\mathbb{Z}^m \times \mathbb{Z}^l$. This is a simple case of group extension that, after doubling and a process of bubble equivalence, the $\mathbb{Z}_2$ generator becomes a decoration by M-building blocks with each mirror plane having $ C^{+}_m = - C^{-}_m $, i.e., an element in $Z^m$, which will be explained more specifically later. 

Among our final results , we remark two points about the TCI classifications in type-4 MSGs are worth mentioning:
\begin{itemize}
	\item  when the MSG has no mirror symmetry, the classification is always $\mathbb{Z}_2$; 
	\item when the MSG has mirror symmetries, the classification is always $\mathbb{Z}^m$, since the $\mathbb{Z}_2$ generator is absorbed into mirror decorations, i.e., decorations by M-building blocks.
\end{itemize}
This means that the $\mathbb{Z}_2$ generator always exists for type-4 MSGs, i.e., no type-4 MSG has trivial classification. 
The above two conclusions, although not proved rigorously, could be understood from the following arguments:
\begin{itemize}
	\item $\mathbb{Z}^l$ generators (which correspond to translation $\mathbb{Z}$ decorations, as discussed below) are prohibited by anti-unitary translations in type-4 MSGs;
	\item The simplest type-4 MSG $P_S 1$, which has anti-unitary half-integer translation in one direction and unitary integer translations in other two directions, has a $\mathbb{Z}_2$ generator of TCI classification (corresponding to a $Z_2$ decoration, as discussed below) which has a nonzero invariant of the anti-unitary half translation. For an arbitrary type-4 MSG, its $\mathbb{Z}_2$ generator can be seen as a complication of the $\mathbb{Z}_2$ generator in $P_S1$, as other crystalline symmetries will add more decorated 2-cells, but cannot trivialize the original $Z_2$ decoration in $P_S1$.
\end{itemize}

\clearpage
\section{Topological invariants}\label{AppendixB}
Topological invariants quantitatively distinguish TCIs from trivial insulators.
Formally, if a state can not be adiabatically deformed into a trivial state preserving symmetry $g$, it has a nontrivial topological invariant protected by $g$. 

To find all MSG symmetries that can independently protect topological invariants, we consider when a single generator of MSG symmetry operation $g$ is present, what TC can be constructed. 
More specifically, we first construct the cell complex with respect to symmetry $g$, and then decorate possible 2D building blocks on 2-cells under the no-open-edge condition. Lastly, we check the effect of bubble equivalence on the decorations.
We say $g$ has (i) trivial invariant if no TC can be constructed, (ii) $\mathbb{Z}$ invariant if there exists a non-trivial TC and can not be trivialized by copying arbitrary times, or (iii) $\mathbb{Z}_n$ invariant if there exists a non-trivial TC and can be trivialized by copying $n$ times.

\begin{table}[ht]
	\centering
	\begin{tabular}{cc}
\hline\hline	
\textbf{MSG Symmetries} & \textbf{Invariant type}
\\
\hline
unitary rotation $ C_n $& \multirow{3}{*}{Trivial} \\
\makecell[c]{anti-unitary improper point group \\
symmetries $P\cdot T$, $M\cdot T$, $S_n\cdot T$} & 
\\
\hline  
unitary improper $ S_n$, $P$, $\{M|\frac{1}{2}\}$ & \multirow{3}{*}{$\mathbb{Z}_2$} \\
anti-unitary translation $ \{ E|t\}\cdot T$ &  \\
anti-unitary proper $C_n\cdot T$, $\{C_n|t\}\cdot T$ & \\
\hline
unitary translation $\{ E | R\}$ & \multirow{4}{*}{$\mathbb{Z}$} \\
unitary mirror $M$ & \\
unitary screw $ \{ C_n | t\} $ &  \\
anti-unitary glide $ \{ M | \frac{1}{2} \}\cdot T$ & \\
\hline\hline			
	\end{tabular}
\caption{\label{table_invariants_appendix} MSG symmetries and their corresponding invariant types. The invariants of the unitary screw and anti-unitary glide are bound to the translation invariant.}
\end{table}

\begin{figure*}
	\centering
	\includegraphics[width=1\textwidth]{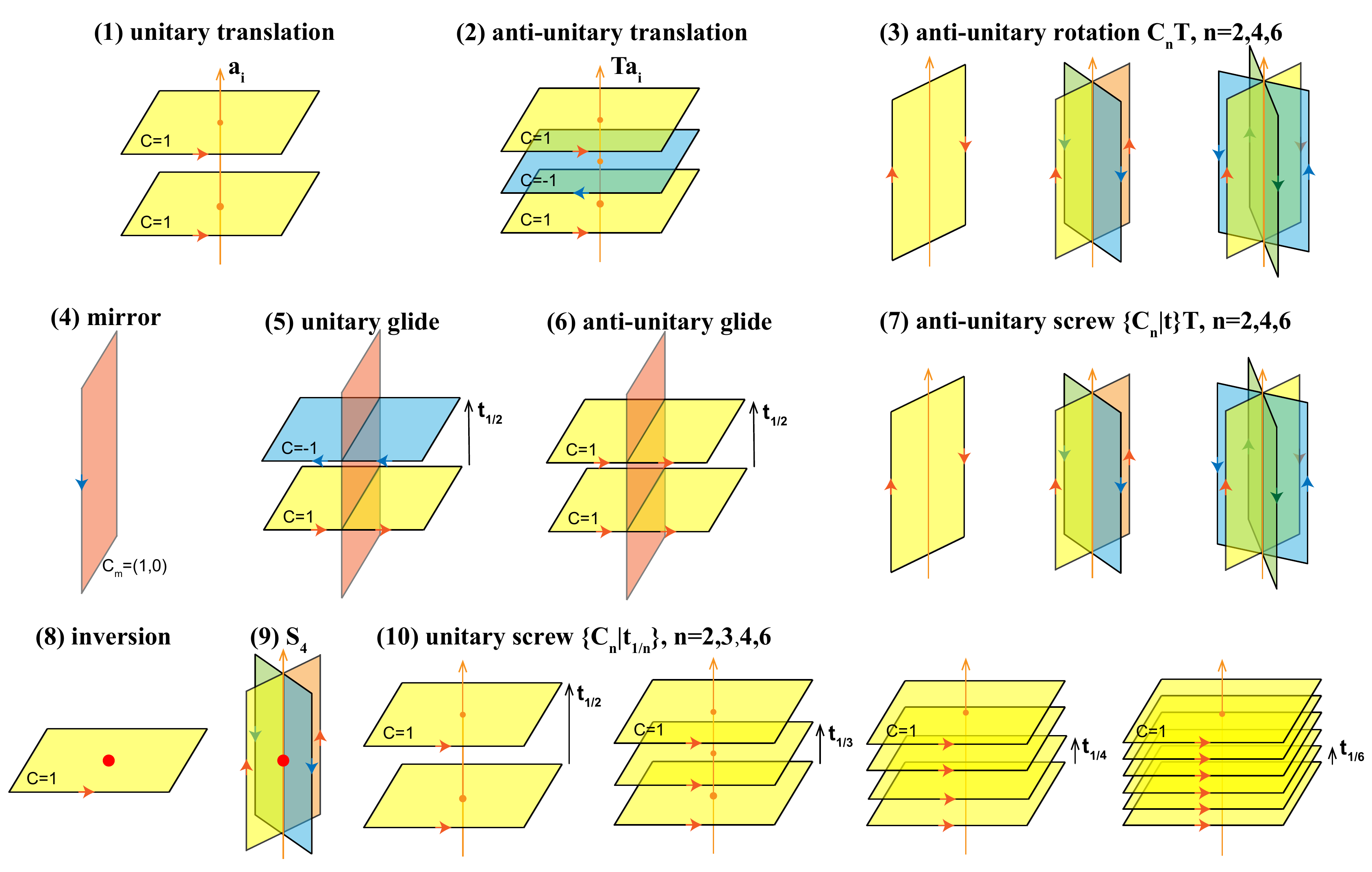}
	\caption{\label{inv_mini_deco}The minimal decoration of the corresponding topological invariant.}
\end{figure*}

We enumerate all topological invariants protected by MSG symmetries in Table.\ref{table_invariants_appendix}, which are either $\mathbb{Z}$-type or $\mathbb{Z}_2$-type, and their corresponding minimal nontrivial decoration, i.e., with invariant $\delta(g)=1$, in Fig.\ref{inv_mini_deco}. 
\begin{enumerate}
    \item Unitary integer translation. 
    Its corresponding invariant is nothing but the weak invariant $\delta_{w,i}$. A minimal decoration with $\delta_{w,i}=1$ is a simple LC by placing $C=1$ layers on integer planes in the $\bm{a}_i$ direction. After doubling $n$ times, this LC is still nontrivial with layers having $C=n$, which confirms the $\mathbb{Z}$-type of weak invariants. 
    
    \item Anti-unitary translation. A minimal decoration with $\delta(\{E|\frac{1}{2}\bm{a}_i\}\cdot T)=1$ is a LC with $C=1$ layers on integer planes and $C=-1$ layers on half-integer planes in the $\bm{a}_i$ direction. Its doubled state can be trivialized by the bubble equivalence, confirming its $\mathbb{Z}_2$-type.
    
    \item Anti-unitary rotations. As shown in Fig.\ref{inv_mini_deco}(3), for $C_2\cdot T$ and $C_6\cdot T$, their minimal decorations have 1 and 3 layers passing the rotation axis, respectively, while for $C_4\cdot T$, there are 4 half-layers that intersect at the rotation axis, forming a non-LC. Their doubled states can all be trivialized, validating their $\mathbb{Z}_2$-type.
    
    \item Mirror. With only mirror symmetry, the 3D space is separated by a single mirror plane (the only 2-cell) into two semi-infinite regions (the two 3-cells), and we need to consider decorations on the mirror plane and the effect of 3D bubbles. Without TRS, the two real-space mirror Chern numbers $( C^{+}_m, C^{-}_m )$ on the mirror plane can take independent values, and they seem to serve as two generators of decorations, i.e., $( C^{+}_m, C^{-}_m )=(1,0)$ and $(0,1)$, forming a $\mathbb{Z}\times \mathbb{Z}$ group. However, as discussed in 
    Sec.\ref{section bubble equivalence}, a decoration with $(C^{+}_m,C^{-}_m)=\pm(1,1)$ can be trivialized by the bubble equivalence. Therefore, there is only one independent generator, which can be chosen as either $(1,0)$ or $(0,1)$, forming a $\mathbb{Z}\times\mathbb{Z}/\{z_1 = z_2 \} = \mathbb{Z}$ group. We plot a minimal decoration of mirror by placing a $(C_m^+,C_m^-)=(1,0)$ state on the mirror plane, as shown in Fig.\ref{inv_mini_deco}(4). 
    
    We can recombine the two mirror Chern numbers into $ C_{\text{diff}} =  C^{+}_m - C^{-}_m $ and $ C_{\text{total}} =  C^{+}_m + C^{-}_m $. Note the bubble equivalence has no effect on $C_{\text{diff}}$, but reduces $C_{\text{total}}$ from $\mathbb{Z}$-valued to $\mathbb{Z}_2$-valued, so the two redefined mirror Chern numbers seem to form a $\mathbb{Z}\times\mathbb{Z}_2$ group. However, the two generators of $\mathbb{Z}$ and $\mathbb{Z}_2$ are not independent, because the $\mathbb{Z}_2$ generator can also be the generator of $\mathbb{Z}$, thus can be absorbed into $\mathbb{Z}$, leading to a $\mathbb{Z}$ group as the final result.

    \item Unitary glide. A minimal decoration with $\delta(\{M|\frac{1}{2}\bm{a}_i\})=1$ is shown in Fig.\ref{inv_mini_deco}(5), which has two layers perpendicular to the glide plane with $C=\pm1$ and connected by the glide half translation in a unit cell. As $(\{M|\frac{1}{2}\bm{a}_i\})^2=\{E|\bm{a}_i\}$, where $\bm{a}_i$ is a lattice vector inside the glide plane, the existence of this non-symmorphic glide symmetry indicates the presence of translation symmetry $\bm{a}_i$. This decoration is similar to the anti-unitary translation decoration, which confirms the $\mathbb{Z}_2$-type of the glide invariant. We remark that this LC of horizontal layers can be deformed into another LC of vertical layers preserving the glide symmetry.
    
    \begin{figure}[ht]
	\centering
	\includegraphics[width=0.4\textwidth]{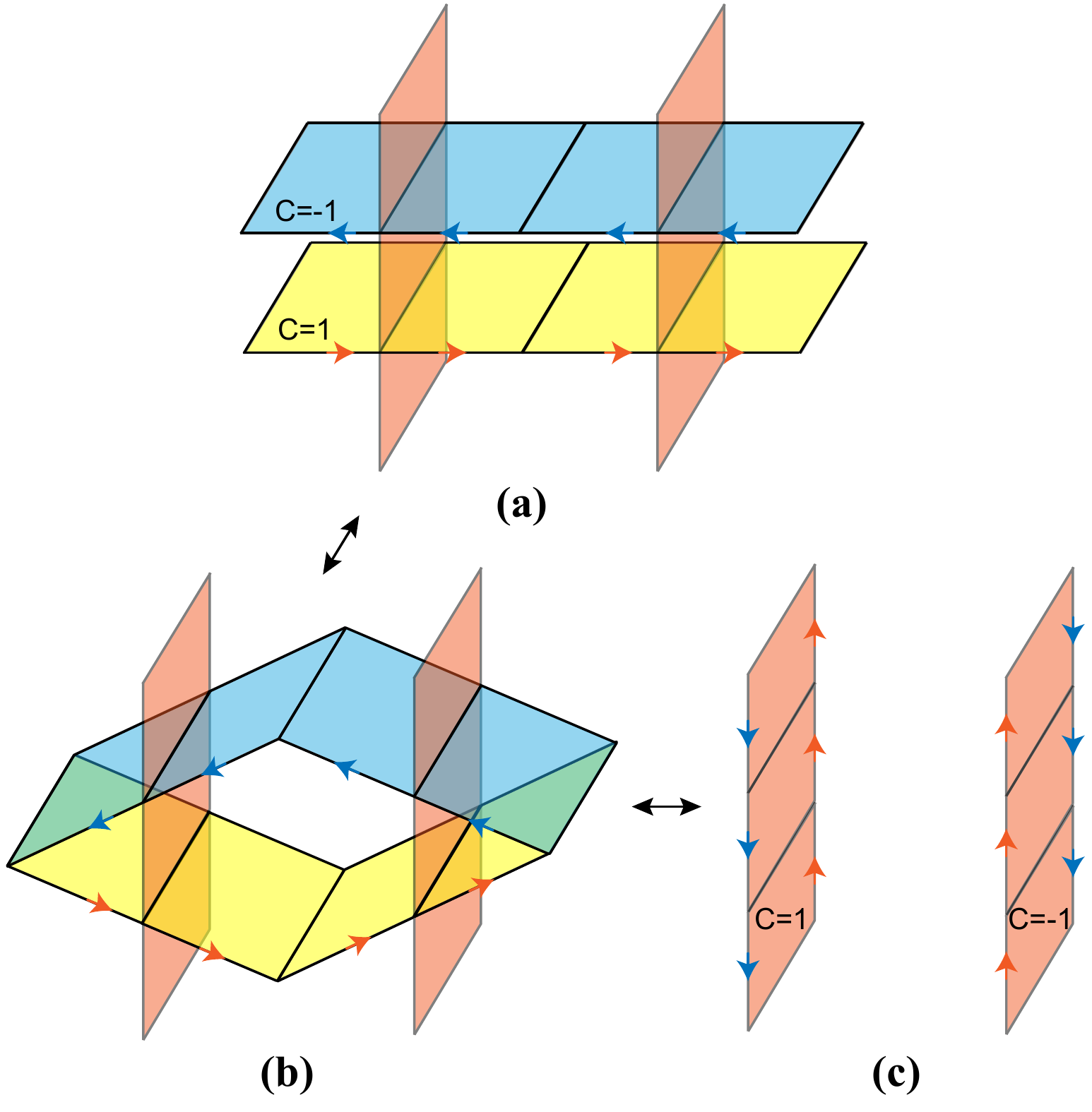}
	\caption{\label{glide_deco}(a)-(c). The process of deforming horizontal layers into vertical layers preserving the glide symmetry.}
    \end{figure}
    
    \item Anti-unitary glide. Its minimal decoration has a similar configuration with the unitary glide, but the Chern number of the two layers in the unit cell are both $C=1$. As $(\{M|\frac{1}{2}\bm{a}_i\}\cdot T)^2=\{E|\bm{a}_i\}$, the existence of the anti-unitary glide also implies the translation symmetry, with their invariants bound together and both being $\mathbb{Z}$-type, i.e., $\delta(\{M|\frac{1}{2}\bm{a}_i\}\cdot T)=\frac{1}{2}\delta_{w,i}$, which enforces $\delta_{w,i}$ to take even numbers. For this minimal decoration, we have $\delta_{w,i}=2$ and $\delta(\{M|\frac{1}{2}\bm{a}_i\}\cdot T)=1$.
    
    \item Anti-unitary screws. The minimal decorations of $ C_n \cdot T$ screws are similar to those of $C_n\cdot T$ rotations. Moreover, these decorations can be transformed into layers vertical to the screw axis with alternating Chern numbers preserving the anti-unitary screw symmetry, as shown in Fig.\ref{au_screw_deco}. We plot the transformation process for $\{C_2|\frac{1}{2}\bm{a}_3\}\cdot T$ in Fig.\ref{au_screw_deco}(4). These invariants are also $\mathbb{Z}_2$-type, as their doubled decorations can be trivialized.
    
    \begin{figure}[ht]
	\centering
	\includegraphics[width=0.48\textwidth]{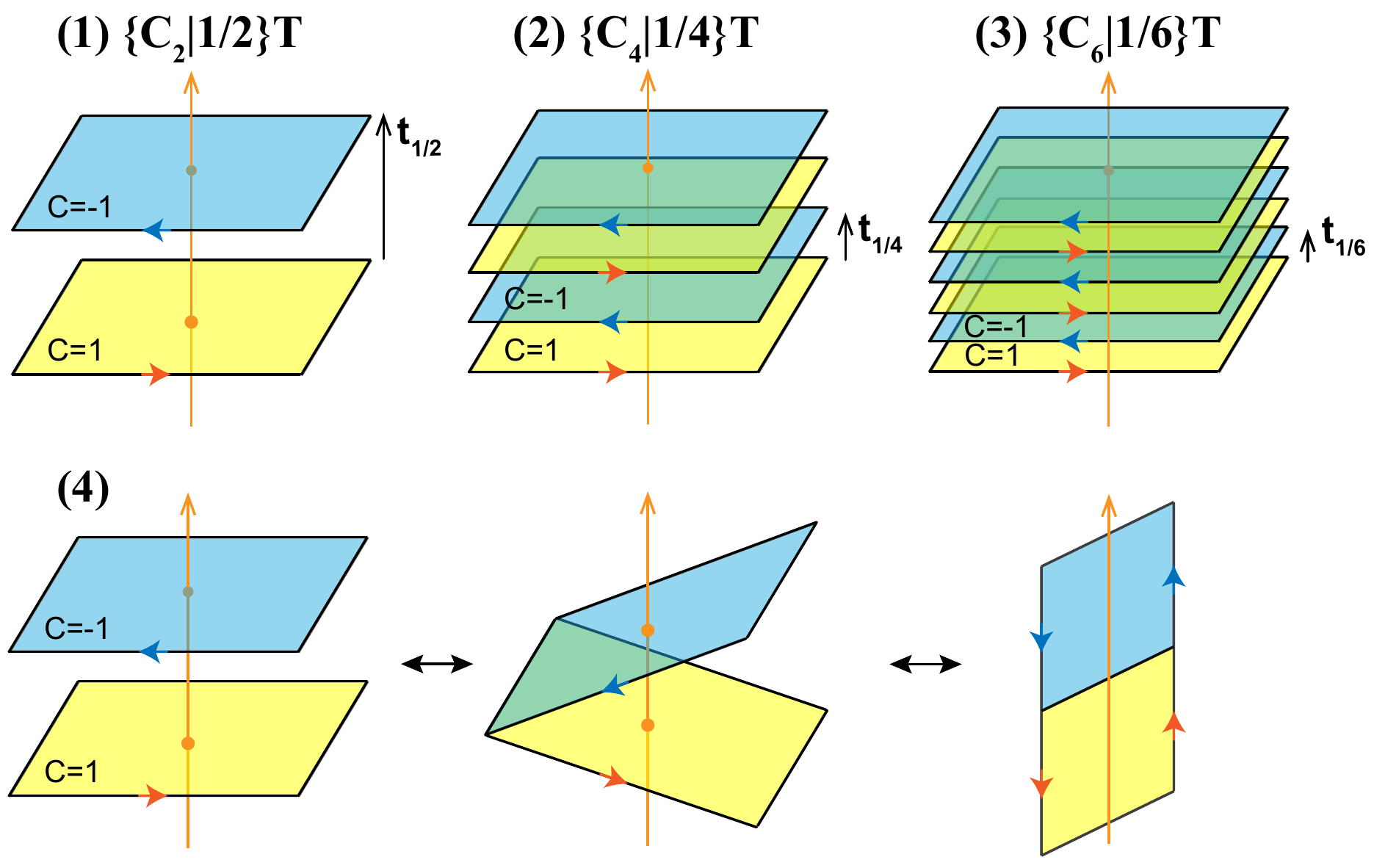}
	\caption{\label{au_screw_deco}(1)-(3). Minimal decorations of anti-unitary screws $\{C_n|\frac{1}{n}\bm{a}_i\}\cdot T,n=2,4,6$, which are equivalent to those in Fig.\ref{inv_mini_deco}(7). (4). The process of deforming two horizontal layers into vertical layers, preserving the $\{C_2|\frac{1}{2}\bm{a}_3\}\cdot T$ symmetry.}
    \end{figure}

    \item Inversion. A minimal decoration with $\delta(P)=1$ is constructed by placing a $C=1$ layer that passes the inversion center. Its doubled state, i.e., a $C=2$ layer, can be trivialized by the bubble equivalence, or equivalently, split into two $C=1$ layers and move to infinity in an inversion-symmetric way. Thus the doubled state has $\delta(P)=0$ and confirms the $\mathbb{Z}_2$-type of $\delta(P)$.
    
    \item $S_4$. A minimal decoration with $\delta(S_4)=1$ can be a non-LC with four half-layers similar to the decoration of $C_4\cdot T$, as shown in Fig.\ref{inv_mini_deco}(9). But unlike $C_4\cdot T$, this non-LC can be transformed into a single layer that passes the $S_4$-center preserving $S_4$ symmetry. $\delta(S_4)$ is $\mathbb{Z}_2$-type as its doubled state can be trivialized. 
    
    Note there are four rotoinverisons $S_n,n=2,3,4,6$, among which $S_2=M, S_3=C_3\cdot P, S_6=C_3\cdot M$, and $S_4$ is the only independent symmetry in the sense that it cannot be decompose into a direct product of smaller point groups. As a result, we only discuss the minimal decoration $S_4$.
    
    \item Unitary screws. In Fig.\ref{inv_mini_deco}(10), we plot minimal decorations of $\{C_n|\frac{1}{n}\bm{a}_i\}$, which have $n$ layers with $C=1$ connected by the screw in a unit cell, and can be seen as a $n$-time copy of the translation decoration. 
    
    In general, a unitary screw can be $\{C_n|\frac{m}{l}\bm{a}_i\}$, where $l,m$ are coprime numbers, e.g., $\frac{3}{4}$ and $\frac{5}{6}$. In these cases, there are also $l$ layers in a unit cell, which has $\delta_{w,i}=l$ and $\delta(\{C_n|\frac{m}{l}\bm{a}_i\})=m$.
\end{enumerate}

As discussed in the main text, these topological invariants can be calculated for TCs in an intuitive way.
For a given symmetry $g$, first, choose a generic point $r$ inside an arbitrary AU, and then use $g$ to transform it to its image point $g \cdot r$. Draw a path connecting these two points without touching any 1- or 0-cells. The invariant $ \delta(g)$ is determined by the decorated 2-cells through the path:
\begin{itemize}
    \item For every symmetry $g$, $\delta(g)$ can be the net Chern number accumulated through the path. 
    Even if $g$ is a mirror symmetry, we can ignore the two mirror sectors and count only the net (total) Chern number. We call this type of invariant defined by the net Chern number ``C-invariants'' and denote them as $\delta_C (g)$.
    \item For a mirror symmetry $M$, we can also consider the Chern numbers for each mirror sectors, i.e., the real-space mirror Chern numbers $(C_m^+,C_m^-)$, which are obtained by counting the mirror Chern insulators decorated on the mirror plane. The momentum space mirror Chern numbers can be calculated from the real space mirror Chern numbers, with procedure similar to that for non-magnetic case shown in Ref.\cite{song2018quantitative}.
\end{itemize}


From the above algorithm for calculating C-invariants for TCs, a homomorphism between the symmetry operations and their invariants naturally arises, as the Chern number accumulated through the path can be superimposed:
\begin{equation}
 \delta_C(g_1) + \delta_C(g_2) = \delta_C(g_1 \cdot g_2 )
 \label{homomorphism}
\end{equation}
With this homomorphism, one only needs to calculate the C-invariants for the generators of an MSG in order to obtain all the invariants.
However, we remark that as some C-invariants are $\mathbb{Z}$-type while others are $\mathbb{Z}_2$-type, it is the original value calculated from the path $r \rightarrow g \cdot r$ before taking module that should be adopted when applying this homomorphism.

In the following text, we sometimes use ``invariants'' to indicate ``C-invariants'' for simplicity, since in most cases they are completely the same except for mirrors. The specific meaning of ``invariants'' can be inferred from the context.

Interestingly, we notice 
the types of C-invariants are closely related to the types of corresponding symmetry operations, especially for those protected by magnetic point group (MPG) symmetries. 
As can been from Table.\ref{table_invariants_appendix}, proper unitary symmetries have trivial C-invariants, while anti-unitary proper symmetries have $\mathbb{Z}_2$ C-invariants (here mirror is special, as strictly speaking its invariant should be $\mathbb{Z}$-type, nevertheless we could regard the total Chern number as its $\mathbb{Z}_2$ C-invariant); unitary improper symmetries have $\mathbb{Z}_2$ C-invariants, while anti-unitary improper symmetries have trivial C-invariants. 


\clearpage
\section{Surface states}\label{AppendixC}

Surface states are in one-to-one correspondence with the topological invariants. In other words, a nontrivial invariant has a corresponding nontrivial surface state, and different surface states can be superimposed. We enumerate all surface states protected by topological invariants in MSGs, which can be seen from their minimal decorations. Note when constructing the minimal decorations of invariants in Fig.\ref{inv_mini_deco}, we ignore the three lattice translations. However, when constructing the surface states of these invariants, we sometimes need to restore the lattice translations and require irrelevant invariants to be zero.

\begin{figure*}
	\centering
	\includegraphics[width=1\textwidth]{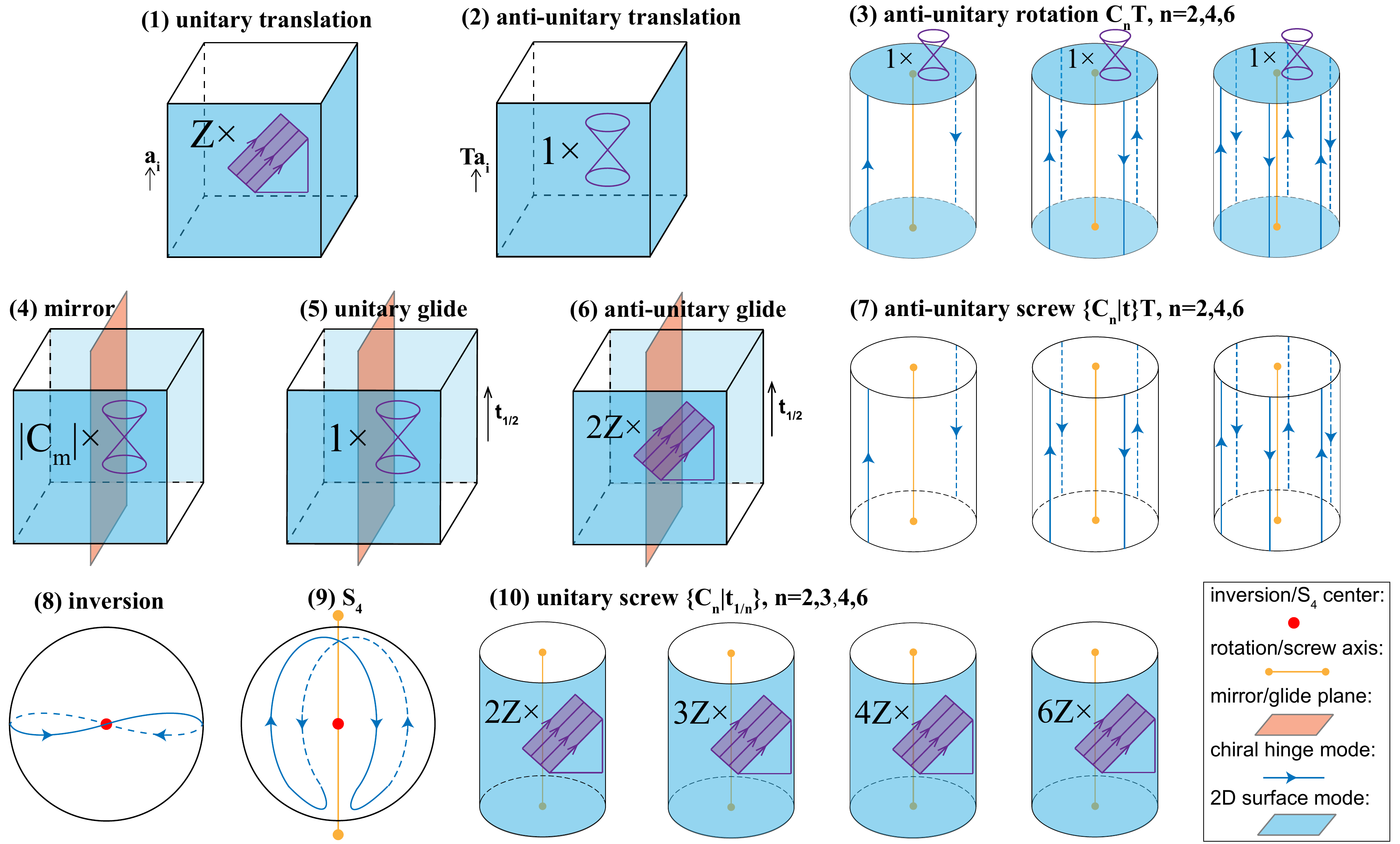}
	\caption{\label{surface_states_appendix}Surface states in MSGs, with the corresponding symmetries marked above the plots. The 2D surface modes can be either sloped-like chiral surface modes in (1),(6),(10) or Dirac cones in (2),(3),(4),(5).}
\end{figure*}

\begin{enumerate}
\item Unitary integer translation 
$\{ E | \bm{a}_{i=1,2,3} \}$. 
The corresponding surface state is known as the 3D quantum anomalous Hall effect. There are $Z=\delta_{w,i}$ chiral surface modes on $\bm{a}_i$-preserving planes, with the surface modes look like a slope in the 2D surface Brillouin zone (BZ).

\item Anti-unitary translation
$ \{ E | \frac{1}{2} \bm{a}_{i=1,2,3} \} \cdot T$. As discussed in Ref.\cite{mong2010antiferromagnetic,fang2013topological}, there are an odd number of Dirac cones on $\bm{a}_{i}$-preserving planes, where $\bm{a}_i$ is the direction of anti-unitary translation. These Dirac cones can only appear at the 4 TRIMs in the 2D surface BZ.

\item Anti-unitary rotation $ C_n\cdot T, n = 2,4,6 $. 
Surface states protected by $ C_2\cdot T$ and $C_4\cdot T$ have been previously discussed in Ref.\cite{fang2015new,shiozaki2014topology,ahn2019symmetry} and Ref.\cite{schindler2018higher}, respectively.

In general, the surface states for $ C_n\cdot T, n = 2,4,6 $ have 2, 4, and 6 chiral hinge modes on the side surface, respectively, while a single Dirac cone on the top surface. For $C_2\cdot T$, the position of the Dirac cone is unpinned\cite{fang2015new}, while for $C_{n=4,6}\cdot T$, the Dirac cone is pinned at $\Gamma$ (or a $C_{n=4,6}\cdot T$-invariant point) in the surface BZ. Note for $C_6\cdot T$, the single Dirac cone can split into three $C_6\cdot T$-symmetric Dirac cones by adding trivial bands.

These surface Dirac cones can be analyzed from the surface Hamiltonian as follows.
Consider a minimal Dirac surface Hamiltonian
\begin{equation}
     H(k) = v_x k_x \sigma_1 + v_y k_y \sigma_2 
\end{equation}
which is invariant under $C_n \cdot T$ symmetry, i.e.,
\begin{equation}
     ( C_n\cdot T) H(k) ( C_n \cdot T)^{-1} = H( (C_n\cdot T)^{-1} k)
\end{equation}
where 
\begin{align}
C_n = e^{i\frac{\pi}{n} \sigma_3 }, \; \;
T = i \sigma_2 \mathcal{K} \end{align}
and $\sigma_{i, i = 1,2,3}$ are the three Pauli matrices.
Mass term proportional to $\sigma_3$ is forbidden by $C_n \cdot T$, i.e., one can not add mass term to gap the Hamiltonian $ H(k)$. 
However, the doubled Hamiltonian $ \tau_0 \otimes H(k)$ can be gapped by adding mass term proportional to $\tau_2 \otimes \sigma_3$, which means a single Dirac cone is stable while two Dirac cones can be gapped. Therefore, the classification of this surface state is $\mathbb{Z}_2$, in accordance with the $\mathbb{Z}_2$ invariant.

\item Mirror. 
As shown in Fig.\ref{surface_states_appendix}(4), when a mirror-preserving termination is made, the mirror-invariant planes in the 3D BZ now become mirror-invariant lines in the 2D surface BZ. Surface states protected by mirror symmetries are determined by the momentum space mirror Chern numbers on each mirror-invariant lines in the surface BZ. In the following, we discuss the surface states on a specific mirror-invariant line of the surface BZ, with $(C_m^+,C_m^-)$ denoting the momentum space mirror Chern numbers on it. If $C^{\pm}_m > 0$, there are $C^{\pm}_m$ right-moving surface modes with mirror eigenvalue $\pm i$, while for $C^{\pm}_m < 0$ there are $|C^{\pm}_m|$ left-moving modes with mirror eigenvalue $\pm i$ (the direction of surface modes is convention-dependent and can be exchanged). As a result,
\begin{enumerate}
    \item when $C_m^{+}= -C_m^{-}$, there are $|C_m|=\frac{1}{2}|C^{+}_m - C^{-}_m|$ Dirac cones, as plotted in Fig.\ref{surface_states_appendix}(4).
    \item when $C_m^{+}\neq -C_m^{-}$:
    \begin{itemize}
        \item If $C_m^{+}C_m^{-} < 0$, there are $N=\min\{|C_m^+|,|C_m^-|\}$ Dirac cones and $||C_m^+|-|C_m^-||$ chiral surface modes.
        \item If $C_m^{+}C_m^{-}\geq 0$, there are $|C_m^+|+|C_m^-|$ chiral surface modes.
    \end{itemize}
\end{enumerate}
Note that the chiral surface modes are protected by not only mirror but also the (total) Chern number of the system. In fact, even if the mirror symmetry is broken, the chiral modes are still preserved due to nonzero Chern number.

Lastly, we remark that there exist hinge modes when a mirror-preserving hinge is made.

\item Unitary glide $ \{ M | \frac{1}{2}\bm{a}_i \} $. Its surface state has been discussed in Ref.\cite{fang2015new, shiozaki2015z, shiozaki2016topology}, and more recently in Ref.\cite{kim2019glide,kim2020glide}, which has an odd number of Dirac cones on the glide-symmetric plane. This surface state can also be understood from the minimal decoration of $ \{ M | \frac{1}{2}\bm{a}_i \} $, which resembles that of the anti-unitary translation. Thus they should share the same type of surface states. These Dirac cones can appear at the glide-invariant lines in the surface BZ, and an even number of Dirac cones can annihilated with each other.

We remark that there exist chiral hinge modes when a glide-preserving hinge is made.

\item Anti-unitary glide $\{  M | \frac{1}{2}\bm{a}_i \}\cdot T$. As $(\{  M | \frac{1}{2}\bm{a}_i \}\cdot T)^2=\{E|\bm{a}_i\}$, the weak invariant $\delta_{w,i}=2Z$ when the invariant $\delta(\{  M | \frac{1}{2}\bm{a}_i \}\cdot T)=Z$, which means there are $2Z$ chiral surface modes on a glide-symmetric plane. 

\item Anti-unitary screw $\{C_n| t\}\cdot T, n = 2,4,6$. 
On a typical cylinder termination, there are 2, 4, and 6 $C_n$-symmetric chiral edge modes on the side surface, respectively. Note the top surface breaks the $\{C_n| t\}\cdot T$ symmetry and thus has no gapless modes in general. 

\item Inversion. There is one inversion-symmetric chiral hinge mode on a surface termination that preserves the inversion symmetry. States with $\delta(P)=1$ have non-trivial axion angle $\theta=\pi$ and are usually referred to as ``axion insulators''. The axion angle $\theta$ is quantized to $0$ or $\pi$ when there exist anti-unitary proper symmetries (including TRS) or unitary improper symmetries, and systems with $\theta=\pi$ can be seen as axion insulators in a generalized sense.

\item $S_4$. There is one $S_4$-symmetric chiral hinge mode on a surface termination that preserves the $S_4$ symmetry.

\item Unitary screw $ \{  C_n | t \}, n = 2,3,4,6 $.
In the surface states shown in Fig.\ref{surface_states_appendix}(10), we set $t = \frac{1}{n}\bm{a}_i$ for simplicity, and there are $n$ chiral surface modes related by $ \{ C_n |\frac{1}{n}\bm{a}_i\} $ on the side surface when $\delta(\{C_n|\frac{1}{n}\bm{a}_i\})=1$.

Generally, the screw vector could be $t=\frac{m}{l}\bm{a}_i$, where $m, l$ are coprime integers.
On the side surface, there are $\delta_{w,i}$ chiral surface modes related by $\{ C_n | \frac{m}{l}\bm{a}_i\}$ per lattice vector, which can be seen from the invariant.
As both $\delta(\{ C_n | \frac{m}{l}\bm{a}_i\})$ and $\delta_{w,i}$ are integers and $ \{ C_n | \frac{m}{l}\bm{a}_i\}^{l} = \{ ( C_n)^l | m\bm{a}_i \} $, $\delta_{w,i}$ should be a multiply of $l$ and $\delta(\{ C_n | \frac{m}{l}\bm{a}_i\})=\frac{m}{l}\delta_{w,i}$. Thus $ \{  C_n | t \}$ shares the same surface state with the unitary translation, similar to the anti-unitary glide.
\end{enumerate}

\clearpage
\section{Decorations, second encounter}\label{AppendixD}
\begin{figure}[ht]
	\centering
	\includegraphics[width=0.48\textwidth]{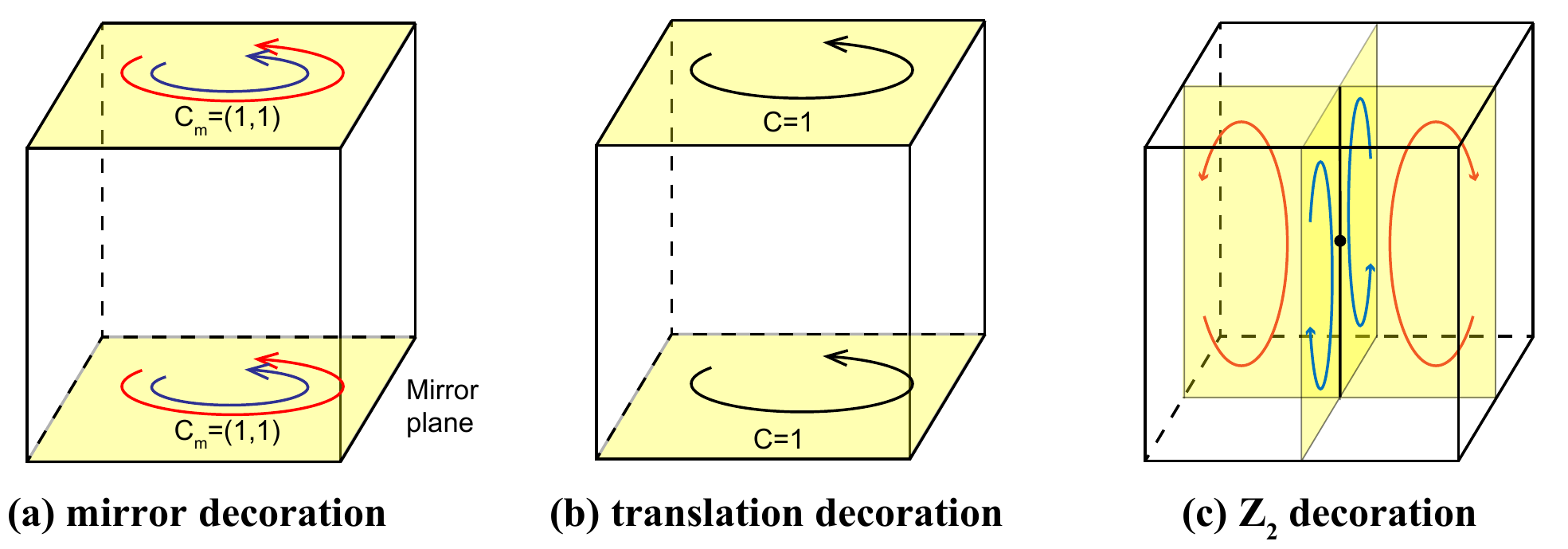}
	\caption{\label{three_type_deco}Three types of decorations. (a). Mirror decoration. (b). $Z$ or translation decorations. (c). $Z_2$ decorations.}
\end{figure}

In this section, we classify the TCs into three types and discuss them in terms of their invariants. 
As can be seen from the general form of TCI classifications, i.e., $ \mathbb{Z}^m \times \mathbb{Z}^l $ or $ \mathbb{Z}^l \times \mathbb{Z}_2 $, there are three types of decorations of TCs:

\begin{enumerate}
\item Mirror Decorations. Mirror Chern insulators with $ ( C^{+}_m, C^{-}_m ) $ = $ (n,-n) $, i.e., zero net Chern numbers, decorated on mirror planes, while all non-mirror 2-cells are not decorated, with an example shown in Fig.\ref{three_type_deco}(a).

\item (Translation) $\mathbb{Z}$ decorations. Decorations with nonzero weak (translation) invariants, which means the Chern number per unit cell is nonzero, as shown in Fig.\ref{three_type_deco}(b).  

\item $Z_2$ decorations. Decorations that are not protected by (unitary) integer translation symmetries (but some are protected by half magnetic translations). They have zero net Chern number per unit cell, i.e., zero weak invariants, and nontrivial $\mathbb{Z}_2$ invariants. As discussed in the main text, these $Z_2$ decorations are axion insulators, characterized by the $\pi$ axion angle. We plot a $Z_2$ decoration of MSG $Pmm2$ in Fig.\ref{three_type_deco}(c).

$Z_2$ decorations can also be further divided into two types depending on whether they have mirror symmetries. When there are no mirror planes, the $Z_2$ decorations become trivial after doubling,
while when there are mirror planes, the doubling of a $Z_2$ decoration followed by bubble equivalence becomes a mirror decoration, where all mirror planes are decorated with $ ( C_m^{+}, C_m^{-} ) = \pm(1,-1) $. 
Caution that we still call it a $Z_2$ decoration for simplicity, which is an abuse of notation, as the decoration actually does not become trivial but transforms into another type of decoration after doubling.

When mirror planes are present, this  
$ Z_2$ decoration must be chosen as a generator of classifications, though its original contribution to the total classification, i.e., the $\mathbb{Z}_2$ factor, is absorbed into $\mathbb{Z}^m$ in a way that $ 2 \bar{Z}_2 \sim \sum^{m}_{i=1} \bar{M}_i $, where
$ \bar{M}_i$'s are generators of mirror decorations and $\sim$ means up to bubble equivalence. 
More specifically, if there are $m$ mirror planes, the $m$ generators for the classification $\mathbb{Z}^m$ should be chosen as $ \{ \bar{Z}_2 , \bar{M}_1 , \bar{M}_2 , \cdots , \bar{M}_{m-1} \} $. 
\end{enumerate}

The three types of decorations can be well characterized by their invariants:
\begin{enumerate}
\item Mirror Decorations. 
As discussed above, this type of decorations are mirror Chern insulators with $ C^{+}_{m} = - C^{-}_{m} $ on all mirror planes. Since the net Chern number is zero for all 2-cells, all invariants except mirror Chern numbers are zero. 

This type of decorations can be defined as those with zero net Chern number on every 2-cell, which is enough to distinguish them from the other two types of decorations.  

\item (Translation) $\mathbb{Z}$ decorations. 
As mentioned before, this type of decorations are protected by (unitary) integer translation symmetries, thus they must have nonzero weak invariants. Besides, other invariants, including the mirror Chern numbers and $\mathbb{Z}_2$ invariants, can also be nonzero.

When mirror planes are present, $\mathbb{Z}$ decorations can have mirror Chern insulators with $C^{+}_m \neq - C^{-}_m $ on mirror planes, which have nonzero net Chern number and do not belong to the mirror decorations. 
Moreover, when the center of a symmetry operation with $\mathbb{Z}_2$ invariant, e.g., $S_4$, is passed by a decorated layer with odd Chern number, the $\mathbb{Z}_2$ invariant is nonzero for this $\mathbb{Z}$ decoration. In fact, a $Z$ decoration just becomes another $\mathbb{Z}$ decoration when added with a mirror or $Z_2$ decoration.

In a word, $\mathbb{Z}$ or translation decorations are those with nonzero net Chern number per unit cell.

\item $Z_2$ decorations. 
Contrary to the $\mathbb{Z}$ decorations, this type of decorations must have zero weak invariants. 
Note the ``weak'' invariants refer to those protected by unitary integer translations, but not magnetic half translations in type-4 MSGs.

For this type of decorations, all $\mathbb{Z}$-type invariants (except mirror Chern numbers), including weak invariants together with unitary screw invariants and anti-unitary glide invariants, are all zero, while all $\mathbb{Z}_2$ invariants must be nonzero, i.e. $\delta( g )_{\mathbb{Z}_2} = 1 $.  

This can be seen with the help of the homomorphism between symmetries and invariants, i.e., Eq.(\ref{homomorphism}).
Firstly, a $Z_2$ decoration must have at least one nonzero $\mathbb{Z}_2$ invariant of some symmetry $g_1$, i.e., $\delta(g_1) = 1$. Then suppose there is another symmetry $g_2$ having $\mathbb{Z}_2$ invariant. 
Their product, denoted as $g_3=g_1\cdot g_2$, must have $\mathbb{Z}$ or trivial invariant. 
This is because $g_1$ and $g_2$, which have $\mathbb{Z}_2$ invariant, must be unitary improper or anti-unitary proper, and their product $g_3$ must be unitary proper or anti-unitary improper, and these two symmetry operations have $\mathbb{Z}$ or trivial invariants, as shown in Table.\ref{table_invariants_appendix}. 
If $\delta(g_2) = 0 $, then $\delta(g_3) = \delta(g_1)+ \delta(g_2) = 1 + 0 = 1 $, which is forbidden in $Z_2$ decorations no matter $g_3$ has $\mathbb{Z}$ or trivial invariant.
Thus we must have $\delta(g_2) =1$. 
Similarly, for any other $g$ with $\mathbb{Z}_2$ invariant, we have $\delta(g) = 1$.

As a result, the values of all $\mathbb{Z}_2$ invariants are bonded and equal to 1 for the $Z_2$ decoration, which can be called as the ``axion invariant'' and take the form of 3D magnetoelectric polarization $P_3$\cite{fang2012bulk}.
This dictates that for any MSG, there is either one independent $Z_2$ decoration (when $\delta(g)_{\mathbb{Z}_2} = 1$), i.e., an axion insulator, or no $Z_2$ decoration (when $\delta(g)_{\mathbb{Z}_2} = 0$).

\begin{figure}[ht]
	\centering
	\includegraphics[width=0.4\textwidth]{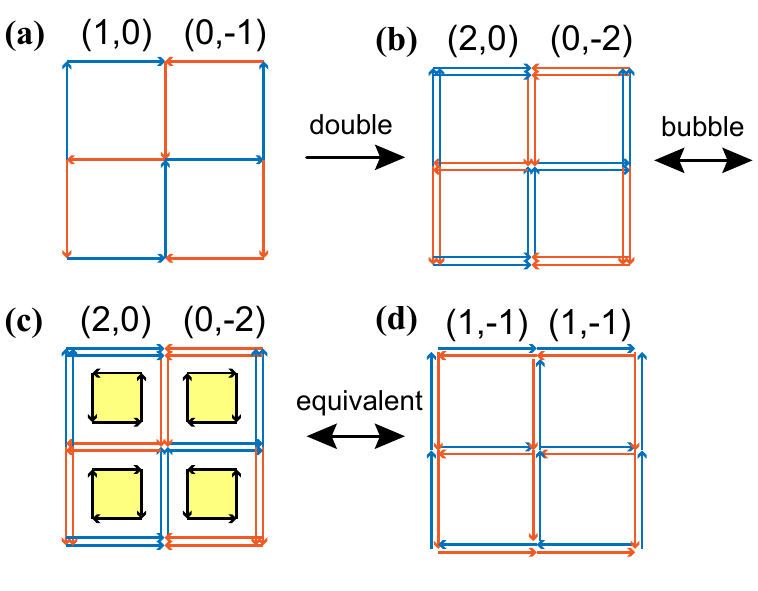}
	\caption{\label{mirrorZ2double}The $Z_2$ decoration becomes a mirror decoration with all mirror planes decorated when doubled and transformed into an LC using the bubble equivalence.}
\end{figure}

There is another interesting and significant property for the $Z_2$ decorations as mentioned previously. When there are mirror planes, the doubling of a $Z_2$ decoration followed by the bubble equivalence becomes a mirror decoration with all mirror planes decorated. This can also be understood from the invariants.
As demonstrated before, the $Z_2$ decoration has all $\mathbb{Z}_2$ invariants nonzero, and mirror symmetries can have the net Chern number as their corresponding $\mathbb{Z}_2$ C-invariants.
This indicates that for a $Z_2$ decoration, each mirror 2-cell is decorated with $C_{\text{total}} = C^{+}_m + C^{-}_m = \pm 1 $ state, i.e., state with $ ( C^{+}_m, C^{-}_m ) = (\pm1, 0)$ or $ ( 0, \pm1 ) $. As shown in Fig.\ref{mirrorZ2double}, after doubling, each mirror 2-cell has $ ( C^{+}_m, C^{-}_m ) = ( \pm 2, 0 )$ or $(0, \pm 2)$. Bubble equivalence changes $ ( C^{+}_m, C^{-}_m ) $ by $ \pm(1,1) $, resulting in $  ( C^{+}_m, C^{-}_m ) = \pm( 1,-1) $ on each mirror 2-cell, which is a mirror decoration with all mirror planes decorated. 

To summarise, $Z_2$ decorations are those with zero net Chern number per unit cell but nonzero net Chern number on some 2-cells. 
\end{enumerate}

\clearpage
\section{Layer and non-layer constructions}\label{AppendixE}

Recall that an overwhelming majority of TCs in non-magnetic SGs are layer constructions(LCs), with exceptions in only 12 SGs, which are called non-layer constructions (non-LCs)\cite{song2019topological,song2018quantitative}. 
Those non-LCs built by 2D TIs have 2D planes that are only partly decorated, namely, the 2-cells decorated with 2D TIs do not extend to the entire 2D plane. On the contrary, there is a vast number of non-LCs in (type-1,3,4) MSGs. It is difficult and also not necessary to exhaustively find all non-LCs in all MSGs, and here we give three representatives for demonstration. 

We remark an interesting property of non-LCs in MSGs. Unlike the non-LCs in non-magnetic SGs which have incomplete 2D decorated planes, here in MSGs all the non-LCs we know have complete 2D decorated planes, which split into small 2D (mirror) Chern insulator pieces with different signs of (mirror) Chern number.

\begin{figure}[ht]
	\centering
	\includegraphics[width=0.48\textwidth]{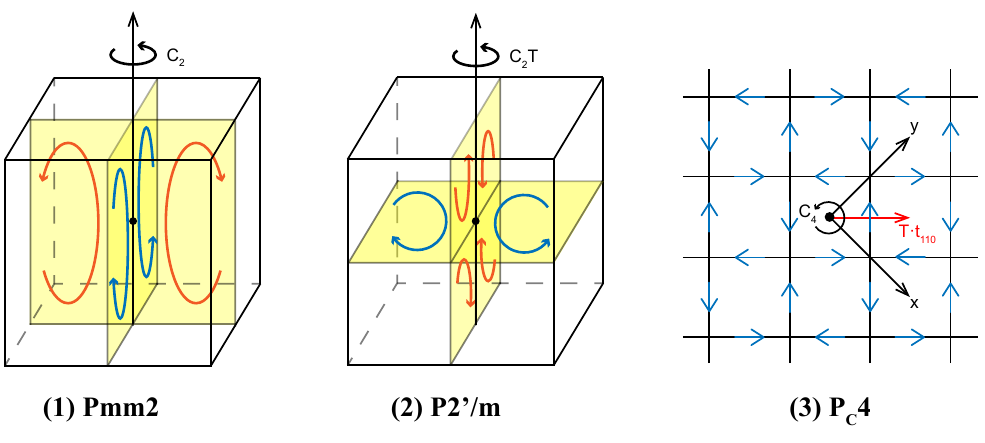}
	\caption{\label{nonLC_example}Three examples of non-layer constructions(non-LCs). (1). The non-LC in $Pmm2$, which has $C_{2z}, M_x$, and $M_y$. (2). The non-LC in $P2'/m$, which has $ C_{2} \cdot T$. (3). The non-LC in $P_C4$, which has $C_{4z}$ and anti-unitary half translation $T\cdot t_{110}$, where $t_{110}=\{E|\frac{1}{2}\frac{1}{2}0\}$. Note in (1) and (2), the side surfaces of the unit cell are also decorated, which we omit for simplicity.}
\end{figure}

\begin{itemize}
\item The first representative non-LC is the $Z_2$ decoration of type-1 MSG $Pmm2$, which has $C_{2z}$ and two orthogonal mirrors $M_x$ and $M_y$ anti-commuting with each other.

Two 2-cells related by $C_2$ symmetry on the same mirror plane have the same directional chiral edge modes on the 1-cell they intersect, i.e., the two red/green curves in Fig.\ref{nonLC_example}(1).
Thus decorating only one mirror plane with C-building blocks breaks the no-open-edge condition, and the other mirror plane must also be decorated to compensate the edge modes on the $C_2$-axis. These decorated 2-cells are fixed on mirror planes and cannot be moved together to trivialize. This decoration could be seen as the simplest one among all non-LCs in MSGs, where only planes in the $x$ and $y$ directions are decorated non-uniformly, with adjacent pieces on a 2D plane having opposite Chern numbers.

\item The second representative non-LC is the $Z_2$ decoration of type-3 MSG $ P 2'/m $ ($P4'/m$ and $P6'/m$ have similar non-LCs), where the crucial symmetry operation that makes it a non-LC is the anti-unitary rotation $ C_{2y} \cdot T$. 
The mirror planes defined by $ M_y $ is split by $ C_{2y} \cdot T$ into two 2-cells, where the two edge modes related by $C_{2y} \cdot T$ have the same direction and can not cancel with each other on the 1-cell they meet. This forces the 2D plane vertical to mirror plane to be decorated as well, in order to satisfy the no-open-edge condition.

\item The third representative non-LC is the $Z_2$ decoration of type-4 MSG 3.6 $P_C 2$. In this case there is no mirror symmetry, but the magnetic translation $T\cdot\{E|\frac{1}{2}\frac{1}{2}0\}$ in the diagonal direction makes the decoration a non-LC.

\end{itemize}

\clearpage
\section{A brief review of the symmetry-based indicator theory}\label{AppendixF}

Symmetry-based indicator (SI) theory uses the symmetry data at high-symmetry-points (HSPs) in the BZ to diagnose the band topology of a system. The symmetry data composes of the irreducible representations (irreps) of the valence bands at HSPs in SGs, or irreducible co-representations (coirreps) in MSGs, and the collection of coirreps at all HSPs is denoted as the band representation (BR). SI theory compares the difference between two linear spaces, i.e., the space of all atomic insulators (AIs) and the space of BRs that satisfy the compatibility relations between all HSPs. A BR having zero SI can be decomposed into an integer combination of AIs, while a BR with nonzero SI cannot. 

AIs are constructed by placing non-interacting atoms in real space, which have flat dispersions. For a specific MSG, the BR of its AIs form a finitely generated Abelian group $\mathbb{Z}^l$, denoted as $\{\text{AI}\}$, where $l$ is the number of generators of this group. We can construct $\{\text{AI}\}$ by first choosing specific Wyckoff positions in the real space and then placing orbits on them, where the orbits have symmetries of the site symmetry group of the Wyckoff positions and can be represented by a coirreps of the site symmetry group, with details shown in Appendix.\ref{AppendixM}.

Similarly, the linear space $\{\text{BS}\}$ composes of the BRs that satisfy the compatibility relations between all HSPs, which ensures no (symmetry-protected) band crossings at high-symmetry lines or planes connecting HSPs, although gapless points could exist at generic momenta. For a given MSG, these two Abelian groups have the same dimension\cite{watanabe2018structure}, and the SI group is defined as the quotient group $X_{\text{BS}}=\{\text{BS}\}/\{\text{AI}\}=\mathbb{Z}_{n_1}\times\mathbb{Z}_{n_2}\times\dots\mathbb{Z}_{n_l}$, which is a finite Abelian group. 

Practically, $\{\text{BS}\}$ can be  derived from $\{\text{AI}\}$ and does not need to be calculated explicitly. For a given MSG, arrange its $\{\text{AI}\}$ bases $\{a_1,a_2,\dots,a_n\}$ into a matrix $A$ and do the Smith normal form decomposition:
\begin{equation}
\begin{aligned}
A_{n\times m}&=
\left(
\begin{matrix}
a_1 \\
a_2 \\
\vdots \\
a_n \\
\end{matrix}
\right) =L_{n\times n}M_{n\times m}R_{m\times m}\\
&=
L\times
\left(
\begin{matrix}
d_1 & & &0\\
& d_2 &  &\\
& & \ddots &\\
0& & & d_l &\\
\vdots &0&0&\vdots
\end{matrix}
\right) 
\left(
\begin{matrix}
b_1 \\
b_2 \\
\vdots \\
b_n \\
\vdots \\
b_m \\
\end{matrix}
\right)\\
&\Rightarrow
L^{-1}\left(
\begin{matrix}
a_1 \\
a_2 \\
\vdots \\
a_n \\
\end{matrix}
\right)=
\left(
\begin{matrix}
d_1b_1 \\
d_2b_2 \\
\vdots \\
d_nb_n\\
\end{matrix}
\right)
\end{aligned}
\end{equation}
where the nonzero diagonal terms $\{d_1,d_2,\dots,d_l\}$ of $M$ are arranged in ascending order and $d_i|d_{i+1}$, with the terms greater than 1 corresponding to the SI group $\mathbb{Z}_{d_k}\times\mathbb{Z}_{d_{k+1}}\times\cdots\mathbb{Z}_{d_l}$. Each row $b_i$ of the right matrix $R$ is a basis of $\{\text{BS}\}$, and $a_i^\prime=d_ib_i$ gives the independent basis of $\{\text{AI}\}$. The rows in $R$ that correspond to the SI group, i.e., $b_k,b_{k+1},\dots,b_l$, are called nontrivial BSs, and must have nontrivial indicators values which generate the SI group.

In the next section, we derive the formulas of SIs in MSGs using physical bases such that their values correspond to some topological invariants. However, we remark here that the formulas of SIs do not have a unique choice, and Smith form decomposition naturally gives a set of SI formula, i.e., the $i$-th column of $R^{-1}$ gives rise to the corresponding $z_i$ indicator, as shown below.

Given an arbitrary band representation $B$, first decompose it using the $\{\text{AI}\}$ bases:
\begin{equation}
B=c\cdot AI=\tilde{c}\cdot M\cdot R,\ \ (\tilde{c}=cL)
\end{equation}
The indicators are then given by
\begin{equation}
z_i=(B\cdot R^{-1})_{i} \bmod n_i
\label{Ind_formula}
\end{equation}
which means the $i$-th column of $R^{-1}$ gives the $z_i$ indicator, because $z_i$ corresponds to the numerator of the fractional number in the decomposition of $B$.

\clearpage
\section{Generating SIs in MSGs}\label{AppendixG}
In Table.\ref{generatingSI_appendix}, we list the generating SIs in 1651 MSGs, and their explicit formulas and correspondence to TCI classifications as well as topological invariants are derived in the following subsections. The word ``generating'' we adopt here means the SIs in all other MSGs can be generated using the combinations of these SIs, with only 10 exceptions denoted as ``corner cases'', which will be introduced in Appendix.\ref{AppendixH}. We also include the type-2 generating MSG $P\overline{4}1'$ here for completeness, which is omitted in the main text where only type-1,3,4 MSGs are considered.

\begin{table}[ht]
	\centering
	\begin{tabular}{c|c|c|c}
		\hline\hline
		\textbf{MSG} & \textbf{MSG type} & $\bm{X}_{\text{BS}}$ & \textbf{SI} \\\hline
		$P\overline{1}$ & 1 & $\mathbb{Z}_{2,2,2,4}$ & $z_{2P,1},z_{2P,2},z_{2P,3},z_{4P}$ \\\hline
		$Pn,n=2,3,4,6$ & 1 & $\mathbb{Z}_n$ & $z_{nC}$ \\\hline
		$Pn/m,n=2,3,4,6$ & 1 & $\mathbb{Z}_{n,n,n}$ & $z_{nm,0}^+,z_{nm,0}^-,z_{nm,\pi}^+$ \\\hline
		$P\overline{4}$ & 1 & $\mathbb{Z}_{2,2,4}$ & $z_{2,S_4}, z_{2,\text{Weyl}}, z_{4C}$ \\\hline
		$Pmmm$ &  1 & $\mathbb{Z}_{2,2,2,4}$ & $z_{2P,1}',z_{2P,2}',z_{2P,3}',z_{4P}'$ \\\hline
		$P4/mmm$ &  1 & $\mathbb{Z}_{2,4,8}$ & $z_{2P,1}^\prime, z_{4m,\pi}^+, z_8$ \\\hline
		$P6/mmm$ &  1 & $\mathbb{Z}_{6,12}$ & $z_{6m,\pi}^+, z_{12}$ \\\hline
		$P\overline{4}1'$ & 2 &  $\mathbb{Z}_{2}$ & $z_{2,S_4}^\prime$ \\\hline
		$Pnc'c',n=2,4,6$ & 3 & $\mathbb{Z}_{n}$ & $z_{nC}^\prime$ \\\hline\hline
	\end{tabular}
	\caption{\label{generatingSI_appendix}Generating SIs and their corresponding MSGs. We use a simplified notation to represent the SI group, i.e., $\mathbb{Z}_{n_1,n_2,\dots}=\mathbb{Z}_{n_1}\times\mathbb{Z}_{n_2}\times\dots$.}
\end{table}

\subsection{Generating SIs in type-1 MSG}
In Ref.\cite{ono2018unified}, Ono and Watanabe derived SIs in a few type-1 key MSGs. Here we do a more detailed investigation, including not only the SI formulas but also their correspondence to topological crystals.

\subsubsection{MSG 2.4 $P \overline{1}$, $X_{BS}=\mathbb{Z}_{2,2,2,4}$, Classification=$\mathbb{Z}^3\times \mathbb{Z}_2$}
MSG 2.4 $P\overline{1}$ has inversion symmetry $P$, and its SIs can be defined using parities at 8 time-reversal invariant momenta (TRIMs):
\begin{equation}
\begin{aligned} 
z_{2P,1} = & \sum_{\mathbf{k} \in \text{TRIM},k_1=\pi} \frac{1}{2}(N_\mathbf{k}^--N_\mathbf{k}^+) \bmod 2 \\ 
z_{2P,2} = & \sum_{\mathbf{k} \in \text{TRIM},k_2=\pi} \frac{1}{2}(N_\mathbf{k}^--N_\mathbf{k}^+) \bmod 2 \\ 
z_{2P,3} = & \sum_{\mathbf{k} \in \text{TRIM},k_3=\pi} \frac{1}{2}(N_\mathbf{k}^--N_\mathbf{k}^+) \bmod 2 \\ 
z_{4P} = & \sum_{\mathbf{k} \in \text{TRIM}} \frac{1}{2}(N_\mathbf{k}^--N_\mathbf{k}^+) \bmod 4 \\ 
\end{aligned}
\end{equation}
where $N_\mathbf{k}^\pm$ is the number of valence bands having positive (negative) parity. 

The first three $z_{2P,i}$ indicators diagnose the Chern number mod 2 on $k_i=0,\pi$ planes, while the $z_{4P}$ indicator can be seen as the sum of Chern number on $k_z=0,\pi$ planes. 

When $\mathbb{Z}_4=1,3$, the Chern number on $k_z=0$ and $k_z=\pi$ plane must differ by an odd number, which indicates an odd number of Weyl points exist between $k_z=0$ and $k_z=\pi$. An equal number of Weyl points of opposite chirality exist between $k_z=0,-\pi$ planes, enforced by the inversion symmetry.

The 4 nontrivial BSs that corresponds to $\mathbb{Z}_{2,2,2,4}$ are
\begin{equation}
\begin{aligned}
b_{1}&=  \overline{V}_{2} - \overline{V}_{3} + \overline{X}_{2} - \overline{X}_{3} + \overline{Y}_{2} - \overline{Y}_{3} - \overline{Z}_{2} + \overline{Z}_{3}  \\ 
b_{2}&=  \overline{Y}_{2} - \overline{Y}_{3} - \overline{Z}_{2} + \overline{Z}_{3}  \\ 
b_{3}&=  \overline{X}_{2} - \overline{X}_{3} - \overline{Z}_{2} + \overline{Z}_{3}  \\ 
b_{4}&=  \overline{Z}_{2} - \overline{Z}_{3}  \\ 
\end{aligned}
\end{equation}
Their indicators are calculated to be (0012), (0110), (1010), (0013), which indeed generate the group $\mathbb{Z}_{2,2,2,4}$. The minus sign in the BSs may be confusing, and one can add proper AIs to them to make the coefficients all positive, which does not change SIs as AIs have zero SIs.

\begin{figure}[ht]
	\centering
	\includegraphics[width=0.48\textwidth]{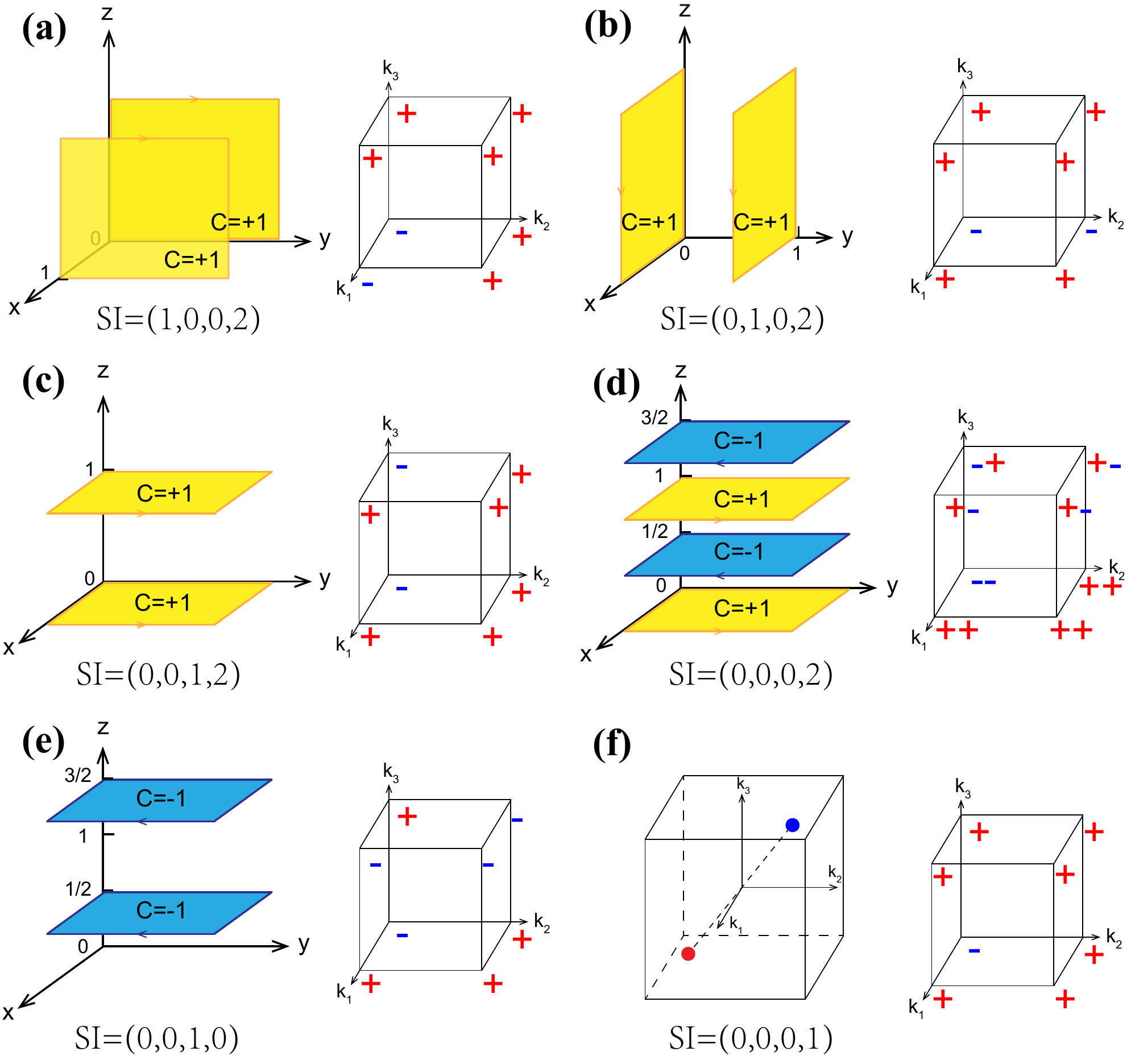}
	\caption{\label{SG2_deco} (a)-(e). Layer constructions and their compatible sets of irreps as well as the corresponding SIs in $P\overline{1}$. (f). The Weyl state and a compatible set of irreps in $P\overline{1}$.}
\end{figure}

The TCI classification of $P\bar{1}$ is $\mathbb{Z}^3\times \mathbb{Z}_2$, where the three $\mathbb{Z}$ indexes correspond to translation decorations, while the $\mathbb{Z}_2$ index is the $Z_2$ decoration protected by inversion. The generators of these decorations have weak and inversion invariants  $(\delta_{w,1},\delta_{w,2},\delta_{w,3},\delta(P))=(100,1),(010,1),(001,0),(000,1)$, respectively. We can attach compatible irreps to these decorations, as shown in Fig.\ref{SG2_deco}(a), (b), (c), and (d).
Their SIs are calculated to be (1002), (0102), (0010), and (0002). The mappings between invariants and SIs are straightforward:
\begin{equation}
z_{2P,i}=\delta_{w,i}\bmod 2,\ \ z_{4P}=2\delta(P)
\end{equation}
which means $z_{4P}=2$ states have $\delta(P)=1$, i.e., they have 1D chiral hinge modes on inversion-preserving surfaces. Note that $z_{4P}=1,3$ are gapless Weyl states and cannot be constructed using the real-space recipe.

\subsubsection{$Pn,\ X_{BS}=\mathbb{Z}_{n}$, Classification=$\mathbb{Z}$}
$Pn, n=2,3,4,6$ have $C_n$ rotations as generators. Because the $C_n$ eigenvalues must be the same along the rotation axis, the Chern number on $k_i=0$ and $k_i=\pi$ plane must be the same, and we take the Chern number on $k_i=0$ plane as the $\mathbb{Z}_n$ indicator, where $i$ denotes the rotation direction. The classification group $\mathbb{Z}$ is generated by a LC with $C=1$ layers on $x_i=n$ planes, which has weak invariant $\delta_{w,i}=1$.

\begin{figure}[ht]
	\centering
	\includegraphics[width=0.35\textwidth]{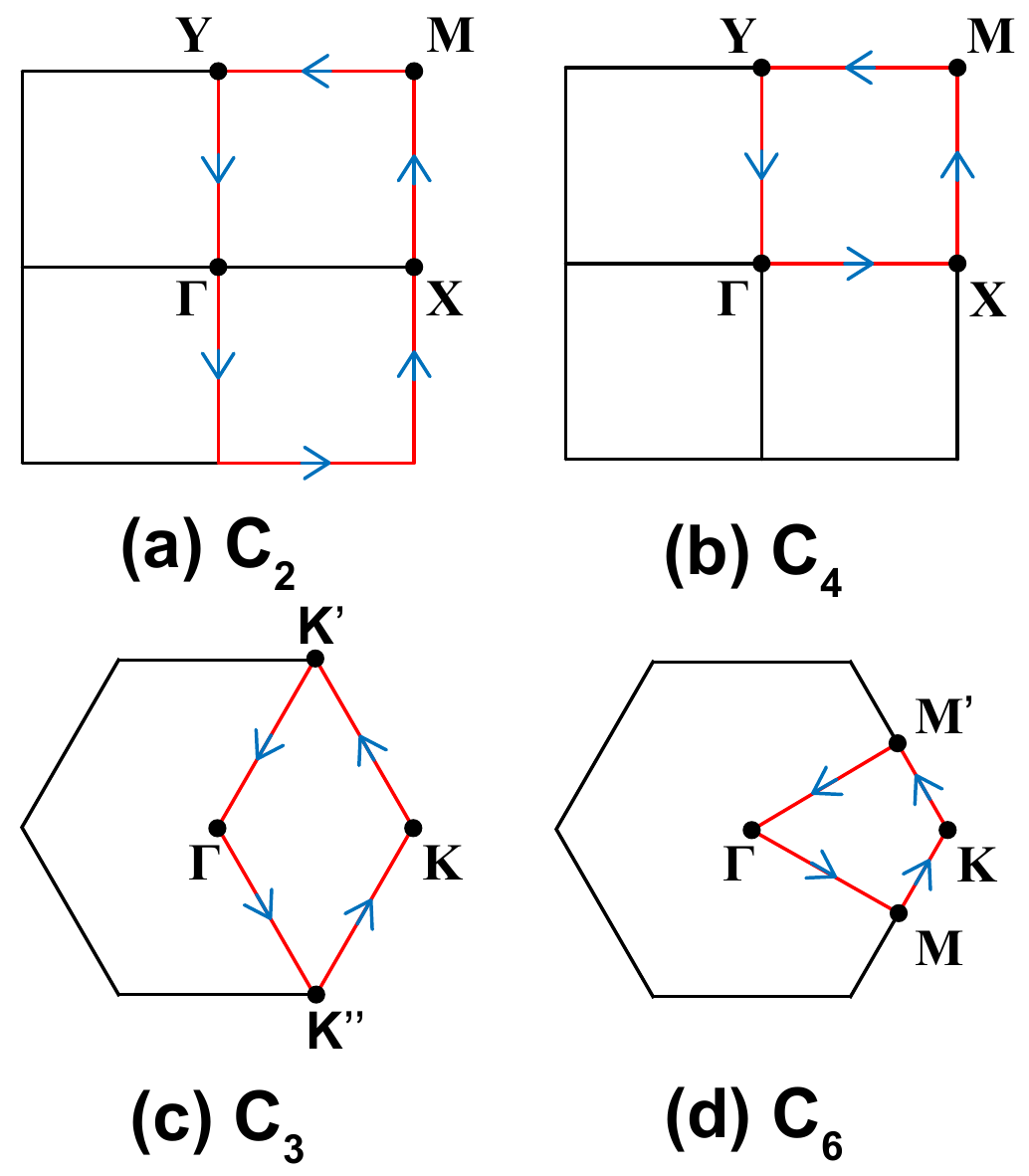}
	\caption{\label{Pn_HSP}HSPs used to calculate Chern number mod n on 2D BZ}
\end{figure}

The formulas for calculating Chern numbers (mod n) using $C_n$ eigenvalues are derived in Ref.\cite{fang2012bulk}:
\begin{equation}
\begin{aligned} 
C_2:&\ \ (-1)^{C}=\prod_{l \in \mathrm{occ} .} B_{C_2}^l(\Gamma) B_{C_2}^l(X) B_{C_2}^l(Y) B_{C_2}^l(M)\\
C_4:&\ \ e^{i \pi C / 2}=\prod_{l \in \mathrm{occ} .} (-1)B_{C_4}^l(\Gamma) B_{C_4}^l(M) B_{C_2}^l(Y)\\
C_3:&\ \ e^{i 2 \pi C / 3} =\prod_{l \in \mathrm{occ} .}(-1) B_{C_3}^l(\Gamma) B_{C_3}^l(K) B_{C_3}^l(K^{\prime}) \\ 
C_6:&\ \ e^{i \pi C / 3} =\prod_{l \in \mathrm{occ} .} (-1)B_{C_6}^l(\Gamma) B_{C_3}^l(K) B_{C_2}^l(M) 
\end{aligned}
\end{equation}
where $B_{C_n}^l(k)$ represent the $C_n$ eigenvalue of the $l$-th band at $C_n$ -invariant point $k$, and the HSPs used in these formulas are plotted in Figure \ref{Pn_HSP}. We can take logarithm to each side of the equation to turn them into the more familiar summation and mod n form, and denote them as $z_{nC}$:
\begin{equation}
\begin{aligned}
z_{2C} &= \sum_{l\in occ} \ln(B_{C_2}^l(\Gamma) B_{C_2}^l(X) B_{C_2}^l(Y) B_{C_2}^l(M))/(i\pi) \bmod 2 \\
&=\sum_{\mathbf{k} \in \Gamma,M,X,Y} \frac{1}{2}(N_\mathbf{k}^--N_\mathbf{k}^+) \bmod 2 \\ 
z_{4C} &= \sum_{l\in occ} \ln(-B_{C_4}^l(\Gamma) B_{C_4}^l(M) B_{C_2}^l(Y)) /(i\frac{\pi}{2})\bmod 4\\
z_{3C} &= \sum_{l\in occ} \ln(-B_{C_3}^l(\Gamma) B_{C_3}^l(K) B_{C_3}^l(K^{\prime})) /(i\frac{2\pi}{3}) \bmod 3\\
z_{6C} &= \sum_{l\in occ} \ln(-B_{C_6}^l(\Gamma) B_{C_3}^l(K) B_{C_2}^l(M)) /(i\frac{\pi}{3}) \bmod 6
\end{aligned}
\end{equation}
These $z_{nC}$ indicators correspond to the weak invariant:
\begin{equation}
	z_{nC}=\delta_{w,i}\bmod n
\end{equation}

In $P\overline{4}$ and $P\overline{3}$, as $S_4=C_4^{-1}$ and $S_3=C_6^{-1}$ on $k_z=0,\pi$ planes, they can also be used to calculate the Chern number. The modified formulas are
\begin{equation}
\begin{aligned}
S_4&:\ z_{4C} = \sum_{l\in occ} \ln(-B_{S_4^{-1}}^l(\Gamma) B_{S_4^{-1}}^l(M) B_{C_2}^l(Y)) /(i\frac{\pi}{2})\bmod 4\\
S_3&:\ z_{6C} = \sum_{l\in occ} \ln(B_{S_3^{-1}}^l(\Gamma) B_{C_3}^l(K) B_{P}^l(M)) /(i\frac{\pi}{3}) \bmod 6
\end{aligned}
\end{equation}

In practice, when calculating these $z_{nC}$ indicators using coirreps of double MSGs, we need to choose the proper representation matrices s.t. the SU(2) part of $C_n$ rotations are aligned, i.e., 
\begin{equation}
\begin{aligned}
&U(C_4)^2=U(C_2),\ U(C_4^{-1})^2=-U(C_2),\\ 
&U(C_6)^2=U(C_3),\ U(C_6^{-1})^2=U(C_3^{-1}),\\
&U(C_6)^3=U(C_2),\ U(C_6^{-1})^3=-U(C_2)
\end{aligned}
\end{equation}
where $U$ denotes the SU(2) matrix. For $S_{3},S_4$, these relations become
\begin{equation}
    U(S_4^{-1})^2=U(C_2), U(S_3^{-1})^2=U(C_3), U(S_3^{-1})^3=U(P)
\end{equation}
Note for $S_3$, there is no $-1$ in the $z_{6C}$ formula, because we are using $P$ instead of $C_2$, which satisfies $P^{-1}=P$ instead of $C_2^{-1}=-C_2$ when SU(2) matrices are considered.

\subsubsection{$Pn/m,\ X_{BS}=\mathbb{Z}_{n,n,n}$, Classification=$\mathbb{Z}^3$}
$Pn/m$ represents four MSGs: $P2/m$, $P4/m$, $P\overline{6}(=P3/m)$, and $P6/m$. All these MSGs have mirror planes vertical to the $C_n$ rotation axis. Assume the rotation is along $x_i$ direction. There are two mirror planes in the BZ, i.e., $k_i=0,\pi$, which allows us to define four Chern numbers: $C_{m,0}^\pm$ and $C_{m,\pi}^\pm$, where $\pm$ means the Chern number defined on the mirror $\pm i$ sector, respectively. It can be proven that only three of them are independent, which correspond to the three $\mathbb{Z}$ indexes in the TCI classification.

The indicators can be chosen as any three independent mirror Chern numbers mod n, and here we take the first three: $z_{nm,0}^+, z_{nm,0}^-, z_{nm,\pi}^+$. The mapping between invariants and indicators are simple:
\begin{equation}
	z_{nm,0/\pi}^\pm=C_{m,0/\pi}^\pm \bmod n
\end{equation}

\paragraph{Example: $P4/m$}
We take $P4/m$ as an example, which has three nontrivial BSs as:
\begin{equation}
\begin{aligned}
b_{1}=& - \overline{\Gamma}_{10} + \overline{\Gamma}_{12} - \overline{M}_{8} + \overline{M}_{12} + \overline{R}_{4} - \overline{R}_{5} - \overline{X}_{5} \\
&+\overline{X}_{6} + 2\overline{Z}_{6} + 3\overline{Z}_{7} -3\overline{Z}_{10} -3\overline{Z}_{11} + \overline{Z}_{12}  \\ 
b_{2}=&  \overline{\Gamma}_{10} - \overline{\Gamma}_{12} - \overline{M}_{6} + \overline{M}_{10} - \overline{R}_{4} + \overline{R}_{5} + \overline{X}_{5} \\
&-\overline{X}_{6} - \overline{Z}_{6} -2\overline{Z}_{7} + 2\overline{Z}_{10} + 2\overline{Z}_{11} - \overline{Z}_{12}  \\ 
b_{3}=&  \overline{Z}_{6} + \overline{Z}_{7} - \overline{Z}_{10} - \overline{Z}_{11}  \\ 
\end{aligned}
\end{equation}
Their SIs are calculated to be $\{z_{4m,0}^+, z_{4m,0}^-, z_{4m,\pi}^+\}= (3,0,3), (0,1,0), (0,0,3)$, which indeed generate the $\mathbb{Z}_{4,4,4}$ group.

\paragraph{Example: $P4_2/m$, $X_{BS}=\mathbb{Z}_{2,4}$}
$P4_2/m$ has a screw symmetry $\{C_{4z}|0,0,1/2\}$ and SI group $\mathbb{Z}_{2,4}$. Its TCI classification is $\mathbb{Z}^2$, whose generators can be taken as a translation decoration (placing $(C_m^+,C_m^-)=(1,0)$ states on $z=1/4,3/4$ planes) and a $Z_2$ decoration, which is a non-LC similar to the non-LC in $Pmmm$, as shown later.

The SIs can be taken as $C_{4m,0}^+, C_{2m,\pi}^+$. The Chern number on $k_z=\pi$ plane can only be calculated by $C_{2z}$, but not $C_{4z}$-screw, as $C_{4z}$-screw does not commute with $M_z$. The nontrivial BSs are
\begin{equation}
\begin{aligned}
b_{1}&=  2\overline{M}_{6} + 2\overline{M}_{8} -2\overline{M}_{10} -2\overline{M}_{12} + \overline{R}_{4} - \overline{R}_{5} + \overline{X}_{4} - \overline{X}_{5}  \\ 
b_{2}&=  \overline{M}_{6} + \overline{M}_{8} - \overline{M}_{10} - \overline{M}_{12}  \\ 
\end{aligned}
\end{equation}
which has SI=(1,0) and (0,1).

Similar argument hold for $P2_1/m$ and $P6_3/m$. $P6_3/m$ has indicator group $\mathbb{Z}_{3,6}$ and can be taken as $C_{6m,0}^+, C_{3m,\pi}^+$, while $P2_1/m$ has indicator group $\mathbb{Z}_{2}$, and can be taken as $z_{2P}^\prime$.

\subsubsection{MSG 81.33 $P\overline{4},\ X_{BS}=\mathbb{Z}_{2,2,4}$, Classification=$\mathbb{Z}\times\mathbb{Z}_2$}
$P\overline{4}$ has $S_4$ symmetry. Its TCI classification is $\mathbb{Z}\times\mathbb{Z}_2$, which correspond to a translation and a $S_4$ decoration, respectively. The translation decoration is placing 2D layers with Chern number $C$ on $z=n+1/2$ planes, while the $S_4$ decoration is composed of 2D layers with Chern number 1 on $z=n$ planes and layers with Chern number -1 on $z=n+\frac{1}{2}$ planes.

The SI group of $P\overline{4}$ is $\mathbb{Z}_{2,2,4}$. Because $S_4$ symmetry is equivalent to $C_4^{-1}$ on $k_z=0,\pi$ planes in the BZ, $S_4$ helps diagnose the Chern number mod 4 on these two planes. As a result, the $\mathbb{Z}$ index of translation decoration turns into the $\mathbb{Z}_4$ factor in the SI group. 

The first $\mathbb{Z}_2$ in the SI group corresponds to Weyl states, which exist when the Chern number on $k_z=0,\pi$ planes differ by $4n+2$, indicating an even number of Weyl points between $k_z=0,\pi$ planes. Note $S_4^2=C_2$, which requires the Chern number on $k_z=0,\pi$ planes to be equal mod 2, and cannot differ by $4n+1$ or $4n+3$.

\begin{figure}[ht]
	\centering
	\includegraphics[width=0.35\textwidth]{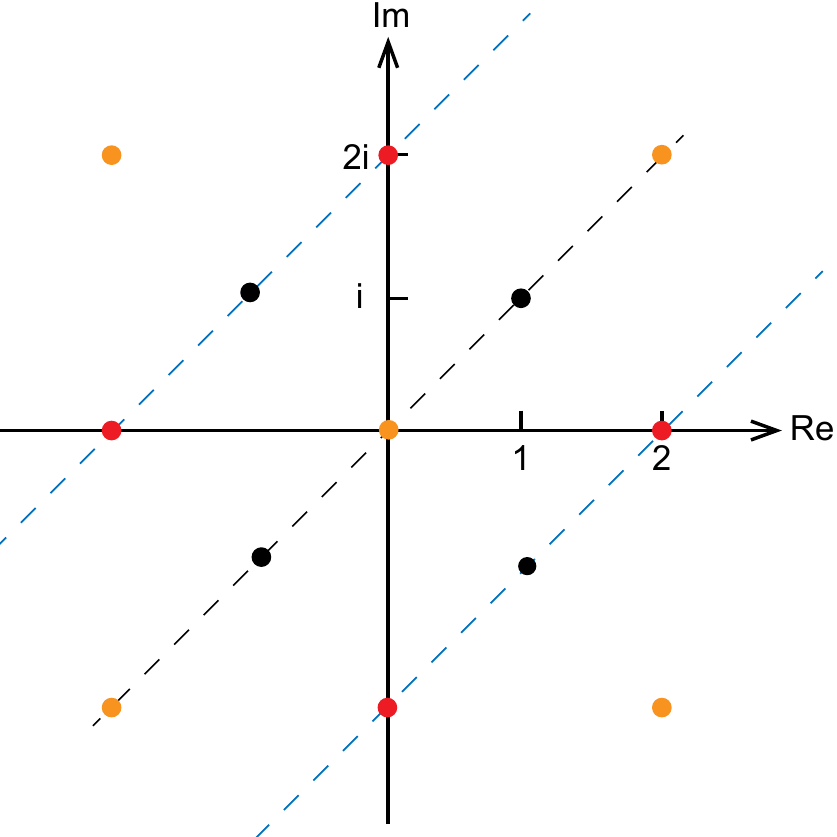}
	\caption{\label{s4z2}.Possible values of $\mu_4$. Orange dots correspond to AIs or the translation decorations, red dots correspond to the $S_4$ decorations, and black dots correspond to Weyl states.}
\end{figure}

The second $\mathbb{Z}_2$ corresponds to the $S_4$ decoration, and can be calculated by the formula:
\begin{equation}
\begin{aligned}
z_{2,S_4} &= \frac{1}{2}(\text{Re}\mu_4-\text{Im}\mu_4) \bmod 2  \\
\mu_4 &= \frac{1}{\sqrt{2}} \sum_{i\in occ.}\sum_{\mathbf{k}\in K_{S_4}}\beta_i(\mathbf{k})\\
\end{aligned}
\end{equation}
where $\beta_i(\mathbf{k})=e^{\alpha\frac{\pi}{4} i}(\alpha=1,3,5,7)$ are the $S_4$ eigenvalues at four $S_4$-invariant momenta $K_{S_4}$. It can be proven that $\mu_4=\pm 2\pm 2i, 0$ for AI or the translation decoration, while $\mu_4=\pm2, \pm2i$ for the $S_4$ decoration. The $z_{2,S_4}$ indicator maps the blue dashed lines in Fig.\ref{s4z2} to $z_2=1$ and the black dashed line to $z_2=0$. This definition is stable against adding AIs to the BS. 

This $z_{2,S_4}$ also maps $\mu_4=\pm(1-i)$ to $z_{2,S_4}=1$, while $\mu_4=\pm(1+i)$ to $z_{2,S_4}=0$. These $\mu_4$ values can only be taken in Weyl states, which means that the $z_2$ indicator classifies Weyl states into two classes. We can define a new $z_{2,\text{Weyl}}$ indicator to represents these Weyl states:
\begin{equation}
	z_{2,\text{Weyl}}=\text{Re}(\mu_4) \bmod 2
\end{equation}

\paragraph{Meaning of the $z_{2,S4}$ and $z_{2,\text{Weyl}}$ indicator.}
From the TCI classification $\mathbb{Z}\times\mathbb{Z}_2$, we can naturally extract $\mathbb{Z}_4\times\mathbb{Z}_2$ as indicators, where the $\mathbb{Z}_2$ indicator is used to identify the $S_4$ decoration.

For the translation decoration, 2D layers with Chern number C=1 are placed on $z=n+1/2$ planes. Therefore, $S_4$ eigenvalues on $k_z=0,\pi$ planes are opposite, which implies that $\mu_4=0$. For the $S_4$ decoration, 2D layers with Chern number C=1 placed on $z=n$ planes have the same $S_4$ eigenvalues on $k_z=0,\pi$ planes, while the 2D layers placed on $z=n+\frac{1}{2}$ planes have opposite $S_4$ eigenvalues on $k_z=0,\pi$ planes. As a result, only the $z=n$ layers contribute to $\mu_4$.

 We tabulate the possible choices of $S_4$ and $C_2$ eigenvalues for a 2D layer with $C=0,1,2,3$ in Table.\ref{table_Chernnumber_S4}:
\begin{table}[ht]
	\centering
	\begin{tabular}{c|c|c|c|c|c}
		\hline
		momenta  & $\Gamma$ & M & X & \multirow{2}{*}{\shortstack{sum of $S_4$\\ eigenvalues $/\sqrt{2}$}} & \multirow{2}{*}{\shortstack{Chern\\ number}} \\\cline{1-4}
		symmetry & $S_4^{-1}$ & $S_4^{-1}$ & $C_2$ &  & \\\hline
		\multirow{8}{*}{eigenvalue} & $e^{i\pi/4}$ & $e^{i3\pi/4}$ & $i$ / $-i$ &$i$ & 
		\multirow{8}{*}{-1 / 1}\\\cline{2-5}
				          & $e^{i3\pi/4}$ & $e^{i\pi/4}$ & $i$ / $-i$ &$i$&\\\cline{2-5}
				          & $e^{-i\pi/4}$ & $e^{-i3\pi/4}$ & $i$ / $-i$ &$-i$&\\\cline{2-5}				          						  
				          & $e^{-i3\pi/4}$ & $e^{-i\pi/4}$ & $i$ / $-i$ &$-i$&\\\cline{2-5}
				          
				          & $e^{i\pi/4}$ & $e^{-i\pi/4}$ & $-i$ / $i$ & $1$&\\\cline{2-5}				          
				          & $e^{-i\pi/4}$ & $e^{i\pi/4}$ & $-i$ / $i$ & $1$&\\\cline{2-5}
				         & $e^{i3\pi/4}$ & $e^{-i3\pi/4}$ &  $-i$ / $i$ & $-1$& \\\cline{2-5}
				         & $e^{-i3\pi/4}$ & $e^{i3\pi/4}$ &  $-i$ / $i$ & $-1$& \\\cline{1-6}
				         
		\multirow{8}{*}{eigenvalue} & $e^{i\pi/4}$ & $e^{i\pi/4}$ & $i$ / $-i$ &$1+i$ & 
		\multirow{8}{*}{2 / 0}\\\cline{2-5}
						 & $e^{-i3\pi/4}$ & $e^{-i3\pi/4}$ & $i$ / $-i$ & $-1-i$&  \\\cline{2-5}    
				         & $e^{i3\pi/4}$ & $e^{-i\pi/4}$ & $i$ / $-i$ &$0$& \\\cline{2-5}
				         & $e^{-i\pi/4}$ & $e^{i3\pi/4}$ & $i$ / $-i$ & $0$& \\\cline{2-5}
				         & $e^{i3\pi/4}$ & $e^{i3\pi/4}$ & $-i$ / $i$ &$-1+i$& \\\cline{2-5}
				         & $e^{-i\pi/4}$ & $e^{-i\pi/4}$ & $-i$ / $i$ & $1-i$& \\\cline{2-5}
				         & $e^{i\pi/4}$ & $e^{-i3\pi/4}$ & $-i$ / $i$ &$0$ & \\\cline{2-5}
				         & $e^{-i3\pi/4}$ & $e^{i\pi/4}$ & $-i$ / $i$ & $0$&  \\\hline 		          				      
	\end{tabular}
	\caption{\label{table_Chernnumber_S4} Possible choices of $S_4$ and $C_2$ eigenvalues for a 2D layer with Chern number $C=0,1,2,3$.}
\end{table}
Note that the SI formula we adopt here is 
$z_{4C} = \sum_{l} \ln(-B_{S_4^{-1}}^l(\Gamma) B_{S_4^{-1}}^l(M) (-1)B_{C_2}^l(Y)) /(i\frac{\pi}{2})\bmod 4$, where the minus sign in front of $B_{C_2}$ comes from the SU(2) matrix alignment for $C_2$, in order to be consistent with the Bilbao irrep convention.

From Table.\ref{table_Chernnumber_S4} we can see that $\mu_4=\pm2,\pm2i$ for the $S_4$ decoration, while $\mu_4=0,\pm2\pm2i$ for AIs. For Weyl states, the Chern number on $k_z=0/\pi$ planes differ by 2. We can enumerate the possible choices of $S_4$ eigenvalues that are compatible with the Weyl states. Note that the $C_2$ eigenvalues on $k_z=0/\pi$ planes need to be the same, with $S_4^{-1}=e^{i\pi/4}, e^{-i3\pi/4}$ corresponding to $C_2=i$ and $S_4^{-1}=e^{-i\pi/4}, e^{i3\pi/4}$ corresponding to $C_2=-i$. We find that all Weyl states has $\mu_4=\pm 1\pm i$. As a result, $z_{2,S_4}$ and $z_{2,\text{Weyl}}$ are well-defined and can successfully identify the $S_4$ decoration and Weyl states.

\paragraph{Nontrivial BSs}
The three nontrivial BSs correspond to $\mathbb{Z}_{2,2,4}$ are:
\begin{equation}
\begin{aligned}
b_{1}&=  \overline{Z}_{5} - \overline{Z}_{6}  \\ 
b_{2}&=  \overline{Z}_{7} - \overline{Z}_{8}  \\ 
b_{3}&=  \overline{\Gamma}_{6} - \overline{\Gamma}_{8} + \overline{Z}_{6} - \overline{Z}_{8}  \\ 
\end{aligned}
\end{equation}
Their indicators $( z_{2,\text{Weyl}}, z_{2,S_4},z_4)$ are calculated to be (1,1,0), (1,0,0), and (0,1,1).

\paragraph{Type-2 MSG}
$P\overline{4}1^\prime$ in type-2 MSG has indicator $\mathbb{Z}_2$ defined as $z_{2,S_4}^\prime=-\frac{1}{2}\mu_4\bmod 2$, which represents STIs, and can be induced from the Weyl states in type-1 MSG. This is because a time-reversal partner of a certain state in type-1 MSG has complex-conjugated $\mu_4$ value. As a result, only $\mu_4=\pm(1\pm i)$ states, which correspond to Weyl states, can have $z_{2,S_4}^\prime=1$, which corresponds to the STI.

\paragraph{MSG 82.39 $I\overline{4},\ X_{BS}=\mathbb{Z}_{2,2,2}$, Classification=$\mathbb{Z}\times\mathbb{Z}_2$.}
The indicator of $I\overline{4}$ is very similar to $P\overline{4}$, and can be taken as $z_{2,\text{Weyl}}, z_{2,S_4}, z_2=z_{4C}/2$. Because $I\bar{4}$ is body-centered, the Chern number on $k_z=0$ plane can only take even number, which enforces $z_{4C}$ to become a $\mathbb{Z}_2$ indicator. 

The $z_{2,\text{Weyl}}$ needs further clarification. Because the TCI classification generators, i.e., translation and $S_4$ decoration, correspond to the $z_{4C}/2$ and $z_{2,S_4}$ separately, the remaining $\mathbb{Z}_2$ indicator must correspond to a gapless state. On the other hand, the $z_{2,\text{Weyl}}$ indicator together with $z_{2,S_4},z_{4C}/2$ successfully mapping AIs to zero and non-trivial BSs to non-zero and generate the $\mathbb{Z}_{2,2,4}$ group, we conclude that the $z_{2,\text{Weyl}}$ indicator is applicable. $z_{2,\text{Weyl}}=1$ corresponds to a Weyl semimetal with Weyl points lying $S_4$-symmetrically around $\Gamma$ at generic momenta, and two Weyl points connected by $S_4$ having opposite chirality.

\subsubsection{MSG 47.249 $Pmmm,\ X_{BS}=\mathbb{Z}_{2,2,2,4}$, Classification=$\mathbb{Z}^6$}
$Pmmm=\{\mathbf{1},C_{2x},C_{2y},C_{2z},P,M_x,M_y,M_z\}$. 

Because there are two mirror planes in each direction, i.e., $x,y,z=0, 1/2$, the TCI classification of $Pmmm$ is $\mathbb{Z}^6$, with each mirror plane corresponding to one $\mathbb{Z}$. Notice three $C_2$ forbid the translation decoration, i.e., a pure stacked C=1 state, on each direction, but only mirror decorations are allowed, i.e., $ ( C_{m}^+,C_m^- )=(n,-n)$ states on $x,y,z=n, n+1/2$ planes. Generators of $\mathbb{Z}^6$ can be taken as five mirror decorations, plus a $Z_2$ generator.

The SI group of $Pmmm$ is $\mathbb{Z}_{2,2,2,4}$. Because $M_x, M_y$, and $M_z$ anti-commute with each other, irreps in $Pmmm$ are all two-dimensional, with the same inversion eigenvalues and opposite mirror eigenvalues, forming effective Kramer pairs. As a result, we can modify the indicator formula in $P\overline{1}$ by replacing $N_k^\pm$ to the number of Kramer pairs:
\begin{equation}
\begin{aligned} 
z_{2P,i}^\prime = & \sum_{\mathbf{k} \in \text{TRIM},k_i=\pi} \frac{1}{4}(N_\mathbf{k}^--N_\mathbf{k}^+) \bmod 2 \\ 
z_{4P}^\prime = & \sum_{\mathbf{k} \in \text{TRIM}} \frac{1}{4}(N_\mathbf{k}^--N_\mathbf{k}^+) \bmod 4 \\ 
\end{aligned}
\end{equation}

The four nontrivial BSs correspond to $\mathbb{Z}_{2,2,2,4}$ are:
\begin{equation}
\begin{aligned}
b_{1}&=  \overline{U}_{5} - \overline{U}_{6} + \overline{X}_{5} - \overline{X}_{6} - \overline{Y}_{5} + \overline{Y}_{6} + \overline{Z}_{5} - \overline{Z}_{6}  \\ 
b_{2}&=  \overline{Y}_{5} - \overline{Y}_{6} - \overline{Z}_{5} + \overline{Z}_{6}  \\ 
b_{3}&=  \overline{X}_{5} - \overline{X}_{6} - \overline{Z}_{5} + \overline{Z}_{6}  \\ 
b_{4}&=  \overline{Z}_{5} - \overline{Z}_{6}  \\ 
\end{aligned}
\end{equation}
Their indicators are calculated to be (0,1,0,2), (0,1,1,0), (1,0,1,0) and (0,0,1,1).

\paragraph{Interpretation of the SI}

\begin{table}[ht]
	\begin{center}
		\begin{tabular}{c|c|cccccccc}
			\hline\hline
			\multicolumn{10}{c}{47.249\ \  $Pmmm$\ \  $ X_{BS}=\mathbb{Z}_{2,2,2,4}$\ \ Classification= $\mathbb{Z}_M^6$}
			\\
			\hline
			deco &  $ Z_{2224} $  & weak & $2^{001}$ & $2^{010}$ & $i$ & $2^{100}$ & $Cm^{100}_{(2)}$ & $Cm^{010}_{(2)}$ & $Cm^{001}_{(2)}$
			\\
			\hline
			$Z_2$ & $0003$ & (000) & 0 & 0 & 1 & 0 & $1\bar{1}00$ & $1\bar{1}00$ & $1\bar{1}00$
			\\
			\hline
			$M(100;0)$ & $1002$ & (000) & 0 & 0 & 0 & 0 & $1\bar{1}1\bar{1}$ & $0000$ & $0000$
			\\
			\hline
			$M(100;\frac{1}{2})$ & $1000$ & (000) & 0 & 0 & 0 & 0 & $1\bar{1}\bar{1}1$ & $0000$ & $0000$
			\\
			\hline
			$M(010;0)$ & $0102$ & (000) & 0 & 0 & 0 & 0 & $0000$ & $1\bar{1}1\bar{1}$ & $0000$
			\\
			\hline
			$M(010;\frac{1}{2})$ & $0100$ & (000) & 0 & 0 & 0 & 0 & $0000$ & $1\bar{1}\bar{1}1$ & $0000$
			\\
			\hline
			$M(001;0)$ & $0012$ & (000) & 0 & 0 & 0 & 0 & $0000$ & $0000$ & $1\bar{1}1\bar{1}$
			\\
			\hline
			$M(001;\frac{1}{2})$ & $0010$ & (000) & 0 & 0 & 0 & 0 & $0000$ & $0000$ & $1\bar{1}\bar{1}1$
			\\
			\hline\hline
		\end{tabular}
	\end{center}
	\caption{\label{Pmmm_mapping}Topological invariants and SIs of $Pmmm$ decoration generators. $M(nml;d)$ denotes an LC with Miller indices $(mnl)$ and the distance $d$ to the origin point. Note we list all 6 mirror decorations for completeness, and any 5 of them plus the $Z_2$ decoration can be chosen as the classification generators.}
\end{table}

We list the topological invariants of decoration generators in Table.\ref{Pmmm_mapping}. The 6 LCs are realized by placing $(C_m^+,C_m^-)=(1,-1)$ states on $x,y,z=n,\frac{1}{2}$ planes, respectively. Their SIs can be calculated by assigning compatible irreps similar to $P\bar{1}$. The correspondence between SIs and invariants is
\begin{equation}
	z_{2P,i}^\prime = C_{m,k_i=\pi}^+\bmod 2,\ z_{4P}^\prime\bmod 2 = \delta(P)
\end{equation}
$z_{4P}^\prime=1,3$ both correspond to $\delta(P)=1$, while $z_{4P}^\prime=2$ corresponds to states with helical hinge modes, similar to the case of type-2 MSG $P\bar{1}1'$ which has the same $\mathbb{Z}_{2,2,2,4}$ SIs.

The $Z_2$ decoration is a non-LC patched using 2D mirror Chern insulators pieces, as discussed in the main text. Although it is not straightforward to get the SI of this decoration, we can understand it by first doubling it and them using bubble equivalence to transform it into an LC, which has mirror Chern number $(1,-1)$ on $x,y,z=n,n+1/2$ planes, i.e., the superposition of the 6 LC generators. In the BZ, the mirror Chern number on $k_x,k_y,k_z=0$ planes is (2,-2), while on $k_x,k_y,k_z=\pi$ plane is (0,0), which has SI=(0,0,0,2). As bubbles has trivial SI, we conclude that the $Z_2$ decoration has SI=(0,0,0,1) or (0,0,0,3). In Appendix.\ref{AppendixL}, we calculated the 3D BHZ model in detail, whose symmetry and mirror Chern numbers are compatible with the $Z_2$ decoration when $M\in(1,3)$. By ignoring irrelevant symmetries, the $Z_2$ decoration must have the same parities at all TRIMs and thus the same $z_{4P}^\prime$ indicator as 3D BHZ model, which means 
\begin{equation}
	z_{4P}^\prime=3
\end{equation}

\subsubsection{MSG 123.339 $P4/mmm,\ X_{BS}=\mathbb{Z}_{2,4,8}$, Classification=$\mathbb{Z}^5$}
$P4/mmm=\{\mathbf{1},C_{2x},C_{2y},C_{2z},C_{4z},C_{2,110},C_{2,1\bar{1}0},P,$ $ M_x,M_y,M_z,M_{110},M_{1\bar{1}0},...\}$. Its TCI classification is $\mathbb{Z}^5$, which can be interpreted as four mirror decorations along x/y and z directions ($C_{4z}$ makes x and y equivalent), each having two mirror planes at 0 and $\frac{1}{2}$, and one mirror decoration along (110)/(1$\bar{1}$0) direction ($C_{4z}$ also makes (110) and (1$\bar{1}$0) equivalent).

The $\mathbb{Z}_2$ indicator corresponds to the weak mirror Chern insulator along x and y direction and can be taken as the $z_{2P,1}^\prime$, i.e., the mirror Chern number mod 2 on $k_x=\pi$ plane.

The $\mathbb{Z}_4$ indicator corresponds to the weak mirror Chern insulator along the $z$ direction and can be taken as the mirror Chern number on the $k_z=\pi$ plane mod 4, i.e.,$z_{4m,\pi}^+$, calculated by the $C_4$ eigenvalues.

The $\mathbb{Z}_8$ indicator can be taken as 
\begin{equation}
\begin{aligned}
z_8&= 2 z_{2,S_4}^\prime - z_{4P}^\prime \bmod 8\\
z_{4P}^\prime &= \sum_{\mathbf{K} \in \operatorname{TRIM}} \frac{1}{4}(
N_\mathbf{k}^--N_\mathbf{k}^+) \\
z_{2,S_4}^\prime & =-\frac{1}{2\sqrt{2}} \sum_{i\in occ.}\sum_{\mathbf{k}\in K_{4}}\beta_i(\mathbf{k})\\
&=\sum_{\mathbf{k}\in K_{S_4}} \frac{1}{2}(n_{\mathbf{k}}^{-\sqrt{2}}-n_{\mathbf{k}}^{\sqrt{2}})\\
\end{aligned}
\end{equation}
where $N_\mathbf{k}^\pm$ are the number of bands with parity $\pm 1$, $z_{2,S_4}^\prime=-\frac{1}{2}\mu_4$ defined by $S_4$ symmetry, and $n_{\mathbf{k}}^{\pm\sqrt{2}}$ is the number of Kramer pairs at $S_4$-invariant TRIMs with $\operatorname{tr}\left[D\left(S_{4}\right)\right]=\pm\sqrt{2}$. Note that $z_{4P}^\prime$ and $z_{2,S_4}^\prime$ in the $z_8$ formula should not mod 4 and mod 2, and their original values are used to calculate $z_8$.

The three nontrivial BSs correspond to $\mathbb{Z}_{2,4,8}$ are:
\begin{equation}
\begin{aligned}
b_{1}&= -2\overline{M}_{7} + 2\overline{M}_{9} - \overline{R}_{5} + \overline{R}_{6}  \\ 
b_{2}&= - \overline{M}_{7} + \overline{M}_{9} - \overline{R}_{5} + \overline{R}_{6} + \overline{Z}_{7} - \overline{Z}_{9}  \\ 
b_{3}&=  \overline{Z}_{7} - \overline{Z}_{9}  \\ 
\end{aligned}
\end{equation}
Their indicators are calculated to be (1,2,0), (0,3,6), (0,1,3).

\paragraph{Interpretation of the SI}

\begin{table*}
	\begin{center}
		\begin{tabular}{c|c|ccccccccc}
\hline\hline
\multicolumn{11}{c}{123.339\ \ $P4/mmm$\ \ $ X_{BS}=\mathbb{Z}_{2,4,8}$\ \  Classification= $\mathbb{Z}_M^5$}
\\
\hline
deco &  $ \mathbb{Z}_{248} $  & weak & $4^{001}$ & $2^{100}$ & $i$ & $2^{110}$ & $\bar{4}^{001}$ & $Cm^{100}_{(2)}$ & $Cm^{001}_{(4)}$ & $Cm^{1\bar{1}0}_{(2)}$
\\
\hline
$Z_2$ & $003$ & ($0$$0$$0$) & $0$ & $0$ & $1$ & $0$ & $1$ & $1\bar{1}00$ & $1\bar{1}00$ & $1\bar{1}$
\\
\hline
M$(100;0)$ & $104$ & ($0$$0$$0$) & $0$ & $0$ & $0$ & $0$ & $0$ & $1\bar{1}1\bar{1}$ & $0000$ & $00$
\\
\hline
M$(100;\frac{1}{2})$ & $100$ & ($0$$0$$0$) & $0$ & $0$ & $0$ & $0$ & $0$ & $1\bar{1}\bar{1}1$ & $0000$ & $00$
\\
\hline
M$(001;0)$ & $016$ & ($0$$0$$0$) & $0$ & $0$ & $0$ & $0$ & $0$ & $0000$ & $1\bar{1}1\bar{1}$ & $00$
\\
\hline
M$(001;\frac{1}{2})$ & $030$ & ($0$$0$$0$) & $0$ & $0$ & $0$ & $0$ & $0$ & $0000$ & $1\bar{1}\bar{1}1$ & $00$
\\
\hline
M$(1\bar{1}0;0)$ & $004$ & ($0$$0$$0$) & $0$ & $0$ & $0$ & $0$ & $0$ & $0000$ & $0000$ & $2\bar{2}$
\\
\hline
\hline
		\end{tabular}
	\end{center}
	\caption{\label{P4mmm_mapping}Topological invariants and SIs of $P4/mmm$ decoration generators.}
\end{table*}

\begin{figure*}
	\centering
	\includegraphics[width=0.9\textwidth]{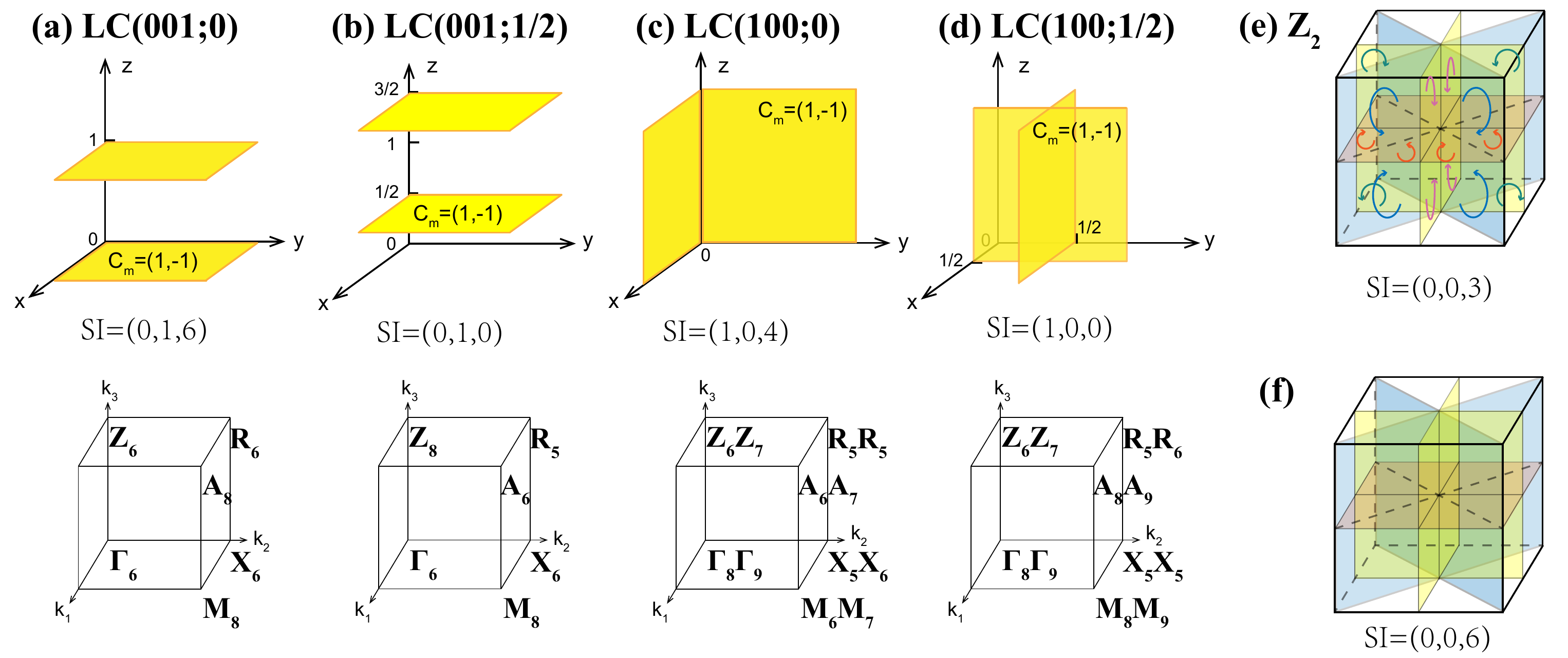}
	\caption{\label{SG123_deco}The decorations of $P4/mmm$. (a)-(d) are the mirror decorations and compatible sets of irreps. (e) is the $Z_2$ decoration and (f) is its doubled mirror decoration, with mirror planes all having mirror Chern number $(1,-1)$.}
\end{figure*}

We list the invariants and SIs of decoration generators in Table.\ref{P4mmm_mapping}, and plot their real space constructions in Fig.\ref{SG123_deco}. We attach possible irreps at HSPs that are compatible with the LCs, with which SIs can be readily calculated.

Similar to $Pmmm$, the $Z_2$ decoration is a non-LC patched by 2D mirror Chern insulator pieces. Its doubled and then bubbled state is an LC with mirror Chern number $(2,-2)$ on $k_x,k_y,k_z,k_{110},k_{1\bar{1}0}=0$ planes and $(0,0)$ on $k_x,k_y,k_z=\pi$ planes, which has SI=$(0,0,6)$. From this we can see that $z_8=3$ or $7$ for the $Z_2$ decoration. In Appendix.\ref{AppendixL}, we calculate the 3D BHZ model, whose symmetry is compatible with this $Z_2$ decoration, and the $z_8$ indicator is calculated to be
\begin{equation}
z_8=3
\end{equation}

The indicator of the $Z_2$ decoration can also be understood in the following way. First, it has inversion invariant $\delta_i=1$, thus has inversion indicator $z_{4P}^\prime=1$ or $3$. On the other hand, it is a two-band system that has $S_4$ invariant $1$, thus it can be seen as a superposition of $\delta(S_4)=1$ and $\delta(S_4)=0$ state in SG81. In SG81, a gapped $\delta(S_4)=0$ state has $\mu_4=\pm 2\pm 2i, 0$ while a $\delta_{S_4}=1$ state has  $\mu_4=\pm2, \pm2i$. By enforcing their sum to be real (because $z_{2,S_4}^\prime$ is always real), we have $z_{2,S_4}^\prime=1 \bmod 2$. Combining these two conditions, $z_8$ can only take odd values for the $Z_2$ decoration.

\subsubsection{MSG 191.233 $P6/mmm$, $X_{BS}=\mathbb{Z}_{6,12}$, Classification=$\mathbb{Z}^4$}
$P6/mmm$ has TCI classification $\mathbb{Z}^4$ due to four sets of independent mirror planes, including two sets of vertical mirrors (in $x$ and $(110)$ direction) and two sets of horizontal mirrors (at $z=0$ and $z=\frac{1}{2}$).

Its SI group is $\mathbb{Z}_{6,12}$. The first $\mathbb{Z}_6$ can be calculated from the mirror Chern insulator in the z direction $z_{6m,k_z=\pi}^+$.

The choice of $\mathbb{Z}_{12}$ is not obvious. We can define $z_{6m,S}=z_{6m,k_z=0}^++z_{6m,k_z=\pi}^+$ and $z_{4P}^\prime$, with $z_{6m,S}=1$ and $z_{4P}^\prime=1$ both correspond to the $Z_2$ decoration, thus they are not independent and merge into one $\mathbb{Z}_{12}$ indicator. Using the same method from Ref.\cite{song2018quantitative}, we have
 \begin{equation}
 \begin{aligned}
 z_{12} \bmod 6 &= z_{6m,S}\\
 z_{12} \bmod 4 &= z_{4P}^\prime
 \end{aligned}
 \end{equation}
 which is equivalent to 
 \begin{equation}
 z_{12}=\left\{z_{6m,S}+3\left[\left(z_{6m,S}-z_{4P}^\prime\right) \bmod 4\right]\right\} \bmod 12
 \end{equation}
We remark that the $z_{12}$ indicator is stable when $z_{6m,S}$ is changed by 6 and $z_{4P}^\prime$ changed by 4, i.e., $z_{6m,S},z_{4P}^\prime$ can first take mod and then insert into the $z_{12}$ formula. 
  
The two non-trivial BSs of $\mathbb{Z}_{6,12}$ are:\\
\begin{equation}
\begin{aligned}
b_{1}&= -2\overline{H}_{7} + 2\overline{H}_{9} -2\overline{K}_{7} + 2\overline{K}_{9} + \overline{L}_{5} - \overline{L}_{6} + \overline{M}_{5} - \overline{M}_{6}  \\ 
b_{2}&=  4\overline{H}_{7} -4\overline{H}_{9} + 4\overline{K}_{7} - \overline{K}_{8} -3\overline{K}_{9} \\
&-2\overline{L}_{5} + 2\overline{L}_{6} - \overline{M}_{5} + \overline{M}_{6}  \\ 
\end{aligned}
\end{equation}
which have $\mathbb{Z}_{6,12}$ indicators (1,2), (4,9).

\begin{table}[ht]
	\begin{center}
		\begin{tabular}{c|c|cccccccc}
\hline\hline
\multicolumn{10}{c}{191.233\ \ $P6/mmm$\ \ $ X_{BS}=\mathbb{Z}_{6,12}$\ \ Classification=$\mathbb{Z}_M^4$}
\\
\hline
deco &  $ Z_{6,12} $  & weak & $6^{001}$ & $2^{010}$ & $i$ & $2^{1\bar{1}0}$ & $Cm^{100}_{(2)}$ & $Cm^{001}_{(6)}$ & $Cm^{\bar{2}10}_{(2)}$
\\
\hline
$Z_2$ & $07$ & ($0$$0$$0$) & $0$ & $0$ & $1$ & $0$ & $1\bar{1}$ & $1\bar{1}00$ & $1\bar{1}$
\\
\hline
$M(100;0)$ & $06$ & ($0$$0$$0$) & $0$ & $0$ & $0$ & $0$ & $2\bar{2}$ & $0000$ & $00$
\\
\hline
$M(001;0)$ & $12$ & ($0$$0$$0$) & $0$ & $0$ & $0$ & $0$ & $00$ & $1\bar{1}1\bar{1}$ & $00$
\\
\hline
$M(001;\frac{1}{2})$ & $50$ & ($0$$0$$0$) & $0$ & $0$ & $0$ & $0$ & $00$ & $1\bar{1}\bar{1}1$ & $00$
\\
\hline
$M(\bar{2}10;0)$ & $06$ & ($0$$0$$0$) & $0$ & $0$ & $0$ & $0$ & $00$ & $0000$ & $2\bar{2}$
\\
\hline
\hline
		\end{tabular}
	\end{center}
	\caption{\label{P6mmm_mapping}Topological invariants and SIs of $P6/mmm$ decoration generators.}
\end{table}

The generators of decorations are listed in Table.\ref{P6mmm_mapping}. For the four mirror decorations, the $z_{12}$ indicator can be calculated by first read the $z_{6m,S}$ from the mirror Chern numbers, and then determine the $z_{4P}^\prime$ by reducing the LC to $Pmmm$. For the $Z_2$ decoration, $z_{12}=7$ is calculated from the BHZ model, which has $z_{6m,S}=1, z_{4P}^\prime=3$.

Type-4 MSG 191.242 has $\mathbb{Z}_{12}$ indicator group and can be defined similarly.

\subsubsection{MSG 147.13 $P\overline{3}$, $X_{BS}=\mathbb{Z}_{2,12}$, Classification=$\mathbb{Z}\times\mathbb{Z}_2$}
MSG $P\overline{3}$ is not a generating MSG, but we also derive its SIs here as they have $\mathbb{Z}_{12}$ indicator, whose formula is not obvious. Its SI group can be rewritten equivalently as 
$\mathbb{Z}_{2}\times\mathbb{Z}_{12}=\mathbb{Z}_{2}\times\mathbb{Z}_{3}\times\mathbb{Z}_4=\mathbb{Z}_{6}\times\mathbb{Z}_{4}$. 

As $S_3$ is equivalent to $C_6$ on $k=0,\pi$ planes, we can take the $\mathbb{Z}_6$ indicator as $z_{6C}$, which corresponds to the translation decoration. The $\mathbb{Z}_4$ indicator is taken as the $z_{4P}$ from $P\bar{1}$, with $z_{4P}=2$ corresponds to $\delta(P)=1$, and odd $z_{4P}$ correspond to Weyl states, where the Chern number on $k_z=0,\pi$ planes differ by 3 mod 6. In the mapping table of $P\overline{3}$, we set the SI group to be $\mathbb{Z}_{6,4}$, and the SI can be directly read from the weak and inversion invariants. $P\overline{3}$ has	TCI classification $\mathbb{Z}\times\mathbb{Z}_2$, where $\mathbb{Z}$ corresponds to the translation decoration and $\mathbb{Z}_2$ is the $Z_2$ decoration, as shown in Table.\ref{147_mapping}.

Four type-3 MSGs 162.77, 163.83, 164.89 and 165.95 with $P\bar{3}$ as subgroup also have similar SIs. 

\begin{table}[ht]
	\begin{center}
		\begin{tabular}{c|c|ccc}
			\hline\hline
			\multicolumn{5}{c}{147.13\ \ $P\bar{3}$\ \ $ X_{BS}=\mathbb{Z}_{6,4}$\ \ Classification=$\mathbb{Z} \times \mathbb{Z}_2$}
			\\
			\hline
			deco &  $ Z_{6,4} $  & weak & $3^{001}$ & $i$
			\\
			\hline
			$Z_2$ & $02$ & ($0$$0$$0$) & $0$ & $1$
			\\
			\hline
			$Z$ & $50$ & ($0$$0$$\bar{1}$) & $0$ & $0$
			\\
			\hline
			\hline
		\end{tabular}
	\end{center}
	\caption{\label{147_mapping}Topological invariants and SIs of $P\overline{3}$ decoration generators.}
\end{table}

\subsection{Generating MSG in type-2 MSGs}
\begin{itemize}
\item MSG 81.34 $P\overline{4}1'$. This type-2 MSG has the following $\mathbb{Z}_2$ indicator:
\begin{equation}
z_{2,S_4}^\prime = -\frac{1}{2}\mu_{S_4}=
\sum_{\mathbf{k}\in K_{S_4}} \frac{1}{2}(n_{\mathbf{k}}^{-\sqrt{2}}-n_{\mathbf{k}}^{\sqrt{2}})\\
\end{equation} 
where $n_{\mathbf{k}}^{\pm\sqrt{2}}$ is the number of Kramer pairs at $S_4$-invariant TRIMs with $\operatorname{tr}\left[D\left(S_{4}\right)\right]=\pm\sqrt{2}$. 
As shown in Ref.\cite{song2018quantitative}, $z_{2,S_4}^\prime=1$ represents a STI.

We remark that all other SIs in type-2 MSGs can be induced from type-1 generating SIs, although their correspondence to topological invariants could change because the definition of invariants is different in type-2 MSGs.
\end{itemize}

\subsection{Generating MSGs in type-3 MSGs}
\subsubsection{$Pnc'c'$, $X_{BS}=\mathbb{Z}_{n}$, Classification=$\mathbb{Z}$}
There are three generating type-3 MSGs and their SIs can be seen as generalizations of SI formulas of $P2,P4$, and $P6$ with an extra $\frac{1}{2}$, because their coirreps are two-dimensional with the same $C_n$ eigenvalues on $k_z=\pi$ plane. 

\begin{itemize}
	\item  MSG 27.81 $Pc'c'2$.
	\begin{equation}
	z_{2C}^\prime = \frac{1}{2}z_{2C} \bmod 2 \\ 
	\end{equation}
	
	\item MSG 103.199  $P4c'c'$
	\begin{equation}
	z_{4C}^\prime = \frac{1}{2}z_{4C} \bmod 4 \\ 
	\end{equation}
	
	\item MSG 184.195 $P6c'c'$.
	\begin{equation}
	z_{6C}^\prime = \frac{1}{2}z_{6C} \bmod 6 \\ 
	\end{equation}
	where $z_{nC}$ on the right hand side has not taken modulo.
\end{itemize}

The TCI classifications of these three MSGs are all $\mathbb{Z}$, i.e., the translation decoration. This decoration has even weak invariant $\delta_{w,3}$ and the Chern number can only take even numbers on $k_z=0/\pi$ planes. The three $Pnc'c'$ all have $\{C_{nz},\{M_x\cdot T|00\frac{1}{2}\},\{M_y\cdot T|00\frac{1}{2}\}\}$, and when there is a layer with Chern number $C=m$ on $z=N$ plane, there will be another plane with the same Chern number on the $z=N+\frac{1}{2}$ plane, which enforces the weak invariant $\delta_{w,3}=2m$.

For MSG 27.81, as $C_{2z}\{M_x\cdot T|00\frac{1}{2}\}=-\{M_x\cdot T|00\frac{1}{2}\}C_{2z}$, where the minus sign comes from the SU(2) matrices, and the eigenvalue of $C_{2z}$ is $\pm i$, which contributes an extra minus when moved outside $T$, the two layers connected by $\{M_x\cdot T|00\frac{1}{2}\}$ must have the same $C_{2z}$ eigenvalues, which validates the $z_{2C}^\prime$ formula.

For MSG 103.199, we have 
$C_{4z}\{M_x\cdot T|00\frac{1}{2}\}=\{M_x\cdot T|00\frac{1}{2}\}C_{4z}^{-1}$. Because the eigenvalues of $C_{4z}$ and $C_{4z}^{-1}$ are conjugated, the two layers connected by $\{M_x\cdot T|00\frac{1}{2}\}$ must have the same $C_{4z}$ eigenvalues, which validates the $z_{4C}^\prime$ formula. Similar argument holds for MSG 184.195. 

As a result, we conclude that for these three MSGs, the $z_{nC}^\prime$ indicator is valid and equals 1 when the weak invariant $\delta_{w,3}=2$, i.e., the mapping between weak invariant and SI is
\begin{equation}
z_{nC}^\prime=\frac{1}{2}\delta_{w,3} \bmod n
\end{equation}

We remark that when there are $C_n,n=2,4,6$ rotation plus two glide-$T$ ($G_x\cdot T,G_y\cdot T$) symmetries, we have $(G_{x/y}\cdot T)^2=-1$ on $k_z=\pi$ plane, and it can be proven that the Berry phase of half of the loop chosen Fig.\ref{Pn_HSP} is quantized to $0$ or $\pi$, which enforces the Chern number of $k_z=\pi$ plane to be even numbers, as shown for SG 27, 103 and 184 in Ref. \cite{song2018diagnosis}.

\paragraph{$z_{nC}^\prime$ in other type-3 MSGs}
The $z_{nC}^\prime, n=2,4$ also apply in a few other MSGs, which are either parent MSGs of $Pnc'c'$, or have a different Bravais lattice.
\begin{itemize}
	\item $z_{2C}^\prime$. MSG 39.199 $Ab'm'2$ and 45.238 $Ib'a'2$ have different lattices from $Pc'c'2$, while	MSG 54.342 $Pc'c'a$ and 56.369 $Pc'c'n$ are parent MSGs of $Pc'c'2$.
	
	\item $z_{4C}^\prime$. MSG 108.237 $I4c'm'$ has a different lattices from $P4c'c'$, while MSG 130.429 $P4/nc'c'$ is a parent MSGs of $P4c'c'$.
\end{itemize}

The $z_{nC}^\prime$ indicator naturally apply in parent MSGs, because we can always use a group's subgroup to calculate its SI, while the case of different Bravais lattice need further clarification, which we leave to the next section.

\subsection{Induce SIs in other MSGs from generating MSGs}
To induce the SIs in all MSGs, we first find the maximum unitary subgroup with nontrivial SI group and then use the unitary subgroup's SI formula to extract effective SI formulas by examining nontrivial BSs. 

Notice that the generating MSGs we considered here are all in a primitive lattice. For an MSG $M$ having the same crystalline symmetries but a complicated lattice, the SI group of $M$ is usually a subgroup of the corresponding primitive lattice MSG $M_0$. One can induce the SI formula of $M$ from $M_0$ by applying the SI formulas of $M_0$ to the nontrivial BSs of $M$ and extract the effective SI formulas. On the other hand, one can also extract the SI formula from the quantitative mappings between TCI classification generators and SIs, as a real space construction of $M$ is also compatible with $M_0$, and they have the same mapping between invariants and SIs. Caution that the HSPs of $M$ and $M_0$ are not the same, but the strong SIs like $z_{4P}^\prime$ remains stable under a change of Bravais lattice, which can be seen from the BZ folding process.

\subsubsection{8 special MSGs with $z_{4P}^\prime$ indicator}
There are 8 type-3 MSGs that have $z_{4P}^\prime$ as SI but their coirreps are not all twofold degenerate to form effective Kramer pairs: MSG 83.45 $P4^\prime/m$, 87.77 $I4^\prime/m$, 124.354 $P4'/mc'c$, 124.355 $P4'/mcc'$, 127.391 $P4'/mbm'$, 128.402 $P4'/mn'c$, 128.403 $P4'/mnc'$, and 140.545 $I4'/mcm'$. We observe that the $z_{4P}^\prime$ formula is valid because (i) they take odd values for nontrivial BSs s.t. they can generate the SI group, and (ii) they equal to zero for all AIs. Note that the $z_{4P}^\prime$ formula in these MSGs is only effective and may become invalid when the origin point is changed. We fix the coordinate system by adopting the convention of Bilbao.

For these 8 MSGs, we need to determine the value of $z_{4P}^\prime$ for the $Z_2$ decorations. Take MSG 83.45 $P4'/m$ as an example. It can be seen that its $Z_2$ decoration is the same as the $Z_2$ decoration in $Pmmm$ if we set the lattice bases $a_1=a_2$. Their common invariants are thus the same with  $C_{m,k_z}=(1,-1,0,0),\delta_i=1$. As a result, they must share the same parities at all TRIMs and thus have $z_{4P}^\prime=3$, as proved using the BHZ model. 

MSG 87.77 $I4^\prime/m$ has the same symmetry operations but a body-centered lattice with $P4'/m$. Because $z_{4P}^\prime$ is a strong index and is stable against the BZ folding, $I4^\prime/m$ must have the same mapping between invariants and $z_{4P}^\prime$. 

The other 6 MSGs all have $P4'/m$ or $I4'/m$ as a subgroup, and their $Z_2$ decorations are compatible with the $Z_2$ decoration in $P4'/m$, i.e., have the same common invariants. As a result, they also take $z_{4P}^\prime=3$.

\subsubsection{The BZ folding process}
\begin{figure*}
	\centering
	\includegraphics[width=0.8\textwidth]{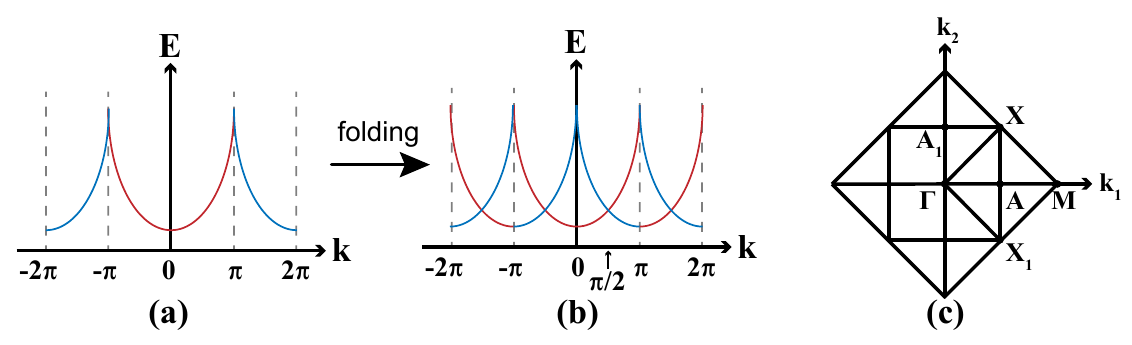}
	\caption{\label{BZfolding}(a), (b). Band folding on the 1D BZ. (c). HSPs of 2D BZ.}
\end{figure*}
\paragraph{Inversion-symmetric BZ}
First, let's consider a 1D BZ with inversion symmetry, as shown in Fig.\ref{BZfolding}. The HSPs are two inversion-invariant points $k=0,\pi$. We plot a schematic energy band, which are symmetric at $k=0,\pi$. When we double the lattice vector, the reciprocal lattice vector is halved and becomes $G=\pi$, leading to a folded BZ. The original TRIMs $k=0,\pi$ becomes equivalent, and $k=\pi/2$ arises as a new TRIM. The band at $k=\pi/2$ must be two-fold degenerate, as the original band is inversion symmetric at $k=\pi$. Assume the parity in the unfolded BZ is $D_0(P),D_\pi(P)$, then the parity at the new TRIMS in the folded BZ is 
\begin{equation}
\begin{aligned}
	D_0^\prime(P)&=
	\left(
	\begin{matrix}
		D_0(P) & 0 \\
		0 & D_\pi(P)\\
	\end{matrix}
	\right),\\
	D_{\pi/2}^\prime(P)&=
	\left(
	\begin{matrix}
		0 & 1 \\
		1 & 0\\
	\end{matrix}
	\right)
	\approxeq
	\left(
	\begin{matrix}
		1 & 0 \\
		0 & -1\\
	\end{matrix}
	\right)
\end{aligned}
\end{equation}
where $\approxeq$ means two matrices are similar. As a result, the summation $\sum_k(N^-_k-N^+_k)$ is unchanged when the BZ is folded.

Next, let's consider a 2D BZ in Fig.\ref{BZfolding}(c), which can be the $k_z=0$ plane of a base-centered or body-centered 3D lattice. The original BZ has 4 TRIMs $\Gamma, M, X, X_1$. After folding, $\Gamma, M$ are folded together, and $X,X_1$ are also folded. There arise two new TRIMs $A,A_1$. Similar to the 1D case, the representation matrices of inversion at $A,A_1$ are off-diagonal and have opposite inversion eigenvalues. As a result, the summation $\sum_k(N^-_k-N^+_k)$ is also unchanged when the 2D BZ is folded.

From the above analysis, we conclude that the $z_{4P}$ indicator is stable when we change the Bravais lattice, so as $z_{4P}^\prime$.

\paragraph{$C_4$-symmetric BZ}
The $z_{4C}$ indicator is also stable when the body-centered lattice is changed to the simple (principal) lattice. In a body-centered lattice BZ in Fig.\ref{BZfolding}(c), the reciprocal lattice vectors are $(1,1,0),(1,0,1),(0,1,1)$. On $k_z=0$ plane, there are two $C_4$-invariant HSPs $\Gamma(0,0,0)$ and $M(1,0,0)$ and a $C_2$-invariant HSP $X(\frac{1}{2},\frac{1}{2},0)$. These three HSPs are used to calculate the $z_{4C}$ indicator. 

When changed to the simple lattice, the reciprocal lattice vectors become the usual $(1,0,0),(0,1,0),(0,0,1)$. On the $k_z=0$ plane, $M$ is folded to $\Gamma$, $X$ becomes a $C_4$-invariant HSP, and there appears a new $C_2$-invariant HSP $A(\frac{1}{2},0,0)$. Assume the three HSPs have the following representations in the body-centered lattice:
\begin{equation}
	D_\Gamma(C_4),\ D_M(C_4),\ D_X(C_2)
\end{equation}
After folding to the simple lattice BZ, the three new HSPs have the following doubled representations:
\begin{equation}
\begin{aligned}
	D_\Gamma^\prime(C_4)&=
	\left(
	\begin{matrix}
	 	D_\Gamma(C_4) & 0 \\
		0 & D_M(C_4)\\
	\end{matrix}
	\right),\ 
	D_X^\prime(C_4)=
	\left(
	\begin{matrix}
		0 & 1 \\
		D_X(C_2) & 0\\
	\end{matrix}
	\right),\\
	D_A^\prime(C_2)&=
	\left(
	\begin{matrix}
		0 & 1 \\
		-1 & 0\\
	\end{matrix}
	\right)
\end{aligned}
\end{equation}

In the $z_{4C}$ formula, the $C_4$ and $C_2$ eigenvalues at $\Gamma,M,X$ are multiplied together. It can be seen that the product of determinant of the three representation matrices remains unchanged, which confirms that the value of $z_{4C}$ remains stable when the lattice is changed.

\clearpage
\section{Corner cases of SI}\label{AppendixH}
There are 10 corner cases of SIs in MSGs that cannot be induced from the generating SIs. We give them SI formulas written specifically in their coirreps. These formulas can be obtained from the Smith normal form decomposition.

\subsection{Corner cases in type-1 MSGs}
MSG 87.75 $I4/m$ has indicator group $\mathbb{Z}_{4,4}$ and two nontrivial BSs:
\begin{equation}
\begin{aligned}
b_1 &= \overline{P}_6  -\overline{P}_8  -\overline{X}_4 + \overline{X}_5 \\  
b_2 &= \overline{M}_6 + \overline{M}_7  -\overline{M}_{10}  -\overline{M}_{11} \\
\end{aligned}
\end{equation}
We can calculate $z_{4m,0}^+, z_{4m,0}^-, z_{2,S_4}, z_{2P}^\prime$ for them, which equal (2,0,1,1) and (3,1,1,1). It can be seen that $z_{4m,0}^+, z_{4m,0}^-$ only form a $\mathbb{Z}_{2,4}$ group, thus cannot be taken directly as the indicator. However, their combination
\begin{equation}
\begin{aligned}
z_{4m,87}^+ & = \frac{1}{2}(z_{4m,0}^+ + z_{4m,0}^-) \bmod 4\\
z_{4m,87}^- & = \frac{1}{2}(z_{4m,0}^+  - z_{4m,0}^-) \bmod 4\\
\end{aligned}
\end{equation}
can generate the $\mathbb{Z}_{4,4}$ group, and the two nontrivial BSs have (1,1) and (2,1) of the new indicators. Note that $z_{4m,0}^+, z_{4m,0}^-$ in the indicator formula have already mod 4.

This MSG has TCI classification $\mathbb{Z}^2$, whose generators can be taken as a translation decoration and a $Z_2$ decoration, which has mirror Chern number (2,0) and (1,-1) respectively, and can be diagnosed by $z_{4m,87}^+$ and $z_{4m,87}^-$. These two indicators are well-defined because they take integer values on any combination of the two generators.

Note that this MSG does not have Weyl states that can be detected by indicator, because it has an extra mirror symmetry compared to SG 81 and 82. Weyl points always appear in $8n$ number and can be pairwise annihilated at generic points on $k_z=0$ plane without changing irreps at HSPs, while in SG 81 and 82, there are $4n$ Weyl points and can only be annihilated at $S_4$ invariant HSPs.

We remark that there are two type-3 MSGs 139.537  $I4/mm'm'$ and 140.547 $I4/mc'm'$, whose unitary halving subgroup is SG 87, also have these two indicators. Type-4 MSG 87.80 $Ic4/m$ has a $\mathbb{Z}_4$ indicator and can be taken as $z_{4m,0}^+$ or $z_{4m,87}^-$, which are equivalent.

\subsection{Corner case in type-2 MSGs}
Type-2 MSG 226.123 $Fm\overline{3}c1'$ has indicator group $\mathbb{Z}_8$, but the $z_8=2z_{2,S_4}^\prime-z_{4P}^\prime$ formula introduced in type-1 SG $P4/mmm$ cannot be used, because some AIs have non-zero $z_8$. As a result, we adopt the SI formula from the Smith normal form: 
\begin{equation}
\begin{aligned}
z_{8, 226.123} &= 3N(\overline{\Gamma}_6) + 3N(\overline{\Gamma}_7) + 4N(\overline{\Gamma}_9) + 2N(\overline{\Gamma}_{10}) + \\
&4N(\overline{L}_4\overline{L}_4) - 4N(\overline{L}_5\overline{L}_6) 
- 3N(\overline{X}_6) + N(\overline{X}_7) \bmod 8
\end{aligned}
\end{equation}

One may wonder why the $z_8$ formula holds in all other MSGs that has $\mathbb{Z}_8$ indicator but only fails in MSG 226.123. This is because 226.123 is the only non-symmorphic MSG with $\mathbb{Z}_8$ indicator, and the inversion and $S_4$ centers in it cannot coincide.

\subsection{Corner cases in type-3 MSGs}
There are 6 corner cases of type-3 MSG whose indicators can not adopt the generating SIs. They can be divided into two classes: five of them have a $\mathbb{Z}_{2}$ indicator and anti-unitary symmetries Glide-$T$ or Mirror-$T$, and the other one with an $\mathbb{Z}_{4}$ indicator group and $C_4\cdot T$. These indicators all represent gapped states, as proved later.

The $\mathbb{Z}_{2}$ class has 5 MSGs:  41.215 $Ab'a'2$, 42.222 $Fm'm'2$, 60.424 $Pb'cn'$, 68.515 $Cc'c'a$, and 110.249 $I4_1c'd'$. The nontrivial BSs of the corner case $\mathbb{Z}_2$ indicator in these MSGs all correspond to topological states with even Chern number on the $k_z=0/\pi$ plane in the conventional BZ, which cannot be diagnosed using the $C_2$ Chern number formula. The $C_2$ Chern number formula with an extra $\frac{1}{2}$ also fails because it gives nonzero values when applied to some AIs. As a result, we adopt the SI formula from the Smith normal form in these MSGs. Note that 60.424 and 68.515 have indicator group $\mathbb{Z}_{2,2}$, with one of the $\mathbb{Z}_2$ being $z_{2P}^\prime$.

\begin{itemize}
\item MSG 41.215 $Ab'a'2$:
\begin{equation}
	z_{2,41.215} = N(\overline{\Gamma}_3) \bmod 2
\end{equation}
\item MSG 42.222 $Fm'm'2$
\begin{equation}
z_{2,42.222} = N(\overline{\Gamma}_3) - N(\overline{A}_3) \bmod 2
\end{equation}
\item MSG 60.424 $Pb'cn'$
\begin{equation}
z_{2,60.424} = N(\overline{\Gamma}_3) \bmod 2
\end{equation}
\item MSG 68.515 $Cc'c'a$
\begin{equation}
z_{2,68.515} = N(\overline{Z}_3\overline{Z}_5) - N(\overline{C}_3\overline{C}_5) \bmod 2
\end{equation}
\item MSG 110.249 $I4_1c'd'$
\begin{equation}
z_{2,110.249} = N(\overline{M}_5\overline{M}_6) \bmod 2 
\end{equation}
\end{itemize}

The $\mathbb{Z}_{4}$ class has only one MSG: 135.487 $P4_2'/mbc'$. The nontrivial BS in this MSG has $z_{4P}=2,\ z_{2m,k_z}=(1,1,0,0)$, which corresponds to the $Z_2$ decoration. Note that $z_{4P}^\prime=2$ for some AIs, which fails to serve as the indicator. We adopt the SI formula from the Smith normal form in this MSG:
\begin{itemize}
\item MSG 135.487 $P4_2'/mbc'$
\begin{equation}
z_{4,135.487} = N(\overline{\Gamma}_5) - 2N(\overline{R}_5\overline{R}_6)
+ N(\overline{S}_5) - 2N(\overline{T}_3) \bmod 4
\end{equation}
\end{itemize}

\subsection{Corner cases in type-4 MSGs}
There are 2 corner cases in type-4 MSGs, i.e., 37.185 $C_acc2$ and 104.209 $P_C4nc$, which both have a $\mathbb{Z}_2$ indicator corresponding to Weyl states, with indicator formulas:
\begin{itemize}
\item MSG 37.185 $C_acc2$
\begin{equation}
z_{2, 37.185} = N(\overline{R}_3) + N(\overline{T}_5\overline{T}_5)
+ N(\overline{Z}_5\overline{Z}_5) \bmod 2
\end{equation}

\item MSG 104.209 $P_C4nc$
\begin{equation}
z_{2,104.209} = 2N(\overline{R}_2\overline{R}_4) \bmod 2
\end{equation}
\end{itemize}

For 37.185, the indicator can be interpreted as $z_2 = N_R^-+(N_T^-+N_Z^-)/2 \bmod 2$, which can be justified using the similar method for the spinless Weyl state in SG 37 shown in Ref.\cite{song2018diagnosis} 

For 104.209, the interpretation of the indicator is similar to the spinless Weyl states in SG 103 in Ref.\cite{song2018diagnosis}, with details showing below. 

\begin{figure}[ht]
	\centering
	\includegraphics[width=0.2\textwidth]{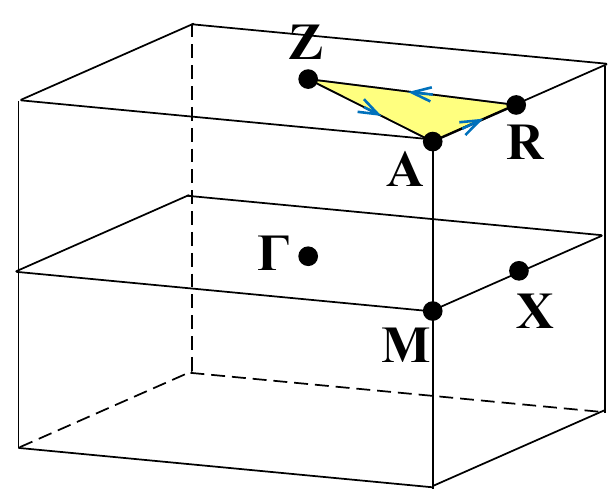}
	\caption{\label{BZ_104}A closed loop in the BZ of MSG 104.209.}
\end{figure}

We choose the loop Z-A-R that encloses one eighth of the 2D BZ of $k_z=\pi$ in Figure.\ref{BZ_104}. The high-symmetry-line Z-A, A-R and R-Z have glide symmetry $G_{1\overline{1}0}, G_y$ and $G_x$, respectively. Starting from $Z$, we choose the two eigenstates of $G_{1\overline{1}0}$ as the bases: $\left|u_1(Z)\right\rangle=\left|u_{1\overline{1}0,+i}(Z)\right\rangle,\left|u_2(Z)\right\rangle=\left|u_1(Z)\right\rangle^*$, and consider the phase they gain after the loop evolution. 

The two bases gain a phase $\theta_1$ after evolving from $Z$ to $A$ and become
\begin{equation}
\left(\begin{array}{cc}
e^{i \theta_1} & 0\\ 
0 & e^{-i \theta_1} 
\end{array}\right)
\left(\begin{array}{l}
\left|u_{1}\right\rangle \\ 
\left|u_{2}\right\rangle
\end{array}\right)
\end{equation}
Thanks to the $C_2T$ symmetry, the phases two bases gain are conjugate to each other. At three HSPs $Z, A, R$, we need to perform a basis transformation, because the three glide symmetries are not simultaneously diagonal. For example, at $A$, we can use 
\begin{equation}
C=\left(\begin{array}{cc}
-i(1+\sqrt{2}) & i(-1+\sqrt{2})\\ 
1 & 1
\end{array}\right)
/\sqrt{-2i\sqrt{2}}
\end{equation}
to transform the eigenstates of $G_{1\overline{1}0}$ to $G_y$. However, there are two irreps $\overline{A}_6$ and $\overline{A}_7$ of SG 104 which have different $C_4$ eigenvalues. For $\overline{A}_6$, $C$ transforms $G_{1\overline{1}0}=i,-i$ to $G_y=-i,i$, while for $\overline{A}_7$, $C$ transforms $G_{1\overline{1}0}=i,-i$ to $G_y=i,-i$. As a result, we can define a representation matrix to distinguish this difference:
\begin{equation}
D_A(C_4)=\left\{
\begin{array}{ll}
\left(\begin{array}{cc}
0 & 1\\ 
1 & 0 
\end{array}\right)
& \text{if } \overline{A}_6 \\ 
\left(\begin{array}{cc}
1 & 0\\ 
0 & 1 
\end{array}\right) 
&\text{if } \overline{A}_7
\end{array}\right.
\end{equation}
Similarly, we can define such representation matrix at $R$ and $Z$:
\begin{equation}
\begin{aligned}
D_R(C_2)&=\left\{
\begin{array}{ll}
\left(\begin{array}{cc}
0 & 1\\ 
1 & 0 
\end{array}\right)
& \text{if } \overline{R}_2\overline{R}_4 \\ 
\left(\begin{array}{cc}
1 & 0\\ 
0 & 1 
\end{array}\right) 
&\text{if } \overline{R}_3\overline{R}_5
\end{array}\right.\\
D_Z(C_4)&=\left\{
\begin{array}{ll}
\left(\begin{array}{cc}
1 & 0\\ 
0 & 1 
\end{array}\right)
& \text{if } \overline{Z}_6 \\ 
\left(\begin{array}{cc}
0 & 1\\ 
1 & 0 
\end{array}\right) 
&\text{if } \overline{Z}_7
\end{array}\right.
\end{aligned}
\end{equation}

Assume the phase gained on A-R and R-Z for $G_y$ and $G_x$ eigenstates are $\theta_2$ and $\theta_3$, respectively, then the total phase gained after the loop evolution is:
\begin{equation}
\begin{aligned}
M&=D_Z(C_4)C^{-1}
\left(\begin{array}{cc}
e^{i \theta_3} & 0\\ 
0 & e^{-i \theta_3} 
\end{array}\right)
D_R(C_2)
\left(\begin{array}{cc}
e^{i \theta_2} & 0\\ 
0 & e^{-i \theta_2} 
\end{array}\right)\\ 
&\cdot D_A(C_4)C^{-1}
\left(\begin{array}{cc}
e^{i \theta_1} & 0\\ 
0 & e^{-i \theta_1} 
\end{array}\right)
\end{aligned}
\end{equation}

The Berry phase of the loop becomes
\begin{equation}
\Phi_B=\det(M)=\det(D_Z(C_4)D_R(C_2)D_A(C_4))
\end{equation}

Note that in type-4 MSG 104.209, there are extra degeneracy, i.e., $\overline{Z}_6\overline{Z}_6$ and $\overline{Z}_7\overline{Z}_7$ (same for $A$). Therefore the determinant $\det(D_{Z/A}(C_4))\equiv 1$, while $\det(D_R(C_2))=-1$ if there are an odd number of $\overline{R}_2\overline{R}_4$. As a result, 
\begin{equation}
\Phi_B=-1\qquad \text{if } N(\overline{R}_2\overline{R}_4)=1 \bmod 2
\end{equation}
and thus $z_{2,104.209}=1$ indicates the existence of the Weyl state.

\subsection{Interpretation of the type-3 corner case indicator formulas}

First, we show the indicators in type-3 corner cases do not represent Weyl states. The possible Weyl points configurations can be determined by looking into the symmetry operations of the MSG, with proper rotations and time-reversal symmetry leave the Weyl point chirality unchanged, while improper operations reverse the chirality.
\begin{enumerate}
\item MSG 41.215 $Ab'a'2$.
This MSG has symmetries $\{1,C_{2z},\{M_x\cdot T|\frac{1}{2}0\frac{1}{2}\},
\{M_y\cdot T|\frac{1}{2}0\frac{1}{2}\}\}$, and $4n$ Weyl points can be created or annihilated at generic points on the $z$-axis without changing irreps at HSPs, and thus cannot be diagnosed using SI.

\item MSG 42.222 $Fm'm'2$.
This MSG has symmetries $\{1,C_{2z},M_x\cdot T,M_y\cdot T\}$. The Weyl points configuration is similar to MSG 42.215 and cannot be diagnosed using SI.

\item MSG 60.424 $Pb'cn'$.
This MSG has generators $\{\{C_{2y}|00\frac{1}{2}\},P,\{M_x\cdot T|\frac{1}{2}\frac{1}{2}0\} \}$. $4n$ Weyl points can be created or annihilated at a generic point $z=z_0$ on the $z$-axis, with the other $4n$ Weyl points at $z=-z_0$. SI fails to diagnose them.

\item MSG 68.515 $Cc'c'a$.
This MSG has generators $\{\{C_{2z}|0\frac{1}{2}0\},P,\{M_x\cdot T|0\frac{1}{2}\frac{1}{2}\} \}$. The analysis is similar to MSG 60.424.

\item MSG 110.249 $I4_1c'd'$.
This MSG has generators $\{\{C_{4z}|0\frac{1}{2}\frac{1}{4}\},\{M_x\cdot T|00\frac{1}{2}\},\{M_y\cdot T|00\frac{1}{2}\} \}$.  Similarly, $8n$ Weyl points can be created or annihilated around a generic point on the $z$-axis.

\item MSG 135.487 $P4_2'/mbc'$.
This MSG has generators $\{\{C_{2x}|\frac{1}{2}\frac{1}{2}0\},P,\{C_{4z}\cdot T|00\frac{1}{2}\} \}$. Similar to 60.424, $8n$ Weyl points can be created or annihilated around two generic point $z=\pm z_0$ on the $z$-axis.
\end{enumerate}

As a result, the SIs in these 6 type-3 MSGs all correspond to the gapped states. Among them, the SI group of MSG 41.215, 42.222 and 110.249 is $\mathbb{Z}_2$, and their TCI classifications are all $\mathbb{Z}$, generated by the translation decoration, which means a nontrivial SI corresponds to all the odd number copies of the translation decoration. 

MSG 60.424 and 68.515 have $\mathbb{Z}_{2,2}$ indicator group, with one $\mathbb{Z}_2$ being $z_{2P}^\prime$, and their TCI classifications are both $\mathbb{Z}\times\mathbb{Z}_2$. Here we calculate the $\mathbb{Z}_{2,2}$ indicators for the TCI classification generators in these two MSGs.

\paragraph{MSG 60.424 $Pb'cn'$}
In MSG 60.424, there are two types of possible LCs, i.e., stacking 2D layers along the $z$ or $y$ directions. 

\begin{itemize}
\item 2D layers on $z=0,\frac{1}{2}$. The layer at $z=0$ has symmetries $\{1,P, \{C_{2x}\cdot T|(\frac{1}{2}\frac{1}{2}0)\},\{M_{x}\cdot T|(\frac{1}{2}\frac{1}{2}0)\}\}$, while the other layer at $z=\frac{1}{2}$ is generated by acting $g=\{C_{2y}|00\frac{1}{2}\}$ on the $z=0$ layer. Assume the 2D layer at $z=0$ has a wave function $W_1(r)$ and representation $D_{k_xk_y}(R)$, then the layer at $z=\frac{1}{2}$ has wave function $W_2(r)=gW_1(r)$. We calculate the representation matrix of $P$ and $g$ for the 3D Bloch function:
\begin{equation}
\begin{aligned}
a_1(k) &=\frac{1}{\sqrt{N}}\sum_n e^{ik_z\cdot n}W_1(r-n)\\
a_2(k) &=\frac{1}{\sqrt{N}}\sum_n e^{ik_z\cdot n}W_2(r-n)\\
\end{aligned}
\end{equation}

As $P\cdot g=\{1|0,0,-1\}g\cdot P$, we have
\begin{equation}
\begin{aligned}
\hat{P}W_1(r-n) &=D_{k_xk_y}(P) W_1(r+n) \\
\hat{P}W_2(r-n) &=D_{k_xk_y}(P) W_2(r+n-1) \\
\Rightarrow
\hat{P}a_1(k) &=D_{k_xk_y}(P) a_1(Pk) \\
\hat{P}a_2(k) &=e^{-ik_z}D_{k_xk_y}(P) a_2(Pk) \\
\Rightarrow
D_k(P)&=
\left(
\begin{matrix}
D_{k_xk_y}(P) & 0 \\
0 & e^{-ik_z}D_{k_xk_y}(P) \\
\end{matrix}
\right)
\end{aligned}
\end{equation}
As a result, $D_k(P)=D_{k_xk_y}(P)\oplus D_{k_xk_y}(P)$  for HSPs on the $k_y=0$ plane, while $D_k(P)=D_{k_xk_y}(P)\oplus -D_{k_xk_y}(P)$ for the $k_y=\pi$ plane.

For $g$, as $g^2=-1$, we have
\begin{equation}
\begin{aligned}
\hat{g}a_1(k) &= a_2(gk) \\
\hat{g}a_2(k) &=-a_1(gk) \\
\Rightarrow
D_k(g)&=
\left(
\begin{matrix}
0 & 1 \\
-1 & 0\\
\end{matrix}
\right)
\end{aligned}
\end{equation}
The representation matrix of $g$ is off-diagonal, which has zero trace. As a result, the numbers of $\pm i$ bands must be the same.

\item 2D layers on $y=0,\frac{1}{2}$. The layer at $y=0$ has symmetries $\{1,P, \{C_{2y}|(00\frac{1}{2})\},\{M_{y}|(00\frac{1}{2})\}\}$, while the other layer at $y=\frac{1}{2}$ is generated by $h=\{C_{2x}\cdot T|\frac{1}{2}\frac{1}{2}0\}$. Assume the 2D layer at $y=0$ has a wave function $W_1(r)$ and representation $D_{k_xk_z}(R)$, then the layer at $y=\frac{1}{2}$ has wave function $W_2(r)=hW_1(r)$. We calculate the representation matrix of $P$ and $g=\{C_{2y}|(00\frac{1}{2})\}$ for the 3D Bloch function.

As $P\cdot h=\{1|-1,-1,0\}h\cdot P$, we have
\begin{equation}
\begin{aligned}
\hat{P}a_1(k) &=D_{k_xk_z}(P) a_1(Pk) \\
\hat{P}a_2(k) &=e^{-i(k_x+k_y)}D_{k_xk_z}(P) a_2(Pk) \\
\Rightarrow
D_k(P)&=
\left(
\begin{matrix}
D_{k_xk_z}(P) & 0 \\
0 & e^{-i(k_x+k_y)}D_{k_xk_z}(P) \\
\end{matrix}
\right)
\end{aligned}
\end{equation}
As a result, $D_k(P)=D_{k_xk_z}(P)\oplus D_{k_xk_z}(P)$  for $\Gamma,C$, while $D_k(P)=D_{k_xk_z}(P)\oplus -D_{k_xk_z}(P)$ for $Y,C$.

For $g$, as $g\cdot h=-\{1|-1,0,1\}h\cdot P$ (where the extra minus sign comes from the SU$_2$ matrix), we have
\begin{equation}
\begin{aligned}
\hat{g}a_1(k) &=D_{k_xk_z}(g) a_1(gk) \\
\hat{g}a_2(k) &=-e^{i(k_x-k_z)}D_{k_xk_z}^*(g) a_2(gk) \\
\Rightarrow
D_k(g)&=
\left(
\begin{matrix}
D_{k_xk_z}(g) & 0 \\
0 & -e^{i(k_x-k_z)}D_{k_xk_z}^*(g) \\
\end{matrix}
\right)
\end{aligned}
\end{equation}
where the complex conjugation is because $h$ is anti-unitary. As a result, $D_k(g)=D_{k_xk_z}(g)\oplus D_{k_xk_z}(g)$  for $\Gamma,Z$, while $D_k(g)=D_{k_xk_z}(g)\oplus -D_{k_xk_z}(g)$ for $Y,C$.
\end{itemize}

MSG 60.424 has the following AIs
\begin{widetext}
\begin{equation}
\begin{aligned}
a_{1}&=  \overline{A}_{2}\overline{A}_{2} + 2\overline{B}_{2} + 2\overline{C}_{5}\overline{C}_{6} + \overline{D}_{2}\overline{D}_{2} + \overline{E}_{2}\overline{E}_{2} + 2\overline{\Gamma}_{5} + 2\overline{\Gamma}_{6} + \overline{Y}_{3}\overline{Y}_{6} + \overline{Y}_{4}\overline{Y}_{5} + \overline{Z}_{3}\overline{Z}_{5} + \overline{Z}_{4}\overline{Z}_{6}  \\ 
a_{2}&=  \overline{A}_{2}\overline{A}_{2} + 2\overline{B}_{2} + 2\overline{C}_{3}\overline{C}_{4} + \overline{D}_{2}\overline{D}_{2} + \overline{E}_{2}\overline{E}_{2} + 2\overline{\Gamma}_{3} + 2\overline{\Gamma}_{4} + \overline{Y}_{3}\overline{Y}_{6} + \overline{Y}_{4}\overline{Y}_{5} + \overline{Z}_{3}\overline{Z}_{5} + \overline{Z}_{4}\overline{Z}_{6}  \\ 
a_{3}&=  \overline{A}_{2}\overline{A}_{2} + 2\overline{B}_{2} + 2\overline{C}_{3}\overline{C}_{4} + \overline{D}_{2}\overline{D}_{2} + \overline{E}_{2}\overline{E}_{2} + 2\overline{\Gamma}_{5} + 2\overline{\Gamma}_{6} + \overline{Y}_{3}\overline{Y}_{6} + \overline{Y}_{4}\overline{Y}_{5} + \overline{Z}_{3}\overline{Z}_{5} + \overline{Z}_{4}\overline{Z}_{6}  \\ 
a_{4}&=  \overline{A}_{2}\overline{A}_{2} + 2\overline{B}_{2} + 2\overline{C}_{5}\overline{C}_{6} + \overline{D}_{2}\overline{D}_{2} + \overline{E}_{2}\overline{E}_{2} + 2\overline{\Gamma}_{3} + 2\overline{\Gamma}_{4} + \overline{Y}_{3}\overline{Y}_{6} + \overline{Y}_{4}\overline{Y}_{5} + \overline{Z}_{3}\overline{Z}_{5} + \overline{Z}_{4}\overline{Z}_{6}  \\ 
a_{5}&=  \overline{A}_{2}\overline{A}_{2} + 2\overline{B}_{2} + \overline{C}_{3}\overline{C}_{4} + \overline{C}_{5}\overline{C}_{6} + \overline{D}_{2}\overline{D}_{2} + \overline{E}_{2}\overline{E}_{2} + 2\overline{\Gamma}_{3} + 2\overline{\Gamma}_{5} + \overline{Y}_{3}\overline{Y}_{6} + \overline{Y}_{4}\overline{Y}_{5} + 2\overline{Z}_{3}\overline{Z}_{5}  \\ 
a_{6}&=  \overline{A}_{2}\overline{A}_{2} + 2\overline{B}_{2} + \overline{C}_{3}\overline{C}_{4} + \overline{C}_{5}\overline{C}_{6} + \overline{D}_{2}\overline{D}_{2} + \overline{E}_{2}\overline{E}_{2} + 2\overline{\Gamma}_{4} + 2\overline{\Gamma}_{6} + \overline{Y}_{3}\overline{Y}_{6} + \overline{Y}_{4}\overline{Y}_{5} + 2\overline{Z}_{4}\overline{Z}_{6}  \\ 
\end{aligned}
\end{equation}
\end{widetext}
and two nontrivial BSs:
\begin{equation}
\begin{aligned}
b_{1}&=  2\overline{\Gamma}_{3} -2\overline{\Gamma}_{6} + \overline{Z}_{3}\overline{Z}_{5} - \overline{Z}_{4}\overline{Z}_{6}  \\ 
b_{2}&=  \overline{\Gamma}_{3} + \overline{\Gamma}_{4} - \overline{\Gamma}_{5} - \overline{\Gamma}_{6}  \\ 
\end{aligned}
\end{equation}
and three Wyckoff positions:
\begin{equation}
\begin{aligned}
4a&=\{(0,0,0),(\frac{1}{2},\frac{1}{2},0),(0,0,\frac{1}{2}),(\frac{1}{2},\frac{1}{2},\frac{1}{2})\}\\
4b&=\{(0,\frac{1}{2},0),(\frac{1}{2},0,0),(0,\frac{1}{2},\frac{1}{2}),(\frac{1}{2},0,\frac{1}{2})\}\\ 4c&=\{(0,y,\frac{1}{4}),(\frac{1}{2},-y+\frac{1}{2},\frac{3}{4}),(0,-y,\frac{3}{4}),(\frac{1}{2},y+\frac{1}{2},\frac{1}{4})\}
\end{aligned}
\end{equation}
$a_{1,2}, a_{3,4}$ and $a_{5,6}$ correspond to Wyckoff positions $4a, 4b$ and $4c$, respectively. The two nontrivial BSs $b_1$ and $b_2$ have indicators $(z_{2P}^\prime,z_{2,60.424})=(1,0), (1,1)$.

It can be seen that $a_{1,2,3,4}$ are consistent with the $z$-directional LC, while $a_{5,6}$ are  consistent with the $y$-directional LC. It can also be seen that $b_2+a_{1,3}, -b_2+a_{2,4}$ satisfies the representation matrices of $z$-directional LC, $b_1+a_6, -b_1+a_5$ satisfies the $y$-directional LC, while other combinations do not satisfy. As a result, the $z$-directional LC of odd Chern number layer has indicator $(1,1)$, while the $y$-directional LC of odd Chern number layer has indicator $(1,0)$, as shown in Table.\ref{table_60424}.

\begin{table}[ht]
	\begin{center}
		\begin{tabular}{c|c|c|c|c}
			\hline
			Decoration & weak & $\delta_i$ & $(z_{2P}^\prime, z_{2,60.424})$ & BS \\ 
			\hline
			z-directional LC($Z_2$) & 000 & 1 & (1,1) & $b_2+$AI \\
			\hline
			y-directional LC(translation) & 020 & 1 & (1,0) & $b_1+$AI\\
			\hline
		\end{tabular}
		\caption{\label{table_60424}Correspondence of invariants and indicators in MSG 60.424}
	\end{center}
\end{table}

\paragraph{MSG 68.515 $Cc'c'a$}
MSG 68.515 has a translation decoration and a $Z_2$ decoration, with the first one being an LC of $C=1$ layers on $z=0,\frac{1}{2}$ planes, and the other one being a non-LC. They have invariants $(\delta_{w,3},\delta_i)=(2,1), (0,1)$, respectively, and their superposition has invariants $(2,0)$. The $\mathbb{Z}_{2,2}$ indicators can be chosen as $(z_{2P}^\prime, z_{2,68.515})$, and we know that the $\delta_i=1$ states have $z_{2P}^\prime=1$. Because any two of the three decorations can be chosen as the generators (e.g., the one with $(\delta_{w,3}, \delta_i) = (0,1)$ and the one with $ (\delta_{w,3}, \delta_i ) =  (2,0) $), none of them can have zero indicator group, i.e., $(z_{2P}^\prime, z_{2,68.515}) = (0,0)$. As a result, the superposition decoration must have $z_{2,68.515}=1$. We then only need to determine the $z_{2,68.515}$ indicator for the translation decoration.

The 2D layer on the $z=0$ plane has symmetries $\{1, P, g=\{C_{2z}|(0\frac{1}{2}0)\},\{M_{z}|(0\frac{1}{2}0)\}\}$, and the layer on the $z=\frac{1}{2}$ plane can be generated by $h=\{C_{2x}\cdot T|(0\frac{1}{2}\frac{1}{2})\}$. Following the derivation in MSG 60.424, with the commutation relation $P\cdot h=\{1|0,-1,-1\}h\cdot P$ and $g\cdot h=-h\cdot g$, the representation matrices of $P$ and $g$ are
\begin{equation}
\begin{aligned}
D_k(P)&=
\left(
\begin{matrix}
D_{k_xk_y}(P) & 0 \\
0 & e^{-i(k_y+k_z)}D_{k_xk_y}(P) \\
\end{matrix}
\right)\\
&=\left\{
\begin{array}{lr}
D_{k_xk_y}(P)\oplus D_{k_xk_y}(P), & \Gamma,Y,B,A \\
D_{k_xk_y}(P)\oplus -D_{k_xk_y}(P), & Z,C,D,E \\
\end{array}
\right.\\
D_k(g)&=
\left(
\begin{matrix}
D_{k_xk_y}(g) & 0 \\
0 & -D_{k_xk_y}^*(g) \\
\end{matrix}
\right)
=D_{k_xk_y}(g)\oplus D_{k_xk_y}(g)
\end{aligned}
\end{equation}

This MSG has the following four AIs (omitting irrelevant AIs) and two $\mathbb{Z}_2$ nontrivial BSs:
\begin{widetext}
\begin{equation}
\begin{aligned}
a_{1}&=  2\overline{A}_{2} + 2\overline{B}_{2} + \overline{C}_{3}\overline{C}_{5} + \overline{C}_{4}\overline{C}_{6} + 2\overline{D}_{2} + 2\overline{E}_{2} + 2\overline{\Gamma}_{5} + 2\overline{\Gamma}_{6} + 2\overline{Y}_{5} + 2\overline{Y}_{6} + \overline{Z}_{3}\overline{Z}_{5} + \overline{Z}_{4}\overline{Z}_{6}  \\ 
a_{2}&=  2\overline{A}_{2} + 2\overline{B}_{2} + \overline{C}_{3}\overline{C}_{5} + \overline{C}_{4}\overline{C}_{6} + 2\overline{D}_{2} + 2\overline{E}_{2} + 2\overline{\Gamma}_{3} + 2\overline{\Gamma}_{4} + 2\overline{Y}_{3} + 2\overline{Y}_{4} + \overline{Z}_{3}\overline{Z}_{5} + \overline{Z}_{4}\overline{Z}_{6}  \\ 
a_{3}&=  2\overline{A}_{2} + 2\overline{B}_{2} + \overline{C}_{3}\overline{C}_{5} + \overline{C}_{4}\overline{C}_{6} + 2\overline{D}_{2} + 2\overline{E}_{2} + 2\overline{\Gamma}_{5} + 2\overline{\Gamma}_{6} + 2\overline{Y}_{3} + 2\overline{Y}_{4} + \overline{Z}_{3}\overline{Z}_{5} + \overline{Z}_{4}\overline{Z}_{6}  \\ 
a_{4}&=  2\overline{A}_{2} + 2\overline{B}_{2} + \overline{C}_{3}\overline{C}_{5} + \overline{C}_{4}\overline{C}_{6} + 2\overline{D}_{2} + 2\overline{E}_{2} + 2\overline{\Gamma}_{3} + 2\overline{\Gamma}_{4} + 2\overline{Y}_{5} + 2\overline{Y}_{6} + \overline{Z}_{3}\overline{Z}_{5} + \overline{Z}_{4}\overline{Z}_{6}  \\ 
b_{1}&=  \overline{Y}_{3} + \overline{Y}_{4} - \overline{Y}_{5} - \overline{Y}_{6}  \\ 
b_{2}&= -2\overline{\Gamma}_{4} + 2\overline{\Gamma}_{5} + \overline{Z}_{3}\overline{Z}_{5} - \overline{Z}_{4}\overline{Z}_{6}  \\ 
\end{aligned}
\end{equation}
\end{widetext}
It can be seen that $a_{1,2,3,4}$ are compatible with the LC of 2D trivial layers on $z=0,\frac{1}{2}$ planes, and $b_2+a_{2,4},-b_2+a_{1,3}$ are compatible with the translation decoration, i.e., placing $C=1$ 2D layers on $z=0,\frac{1}{2}$ planes. Note that combinations of AIs with $b_1$ are not compatible.

Two nontrivial BSs $b_1,b_2$ have $(z_{2P}^\prime, z_{2,68.515})=(1,0),(1,1)$, respectively. As a result, the translation decoration has indicator $(1,1)$, as shown in Table.\ref{table_68515}.

\begin{table}[ht]
	\begin{center}
		\begin{tabular}{c|c|c|c|c}
			\hline
			Decoration & weak & $\delta_i$ & $(z_{2P}^\prime, z_{2,68.515})$ & possible BS \\ 
			\hline
			Translation & 002 & 1 & (1,1) & $b_2+$AI \\
			\hline
			$Z_2$ & 000 & 1 & (1,0) & $b_1+$AI\\
			\hline
			Translation+$Z_2$ & 002 & 0 & (0,1) & $b_1+b_2+$AI\\
			\hline
		\end{tabular}
		\caption{\label{table_68515}Correspondence of invariants and indicators in MSG 68.515}
	\end{center}
\end{table}

\begin{figure}[ht]
\centering
\includegraphics[width=0.2\textwidth]{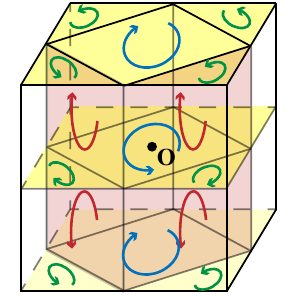}
\caption{\label{135_487Z2}The $Z_2$ decoration of MSG 135.487. Note we do not plot every edge modes for simplicity, which can be inferred from the symmetries.}
\end{figure}

\paragraph{MSG 135.487 $P4_2'/mbc'$}
The $Z_2$ decoration of MSG 135.487 has $C_{m,k_z}=(1,-1,0,0), \delta_i=1$, and we plot it in Fig.\ref{135_487Z2}. It can be seen that this decoration is identical to the $Z_2$ decoration of $Pmmm$ if we first shift the origin point by half of the lattice vector in the $x$ direction and then rotate it by $\pi/4$ in the $z$ direction, which makes the BHZ model still applicable. 

The BHZ model used for describing $Pmmm$ has $(z_{2P,1}^\prime,z_{2P,2}^\prime,z_{2P,3}^\prime,z_{4P}^\prime)=(0,0,0,3)$. Because the three weak indicators all equal zero, the $z_{4P}^\prime$ indicator is stable when the origin point is changed to any other inversion centers. Moreover, rotating the $Z_2$ decoration by $\pi/4$ in the $z$ direction is equivalent to change the base-centered lattice to a simple(principal) lattice, which again does not change the $z_{4P}^\prime$ indicator. As a result, we conclude that the $Z_2$ decoration of MSG 135.487 also has $z_{4P}^\prime=3$. On the other hand, as the BHZ model has time-reversal symmetry and all 8 TRIMs have Kramer pair of opposite $C_{2z}$ eigenvalues, the $C_{2z}$ eigenvalues at TRIMS in the $Z_2$ decoration must also appear in opposite pairs when the time-reversal symmetry is broken. Based on these observations, we find a compatible set of coirreps for this $Z_2$ decoration as $(2\bar{\Gamma}_5,\bar{S}_5,\bar{S}_6,\bar{T}_3,\bar{T}_4,\bar{Y}_3,\bar{Y}_4,\bar{U}_3,\bar{U}_4,\bar{X}_3,\bar{X}_4,\bar{R}_5\bar{R}_6,\bar{Z}_5\bar{Z}_6)$. Substituting into the $z_{4,135.487}$, we have
\begin{equation}
z_{4,135.487}=3
\end{equation}
for the $Z_2$ decoration.

\clearpage
\section{Weyl semimetals in MSGs}\label{AppendixI}
There are two types of Weyl semimetals in MSGs that can be diagnosed by indicators:
\begin{itemize}
	\item inter-plane Weyl points: lie between $k_i=0$ and $k_i=\pi$ plane;
	\item in-plane Weyl points: lie on $k_i=0$ or $\pi$ plane.
\end{itemize} 
where $i$ denotes the main rotation axis of the MSG. Although these Weyl points exist at generic momenta, their creation and annihilation must happen at HSPs, which makes it possible to detect them using indicators. Note all SIs of type-2 spinful MSG cannot represent gapless states, as shown in Ref.\cite{song2018quantitative}.

\subsection{Type-1 MSGs}
Inter-plane Weyl points exist when the Chern number on $k_i=0$ and $\pi$ plane are different. In type-1 MSGs, inversion and $S_4$ symmetry can help diagnose inter-plane Weyl points.

MSG 2.4 $P\overline{1}$, 147.13 $P\overline{3}$, and 148.17 $R\overline{3}$ all have $z_{4P}$ indicator, and its odd values represent inter-plane Weyl states, as the Chern numbers on $k_z=0/\pi$ planes are different.

MSG 81.33 $P\overline{4}$ and 82.39 $I4_1$ have $z_{2,\text{Weyl}}$ indicator, which also indicates inter-plane Weyl states.

The Weyl points configuration of MSG 2.4 and 81.33 are shown in Fig.\ref{weyl_config}(1)(2).

\subsection{Type-4 MSGs}
 Type-4 MSGs have anti-unitary half translation $T\tau$, which forbids the inter-plane Weyl points indicated by inversion and $S_4$ symmetry in type-1 MSGs. However, type-4 MSGs can host in-plane Weyl points, thanks to the $C_2T$ symmetry with reasons given below. 
 
$C_2T$ symmetry transfers the Berry curvature $F(k)$ to $-F(k)$ on $k_z=0/\pi$ plane, thus the Berry phase along any loop on $k_i=0$ plane is quantized to $0$ or $\pi$, with $\pi$-Berry phase denoting an odd number of Weyl points (of the same chirality) inside the loop. This $\pi$-Berry phase can be readily calculated using $C_n$ eigenvalues at HSPs.

\begin{figure*}
	\centering
	\includegraphics[width=1\textwidth]{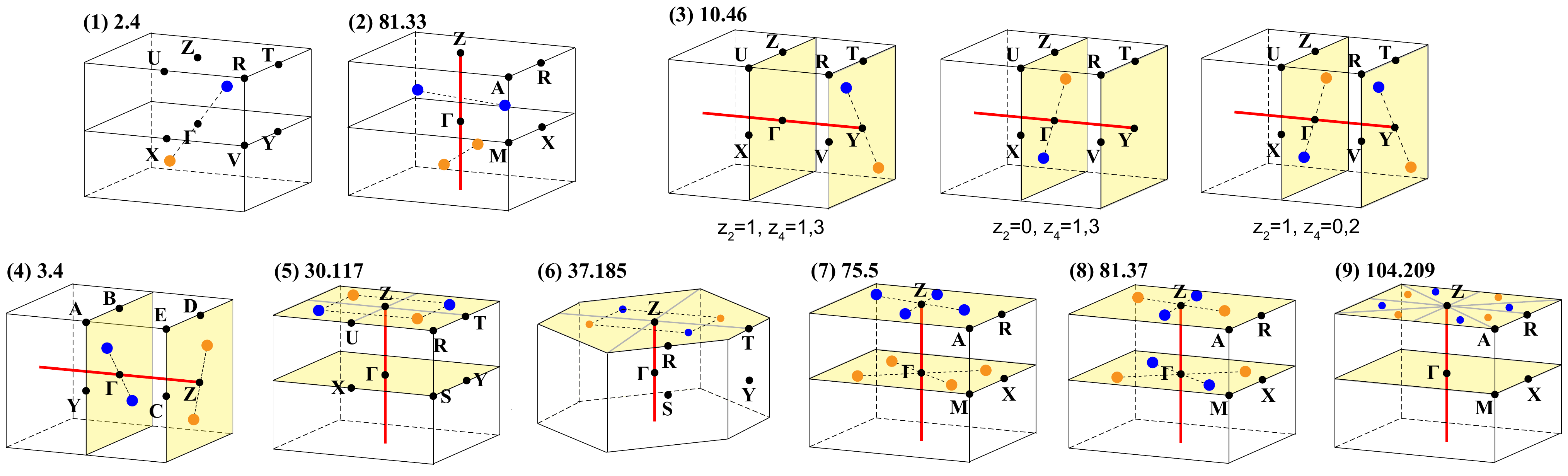}
	\caption{\label{weyl_config} Minimal configurations for Weyl states in MSGs, with the blue and orange dots represent Weyl points of opposite monopole charges, and the red lines denote the main rotation axis. (1) MSG 2.4. (2) MSG 81.33. (3) 10.46. (4) MSG 3.4. (5) 30.117 and 34.163. (6) 37.185. (7) 75.5 and 77.17. (8) 81.37. (9) 104.209. }
\end{figure*}

There are 8 type-4 MSGs that can host in-plane Weyl points, all having a $\mathbb{Z}_2$ indicator group:
\begin{itemize}
\item MSG 3.4 $P_a2$. 
We can take the Berry phase formula using $C_2$ eigenvalues on $k_y=0$ plane as the indicator. $z_{2C}=1$ indicates the Berry phase around half BZ is $\pi$, implying an odd number of Weyl points. $C_{2y}$ dictates a same number of Weyl points of the same chirality on the other half $k_y=0$ BZ. The same argument holds for the $k_y=\pi$ plane, where Weyl points have opposite chirality from the $k_y=0$ plane, to make sure the total chirality is zero. The Weyl points configuration is shown in Fig.\ref{weyl_config}(4).

\item MSG 30.117 $P_bnc2$ and 34.163 $P_Cnn2$. These two MSGs can host $4n$ Weyl points on $k_z=\pi$ plane, with Weyl points connected by glide symmetry $G_x/G_y$ having opposite chirality, as shown in Fig.\ref{weyl_config}(5). The indicator formula can be taken as the Berry phase formula using $C_2$ eigenvalues on $k_z=\pi$ plane divided by 2 because the coirreps on $k_z=\pi$ plane are all two-fold degenerate with the same $C_2$ eigenvalue. The justification of this indicator is very similar to the Weyl states in SG 27 in Ref.\cite{song2018diagnosis}. We remark here that although the Weyl points on $k_z=\pi$ plane have opposite chirality, they cannot annihilate with each other without changing the irreps on HSPs, as they are also connected by $G_{x/y}T$ with $(G_{x/y}T)^2=-1$, as shown in Ref.\cite{fang2016topological}. On $k_z=0$ plane we have $(G_{x/y}T)^2=1$, which means the Weyl points are not protected.

\item MSG 37.185 $C_acc2$. This MSG can also host $4n$ Weyl points on $k_z=\pi$ plane, with Weyl points connected by glide symmetry $G_x/G_y$ having opposite chirality, as shown in Fig.\ref{weyl_config}(6). Its SI takes the corner case formula $z_{2,37.185}$.

\item MSG 75.5 $P_C4$ and 77.17 $P_C4_2$. These two MSGs are very similar to MSG 3.4, and we can take the Berry phase formula using $C_4$ eigenvalues on the $k_z=0$ plane as the indicator, with $z_4=2$ dictating an odd number of Weyl points in a quarter of the 2D BZ, i.e., $4n$ Weyl points of the same chirality on the $k_z=0$ plane. There are also a same number of Weyl points of the opposite chirality on the $k_z=\pi$ plane. The Weyl points configuration is shown in Fig.\ref{weyl_config}(7).

\item MSG 81.37 $P_C\overline{4}$. This MSG can host $4n$ Weyl points on both $k_z=0/\pi$ plane, with Weyl points connected by $S_4$ symmetry having opposite chirality, as shown in Fig.\ref{weyl_config}(8). It can be seen from AI and nontrivial BSs that $z_{4C,k_z=0}=z_{4C,k_z=\pi}$ always holds, which means the Weyl points will appear simultaneously on $k_z=0,\pi$ planes. This Weyl configuration can be detected by Berry phase formula using $S_4$ eigenvalues on the $k_z=0$ plane, as $S_4=C_4^{-1}$ on $k_z=0/\pi$ plane. The justification of this indicator is very similar to the Weyl states in SG 81 in Ref.\cite{song2018diagnosis}. Notice the Weyl points of opposite chirality are connected by $S_4$ symmetry, and they can only be annihilated at HSPs on $k_z=0/\pi$ plane.

\item MSG 104.209 $P_C4nc$. This MSG can host $8n$ Weyl points on $k_z=\pi$ plane, with Weyl points connected by glide symmetry $G_x/G_y$ having opposite chirality, as shown in Fig.\ref{weyl_config}(9). Its SI takes the corner case formula $z_{2,104.209}$.
\end{itemize}

\begin{table*}
\begin{center}
\begin{tabular}{c|c|c|c}
\hline
\multicolumn{4}{c}{Weyl states in type-1 MSGs}\\	
\hline
MSG & SI & SI formula & Weyl type  \\ 
\hline
 \makecell[c]{2.4 $P\overline{1}$ \\ 147.13 $P\overline{3}$, 148.17 $R\overline{3}$}
 	 & $\mathbb{Z}_{4}\times\mathbb{Z}_*$ & $z_{4P}=1,3$ & inter-plane \\
\hline
 \makecell[c]{81.33 $P\overline{4}$, 82.39 $I\overline{4}$} & $\mathbb{Z}_{2}\times\mathbb{Z}_*$ & $z_{2,\text{Weyl}}=1$ & inter-plane \\
\hline
\multicolumn{4}{c}{Weyl states in type-3 MSGs}\\
\hline
\makecell[c]{10.46 $P2'/m'$ \\ 13.69 $P2'/c'$}
 & $\mathbb{Z}_{2,4}\times\mathbb{Z}_*$ & $z_{2P,2}=1, z_{4P}=1,3$ & in-plane \\
 \hline
175.141 $P6'/m'$ & $\mathbb{Z}_{2,4}$ & $z_{6C}=3, z_{4P}=1,3$ & in-plane \\
\hline
\makecell[c]{11.54 $P2_1'/m'$, 14.79 $P2'_1/c'$\\
12.62 $C2'/m'$, 15.89 $C2'/c'$} & $\mathbb{Z}_4\times\mathbb{Z}_*$ & $z_{4P}=1,3$ & in-plane \\
\hline
\makecell[c]{162.77, 163.83, 164.89 \\ 
	165.95, 166.101, 167.107\\ 176.147} & $\mathbb{Z}_4\times\mathbb{Z}_*$ & $z_{4P}=1,3$ & inter-plane \\
\hline
\makecell[c]{111.255, 112.263, 113.271 \\ 114.279, 115.287, 116.295 \\ 
	117.303, 118.311, 119.319 \\ 120.325, 121.331, 122.337}
& $\mathbb{Z}_2\times\mathbb{Z}_*$ & $z_{2,\text{Weyl}}=1$ & inter-plane \\
\hline
\multicolumn{4}{c}{Weyl states in type-4 MSGs}\\
\hline
3.4 $P_a2$ & $\mathbb{Z}_2$ & $z_{2C}=1$ & in-plane \\
\hline
30.117 $P_bnc2$, 34.163 $P_Cnn2$  & $\mathbb{Z}_2$ &  $z_{2C}^\prime=1$   & in-plane \\
\hline
37.185 $C_acc2$ & $\mathbb{Z}_2$ &  $z_{2,37.185}=1$   & in-plane \\
\hline
\makecell[c]{75.5 $P_C4$, 77.17 $P_C4_2$\\ 81.37 $P_C\overline{4}$ } & $\mathbb{Z}_2\times\mathbb{Z}_*$ & $z_{4C}=2$  & in-plane \\
\hline
104.209 $P_C4nc$ & $\mathbb{Z}_2$ &  $z_{2,104.209}=1$   & in-plane \\
\hline
\end{tabular}
\caption{\label{weyl_table} Summary of Weyl states in MSGs. $\mathbb{Z}_*$ stands for extra $\mathbb{Z}_n$ factors.}
\end{center}
\end{table*}

\subsection{Type-3 MSGs}
Most of the Weyl states in type-3 MSGs can be induced from type-1 MSGs, calculated by inversion or $S_4$ eigenvalues, as listed in the following table. 

However, when the MSG also has $C_{2i}\cdot T$ symmetry, the inter-plane Weyl points could be pinned on $k_i=0/\pi$ planes and become in-plane Weyl. There are three special type-3 MSGs that have two indicators to specify the Weyl points configuration:

\begin{itemize}
\item MSG 10.46 $P2'/m'$ and 13.69 $P2'/c'$. These two MSGs all have $C_{2y}\cdot T$, and inversion can be used to calculate the Berry phase of half $k_y=\pi$ plane using $z_{2P,2}$, with $\pi$-Berry phase corresponding to the Weyl points of opposite chirality on the $k_y=\pi$ plane. We plot the Weyl point configurations of MSG 10.46 in Fig.\ref{weyl_config}(3):
\begin{itemize}
    \item If $z_{2P,2}=1$, when (i) $z_{4P}=1,3$, the Berry phase of half of $k_y=0/\pi$ plane differs by $\pi$ and only the $k_y=\pi$ plane has Weyl points, while (ii) $z_{4P}=0,2$ mean both $k_y=0/\pi$ planes host in-plane Weyl points.
    \item If $z_{2P,2}=0$, when (i) $z_{4P}=1,3$, only $k_y=0$ plane hosts in-plane Weyl points, while (ii) $z_{4P}=0,2$ mean no plane has Weyl points.
\end{itemize}

Note that in MSG 11.54 $P2_1'/m'$, 14.79 $P2'_1/c'$, 12.62 $C2'/m'$, and 15.89 $C2'/c'$, which also have $C_2T$, there is only one $z_4$ indicator to specify the Weyl points configuration. This is because in the first two MSGs the coirreps are two-fold degenerate of opposite parity on the $k_y=\pi$ plane such that Weyl points can only appear on the $k_y=0$ plane, while the last two MSGs are $C$-face-centered lattice, with $z_{4P}=1,3$ effectively diagnosing the $\pi$-Berry phase of half $k_y=0$ plane of the conventional BZ and corresponding to an odd number of Weyl points on it.

\item MSG 175.141 $P6'/m'$. $S_3$ symmetry in this MSG can be used to calculate the $\pi$-Berry phase of one sixth of the $k_z=0$ plane, as $S_3=C_6^{-1}$ on $k_z=0$ plane. $k=0/\pi$ planes can both host in-plane Weyl points, with Weyl points connected by $S_3$ symmetry having opposite chirality. 
\begin{itemize}
    \item If $z_{6C}=3$, when (i) $z_{4P}=1,3$, the Berry phase of half of $k_z=0/\pi$ plane differs by $\pi$ and only the $k_z=\pi$ plane has Weyl points, while (ii) $z_{4P}=0,2$ mean both $k_z=0/\pi$ planes host in-plane Weyl points.
    \item If $z_{6C}=0$, when (i) $z_{4P}=1,3$, only $k_z=0$ plane hosts in-plane Weyl points, while (ii) $z_{4P}=0,2$ mean no plane has Weyl points.
\end{itemize}
\end{itemize}

\clearpage
\section{Summary of generating SIs in MSGs}\label{AppendixJ}
We summarize the generating SI formulas in this section. The first subscript number in the SI name represents its order, and the prime in some indicators represents they are $\frac{1}{2}$ of the original indicators without prime.

\subsection{Generating SIs in type-1 \& 2 MSGs}
\begin{itemize}
\item MSG 2.4 $P\overline{1}$. 
\begin{equation}
\begin{aligned} 
z_{2P,i} = & \sum_{\mathbf{k} \in \text{TRIM},k_i=\pi} \frac{1}{2}(N_\mathbf{k}^--N_\mathbf{k}^+) \bmod 2 \\ 
z_{4P} = & \sum_{\mathbf{k} \in \text{TRIM}} \frac{1}{2}(N_\mathbf{k}^--N_\mathbf{k}^+) \bmod 4 \\ 
\end{aligned}
\end{equation}
where $N_\mathbf{k}^\pm$ is the number of valence bands having positive (negative) parity. 

\item MSG 11.50 $P2_1/m$.
\begin{equation}
\begin{aligned} 
z_{2P}^\prime = \frac{1}{2}z_{4P} = \sum_{\mathbf{k} \in \text{TRIM}} \frac{1}{4}(N_\mathbf{k}^--N_\mathbf{k}^+) \bmod 2 
\end{aligned}
\end{equation}
Note this SI is not included as a generating SI in the first section, because it can be directly induced from $z_{4P}$ by taking only even values. We list it here for ease of consulting.

\item MSG 47.249 $Pmmm$.
\begin{equation}
\begin{aligned} 
z_{2P,i}^\prime = & \sum_{\mathbf{k} \in \text{TRIM},k_i=\pi} \frac{1}{4}(N_\mathbf{k}^--N_\mathbf{k}^+) \bmod 2 \\ 
z_{4P}^\prime = & \sum_{\mathbf{k} \in \text{TRIM}} \frac{1}{4}(N_\mathbf{k}^--N_\mathbf{k}^+) \bmod 4 \\ 
\end{aligned}
\end{equation}

\item Indicators defined by $C_n$ rotations:

MSG 3.1 $P2$.
\begin{equation}
 \begin{aligned}
 z_{2C} &= \sum_{l\in occ} \ln(B_{C_2}^l(\Gamma) B_{C_2}^l(X) B_{C_2}^l(Y) B_{C_2}^l(M))/(i\pi) \bmod 2 \\
&=\sum_{\mathbf{k} \in \Gamma,M,X,Y} \frac{1}{2}(N_\mathbf{k}^--N_\mathbf{k}^+) \bmod 2 \\ 
\end{aligned}
\end{equation} 
 
MSG 75.1 $P4$.
\begin{equation}
 z_{4C} = \sum_{l\in occ} \ln(-B_{C_4}^l(\Gamma) B_{C_4}^l(M) B_{C_2}^l(Y)) /(i\frac{\pi}{2})\bmod 4
\end{equation}

MSG 143.1 $P3$.
\begin{equation}
 z_{3C} = \sum_{l\in occ} \ln(-B_{C_3}^l(\Gamma) B_{C_3}^l(K) B_{C_3}^l(K^{\prime})) /(i\frac{2\pi}{3}) \bmod 3
\end{equation}

MSG 168.109 $P6$.
\begin{equation}
 z_{6C} = \sum_{l\in occ} \ln(-B_{C_6}^l(\Gamma) B_{C_3}^l(K) B_{C_2}^l(M)) /(i\frac{\pi}{3}) \bmod 6
\end{equation}
where $C$ stands for Chern number or $C_n$ rotation, and $B_{C_n}^i(\mathbf{k})$ represents the  $C_n$ eigenvalue of the $i$-th band at HSP $\mathbf{k}$. This Chern number is calculated on the $k_i=0$ plane by default, and will be changed to $z_{nC,\pi}$ if calculated on the $k_i=\pi$ plane.

Note that these SIs are also adopted in some other MSGs and take only even values, for example, the $z_{4C}$ can only take $0,2$ in MSG $I4$, and we denote them as $z_{nC}/2$ for simplicity.

\item Indicators defined by $C_n$ rotations and mirror symmetry to calculate the mirror Chern numbers mod n, which can be calculated by $z_{nC}$ formula using $M=\pm i$ bands on $k_i=0/\pi$ plane. We name these indicators as $z_{nm,0/\pi}^\pm$, where $0/\pi$ denotes which $k$ plane and $\pm$ which mirror sector.

MSG 10.41 $P2/m$: $z_{2m,0}^+,z_{2m,0}^-, z_{2m,\pi}^+$.

MSG 83.43 $P4/m$: $z_{4m,0}^+,z_{4m,0}^-, z_{4m,\pi}^+$.

MSG 174.133 $P\overline{6}=P3/m$: $z_{3m,0}^+$,$z_{3m,0}^-$, $z_{3m,\pi}^+$.

MSG 175.137 $P6/m$: $z_{6m,0}^+,z_{6m,0}^-, z_{6m,\pi}^+$.

\item MSG 81.33 $P\overline{4}$. In this MSG, aside from $z_{4C}$, we can also define two $\mathbb{Z}_2$ indicators $z_{2,S_4}$ and $z_{2,\text{Weyl}}$ to represent the $S_4$ decoration and inter-plane Weyl states:
\begin{equation}
\begin{aligned}
\mu_{S_4} &= \frac{1}{\sqrt{2}} \sum_{i\in occ.}\sum_{\mathbf{k}\in K_{S_4}}\beta_i(\mathbf{k})\\
z_{2,S_4} &= \frac{1}{2}(\text{Re}\mu_{S_4}-\text{Im}\mu_{S_4}) \bmod 2  \\
z_{2,\text{Weyl}} &= \text{Re}(\mu_{S_4}) \bmod 2
\end{aligned}
\end{equation}
where $\beta_i(\mathbf{k})=e^{\alpha\frac{\pi}{4} i}(\alpha=1,3,5,7)$ are the $S_4$ eigenvalues at $S_4$-invariant points $K_{S_4}$.

\item MSG 81.34 $P\overline{4}1'$. This type-2 MSG has the following $z_{2,S_4}^\prime$ $\mathbb{Z}_2$ indicator:
\begin{equation}
z_{2,S_4}^\prime = -\frac{1}{2}\mu_{S_4}=
\sum_{\mathbf{k}\in K_{S_4}} \frac{1}{2}(n_{\mathbf{k}}^{-\sqrt{2}}-n_{\mathbf{k}}^{\sqrt{2}})
\end{equation}
where $n_{\mathbf{k}}^{\pm\sqrt{2}}$ is the number of Kramer pairs at $S_4$-invariant TRIMs with $\operatorname{tr}\left[D\left(S_{4}\right)\right]=\pm\sqrt{2}$. $z_{2,S_4}^\prime=1$ represents a STI.

\item MSG 123.339 $P4/mmm$. This MSG has an $\mathbb{Z}_8$ indicator which is taken as 
\begin{equation}
z_8=2 z_{2,S_4}^\prime - z_{4P}^\prime \bmod 8
\end{equation}
 Note that $z_{4P}^\prime$ and $z_{2,S_4}^\prime$ in $z_8$ formula should not take mod, and their original values are used to calculate $z_8$.

\item MSG 191.233 $P6/mmm$. This MSG has an $\mathbb{Z}_{12}$ indicator which be taken as 
 \begin{equation}
z_{12}=\left\{z_{6m,S}+3\left[\left(z_{6m,S}-z_{4P}^\prime\right) \bmod 4\right]\right\} \quad \bmod 12
\end{equation}
where $z_{6m,S}$ has more than one choices, and here we choose it as $z_{6m,S}=z_{6m,0}^++z_{6m,\pi}^+$.

In type-2 MSGs, there are similar $z_{12}$ indicators, where in MSG 175.138, 191.234 and 192.244 the definition of $z_{6m,S}$ is the same, while in MSG 176.144, 193.254 and 194.264, $z_{6m,S}=z_{6m,0}$ because they have $C_6$-screw which forces the mirror Chern number on $k_z=\pi$ plane to be zero. 

\end{itemize}

\subsection{Generating SIs in type-3 MSG}
\begin{itemize}
\item  MSG 27.81 $Pc'c'2$.
\begin{equation}
	z_{2C}^\prime = \frac{1}{2}z_{2C} \bmod 2 \\ 
\end{equation}

MSG 103.199  $P4c'c'$
\begin{equation}
z_{4C}^\prime = \frac{1}{2}z_{4C} \bmod 4 \\ 
\end{equation}

MSG 184.195 $P6c'c'$.
\begin{equation}
z_{6C}^\prime = \frac{1}{2}z_{6C} \bmod 6 \\ 
\end{equation}
where $z_{nC}$ on the RHS has not taken mod.

\end{itemize}

\subsection{Corner cases SI}
Type-1 MSG:
\begin{itemize}
\item 87.75 $I4/m$
\begin{equation}
\begin{aligned}
z_{4m,87}^+ & = \frac{1}{2}(z_{4m,0}^+ + z_{4m,0}^-) \bmod 4\\
z_{4m,87}^- & = \frac{1}{2}(z_{4m,0}^+  - z_{4m,0}^-) \bmod 4\\
\end{aligned}
\end{equation}
\end{itemize}

Type-2 MSG 
\begin{itemize}
	\item 226.123 $Fm\overline{3}c1'$ 
    \begin{equation}
    \begin{aligned}
    z_{8, 226.123} &= 3N(\overline{\Gamma}_6) + 3N(\overline{\Gamma}_7) + 4N(\overline{\Gamma}_9) + 2N(\overline{\Gamma}_{10}) + \\
    &4N(\overline{L}_4\overline{L}_4) - 4N(\overline{L}_5\overline{L}_6) 
    - 3N(\overline{X}_6) + N(\overline{X}_7) \bmod 8
    \end{aligned}
    \end{equation}
\end{itemize}

Type-3 MSGs:
\begin{itemize}
	\item MSG 41.215 $Ab'a'2$
	\begin{equation}
	z_{2,41.215} = N(\overline{\Gamma}_3) \bmod 2
	\end{equation}
	\item MSG 42.222 $Fm'm'2$
	\begin{equation}
	z_{2,42.222} = N(\overline{\Gamma}_3) - N(\overline{A}_3) \bmod 2
	\end{equation}
	\item MSG 60.424 $Pb'cn'$
	\begin{equation}
	z_{2,60.424} = N(\overline{\Gamma}_3) \bmod 2
	\end{equation}
	\item MSG 68.515 $Cc'c'a$
	\begin{equation}
	z_{2,68.515} = N(\overline{Z}_3\overline{Z}_5) - N(\overline{C}_3\overline{C}_5) \bmod 2
	\end{equation}
	\item MSG 110.249 $I4_1c'd'$
	\begin{equation}
	z_{2,110.249} = N(\overline{M}_5\overline{M}_6) \bmod 2 
	\end{equation}
	
	\item MSG 135.487 $P4_2'/mbc'$
	\begin{equation}
	z_{4,135.487} = N(\overline{\Gamma}_5) - 2N(\overline{R}_5\overline{R}_6)
	+ N(\overline{S}_5) - 2N(\overline{T}_3) \bmod 4
	\end{equation}
\end{itemize}

Type-4 MSGs (both are Weyl states):
\begin{itemize}
	\item MSG 37.185 $C_acc2$
	\begin{equation}
	z_{2, 37.185} = N(\overline{R}_3) + N(\overline{T}_5\overline{T}_5)
	+ N(\overline{Z}_5\overline{Z}_5) \bmod 2
	\end{equation}
	
	\item MSG 104.209 $P_C4nc$
	\begin{equation}
	z_{2,104.209} = 2N(\overline{R}_2\overline{R}_4) \bmod 2
	\end{equation}
\end{itemize}

\clearpage
\section{The Berry phase of closed loops in 2D BZ}\label{AppendixK}
\begin{figure}[ht]
	\centering
	\includegraphics[width=0.35\textwidth]{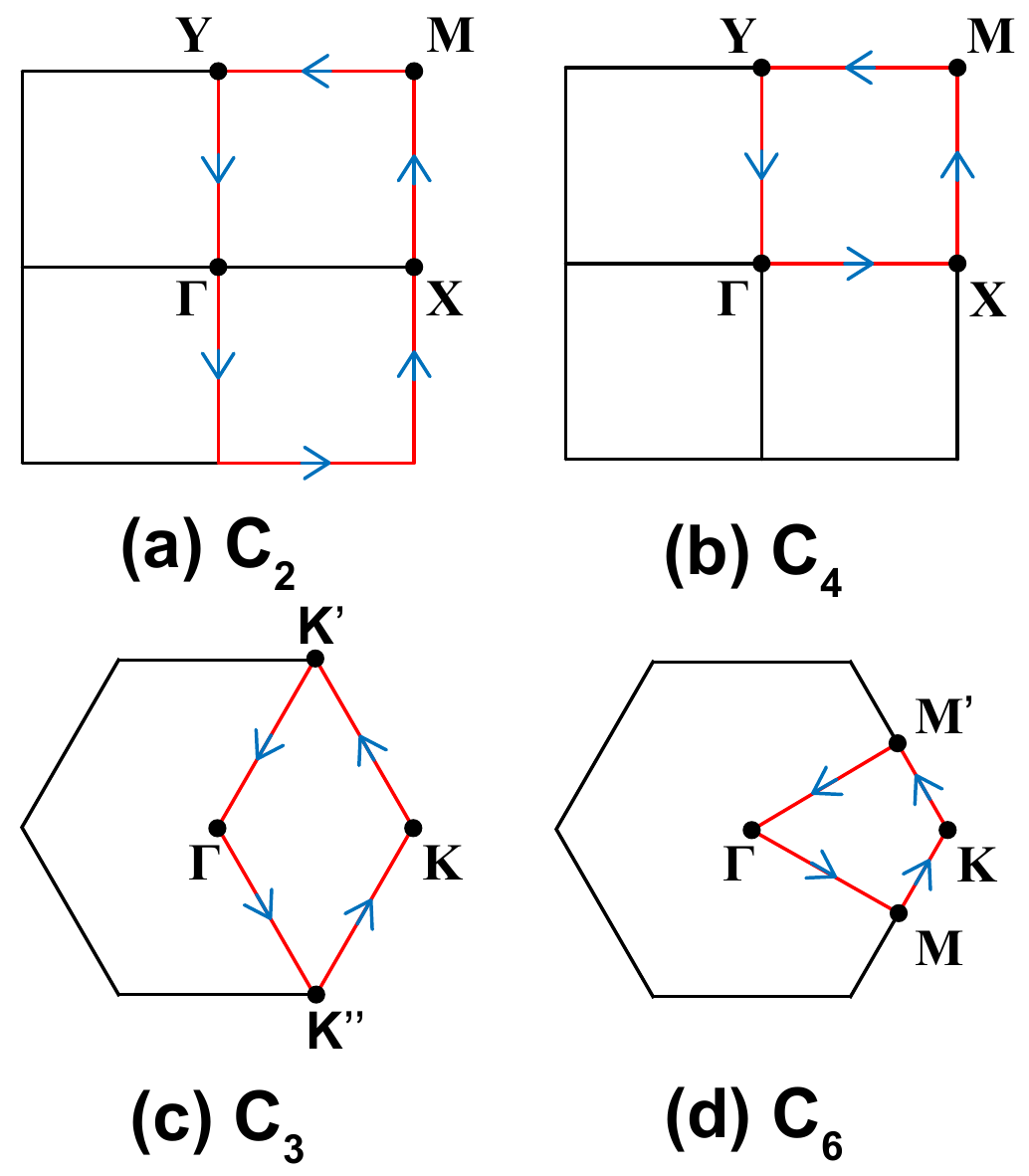}
	\caption{\label{loops}Loops and HSPs used to calculate the Berry phase on 2D BZs}
\end{figure}

The formulas for calculating the Berry phase  $\Phi_B$ around a closed loop in Fig.\ref{loops} using $C_n$ eigenvalues are derived in Ref.\cite{fang2012bulk}:
\begin{equation}
\begin{aligned} 
C_2:&\ \ e^{i\Phi_B}=\prod_{l \in \mathrm{occ} .} B_{C_2}^l(\Gamma) B_{C_2}^l(X) B_{C_2}^l(Y) B_{C_2}^l(M)\\
C_4:&\ \ e^{i\Phi_B}=\prod_{l \in \mathrm{occ} .} (-1)B_{C_4}^l(\Gamma) B_{C_4}^l(M) B_{C_2}^l(Y)\\
C_3:&\ \ e^{i\Phi_B} =\prod_{l \in \mathrm{occ} .}(-1) B_{C_3}^l(\Gamma) B_{C_3}^l(K) B_{C_3}^l(K^{\prime}) \\ 
C_6:&\ \ e^{i\Phi_B}=\prod_{l \in \mathrm{occ} .} (-1)B_{C_6}^l(\Gamma) B_{C_3}^l(K) B_{C_2}^l(M) 
\end{aligned}
\end{equation}
where $B_{C_n}^l(k)$ represent the $C_n$ eigenvalue of the $l$-th band at $C_n$ -invariant point $k$.

These formulas can be used to diagnose Weyl points. For MSGs with $PT$ symmetry, Berry curvature is forced to be zero (for gapped momenta). Also for MSGs with $C_{2i}T$, the Berry curvature on $k_i=0,\pi$ planes are forced to be zero (for gapped momenta). As a result, in these MSGs, the Berry phase around any closed loops equals zero, or equals $\pi$ when there are an odd number of Weyl points lie in these loops. These formulas using $C_n$ eigenvalues can be used to diagnose the $\pi$ Berry phase.

\clearpage
\section{The BHZ model for describing non-layer constructions}\label{AppendixL}
Bernevig-Hughes-Zhang(BHZ) model was first introduced in the 2D HgTe system to characterize the quantum-spin Hall effect (QSHE), which can be generalized to 3D to describe strong TIs. In this appendix, we calculate the mirror Chern numbers and SIs of the BHZ model.

\subsection{2D Chern insulator model}
We first introduce a 2D Chern insulator model, which is known as the half-BHZ model or the Qi-Wu-Zhang model\cite{qi2006topological}:
\begin{equation}
\begin{aligned}
&H=\bm{H(k)}\cdot\bm{\sigma}\\ 
&H_x(k)=\sin(k_x)\\ 
&H_y(k)=\sin(k_y)\\ 
&H_z(k)=M-\cos(k_x)-\cos(k_y)
\end{aligned}
\end{equation}
which has eigenvalues $E_{\pm}=\pm|\bm{H(k)}|$. The eigenvector of $E_-$ has two different forms
\begin{equation}
\begin{aligned}
u_-^1(k)&=\frac{1}{N_-^1}
\left(\begin{array}{c}
H_z-|H|\\ 
H_x+iH_y
\end{array}\right),\\
u_-^2(k)&=\frac{1}{N_-^2}
\left(\begin{array}{c}
-H_x+iH_y\\ 
H_z+|H|
\end{array}\right)
\end{aligned}
\end{equation}
where $N_-^{1,2}$ are normalization factors. These two eigenvectors differ by a phase
\begin{equation}
\begin{aligned}
u_-^2&=u_-^1e^{i\phi} \\
\Rightarrow e^{i\phi}&=\frac{|H_x+iH_y|}{H_x+iH_y} \approx \frac{|k_x+ik_y|}{k_x+ik_y}=e^{-i\theta}
\end{aligned}
\end{equation}
where we use the approximation $\sin(k)\approx k$ if $k\rightarrow 0$, and $e^{i\theta}=(k_x+ik_y)/k$. The Berry connection satisfies $A_-^2=A_-^1+\nabla_k\phi$. 
When $H_x=H_y=0$, $u_-^1=0$ if $H_z>0$, while $u_-^2=0$ if $H_z<0$. As a result, we need to choose $u_-^{1,2}$ properly depending on whether $H_z(k)$ is positive or negative. 

The upper band also has two forms of eigenvectors
\begin{equation}
\begin{aligned}
u_+^1(k)&=\frac{1}{N_+^1}
\left(\begin{array}{c}
H_z+|H|\\ 
H_x+iH_y
\end{array}\right),\\ 
u_+^2(k)&=\frac{1}{N_+^2}
\left(\begin{array}{c}
-H_x+iH_y\\ 
H_z-|H|
\end{array}\right)
\end{aligned}
\end{equation}
These two eigenvectors differ by a phase
\begin{equation}
\begin{aligned}
u_+^2&=u_+^1e^{i\phi} \\
\Rightarrow e^{i\phi}&=\frac{-H_x+iH_y}{|-H_x+iH_y|} \approx \frac{-k_x+ik_y}{|-k_x+ik_y|}=e^{-i\theta}
\end{aligned}
\end{equation}
When $H_x=H_y=0$, $u_+^1=0$ if $H_z<0$, while $u_+^2=0$ if $H_z>0$. Note that the choice of $u_+^{1,2}$ is opposite to $u_-^{1,2}$, which ensures $C_+=-C_-$. The zero total Chern number is also enforced by the time-reversal symmetry (TRS).

The topological property of the this 2D BHZ model depends on the sole parameter $M$. If $M=0,\pm2$, the model is gapless, which are topological phase transition points. If $M>2$ or $M<-2$, the model is equivalent to $M\rightarrow\pm\infty$, which is trivial atomic insulator. If $0<M<2$, the model is topological and has $C_\pm=\mp1$, where $C_\pm$ is the Chern number of $E_\pm$ band, while $C_\pm=\pm1$ if $-2<M<0$.

If $0<M<2$, we have $H_z<0$ when $k=(0,0)$, while $H_z>0$ when $k=(\pi,0),(0,\pi),(\pi,\pi)$. We can choose $u_-^1$ in the vicinity of $\Gamma$ and $u_-^2$ in other momenta. The Chern number $C_-$ can be calculated as
\begin{equation}
\begin{aligned}
2\pi C_-
&=\int\int_{D_1}\nabla_k\times A_-^1+\int\int_{D_2}\nabla_k\times A_-^2\\
&=\int_{\partial D_1}A_-^1+\int_{\partial D_2}A_-^2 =\int_{\partial D_1}(A_-^1-A_-^2)\\
&=\int_{\partial D_1} (-\nabla_k\phi)=-(\phi(2\pi)-\phi(0))\\
&=2\pi\\
\Rightarrow\ C_-&=1
\end{aligned}
\end{equation}

If $-2<M<0$, we have $H_z<0$ when $k=(0,0),(\pi,0),(0,\pi)$ and $H_z>0$ when $k=(\pi,\pi)$. As a result, we can choose $u_-^2$ in the vicinity of $k=(\pi,\pi)$ and $u_-^1$ in other momenta. The phase $e^{i\phi}=e^{i\theta}$ near $k=(\pi,\pi)$, where $e^{i\theta}=(k_x-\pi+i(k_y-\pi))/|k-(\pi,\pi)|$. The Chern number $C_-=-1$ can be calculated similarly, where the minus sign is due to the clockwise orientation of $\partial D_2$.

\begin{figure}
	\centering
	\includegraphics[width=0.4\textwidth]{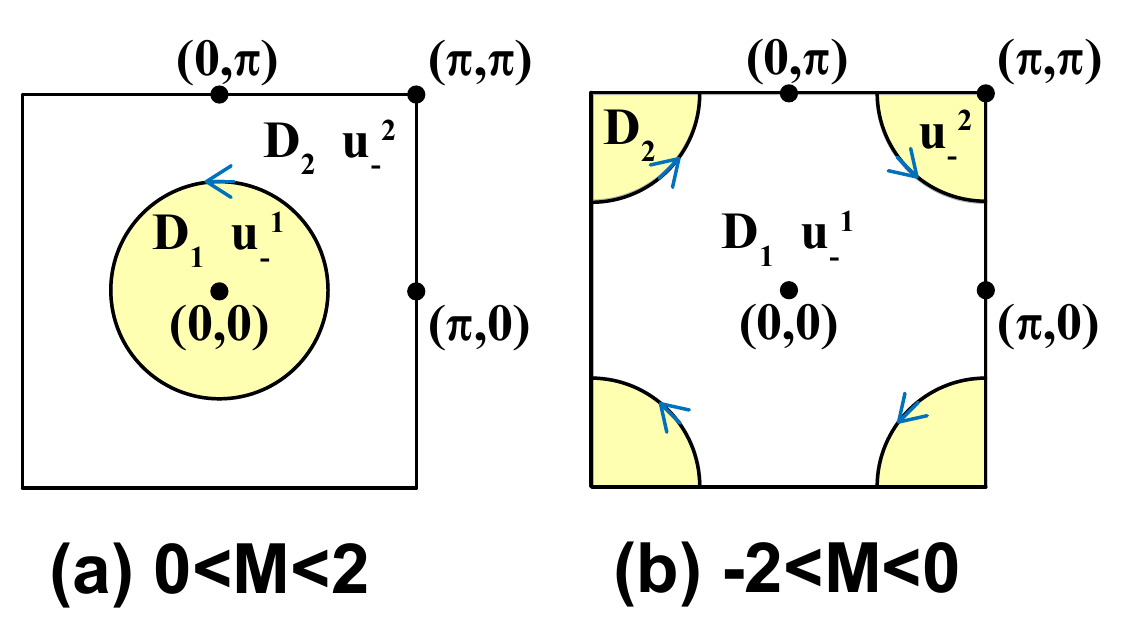}
	\caption{\label{BHZ_BZ}The choice of eigenvectors in 2D BZ}
\end{figure}

\paragraph{When $H_z$ adds a minus sign}
Here we summarize the result of 2D BHZ model when $H_z$ adds a minus sign, i.e.,
\begin{equation}
	H_z=-(M-\cos(k_x)-\cos(k_y))
\end{equation}
In this case, the eigenvalues and the form of eigenvectors are not changed, but the choice of eigenvectors change:
\begin{itemize}
\item If $0<M<2$, we have $H_z>0$ when $k=(0,0)$, while $H_z<0$ when $k=(\pi,0),(0,\pi),(\pi,\pi)$. We can choose $u_-^2$ in the vicinity of $\Gamma$ and $u_-^1$ in other momenta. As a result, we have $C_\pm=\pm 1$.
\item If $-2<M<0$, we have $H_z>0$ when $k=(0,0),(\pi,0),(0,\pi)$ and $H_z<0$ when $k=(\pi,\pi)$. As a result, we can choose $u_-^1$ in the vicinity of $k=(\pi,\pi)$ and $u_-^2$ in other momenta. As a result, we have $C_\pm=\mp 1$.
\end{itemize}

\subsection{3D BHZ model}
A simplified version of 3D BHZ model can be written as
\begin{equation}
H(k)=(M-\sum_{i=1,2,3} \cos(k_i))\Sigma_{30}+\sum_{i=1,2,3} \sin(k_i)\Sigma_{1i}
\end{equation}
where $\Sigma_{ij}=\sigma_i\otimes\sigma_j$ and M is the sole parameter. This model has $O_h$ and time-reversal symmetry. Assume a symmetry operation $R$ has representation matrix $D(R)$. The Hamiltonian with symmetry $D(R)$ satisfies $H(k)=D^{-1}(R)H(Rk)D(R)$. We can calculate the representation matrix for the following symmetries:
\begin{itemize}
\item Inversion $P$: $D(P)=\Sigma_{30}$.
\item $M_x$: $D(M_x)=i\Sigma_{31}$.
\item $M_y$: $D(M_y)=i\Sigma_{32}$.
\item $M_z$: $D(M_z)=i\Sigma_{33}$.
\item $D(M_{1\bar{1}0})$: $D(M_{1\bar{1}0})=\frac{i}{\sqrt{2}}(\Sigma_{32}-\Sigma_{31})$.
\item $S_4$: $D(S_4)=\frac{1}{\sqrt{2}}(\Sigma_{30}-i\Sigma_{33})$.
\item $T$: $D(T)=i\Sigma_{02}\hat{K}$, where $\hat{K}$ is the complex conjugation.
\end{itemize}

Here we mainly discuss the $M-1\in(0,2)$ phase and calculate the mirror Chern numbers along with SI. The $M-1\in(-2,0)$ phase can be derived similarly. 

\paragraph{Mirror Chern number on $k_z=0,\pi$ planes}
When $k_z=0$, the Hamiltonian reduces to 
\begin{equation}
\begin{aligned}
	H(k_z=0)=&(M-1-\cos(k_x)-\cos(k_y))\Sigma_{30}\\
	&+ \sin(k_x)\Sigma_{11}+ \sin(k_y)\Sigma_{12}
\end{aligned}
\end{equation}
We can block-diagonalize $H(k)$ using
\begin{equation}
S=\left(\begin{array}{cccc}
1 & 0 & 0 & 0  \\
0 & 0 & 0 & 1  \\
0 & 0 & 1 & 0  \\
0 & 1 & 0 & 0  \\
\end{array}\right)
\end{equation}

\begin{equation}
\begin{aligned}
\Rightarrow 
H^\prime(k_z=0)&=S^{-1}H(k_z=0)S\\
&=[\sin(k_x)\sigma_1+\sin(k_y)\sigma_2+M(k)\sigma_3]\\
&\oplus [\sin(k_x)\sigma_1+\sin(k_y)\sigma_2-M(k)\sigma_3]\\
\end{aligned}
\end{equation}
where $M(k) = M-1-\cos(k_x)-\cos(k_y)$. The representation matrix of $P$ and $M_z$ becomes
\begin{equation}
\begin{aligned}
D^\prime(P)&=S^{-1}D(P)S=
\left(\begin{array}{cccc}
1 & 0 & 0 & 0  \\
0 & -1 & 0 & 0  \\
0 & 0 & -1 & 0  \\
0 & 0 & 0 & 1  \\
\end{array}\right),\\
D^\prime(M_z)=S^{-1}D(M_z)S&=
\left(\begin{array}{cccc}
i & 0 & 0 & 0  \\
0 & i & 0 & 0  \\
0 & 0 & -i & 0  \\
0 & 0 & 0 & -i  \\
\end{array}\right)
\end{aligned}
\end{equation}

We arrange the eigenvalues as $E=(E_{+}(k), E_{-}(k))\oplus(E_{+}(k), E_{-}(k))$, and calculate the corresponding eigenvectors at different TRIMs. At $\Gamma$, we have
\begin{equation}
|\psi_\Gamma\rangle=(u_+^2(\Gamma), u_-^1(\Gamma))\oplus(u_+^1(\Gamma),u_-^2(\Gamma))=
\left(\begin{array}{cccc}
0 & 1 & 0 & 0  \\
1 & 0 & 0 & 0  \\
0 & 0 & 1 & 0  \\
0 & 0 & 0 & 1  \\
\end{array}\right)
\end{equation}
Under this basis, the representation matrix of $P$ and $M_z$ becomes
\begin{equation}
\begin{aligned}
D^{\prime\prime}(P)&=\langle\psi_\Gamma|D^\prime(P)|\psi_\Gamma\rangle=
\left(\begin{array}{cccc}
-1 & 0 & 0 & 0  \\
0 & 1 & 0 & 0  \\
0 & 0 & -1 & 0  \\
0 & 0 & 0 & 1  \\
\end{array}\right),\\
D^{\prime\prime}(M_z)&=\langle\psi_\Gamma|D^\prime(M_z)|\psi_\Gamma\rangle=
\left(\begin{array}{cccc}
i & 0 & 0 & 0  \\
0 & i & 0 & 0  \\
0 & 0 & -i & 0  \\
0 & 0 & 0 & -i  \\
\end{array}\right)
\end{aligned}
\end{equation}
As a result, we conclude that the $E_-(\Gamma)$ bands is a Kramers pair with inversion eigenvalue $+1$, while the  $E_+(\Gamma)$ bands a another with inversion eigenvalue $-1$. 

At other three TRIMs, $k=(\pi,0,0),(0,\pi,0),(\pi,\pi,0)$, the eigenvectors can be choose as
\begin{equation}
|\psi_k\rangle=(u_+^2(k), u_-^1(k))\oplus(u_+^1(k),u_-^2(k))=
\left(\begin{array}{cccc}
1 & 0 & 0 & 0  \\
0 & 1 & 0 & 0  \\
0 & 0 & 0 & 1  \\
0 & 0 & 1 & 0  \\
\end{array}\right)
\end{equation}
Under this basis, the representation matrix of $P$ and $M_z$ becomes
\begin{equation}
\begin{aligned}
D^{\prime\prime}(P)&=\langle\psi_k|D^\prime(P)|\psi_k\rangle=
\left(\begin{array}{cccc}
1 & 0 & 0 & 0  \\
0 & -1 & 0 & 0  \\
0 & 0 & 1 & 0  \\
0 & 0 & 0 & -1  \\
\end{array}\right),\\
D^{\prime\prime}(M_z)&=\langle\psi_k|D^\prime(M_z)|\psi_k\rangle=
\left(\begin{array}{cccc}
i & 0 & 0 & 0  \\
0 & i & 0 & 0  \\
0 & 0 & -i & 0  \\
0 & 0 & 0 & -i  \\
\end{array}\right)\\
\end{aligned}
\end{equation}
The inversion eigenvalues of these three TRIMs are opposite from $\Gamma$, i.e., $E_-(k)$ band being $-1$ and $E_+(k)$ band being $+1$. The $E_-(\Gamma)$ bands form a Kramers pair whose mirror Chern number are $C_{m,k_z=0}^+=1, C_{m,k_z=0}^-=-1$, while the  $E_+(\Gamma)$ bands form another Kramers pair whose mirror Chern numbers are $C_{m,k_z=0}^+=-1, C_{m,k_z=0}^-=1$ 

On $k_z=\pi$ plane, the Hamiltonian can be block-diagonalized similarly, with $M(k)=M+1-\cos(k_x)-\cos(k_y)$. $M-1\in(0,2)$ implies that $M+1>2$, i.e., the model is in the trivial phase, and we can choose the same eigenvector on 4 TRIMs on $k_z=\pi$ plane:
\begin{equation}
|\psi_k\rangle=(u_+^1(k), u_-^2(k))\oplus(u_+^2(k),u_-^1(k))=
\left(\begin{array}{cccc}
1 & 0 & 0 & 0  \\
0 & 1 & 0 & 0  \\
0 & 0 & 0 & 1  \\
0 & 0 & 1 & 0  \\
\end{array}\right)
\end{equation}
which is the same as the three TRIMs on $k_z=0$ plane except $\Gamma$. As a result, we have the parity of $E_-(k)$ band being $-1$ and $E_+(k)$ band being $+1$. The mirror Chern numbers are both zero for $M_z=\pm i$ sectors.

From the analysis above, we conclude that when $M\in(1,3)$, the occupied band $E_-$ has mirror Chern number $(C_{m,0}^+,C_{m,\pi}^-,C_{m,\pi}^+,C_{m,\pi}^-)=(1,-1,0,0)$, and the parity being $+1$ for $\Gamma$ and $-1$ for other seven TRIMs. Based on this result, we know that the 3D BHZ model has
\begin{equation}
	z_{4P}^\prime=3
\end{equation}

The mirror Chern numbers in the $k_x$ and $k_y$ direction are the same as $k_z$ direction as the model has $C_3$ symmetry in the $(111)$ direction, which makes the three directions equivalent.

\paragraph{Mirror Chern number in the $(1,\bar{1},0)$ direction.}
For the $O_h$ group, mirror $x,y,z$ are in the same conjugacy class, while mirror $(1,\pm1,0),(1,0,\pm1),(0,1,\pm1)$ are in another conjugacy class. To calculate the mirror Chern number of the second conjugacy class, we consider the $M_{1\bar{1}0}$ symmetry:
$ M_{1\bar{1}0}=
\left(\begin{array}{ccc}
0 & 1 & 0  \\
1 & 0 & 0  \\
0 & 0 & 1  \\
\end{array}\right) $.
The representation matrix of $M_{1\bar{1}0}$ satisfies
\begin{equation}
\begin{aligned}
\Sigma_{30}D(M_{1\bar{1}0})&=D(M_{1\bar{1}0})\Sigma_{30},\ \Sigma_{13}D(M_{1\bar{1}0})=D(M_{1\bar{1}0})\Sigma_{13},\ \\
\Sigma_{11}D(M_{1\bar{1}0})&=D(M_{1\bar{1}0})\Sigma_{12},\ \Sigma_{12}D(M_{1\bar{1}0})=D(M_{1\bar{1}0})\Sigma_{11},\  \\
&\Rightarrow \
D(M_{1\bar{1}0})=\frac{i}{\sqrt{2}}(\Sigma_{32}-\Sigma_{31})
\end{aligned}
\end{equation}
We can diagonalize it using:
\begin{equation}
\begin{aligned}
S_{1}&=\frac{1}{\sqrt{2}}
\left(\begin{array}{cccc}
\sqrt{2} & 1+i & 0 & 0  \\
-1+i & \sqrt{2}  & 0 & 0  \\
0 & 0 & \sqrt{2}  & 1+i  \\
0 & 0 & -1+i & \sqrt{2}   \\
\end{array}\right)\
\Rightarrow\\
D^\prime(M_{1\bar{1}0})&=S_1^{-1}D(M_{1\bar{1}0})S_1=
\left(\begin{array}{cccc}
i & 0 & 0 & 0  \\
0 & -i & 0 & 0  \\
0 & 0 & -i & 0  \\
0 & 0 & 0 & i  \\
\end{array}\right)\\
\end{aligned}
\end{equation}

The $k_{1\bar{1}0}=0$ plane is defined by $k_x=k_y$. On this plane, the Hamiltonian becomes
\begin{equation}
\begin{aligned}
H(k_{1\bar{1}0}=0)&=(M-2\cos(k_x)-\cos(k_z))\Sigma_{30}+\\ &\sin(k_x)(\Sigma_{11}+\Sigma_{12})+\sin{k_z}\Sigma_{13}
\end{aligned}
\end{equation}
We first change basis using $S_1$, where the Hamiltonian becomes very similar to $H(k_z=0)$, and them we block-diagonalize it using $S$:
\begin{equation}
\begin{aligned}
H^\prime(k_{1\bar{1}0}=0)&=(S_1S)^{-1}H(k_{1\bar{1}0}=0)S_1S\\
&=[t_2\sigma_1+t_1\sigma_2+M(k)\sigma_3]\\
&\oplus [t_2\sigma_1+t_1\sigma_2-M(k)\sigma_3]\\
\end{aligned}
\end{equation}
where $t_1=\sin(k_x)-\frac{1}{\sqrt{2}}\sin{k_z}, t_2=\sin(k_x)+\frac{1}{\sqrt{2}}\sin{k_z}, M(k) = M-2\cos(k_x)-\cos(k_z)$. The representation matrix of $P$ and $M_{1\bar{1}0}$ becomes
\begin{equation}
\begin{aligned}
D^\prime(P)&=(S_1S)^{-1}D(P)S_1S=
\left(\begin{array}{cccc}
1 & 0 & 0 & 0  \\
0 & -1 & 0 & 0  \\
0 & 0 & -1 & 0  \\
0 & 0 & 0 & 1  \\
\end{array}\right),\\
D^{\prime\prime}(M_{1\bar{1}0})&=S^{-1}D^\prime(M_{1\bar{1}0})S=
\left(\begin{array}{cccc}
i & 0 & 0 & 0  \\
0 & i & 0 & 0  \\
0 & 0 & -i & 0  \\
0 & 0 & 0 & -i  \\
\end{array}\right)\\
\end{aligned}
\end{equation}

The mirror Chern number can be calculated similarly as $C_{m,1\bar{1}0}=(-1,1)$, where the minus sign comes from the orientation of $t_1$ and $t_2$.

\paragraph{$S_4$ eigenvalues at $S_4$-invariant points}
The $S_4$ symmetry in $z$ direction has $O(3)$ rotation matrix
$ S_4=
\left(\begin{array}{ccc}
0 & 1 & 0  \\
-1 & 0 & 0  \\
0 & 0 & -1  \\
\end{array}\right) $.
The representation matrix of $S_4$ satisfies
\begin{equation}
\begin{aligned}
\Sigma_{30}D(S_4)&=D(S_4)\Sigma_{30},\ \Sigma_{13}D(S_4)=-D(S_4)\Sigma_{13},\ \\
\Sigma_{11}D(S_4)&=D(S_4)\Sigma_{12},\ \Sigma_{12}D(S_4)=-D(S_4)\Sigma_{11},\  \\
&\Rightarrow \
D(S_4)=\frac{1}{\sqrt{2}}(\Sigma_{30}-i\Sigma_{33})
\end{aligned}
\end{equation}

There are four $S_4$-invariant momenta $\Gamma$, $M(\pi,\pi,0)$, $X(0,\pi,0)$, and $R(\pi,\pi,\pi)$. Using the eigenvectors derived before, we have the corresponding representation matrices:
\begin{equation}
\begin{aligned}
	D^{\prime}_\Gamma(S_4)&=\langle\psi_\Gamma|S^{-1}D(S_4)S|\psi_\Gamma\rangle\\
	&=\frac{1}{\sqrt{2}}
	\left(\begin{array}{cccc}
	-1-i & 0 & 0 & 0  \\
	0 & 1-i & 0 & 0  \\
	0 & 0 & -1+i & 0  \\
	0 & 0 & 0 & 1+i  \\
	\end{array}\right) \\
D^{\prime}_{M,X,R}(S_4)&=\langle\psi_{M,X,R}|S^{-1}D(S_4)S|\psi_{M,X,R}\rangle\\
&=\frac{1}{\sqrt{2}}
\left(\begin{array}{cccc}
1-i & 0 & 0 & 0  \\
0 & -1-i & 0 & 0  \\
0 & 0 & 1+i & 0  \\
0 & 0 & 0 & -1+i  \\
\end{array}\right)	
\end{aligned}
\end{equation}
As a result, the $S_4$ eigenvalue of the $E_-$ Kramer pair is $\sqrt{2}$ at $\Gamma$ and $-\sqrt{2}$ at $M,X,R$. The $S_4$ indicator can be calculated as
\begin{equation}
z_{2,S_4}^\prime = 
\sum_{\mathbf{k}\in K_{S_4}} \frac{1}{2}(n_{\mathbf{k}}^{-\sqrt{2}}-n_{\mathbf{k}}^{\sqrt{2}})=1
\end{equation}

The $z_8$ indicator can be calculated as
\begin{equation}
	z_8=2z_{2,S_4}^\prime-z_{4P}^\prime \bmod 8 = 7
\end{equation}

\paragraph{$z_8$ indicator for the $Z_2$ decoration in MSG $Pm\bar{3}m$}
The $Z_2$ decoration in MSG 221.92 $Pm\bar{3}m$ has mirror Chern number $C_{m,k_z}=(1,-1,0,0)$ and $C_{m,1\bar{1}0}=(1,-1)$. It can be seen as a superposition of 3D BHZ model with $M\in(1,3)$, whose mirror Chern numbers are $C_{m,k_z}=(1,-1,0,0)$, $C_{m,1\bar{1}0}=(-1,1)$, and a $(1\bar{1}0,0)$-LC which has mirror Chern numbers $C_{m,k_z}=(0,0,0,0)$, $C_{m,1\bar{1}0}=(2,-2)$ and $z_8=4$. As a result, the $Z_2$ decoration has $z_8=3$.

\paragraph{Verification of the $z_{4C}$ formula}
For the BHZ model, we can calculate its $C_4$ and $C_2$ eigenvalues on HSPs. When $1<M<3$, as $C_4=S_4\cdot P, C_2=M_z\cdot P$, we have
\begin{equation}
\begin{aligned}
D_\Gamma(C_4)&=\frac{1}{\sqrt{2}}
\left(\begin{array}{cccc}
1+i & 0 & 0 & 0  \\
0 & 1-i & 0 & 0  \\
0 & 0 & 1-i & 0  \\
0 & 0 & 0 & 1+i  \\
\end{array}\right),\\
D_M(C_4)&=\frac{1}{\sqrt{2}}
\left(\begin{array}{cccc}
1-i & 0 & 0 & 0  \\
0 & 1+i & 0 & 0  \\
0 & 0 & 1+i & 0  \\
0 & 0 & 0 & 1-i  \\
\end{array}\right)\\
D_X(C_2)&=
\left(\begin{array}{cccc}
i & 0 & 0 & 0  \\
0 & -i & 0 & 0  \\
0 & 0 & -i & 0  \\
0 & 0 & 0 & i  \\
\end{array}\right)
\end{aligned}
\end{equation}
For the lower band with $M_z=+i$, we have $B_{C_4}(\Gamma)=e^{-i\pi/4}, B_{C_4}(M)=e^{i\pi/4}, B_{C_2}(X)=-i$, which leads to 
\begin{equation}
	 z_{4C} = \sum_{l\in occ} \ln(-B_{C_4}^l(\Gamma) B_{C_4}^l(M) B_{C_2}^l(Y)) /(i\frac{\pi}{2})\bmod 4=1
\end{equation}

\clearpage
\section{The construction of atomic insulator basis in MSGs}\label{AppendixM}
In this Appendix, following Ref.\cite{cano2018building}, we show the procedures for constructing the band representation of atomic insulators in SGs and MSGs.

\subsection{General procedures}
\paragraph{Wyckoff positions}
Wyckoff positions (WPs) are specific points in the unit cell of an (M)SG $G$, classified by their site symmetry groups (SSGs). The SSG of a given WP $\mathbf{q}$ is defined as the MSG elements that keep $\mathbf{q}$ fixed, i.e.
$$
G_q=\left\{g \in G \mid  g \mathbf{q}=\mathbf{q}\right\}
$$
Caution that here we require that $g \mathbf{q}=\mathbf{q}$, but not the more usual $g\mathbf{q}=\mathbf{q}+\mathbf{R}$, where $\mathbf{R}$ is a lattice vector. This convention is more convenient as the symmetry operations will not change the center of Wannier functions to another unit cell.

The representatives $g_{\alpha}$ of the quotient group $G / G_q$ are used to generate the equivalent positions of $\mathbf{q}$: 
$$\mathbf{q}_{\alpha}=g_{\alpha} \mathbf{q}, \quad g_{\alpha} \in G / G_q$$
where $g_\alpha$ may also contain lattice translations s.t  $\mathbf{q}_{\alpha}$ locates in the home cell. The SSG of $\mathbf{q}_\alpha$ is 
$$G_{\mathbf{q}_{\alpha}} \equiv\left\{g_{\alpha} h g_{\alpha}^{-1} \mid h \in G_{\mathbf{q}}\right\}$$

\paragraph{Wannier functions}
Assume a given set of equivalent WPs $\{\mathbf{q},\mathbf{q}_2, ...\mathbf{q}_{n}\}$ has SSG $G_q$. We can place localized Wannier functions $W_{i1}(r)$ on $\mathbf{q}$, which form the bases of an m-dimensional representation $D_{ij}(g)$ of $G_q$:
\begin{equation}
gW_{i1}(r)= W_{i1}(g^{-1}r)=\sum_j W_{j1}(r)D_{ji}(g),\ \ g\in G_q
\end{equation}
where $i=1,...,m$ indicate m bases, and the second index indicates n sites. Notice that $g\in G_q$ may contain some integer translations s.t $W_{i1}(g^{-1}r)$ lies in the same unit cell as $W_{i1}(r)$, and in more general case, 
$W_{i1}(g^{-1}r)=\sum_j W_{j1}(r-R_\mu)D_{ji}(g)$, where $R_\mu=g\mathbf{q}_1-\mathbf{q}_1$.

The Wannier functions on $\mathbf{q}_\alpha$ are
\begin{equation}
W_{i\alpha}(r)=g_\alpha W_{i1}(r) = W_{i1}(g_\alpha^{-1}r)
\end{equation}
which form the bases of the conjugate representation $D^\alpha(g)=D(g_\alpha^{-1}gg_\alpha)$.

Wannier functions in other unit cells transform similarly:
\begin{equation}
\begin{aligned}
gW_{i1}(r-R_\mu)&= g\{1|R_\mu\}W_{i1}(r)=\{1|gR_\mu\}gW_{i1}(r)\\
&=	\{1|gR_\mu\}W_{i1}(g^{-1}r)\\
&=\sum_j W_{j1}(r-gR_\mu)D_{ji}(g),\ \ g\in G_q
\end{aligned}
\end{equation}

For other symmetry operations $h\notin G_q$, we can derive the following relation using coset representatives:
\begin{equation}
hg_\alpha=\{1|R_{\beta\alpha}\}g_\beta g,\ \ R_{\beta\alpha}=h\mathbf{q}_\alpha-\mathbf{q}_\beta
\label{coset_rel}
\end{equation}
where $g_\alpha,g_\beta$ are coset representatives and $g\in G_q$. For given $h$ and $g_\alpha$, we can find $g_\beta$ by enumerating coset representatives s.t. $g_\beta hg_\alpha\in G_q$. The translation $R_{\beta\alpha}$ exists because we set Wannier functions $W_{i\alpha}$ in the same unit cell and can be derived as:
\begin{equation}
\begin{aligned}
hg_\alpha \mathbf{q}&=h\mathbf{q}_\alpha=\{1|R_{\beta\alpha}\}g_\beta g \mathbf{q}=\{1|R_{\beta\alpha}\}g_\beta \mathbf{q}=\{1|R_{\beta\alpha}\}\mathbf{q}_\beta\\
&\Rightarrow R_{\beta\alpha}=h\mathbf{q}_\alpha-\mathbf{q}_\beta
\end{aligned}
\end{equation}
where we have used $g\mathbf{q}=\mathbf{q},\ g_i\mathbf{q}=\mathbf{q}_i$. These relations hold if we take the convention that integer translations are absorbed into $g,g_i$ s.t. $\mathbf{q},\mathbf{q}_i$ lies in the home cell. In more general cases, 
\begin{equation}
R_{\beta\alpha}=t_{hg_\alpha-g_\beta g}
\end{equation}
where $t$ denotes the translation part of the operation. 

With Eq.\ref{coset_rel}, we consider how $W_{i\alpha}$ transform under arbitrary $h=\{R|\tau\}\notin G_q$, i.e., lifting the representation of $G_q$ to a representation of $G$:
\begin{equation}
\begin{aligned}
hW_{i\alpha}(r-R_\mu)
&=h\{1|R_\mu\}g_\alpha W_{i1}(r)=\{1|RR_\mu\}hg_\alpha W_{i1}(r)\\
&=\{1|RR_\mu+R_{\beta\alpha}\}g_\beta g W_{i1}(r)\\
&=\{1|RR_\mu+R_{\beta\alpha}\} \sum_j W_{j\beta}(r)D_{ji}(g)\\
&=\sum_jW_{j\beta}(r-RR_\mu-R_{\beta\alpha})D_{ji}(g)
\end{aligned}
\end{equation}

\paragraph{Bloch functions}
The Bloch functions are generated from the Wannier functions: 
\begin{equation}
a_{i\alpha}(k,r)=\frac{1}{\sqrt{N}}\sum_{\mu} e^{ik\cdot R_\mu}W_{i\alpha}(r-R_\mu)
\end{equation}
For $h\in G$, they transform as 
\begin{equation}
\begin{aligned}
ha_{i\alpha}(k,r)
=&\frac{1}{\sqrt{N}}\sum_{\mu} e^{ik\cdot R_\mu} hW_{i\alpha}(r-R_\mu)\\
=&\frac{1}{\sqrt{N}}\sum_{\mu} e^{ik\cdot R_\mu}
\sum_jW_{j\beta}(r-RR_\mu-R_{\beta\alpha})D_{ji}(g)\\
=&e^{-iRk\cdot R_{\beta\alpha}}\sum_jD_{ji}(g)\\
&\frac{1}{\sqrt{N}}\sum_{\mu} e^{iRk\cdot (RR_\mu+R_{\beta\alpha})} W_{j\beta}(r-(RR_\mu+R_{\beta\alpha}))\\
=&e^{-iRk\cdot R_{\beta\alpha}}\sum_jD_{ji}(g) a_{j\beta}(Rk,r)
\end{aligned}
\end{equation}

As a result, the band representation $D_G^k(h)$ is:
\begin{equation}
\begin{aligned}
D_G^{k}(h)_{j\beta,i\alpha}
&=\langle a_{j\beta}(Rk,r) |h| a_{i\alpha}(k,r)\rangle\\
&=e^{-iRk\cdot R_{\beta\alpha}}D_{ji}(g)
\end{aligned}
\end{equation}
where $g=g_\beta^{-1}\{1|-R_{\beta\alpha}\}hg_\alpha,\  R_{\beta\alpha}=t_{hg_\alpha-g_\beta g}$.

\paragraph{Note: non-orthogonal rotations in trigonal and hexagonal lattices}
The formula above failed in trigonal and hexagonal lattice, because the 3D rotation matrices are not orthogonal, i.e., $R^{T}\ne R^{-1}\Rightarrow k\cdot R_\mu\ne Rk\cdot RR_\mu$. Instead, $(k,R_\mu)=k^T\cdot R_\mu=k^T\cdot R^{-1}RR_\mu=k^TR^{-1}\cdot RR_\mu=((R^{-1})^Tk,RR_\mu)$. As a result, we have
\begin{equation}
\begin{aligned}
D_G^{k}(h)_{j\beta,i\alpha}
&=\langle a_{j\beta}(kR^{-1},r) |h| a_{i\alpha}(k,r)\rangle\\
&=e^{-ikR^{-1}\cdot R_{\beta\alpha}}D_{ji}(g)\\
\end{aligned}
\end{equation}

\paragraph{Note: Bilbao irrep convention}
$D^k_G$ is a ``small representation" when restricted to the little group $G_\mathbf{k}=\{h=\{R|\tau\}\in G|R\mathbf{k} = \mathbf{k}\}$, as the representation matrices of two operations that differ by a lattice translation differ by a phase factor $e^{-iRk\cdot R_\mu}=e^{-ik\cdot R_\mu}$.

However, Bilbao irrep matrices take the convention that a lattice translation is represented by $e^{ik\cdot R_\mu}$. As a result, if we want to decompose the band representation $D^k_G$ into Bilbao irreps, we need to take the convention that
\begin{equation}
\begin{aligned}
D_G^{k}(h)_{j\beta,i\alpha}
&=e^{ikR^{-1}\cdot R_{\beta\alpha}}D_{ji}(g)
\end{aligned}
\end{equation}
This is legitimate because it is equivalent to modify the rule of action $gW(r)=W(g^{-1}r)$ to $gW(r)=W(gr)$.

\paragraph{Decompose into small representations}
For a fixed momenta $k$, $D_G^k(h)$ is a $nm\times nm$-dimensional matrix and can be block-diagonalized into the small representations of the little group $G_\mathbf{k}$:
\begin{equation}
(D \uparrow G) \downarrow G_{\mathbf{k}} \cong \bigoplus_{i} m_{i}^{\mathbf{k}} \sigma_{i}^{\mathbf{k}}
\end{equation}
where $\sigma_{i}^{\mathbf{k}}$ are irreps of $G_\mathbf{k}$.

The characters of $D_G^k(h)$ are
\begin{equation}
\begin{aligned}
\chi_G^k(h)
&=\sum_{i,\alpha}D_G^{k}(h)_{i\alpha,i\alpha}
=\sum_\alpha e^{-iRk\cdot R_{\alpha\alpha}}\chi[D(g)]
\end{aligned}
\end{equation}
The decomposition coefficients $m_{i}^{\mathbf{k}}$ can be readily calculated using the orthogonal theorem of characters.

\paragraph{Summary of the procedures for deriving AI}
We summarize here the general procedure for deriving the BS of AIs:
\begin{enumerate}
	\item Choose a Wyckoff point $\mathbf{q}$ and find the coset decomposition
	$G=\bigcup_{\alpha=1} g_{\alpha}G_{\mathbf{q}}$.
	\item Choose an irrep $D$ of $G_\mathbf{q}$. For a given HSP $\mathbf{k}$, calcuate the character $\chi_G^k(h)$:
	\begin{enumerate}
		\item for each $h\in G_\mathbf{k}$ and coset representative $g_\alpha$, find $g=g_\alpha^{-1}\{1|-R_{\alpha\alpha}\}hg_\alpha$
		\item calculate the summation 
		$\chi_G^k(h)=\sum_\alpha e^{ikR^{-1}\cdot R_{\alpha\alpha}}\tilde{\chi}[D(g)]$.
	\end{enumerate}
	\item Use the orthogonal theorem of characters to decompose $\chi_G^k(h)$ into the irreps of $G_\mathbf{k}$: $$\chi_{G}^{\mathbf{k}}(h)=\sum_{i} m_{i}^{\mathbf{k}} \chi_{\sigma_{i}}^{\mathbf{k}}(h)$$
\end{enumerate} 


\subsection{Modification of the procedures for anti-unitary symmetries}
In the above derivation, we have only considered unitary symmetries, i.e., the derived AIs belong to type-1 MSG. For anti-unitary symmetries, we make the following modification:
\begin{enumerate}
\item In the relation 
\begin{equation}
hg_\alpha=\{1|R_{\beta\alpha}\}g_\beta g,\ \ 
\Rightarrow g=g_\beta^{-1}\{1|-R_{\beta\alpha}\}hg_\alpha
\end{equation}
the $\pm 1$ phase correction for SU(2) matrix of $g$ need to consider complex conjugation if the symmetries are anti-unitary.

\item When $g_\beta$ is anti-unitary, $D_{ji}(g)$ needs to take complex conjugation: 
\begin{equation}
\begin{aligned}
hW_{i\alpha}(r-R_\mu)
&=h\{1|R_\mu\}g_\alpha W_{i1}(r)=\{1|RR_\mu\}hg_\alpha W_{i1}(r)\\
&=\{1|RR_\mu+R_{\beta\alpha}\}g_\beta g W_{i1}(r)\\
&=\{1|RR_\mu+R_{\beta\alpha}\} \sum_j W_{j\beta}(r)D^*_{ji}(g)\\
&=\sum_jW_{j\beta}(r-RR_\mu-R_{\beta\alpha})D^*_{ji}(g)
\end{aligned}
\end{equation}

\item When $g_\beta$ is anti-unitary ($h$ unitary), the Bloch function transforms as:
\begin{equation}
\begin{aligned}
ha_{i\alpha}(k,r)
=&\frac{1}{\sqrt{N}}\sum_{\mu} e^{ik\cdot R_\mu} hW_{i\alpha}(r-R_\mu)\\
=&\frac{1}{\sqrt{N}}\sum_{\mu} e^{ik\cdot R_\mu}
\sum_jW_{j\beta}(r-RR_\mu-R_{\beta\alpha})D^*_{ji}(g)\\
=&e^{-iRk\cdot R_{\beta\alpha}}\sum_j D^*_{ji}(g)\\
&\frac{1}{\sqrt{N}}\sum_{\mu} e^{iRk\cdot (RR_\mu+R_{\beta\alpha})} W_{j\beta}(r-(RR_\mu+R_{\beta\alpha}))\\
=&e^{-iRk\cdot R_{\beta\alpha}}\sum_j D^*_{ji}(g) a_{j\beta}(-Rk,r)
\end{aligned}
\end{equation}

\item When $g_\beta$ is anti-unitary, the band representation becomes:
\begin{equation}
\begin{aligned}
D_G^{k}(h)_{j\beta,i\alpha}
&=\langle a_{j\beta}(-Rk,r) |h| a_{i\alpha}(k,r)\rangle\\
&=e^{-iRk\cdot R_{\beta\alpha}}D^*_{ji}(g)
\end{aligned}
\end{equation}
\end{enumerate}

To sum up, under Bilbao convention, the band representation $D_G^{k}$ is
\begin{equation}
\begin{aligned}
D_G^{k}(h)_{j\beta,i\alpha}
&=e^{+iRk\cdot R_{\beta\alpha}}D_{ji}(g),\ \ h\in G,\ g_\beta \text{ unitary}\\
D_G^{k}(h)_{j\beta,i\alpha}
&=e^{+iRk\cdot R_{\beta\alpha}}D^*_{ji}(g),\ \ h\in G,\ g_\beta \text{ anti-unitary}\\
\end{aligned}
\end{equation}

\clearpage
\newpage
\section{Table: TCI classifications of MSGs}\label{AppendixN}

\begin{longtable*}{c|c||c|c||c|c||c|c||c|c}
\caption{TCI classifications of MSGs}
\label{table_classification}
\endfirsthead
\hline
1.1 & $\mathbb{Z}^3$ & 
1.3 & $\mathbb{Z}_2$ & 
2.4 & $\mathbb{Z}^3 \times \mathbb{Z}_2$ & 
2.6 & $\text{N/A}$ & 
2.7 & $\mathbb{Z}_2$
\\
\hline
3.1 & $\mathbb{Z}$ & 
3.3 & $\mathbb{Z}^2 \times \mathbb{Z}_2$ & 
3.4 & $\mathbb{Z}_2$ & 
3.5 & $\mathbb{Z}_2$ & 
3.6 & $\mathbb{Z}_2$
\\
\hline
4.7 & $\mathbb{Z}$ & 
4.9 & $\mathbb{Z}^2 \times \mathbb{Z}_2$ & 
4.10 & $\mathbb{Z}_2$ & 
4.11 & $\mathbb{Z}_2$ & 
4.12 & $\mathbb{Z}_2$
\\
\hline
5.13 & $\mathbb{Z}$ & 
5.15 & $\mathbb{Z}^2 \times \mathbb{Z}_2$ & 
5.16 & $\mathbb{Z}_2$ & 
5.17 & $\mathbb{Z}_2$ & 
6.18 & $\mathbb{Z}_M^2 \times \mathbb{Z}$
\\
\hline
6.20 & $\mathbb{Z}^2$ & 
6.21 & $\mathbb{Z}_M^2$ & 
6.22 & $\mathbb{Z}_M$ & 
6.23 & $\mathbb{Z}_M$ & 
7.24 & $\mathbb{Z} \times \mathbb{Z}_2$
\\
\hline
7.26 & $\mathbb{Z}^2$ & 
7.27 & $\mathbb{Z}_2$ & 
7.28 & $\mathbb{Z}_2$ & 
7.29 & $\mathbb{Z}_2$ & 
7.30 & $\mathbb{Z}_2$
\\
\hline
7.31 & $\mathbb{Z}_2$ & 
8.32 & $\mathbb{Z}_M \times \mathbb{Z}$ & 
8.34 & $\mathbb{Z}^2$ & 
8.35 & $\mathbb{Z}_M$ & 
8.36 & $\mathbb{Z}_M$
\\
\hline
9.37 & $\mathbb{Z} \times \mathbb{Z}_2$ & 
9.39 & $\mathbb{Z}^2$ & 
9.40 & $\mathbb{Z}_2$ & 
9.41 & $\mathbb{Z}_2$ & 
10.42 & $\mathbb{Z}_M^2 \times \mathbb{Z}$
\\
\hline
\end{longtable*}

\clearpage
\newpage
\section{Table: Invariants and SIs of TCI classification generators for MSGs with non-trivial SI group}\label{AppendixO}

\subsection{Notations in the table of mapping}

The quantitative mappings between invariants and SIs for all MSGs are tabulated in Table.\ref{table_nontivial}.
For each MSG with a nontrivial SI group, we list all the independent decorations as generators for TCI classification. However, when the MSG has mirror symmetries, we give one extra generator, that is, if the classification is $\mathbb{Z}^m$, we give $m$ mirror decorations plus a $\mathbb{Z}_2$ decoration, and one can remove any one of the $m$ mirror decorations in order to obtain $m$ independent decorations. For each decoration, we show its invariants and SIs.

For MSGs without SI, we also list their independent decorations corresponding to the classification in Table.\ref{table_tivial}.
 
Here we explain the notations used in these two tables.

\paragraph{The top row}
The top row gives the general information of the MSG. 
The MSG number and MSG label are given in Belov-Neronova-Smirnova (BNS) setting, followed by the TCI classification, where we separate the mirror decorations and translations into $\mathbb{Z}_M$ and $\mathbb{Z}$, with the subscript $M$ denotes mirror.
The SI group is given subsequently as $X_{\text{BS}}$, followed by the corresponding generating SI name.

\paragraph{Decorations}
The first column ``deco'' denotes the type of decoration.
As $Z$ (translation) and $Z_2$ decorations are generically not constructed by LCs, we just use ``$Z$'' and ``$Z_2$'' to denote them, respectively, while the mirror decorations are all constructed by layers, which we use ``$M (m n l; d)$'' to denote the mirror decorations, with $(m n l; d)$ representing Miller indices with respect to the conventional lattice.
The second column, ``$\mathbb{Z}_{n_1, n_2, \cdots }$'' gives the SI values for each generator.
The third column shows the name of MSG symmetries and their corresponding invariants of the generators.

\paragraph{Invariants}
The first set ``weak=$(w_1 w_2 w_3)$'' gives the weak invariants along the three lattice bases of the primitive cell, followed by other invariants defined in real-space, including those of inversion, $S_4$, rotation, screw, glide, and their combinations with TRS.
For type-4 MSGs, the invariant of half magnetic translation, $\{ E | \frac{1}{2} \bm{a} \} \cdot T $ is given after the weak invariants.
Note the mirror invariants are given not by the real-space ones, but the momentum space mirror Chern numbers, which appear at the end of the row for MSGs with mirror symmetries.

\paragraph{MSG elements}
We represent each MSG symmetry operation in the following short symbols:
\begin{itemize}
    \item rotation: $n^{hkl}$, where $n$ is the rotation angle and $(hkl)$ are the Miller indices of the rotation axis in the conventional lattice.
    \item screw: $n^{hkl}_{t_1 t_2 t_3}$, where $(hkl)$ denote the screw axis and $(t_1 t_2 t_3)$ are the indices for the screw vector, both given in the conventional lattice.
    
    \item inversion: $i$. In some MSGs, inversion could carry  fractional translations and becomes $i_{ t_1 t_2 t_3 }$, where $(t_1 t_2 t_3)$ is given in the conventional lattice.
    \item $S_n$: $\bar{n}^{hkl}$, where $(hkl)$ denote the rotoinversion axis in the conventional lattice.
    Similar to inversion,  it could become $\bar{n}^{h k l}_{ t_1 t_2 t_3 }$ in some MSGs.
    
    \item glide: $g^{hkl}_{t_1 t_2 t_3}$, where $(hkl)$ denote the glide plane and $(t_1 t_2 t_3)$ are the indices for the glide vector, both given in the conventional lattice.
    \item anti-unitary symmetries: $ \widehat{O}* $, where $*$ denote the TRS and $\widehat{O}$ can be symmetry operations mentioned above plus the half lattice translation, e.g., $\{1| \frac{1}{2}00\}$.
\end{itemize}

\paragraph{Mirror Chern number}
Mirror Chern numbers are denoted as $Cm^{hkl}_{(n)}$. 
Here $(hkl)$ denote the Miller indices of the mirror plane and $n$ denotes the highest (unitary) $C_n$ rotation along the direction of the mirror, which is set to be $0$ when there is no such rotation.

There are two kinds of $C_m$ due to different Bravais lattices, i.e., $C_m = ( 
C^{+}_{k_{\perp} =0}, C^{-}_{k_{\perp}=0}, C^{+}_{k_{\perp}=\pi}, 
C^{-}_{k_{\perp}= \pi} )$ or 
$ C_m = ( 
C^{+}_{k_{\perp} =0}, C^{-}_{k_{\perp}=0} ) $,
where $k_{\perp}$ is in the normal direction of mirror.
This depends on whether there are two mirror-symmetric planes in the BZ, i.e., $k_{\perp}=0$ and $k_{\perp}=\pi$, or there is only one mirror-symmetric plane in BZ, i.e., $k_{\perp}=0$.

\clearpage
\newpage
\begin{longtable*}{c|c|cccccccccccccccccc}
\caption{MSGs with nontrivial SI group}
\label{table_nontivial}
\endfirsthead
\hline
\multicolumn{20}{c}{2.4\;\;\; $P\bar{1}$\;\;\; Classification = $\mathbb{Z}^3 \times \mathbb{Z}_2$\;\;\;$ \text{X}_{\text{BS}}= \mathbb{Z}_{2,2,2,4}$\;\;\;SI = $z_{2P,1},z_{2P,2},z_{2P,3},z_{4P}$}
\\
\hline
deco &  $ \mathbb{Z}_{2,2,2,4} $  & weak & $i$
\\
\hline
$Z_2$ & $0002$ & ($0$$0$$0$) & $1$
\\
\hline
$Z$ & $1002$ & ($\bar{1}$$0$$0$) & $1$
\\
\hline
$Z$ & $0102$ & ($0$$1$$0$) & $1$
\\
\hline
$Z$ & $0010$ & ($0$$0$$1$) & $0$
\\
\hline
\hline
\multicolumn{20}{c}{2.7\;\;\; $P_{S}\bar{1}$\;\;\; Classification = $\mathbb{Z}_2$\;\;\;$ \text{X}_{\text{BS}}= \mathbb{Z}_{2}$\;\;\;SI = $z_{2P}^\prime$}
\\
\hline
deco &  $ \mathbb{Z}_{2} $  & weak &  $ \{ 1 |0 0 \frac{1}{2}\} * $  & $i$
\\
\hline
\end{longtable*}

\newpage
\clearpage
\section{Table: Invariants of TCI classification generators for MSGs with trivial SI group}\label{AppendixP}

\begin{longtable*}{c|c|cccccccccccccccc}
\caption{MSGs with trivial SI group}
\label{table_tivial}
\endfirsthead
\hline
\multicolumn{18}{c}{1.1\;\; $P1$\;\; Classification = $\mathbb{Z}^3$}
\\
\hline
deco & weak & 
\\
\hline
$Z$ & ($0$$0$$\bar{1}$) & 
\\
\hline
$Z$ & ($0$$1$$0$) & 
\\
\hline
$Z$ & ($\bar{1}$$0$$0$) & 
\\
\hline
\hline
\multicolumn{18}{c}{1.3\;\; $P_{S}1$\;\; Classification = $\mathbb{Z}_2$}
\\
\hline
deco & weak &  $ \{ 1 |0 0 \frac{1}{2}\} * $  & 
\\
\hline
$Z_2$ & ($0$$0$$0$) & $1$ & 
\\
\hline
\end{longtable*}

\end{document}